\documentclass[10pt]{article} 

\usepackage{amssymb,amsmath,amsfonts,theorem,subfigure} 
\usepackage[usenames]{color}
\usepackage{pstricks}
\usepackage{euler,palatino}

\newcommand{\be}{\begin{equation}}
\newcommand{\ee}{\end{equation}}

\newcommand{\spec}{\text{\rm spec\,}}
\newcommand{\tr}{\text{\rm tr\,}}
\newcommand{\lw}{\psset{linewidth=0.5pt}}
\newcommand{\unlw}{\psset{linewidth=1pt}}
\newcommand{\pmm}{P_{\{m^m\}}^d}
\newcommand{\upto}{{UpTo}}
\newrgbcolor{myc}{1 0 0}
\newrgbcolor{myc2}{0 0 1}
\newrgbcolor{myc3}{1 0 1}

\newtheorem{Lemme}{Lemma}[section]

\newtheorem{Proposition}[Lemme]{Proposition}

\newtheorem{Definition}{Definition}[section]

\title{{\huge The Jordan Structure\\ of Two Dimensional Loop Models}}
\author{
{\Large Alexi Morin-Duchesne}\footnote{\ttfamily alexi.morin-duchesne{\char'100}umontreal.ca},\\
\it D\'epartement de physique\\ 
\it Universit\'e de Montr\'eal, C.P.\ 6128, succ.\ centre-ville, Montr\'eal\\ 
\it Qu\'ebec, Canada, H3C 3J7\\[10pt]
{\Large Yvan Saint-Aubin}\footnote{\ttfamily saint{\char'100}dms.umontreal.ca}\\
\it D\'epartement de math\'ematiques et de statistique\\ 
\it Universit\'e de Montr\'eal, C.P.\ 6128, succ.\ centre-ville, Montr\'eal\\ 
\it Qu\'ebec, Canada, H3C 3J7\\[10pt]}

\begin{document} 
\maketitle

%
%
 
\begin{abstract}
We show how to use the link representation of the transfer matrix $D_N$ of loop models on the lattice to calculate partition functions, at criticality, of the Fortuin-Kasteleyn model with various boundary conditions and parameter $\beta = 2 \cos( \pi(1-a/b)), a,b\in \mathbb N$ and, more specifically, partition functions of the corresponding $Q$-Potts spin models, with $Q=\beta^2$. The braid limit of $D_N$ is shown to be a central element $F_N(\beta)$ of the Temperley-Lieb algebra $TL_N(\beta)$, its eigenvalues are determined and, for generic $\beta$, a basis of its eigenvectors is constructed using the Wenzl-Jones projector. To any element of this basis is associated a number of defects $d$, $0\le d\le N$, and the basis vectors with the same $d$ span a sector. Because components of these eigenvectors are singular when $b \in \mathbb{Z}^*$ and $a \in 2 \mathbb{Z} + 1$, the link representations of $F_N$ and $D_N$ are shown to have Jordan blocks between sectors $d$ and $d'$ when $d-d' < 2b$ and $(d+d')/2 \equiv  b-1 \  \textrm{mod}  \ 2b$ ($d>d'$). When $a$ and $b$ do not satisfy the previous constraint, $D_N$ is diagonalizable.\\


\end{abstract}

%
%

\tableofcontents

%
%

\section{Introduction}
\label{sec:intro}

Which are the representations of the Virasoro algebra involved in the thermodynamical limit of the Fortuin-Kasteleyn description of lattice spin models? This is such a natural question that it is somewhat surprising that the answer is not known precisely. 

In the spin description of various two dimensional lattice models, like the Ising model, the answer is known. Thanks to early work by Onsager \cite{Onsager} and others \cite{Yang,Baxter,OPW}, the spectrum of the transfer matrix of many models is known for various boundary conditions. To probe the hypothesis of conformal invariance of critical phenomenon \cite{BPZ} put forward in 1984, it was imperative to tie the lattice description to one of the conformal field theories (CFT). For that the transfer matrix, properly scaled, was identified to the generator $L_0$ of the Virasoro algebra. The possibility to use the characters of its three irreducible representations at $c=\frac12$, with conformal dimensions $0$, $1/2$ and $1/16$, to reproduce Onsager's spectrum of the Ising model was maybe the most compelling evidence for this hypothesis. The CFT description of rational models, denoted $\mathcal{M}(a,b)$, with $a,b \in \mathbb{N}$ ($b>a$), and central charge given by
\begin{equation} c = c_{a,b}= 1- \frac{6(b-a)^2}{ab} \label{eq:centralcharge}\end{equation}
is of particular interest, as the Ising model and $3$-Potts model belong to this family, as $\mathcal{M}(3,4)$ and $\mathcal{M}(4,5)$ respectively. For models whose spectrum was not found analytically, the hypothesis of conformal invariance was later supported by strong numerical evidence, leading to many conjectures (\cite{LPSA}, \cite{Cardy1}, \cite{Watts}, \cite{Nienhuis}), some of which were proven recently (\cite{Smirnov1}, \cite{Duminil}).

In the Fortuin-Kasteleyn description (FK) of lattice models \cite{Essam} (or simply loop models), the partition function is a sum over graphs instead of a sum over spin configurations like in the spin description. Each graph corresponds to many configurations, each spin configuration to many graphs. A transfer matrix $D_N$ for the loop models is also available (see for example \cite{Zhou}) and much effort has been dedicated to the understanding of its spectrum and ground-state in particular situations (\cite{PDF}, \cite{PR}). But, in standard representations, the spin and loop transfer matrices of the same model on a given finite lattice do not have the same size. How can they possibly lead to the same partition function? Will they have the same spectrum in the thermodynamical limit? If not, the Virasoro representations involved in either limit are likely to be different.
In the continuum limit, a similar problem arises and Read and Saleur \cite{ReadSaleur} wrote partition functions on the torus in terms of partition functions of free fields. Jacobsen and Richard \cite{Jacobsen} later proved that similar formulae hold for finite lattices. But partition functions reveal only a small part of the underlying representations.

It has been argued that multi-point correlation functions of these loop models exhibit logarithmic behavior \cite{ZiffCardy}. This new feature does not fit well with the standard description of rational CFT's. Indeed, the differential equations obtained from singular vectors have logarithmic solutions, but the correlation functions of rational theories do not have logarithmic behavior. Of course, an extended paradigm, logarithmic conformal field theories (LCFT), is already available and allows these logarithmic functions. In these theories, a new field, the logarithmic partner of the stress energy tensor $T$, appears and is coupled to it in the operator product expansion, yielding the required logarithmic dependence for correlation functions \cite{Gurarie}. At the level of the Virasoro algebra, this translates into a generator $L_0$ that is no longer diagonalizable, but whose restriction to a fixed conformal weight subspace has Jordan blocks. One might also want to characterize the properties of a CFT based on indecomposable Virasoro representations, independently of the existence of such a logarithmic partner of $T$. 

To probe the question raised at the very beginning, we shall use the loop transfer matrix $D_N, N\ge 1$, as introduced in \cite{PRZ}. This matrix was introduced earlier, of course, but the authors' observations indicate that the answer might be quite subtle for this particular form. In their notation $D_N$ is an element of the Temperley-Lieb algebra $TL_N(\beta)$ whose parameter $\beta=2\cos(\lambda)$ is related to the central charge $c_{a,b}$ by $\lambda= (b-a) \pi / b$. In the given representation, the link representation, the element $D_N$ acts on a vector space naturally broken into sectors. Pearce, Rasmussen and Zuber first provided numerical evidence that the eigenvalues of $D_N$, restricted to any of these sectors, reproduce the first highest weights of some highest weight representations of the Virasoro algebra. And second they showed that, on some examples and for some boundary conditions, the matrix $D_N$ has Jordan blocks between some of the sectors, a telltale indication of logarithmic structure. Consequently they named {\em logarithmic minimal models} $\mathcal{LM}(a,b)$ the thermodynamical limit of these lattice loop models. We shall use simpler boundary conditions than theirs. Ours will be shown to correspond to free boundary conditions on a strip in the corresponding spin description.

Not surprisingly, the formulation in term of loop models owes much to the Temperley-Lieb algebra $TL_N(\beta)$ and its representations. This algebra is not semi-simple for all values of $\beta$. Actually it fails to be semi-simple precisely for the $\beta$ corresponding to the rational models. Its representation theory, at these values, is much more involved than for generic $\beta$ and its principal indecomposable modules were classified by  Goodman and Wenzl \cite{GoodmanWenzl}, and Martin \cite{Martin}. We shall use some of their tools in our study.

The goal of this paper is two-fold: to show how the spectrum of $D_N$ relates to the spin models and to describe the Jordan structure of the double-row matrix, for all $N$, and at the critical values of $\beta$ (or $\lambda$). Would $D_N$ contain Jordan blocks, this would be a clear indication that the thermodynamical limit of the FK description requires indecomposable representations of the Virasoro algebra.

The introductory question has also been attacked from a different standpoint, that of conformal field theory. Percolation, the $Q$-Potts model with $Q=1$, can be used as an example. It is believed that the thermodynamical limit of percolation is described by a theory with central charge $c=0$. (A historical presentation of the difficulties arising in setting percolation within the context of CFT is given in the introduction of \cite{Ridout2}.) However, if only irreducible representations of the Virasoro algebra are used, the theory is trivial. One {\em must} introduce indecomposable representations, leading naturally to a logarithmic conformal field theory \cite{Gurarie}. The operator product expansion is then used to probe which representations should appear. However there exist more than one well-defined CFT at $c=0$ and physical arguments will be needed to pick the right one for percolation. Studying the original problem on a finite lattice, as here, is a first step. But, to describe crossing probabilities, changes in boundary conditions will need to be included, a step that we have not considered. 

Here is the layout of the paper. Section \ref{sec:TL} makes explicit the relation between $Q$-Potts models at criticality and the double-row matrix $D_N$. After a quick reminder of definitions of the Temperley-Lieb algebras $TL_N$ and its link representation $\rho$, we introduce a trace $\tau$ on $TL_N$ and show that it is equivalent to a weighted trace on diagonal blocks of the representation $\rho$. Partition functions for the $Q$-Potts spin models on the cylinder is then expressed in terms of $\rho(D_N)$ on the cylinder. Extensions to various boundary conditions for the spin models are also given. In section \ref{sec:fn}, we study $C_{2N}$, the top Fourier coefficient of $D_N$, and show that it is central in $TL_N$.  (Appendix \ref{app:b} provides a simple method for calculating elements of $\rho(C_{2N})$ using $2\times 2$  matrices.) Using the Wenzl-Jones projector, we find a basis of eigenvectors of $\rho(C_{2N})$ for non-critical $\lambda$. The critical $\lambda$'s are studied in section {\ref{sec:jordan}}. The singularities of the eigenvectors obtained in the previous section are identified. (Appendix \ref{app:a} gives the main lemmas with their technical proofs.) This singular behavior translates to a non-trivial Jordan structure of $\rho(C_{2N})$. More precisely, we find that when $\lambda = \frac{(b-a)\pi}{b}$, with $b \in \mathbb{Z}^*$ and $a \in 2 \mathbb{Z} + 1$, the matrix $\rho(C_{2N})$ has Jordan blocks linking sectors $d$ and $d'$ ($d>d'$) when $d-d' < 2b$ and $\frac{d+d'}{2} \equiv  b-1 \  \textrm{mod}  \ 2b$. An extension of this to $\rho(D_N)$ exists and is our main result.

%
%

\section{Transfer matrices of $Q$-Potts and loop models}
\label{sec:TL}

\subsection{The Temperley-Lieb algebra as an algebra of connectivities}
\label{sec:TLconnect}

We introduce the Temperley-Lieb algebra $TL_N(\beta)$ in a slightly uncommon way. (See \cite{KauffmanLins} for a longer description of this approach.) Let $N$ be a positive integer and draw a rectangle with $2N$ marked points on it, $N$ on its upper side, $N$ on the bottom. A {\em connectivity diagram}, or simply a {\em connectivity}, is a pairwise pairing of all points by non-crossing curves drawn within the rectangular box. There are $C_N$ distinct connectivities, where $C_N=\frac1{N+1}\left(\begin{smallmatrix}2N\\ N\end{smallmatrix}\right)$ is the Catalan number. Let $\mathcal A_N$ be the set of formal linear combinations over $\mathbb C$ of these connectivities. It is a vector space of dimension $C_N$.

Choose now $\beta\in \mathbb C$. We define a product between two connectivities. If $c$ and $d$ are two such connectivities, their $\beta$-product $cd$ is obtained by drawing their two rectangles one above the other, $d$ being on top, and looking how the points of the top side of $d$ and those on the bottom side of $c$ are connected by the curves of $c$ and $d$. The $\beta$-product $cd$ is the resulting connectivity multiplied by a factor of $\beta$ for each closed loop in the diagram with the two rectangles on top of each other. For example here is the product of two connectivities with $N=6$. 
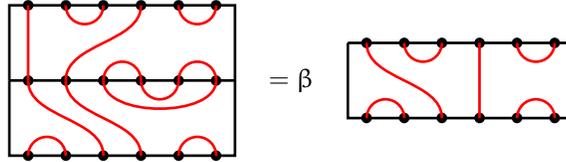
\begin{equation*}
\begin{pspicture}(3,-1)(1.5,1)
\psset{unit=0.5}
\psline[linewidth=1pt]{-}(0.5,0)(6.5,0)(6.5,2)(0.5,2)(0.5,0)
\psline[linewidth=1pt]{-}(6.5,0)(6.5,-2)(0.5,-2)(0.5,0)
\psset{linewidth=1pt}
\psdots(1,2)(2,2)(3,2)(4,2)(5,2)(6,2)(1,0)(2,0)(3,0)(4,0)(5,0)(6,0)(1,-2)(2,-2)(3,-2)(4,-2)(5,-2)(6,-2)
\psset{linecolor=myc}
\psline{-}(1,0)(1,2)
\psarc(2.5,2){0.5}{180}{360}
\psarc(5.5,2){0.5}{180}{360}
\psarc(3.5,0){0.5}{0}{180}
\psarc(5.5,0){0.5}{0}{180}
\psbezier{-}(4,2)(4,1)(2,1)(2,0)
\psbezier{-}(1,0)(1,-1)(3,-1)(3,-2)
\psbezier{-}(2,0)(2,-1)(4,-1)(4,-2)
\psarc(4.5,0){0.5}{180}{360}
\psarc(1.5,-2){0.5}{0}{180}
\psarc(5.5,-2){0.5}{0}{180}
\psbezier{-}(3,0)(3,-1)(6,-1)(6,0) \rput(8,0){$=\beta$}
\end{pspicture}
\begin{pspicture}(-1.5,-1)(1.5,0.5)
\psset{unit=0.5}
\psline[linewidth=1pt]{-}(0.5,1)(6.5,1)(6.5,-1)(0.5,-1)(0.5,1)
\psset{linewidth=1pt}
\psdots(1,1)(2,1)(3,1)(4,1)(5,1)(6,1)(1,-1)(2,-1)(3,-1)(4,-1)(5,-1)(6,-1)
\psset{linecolor=myc}
\psline{-}(4,1)(4,-1)
\psarc(2.5,1){0.5}{180}{360}
\psarc(5.5,1){0.5}{180}{360}
\psarc(1.5,-1){0.5}{0}{180}
\psarc(5.5,-1){0.5}{0}{180}
\psbezier{-}(1,1)(1,0)(3,0)(3,-1)
\end{pspicture}
\end{equation*}
This product between connectivities is extended to $\mathcal A_N$ linearly on both entries. It is associative, non-commutative and it has a unit element, namely the connectivity that pairs points on the same vertical line. This element is written as $id$.

\begin{Definition}The Temperley-Lieb algebra $TL_N(\beta)$ is the vector space $\mathcal A_N$ endowed with the $\beta$-product just defined.\end{Definition}

The connectivities $e_i$, $i=1, \dots, N-1$, play an important role. Let the points be labeled from left to right. The connectivity $e_i$ pairs all $2N$ points like the unit $id$ does, except those on positions $i$ and $i+1$ on both top and bottom sides; the points $i$ and $i+1$ on the top are paired together, so are those on the bottom. It is easily shown that, as an algebra, $TL_N(\beta)$ is generated by the connectivities $e_i$, $i=1, \dots, N-1$. It is usual in the literature to define $TL_N(\beta)$ using the generators $e_i$ together with the relations between them, namely
\begin{alignat*}{3}
e_i^2&=\beta e_i,&\qquad &\\
e_ie_j&=e_je_i, &&\text{\rm for }|i-j|>1\\
e_ie_{i\pm 1}e_i&=e_i&& \text{\rm  when $i, i\pm 1 \in \{1,2, \dots, N-1\}$.}
\end{alignat*}

For every $\beta\in(0,2)$, there is a one-parameter family of elements in $TL_N(\beta)$ that has been studied by many because of its relationship with statistical physics. We use the notation of the double-row transfer matrix $D_N(\lambda,u)$ introduced in \cite{PRZ}. It is defined graphically by
\begin{equation*}
\psset{linewidth=1pt}
\begin{pspicture}(-2,0)(0,2.2)
\rput(-1.7,1){$D_N(\lambda,u) = $}
\end{pspicture}
\overbrace{
\begin{pspicture}(-0,0)(5,2.2)
\psdots(0.5,0)(1.5,0)(4.5,0)
\psdots(0.5,2)(1.5,2)(4.5,2)
\lw
\psline{-}(0,0)(1,0)(1,1)(0,1)(0,0)\psarc[linewidth=0.5pt]{-}(0,0){0.25}{0}{90}\rput(0.5,0.5){$u$}
\psline{-}(1,0)(2,0)(2,1)(1,1)(1,0)\psarc[linewidth=0.5pt]{-}(1,0){0.25}{0}{90}\rput(1.5,0.5){$u$}
\psline{-}(4,0)(5,0)(5,1)(4,1)(4,0)\psarc[linewidth=0.5pt]{-}(4,0){0.25}{0}{90}\rput(4.5,0.5){$u$}
\psline{-}(0,1)(1,1)(1,2)(0,2)(0,1)\psarc[linewidth=0.5pt]{-}(0,1){0.25}{0}{90}\rput(0.5,1.5){$\lambda-u$}
\psline{-}(1,1)(2,1)(2,2)(1,2)(1,1)\psarc[linewidth=0.5pt]{-}(1,1){0.25}{0}{90}\rput(1.5,1.5){$\lambda-u$}
\psline{-}(4,1)(5,1)(5,2)(4,2)(4,1)\psarc[linewidth=0.5pt]{-}(4,1){0.25}{0}{90}\rput(4.5,1.5){$\lambda-u$}
\psline{-}(2,0)(2.5,0)\psline[linestyle=dashed,dash=2pt 2pt]{-}(2.5,0)(3.5,0)\psline{-}(3.5,0)(4,0)
\psline{-}(2,1)(2.5,1)\psline[linestyle=dashed,dash=2pt 2pt]{-}(2.5,1)(3.5,1)\psline{-}(3.5,1)(4,1)
\psline{-}(2,2)(2.5,2)\psline[linestyle=dashed,dash=2pt 2pt]{-}(2.5,2)(3.5,2)\psline{-}(3.5,2)(4,2)
\psset{linecolor=myc}\unlw
\psarc{-}(0,1){0.5}{90}{270}
\psarc{-}(5,1){0.5}{270}{450}
\end{pspicture}}^N
\end{equation*}
where each box stands for the sum
\begin{equation*}
\psset{linewidth=1pt}
\begin{pspicture}(-0.5,-0.1)(0.5,0.5)
\psline{-}(-0.5,-0.5)(0.5,-0.5)(0.5,0.5)(-0.5,0.5)(-0.5,-0.5)
\psarc[linewidth=0.5pt]{-}(-0.5,-0.5){0.25}{0}{90}
\rput(0,0){$u$}
\end{pspicture}\ =\ \sin(\lambda-u)\ \ 
\begin{pspicture}(-0.5,-0.1)(0.5,0.5)
\psline{-}(-0.5,-0.5)(0.5,-0.5)(0.5,0.5)(-0.5,0.5)(-0.5,-0.5)
\psset{linecolor=myc}
\psarc{-}(0.5,-0.5){0.5}{90}{180}
\psarc{-}(-0.5,0.5){0.5}{270}{360}
\end{pspicture}\ +\ \sin u\ \ 
\begin{pspicture}(-0.5,-0.1)(0.5,0.5)
\psline{-}(-0.5,-0.5)(0.5,-0.5)(0.5,0.5)(-0.5,0.5)(-0.5,-0.5)
\psset{linecolor=myc}
\psarc{-}(-0.5,-0.5){0.5}{0}{90}
\psarc{-}(0.5,0.5){0.5}{180}{270}
\end{pspicture}
\ \ =\ \ 
\begin{pspicture}(-0.5,-0.1)(0.5,0.5)
\psline{-}(-0.5,-0.5)(0.5,-0.5)(0.5,0.5)(-0.5,0.5)(-0.5,-0.5)
\psarc[linewidth=0.5pt]{-}(0.5,-0.5){0.25}{90}{180}
\rput(0,0){$\lambda-u$}
\end{pspicture}
\end{equation*}
\\
with $\beta=2\cos \lambda$, $\lambda\in(0,\frac{\pi}2)$ and $u\in[0,\lambda]$ is the anisotropy parameter and will be given a physical interpretation in section \ref{sec:QPotts}. (A global factor to the weights of the two states does not change the physics; our factor differs from that of \cite{PRZ}.) The matrix $D_N(\lambda,u)$ is therefore defined by $2N$ sums over the two states. As for the $\beta$-product of connectivities, any closed loop gives rise to a factor $\beta$. The simplest case is for $N=2$. There are $2^4$ configurations, each contributing to one of the two connectivities:
\begin{align*}
D_2(\lambda,u)=&
\left(
2\beta (\sin^3u\sin(\lambda-u)+\sin u\sin^3(\lambda-u))+(4+\beta^2)\sin^2u\sin^2(\lambda-u)\right)
\ 
\psset{unit=0.5}
\begin{pspicture}(-0.5,-0.25)(1.5,0.5)
\psline[linewidth=1pt]{-}(-0.5,-0.5)(1.5,-0.5)(1.5,0.5)(-0.5,0.5)(-0.5,-0.5)
\psdots(0,-0.5)(1,-0.5)(0,0.5)(1,0.5)
\psset{linecolor=myc}
\psbezier{-}(0,-0.5)(0,0)(1,0)(1,-0.5)
\psbezier{-}(0,0.5)(0,0)(1,0)(1,0.5)
\end{pspicture}\\ 
+&\left(\beta(\sin^4u+\sin^2u\sin^2(\lambda-u)+\sin^4(\lambda-u))+2(\sin^3u\sin(\lambda-u)+\sin u\sin^3(\lambda-u))\right)
\ 
\psset{unit=0.5}
\begin{pspicture}(-0.5,-0.25)(1.5,0.5)
\psline[linewidth=1pt]{-}(-0.5,-0.5)(1.5,-0.5)(1.5,0.5)(-0.5,0.5)(-0.5,-0.5)
\psdots(0,-0.5)(1,-0.5)(0,0.5)(1,0.5)
\psset{linecolor=myc}
\psline{-}(0,-0.5)(0,0.5)
\psline{-}(1,-0.5)(1,0.5)
\end{pspicture} \ . \ 
\end{align*}
The most important property of the $D_N(\lambda,u)$ is the following (for the proof, see \cite{Behrend}):
\begin{equation}
[D_N(\lambda,u),D_N(\lambda, v)]=0, \qquad u,v\in [0,\lambda].
\label{eq:commu}
\end{equation}
It is a direct consequence of the Yang-Baxter equation. If some coefficient of $D_N(\lambda,u)$, for a fixed $\lambda$, is interpreted as a Hamiltonian, then the above equation leads to a family of constants of motion for this dynamical system. When there is no confusion, the dependence upon $\lambda$ and $u$ of $D_N(\lambda,u)$ and the $\beta$ in $TL_N(\beta)$ will be omitted. 

\subsection{The link representation of $TL_N(\beta)$}

Several representations are useful to study $TL_N(\beta)$. Since $TL_N(\beta)$ is itself a vector space, it comes with a natural representation on $\mathbb C^{C_N}$. Another one that appears naturally in spin systems acts on the tensor product $(\mathbb C^2)^{\otimes N}$ and Goodman and Wenzl \cite{GoodmanWenzl} have shown that it is a faithful representation.  Following \cite{PRZ}, we will use yet another representation, that on link states. (See also \cite{Westbury}.) A {\em $N$-link state} is a set of non-crossing curves, drawn above a horizontal segment, pairing $N$ points among themselves or to infinity (more than one point can be connected to infinity). In the latter case, we draw the curve as a vertical segment and call  {\em defects} such  pairings to infinity. The number of defects of a link state $u$ will be denoted $d(u)$.  The set of all $N$-link states is denoted by $B_N$ and we shall order $B_N$ such that the number of defects is increasing. (The actual order among link states of a given defect number will not play a role.) Formal linear combinations over $\mathbb C$ of elements of $B_N$ form the vector space $V_N$. It is of dimension $\left(\begin{smallmatrix}N\\ \lfloor N/2 \rfloor \end{smallmatrix}\right)$. Subspaces $V_N^d$ are those spanned by link states with $d$ defects. Their dimension is $\dim V_N^d=\left(\begin{smallmatrix}N\\ (N-d)/2\end{smallmatrix}\right)-
\left(\begin{smallmatrix}N\\ (N-d)/2-1\end{smallmatrix}\right)$ and $V_N=\oplus_{0\le d\le N}V_N^d$. Note that $V_N^d$ is non-trivial only if $d$ and $N$ have the same parity. As an example, there are six link states for $N=4$:
\begin{equation*}
B_4=\bigg\{
\psset{unit=1}
\psset{linewidth=1pt}
\begin{pspicture}(-0.1,0.1)(11.4,0.9)
\psset{linecolor=myc2}
\psarc{-}(0.2,0){0.2}{0}{180}
\psarc{-}(1,0){0.2}{0}{180}
\psset{linecolor=black}
\psdots(0.0,0)(0.4,0)(0.8,0)(1.2,0)
\psdots(2.0,0)(2.4,0)(2.8,0)(3.2,0)
\rput(1.6,0){,}
\psset{linecolor=myc2}
\psarc{-}(2.6,0){0.2}{0}{180}
\psbezier{-}(2,0)(2,0.4)(3.2,0.4)(3.2,0)
\psset{linecolor=black}
\psdots(4.0,0)(4.4,0)(4.8,0)(5.2,0)
\rput(3.6,0){,}
\psset{linecolor=myc2}
\psline{-}(4,0)(4,0.55)
\psline{-}(4.4,0)(4.4,0.55)
\psarc{-}(5,0){0.2}{0}{180}
\psset{linecolor=black}
\rput(5.6,0){,}
\psdots(6,0)(7.2,0)(6.4,0)(6.8,0)
\psset{linecolor=myc2}
\psline{-}(6,0)(6,0.55)
\psline{-}(7.2,0)(7.2,0.55)
\psarc{-}(6.6,0){0.2}{0}{180}
\psset{linecolor=black}
\rput(7.6,0){,}
\psdots(8.0,0)(8.4,0)(8.8,0)(9.2,0)
\psset{linecolor=myc2}
\psarc{-}(8.2,0){0.2}{0}{180}
\psline{-}(8.8,0)(8.8,0.55)
\psline{-}(9.2,0)(9.2,0.55)
\psset{linecolor=black}
\rput(9.6,0){,}
\psdots(10,0)(10.4,0)(10.8,0)(11.2,0)
\psset{linecolor=myc2}
\psline{-}(10,0)(10,0.55)
\psline{-}(10.4,0)(10.4,0.55)
\psline{-}(10.8,0)(10.8,0.55)
\psline{-}(11.2,0)(11.2,0.55)
\psset{linecolor=black}
\end{pspicture}
\bigg\}.
\label{eq:baseB2}
\end{equation*}

To each $N$-connectivity $c$ corresponds a matrix $\rho(c)\in \text{\rm End}(V_N)$. Let $v\in B_N$. To determine $\rho(c)$, we draw $v$ on top of $c$ (and denote the resulting diagram $cv$) and read how the bottom sites of $c$ are connected among each other or to infinity. The result is a link state $w$ in $B_N$. Any closed loop gives a factor of $\beta$. The column $v$ in $\rho(c)$ contains therefore a single non-zero matrix element at position $w$ and its value is the product of all $\beta$ factors.
\begin{equation}
\begin{pspicture}(-0.5,-0.3)(9.5,0.8)
\psset{unit=0.5}
\psline[linewidth=1pt]{-}(0.5,1)(8.5,1)(8.5,-1)(0.5,-1)(0.5,1)
\psset{linewidth=1pt}
\psdots(1,1)(2,1)(3,1)(4,1)(5,1)(6,1)(7,1)(8,1)(1,-1)(2,-1)(3,-1)(4,-1)(5,-1)(6,-1)(7,-1)(8,-1)
\psset{linecolor=myc}
\psline{-}(1,1)(1,-1)
\psarc(6.5,1){0.5}{180}{360}
\psarc(2.5,-1){0.5}{0}{180}
\psarc(6.5,-1){0.5}{0}{180}
\psbezier{-}(2,1)(2,0)(4,0)(4,-1)
\psbezier{-}(3,1)(3,0)(5,0)(5,-1)
\psbezier{-}(5,1)(5,0)(8,0)(8,1)
\psbezier{-}(4,1)(4,0)(8,0)(8,-1)
\psset{linecolor=myc2}
\psline{-}(3,1)(3,2)
\psline{-}(4,1)(4,2)
\psarc(1.5,1){0.5}{0}{180}
\psarc(6.5,1){0.5}{0}{180}
\psbezier{-}(5,1)(5,2)(8,2)(8,1)
\psset{linecolor=black}
\rput(9.7,0){$= \beta^2$}
\psdots(17.7,-0.5)(16.7,-0.5)(15.7,-0.5)(14.7,-0.5)(13.7,-0.5)(12.7,-0.5)(11.7,-0.5)(10.7,-0.5)
\psset{linecolor=myc2}
\psarc(12.2,-0.5){0.5}{0}{180}
\psbezier{-}(10.7,-0.5)(10.7,0.5)(13.7,0.5)(13.7,-0.5)
\psline{-}(14.7,-0.5)(14.7,0.5)
\psline{-}(17.7,-0.5)(17.7,0.5)
\psarc(16.2,-0.5){0.5}{0}{180}
\end{pspicture}
\label{eq:prodcv}
\end{equation}
Note that the action of $c$ on a link state cannot increase its number of defects and, if $v\in V_N^d$, then $\rho(c)v\in  \oplus_{d'\le d}V_N^{d'}$. We extend $\rho$ linearly to the algebra $TL_N(\beta)$.

\begin{Lemme} $\rho:TL_N(\beta)\rightarrow \text{\rm End\,}{\mathbb C}^{\left(\begin{smallmatrix}N\\ \lfloor N/2\rfloor \end{smallmatrix}\right)}$ is a representation.
\end{Lemme}

\noindent{\scshape Proof\ \ }The only thing to check is whether $\rho(cd)=\rho(c)\rho(d)$ for all connectivities $c,d$. The product of $cd$ is, up to a power of $\beta$, a connectivity. As noted before, a single element is non-zero in each column of $\rho(c)$, $\rho(d)$ and $\rho(cd)$. Fix a link state $u\in B_N$ and let $v,w\in B_N$ be such that $(\rho(d))_{vu}$ and $(\rho(cd))_{wu}$ are non-zero. Because the process of reading the pairs connected in the diagram $cdu$ is associative, that is $(cd)u=c(du)$, then $(\rho(c))_{wv}$ must be non-zero. Therefore the only thing to check is whether the factors $\beta$ in $(\rho(cd))_{wu}$ and $(\rho(c))_{wv}(\rho(d))_{vu}$ are the same. (Note that there is no sum on $v$.) These factors are indeed equal. They are both the number of closed loops in the diagram with, from top to bottom, the link state $u$, the connectivity $d$ and then $c$. The two expressions count the number of loops in a different order however. For example, the term $(\rho(cd))_{wu}$ counts first the loops that are closed when the connectivities $c$ and $d$ are glued, and then the additional loops when the link $u$ is added.\hfill$\square$

\smallskip
We shall refer to $\rho$ as the link representation.

\subsection{Traces in the link representation}

We will be interested in making contacts between elements of $TL_N(\beta)$ and partition functions $\tilde\Lambda$ of $Q$-Potts models. In section \ref{sec:QPotts}, the partition function $Z_{N\times M}$ is obtained as the trace of some powers of the transfer matrix $\tilde\Lambda$. Similar questions were already addressed for some models and representations of $TL_N$. Our argument to answer this question is similar to that of Jacobsen and Richard \cite{Jacobsen}. Their computation is however done for a representation of dimension $C_N$, much larger than ours, and has to identify some of the degeneracies among the representations that constitute the starting one. The link representation avoids some of these difficulties and it is therefore useful to give the shorter computation for this case. A similar calculation, in the link basis, can be found in \cite{JacobsenSaleur}.

We now introduce an operation akin the trace on the algebra $TL_N(\beta)$. 

\begin{Definition} If $c$ is an $N$-connectivity, the trace of $c$, noted $\tau(c)$, is $\beta^{\#(c)}$ where $\#(c)$ is the number of loops closed by the process of identifying the points on the top with those on the bottom of the rectangle. The trace $\tau:TL_N(\beta)\rightarrow \mathbb C$ is the linear extension of the trace of connectivities to the whole algebra.
\end{Definition}
Note that $\tau$ is not a representation. 

The double-row matrix $D_N(\lambda, u)$ is an element of $TL_N(\beta)$ and can then be written as $D_N=\sum_c \alpha_c c$ where the sum is over connectivities and $\alpha_c\in \mathbb C$. Note that any power $D_N^M$, seen as an element of $TL_N(\beta)$, is also of the same form $D_N^M=\sum \alpha_{c,M}c$ with $\alpha_{c,M}\in\mathbb C$. It will be shown in section \ref{sec:QPotts} that, for the Potts models, $\beta^2$ equals $Q=1,2,3$ or $4$ and their partition function is related to $\tau(D_N^M)$ when $N$ is even by
\begin{equation}Z_{N \times M} \propto \sum_c \alpha_{c,M}\beta^{\#(c)}=\sum_c\alpha_{c,M}\tau(c)=\tau(D_N^M).\label{eq:partitionFunction}
\end{equation}
A natural question is then to relate $\tau(D_N^M)$ and $\tr\rho(D_N^M)$ or, for any $C\in TL_N(\beta)$, to relate $\tau(C)$ and $\tr\rho(C)$. We therefore turn to the (usual) trace $\text{\text tr\,}\rho(e)$, for a connectivity $e\in TL_N(\beta)$. Because $\rho(e)$ has a single element per column, its trace will be the sum of eigenvalues of link states that are eigenvectors of $\rho(e)$. Another important number characterizing connectivities is the following. 
\begin{Definition} Let $\delta(c)$ be the maximum number of defects of link states among those that are eigenvectors of $\rho(c)$:
$$\delta(c)=\max_{u\in B_N, \rho(c)_{uu}\neq 0}\left\{d(u)\right\}.$$
Note that $\#(c)\ge \delta(c)$.
\end{Definition}
This allows to count the number of eigenstates of $\rho(c)$ in $B_N$.
\begin{Lemme}The number of link states with $d$ defects that are eigenstates of $\rho(c)$ is
$$\dim V^d_{\delta(c)}=\begin{pmatrix}\delta(c)\\ (\delta(c)-d)/2\end{pmatrix}-
\begin{pmatrix}\delta(c)\\ (\delta(c)-d)/2-1\end{pmatrix}.$$
\end{Lemme}
\noindent{\scshape Proof\ \ }The unit element $id$ has $\delta(id)=N$ and all link states are eigenvectors of $\rho(id)= id_{\textrm{dim}V_N}$. The numbers given in the statement are then the dimensions of the $V_N^d$.

Let c be a connectivity that is not the unit. We construct its link eigenstates as follows. Since $c\neq id$, there are for sure two points on the bottom of the rectangle that are paired. Among all pairs of joined points on the bottom, choose one such that the points are contiguous. Such a pairing must also occur in any eigenstate of $\rho(c)$. Consider the diagram of the connectivity $c$ to which an arc has been added to its top connecting the same pair. One of the following diagrams, where the original pairing is in gray, represents the extended curve thus created.
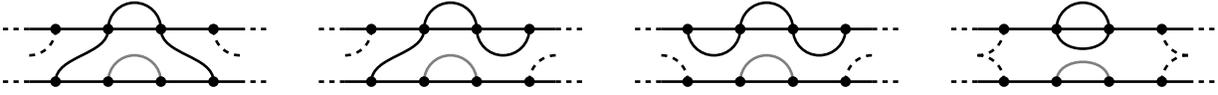
\begin{equation*}
\psset{unit=0.7}
\begin{pspicture}(-1,0)(4,1.5)
\psline[linewidth=1pt,linestyle=dashed,dash=2pt 2pt]{-}(-1,0)(-0.5,0)
\psline[linewidth=1pt,linestyle=dashed,dash=2pt 2pt]{-}(-1,1)(-0.5,1)
\psline[linewidth=1pt,linestyle=dashed,dash=2pt 2pt]{-}(3.5,0)(4,0)
\psline[linewidth=1pt,linestyle=dashed,dash=2pt 2pt]{-}(3.5,1)(4,1)
\psarc[linewidth=1pt,linestyle=dashed,dash=2pt 2pt](-0.5,1){0.5}{270}{360}
\psarc[linewidth=1pt,linestyle=dashed,dash=2pt 2pt](3.5,1){0.5}{180}{270}
\psline[linewidth=1pt]{-}(-0.5,0)(3.5,0)
\psline[linewidth=1pt]{-}(-0.5,1)(3.5,1)
\psset{linewidth=1pt}
\psarc(1.5,1){0.5}{0}{180}
\psarc[linecolor=gray](1.5,0){0.5}{0}{180}
\psbezier{-}(0,0)(0,0.5)(1,0.5)(1,1)
\psbezier{-}(3,0)(3,0.5)(2,0.5)(2,1)
\psdots(0,0)(1,0)(2,0)(3,0)(0,1)(1,1)(2,1)(3,1)
\end{pspicture}\qquad
\begin{pspicture}(-1,0)(4,1.5)
\psline[linewidth=1pt,linestyle=dashed,dash=2pt 2pt]{-}(-1,0)(-0.5,0)
\psline[linewidth=1pt,linestyle=dashed,dash=2pt 2pt]{-}(-1,1)(-0.5,1)
\psline[linewidth=1pt,linestyle=dashed,dash=2pt 2pt]{-}(3.5,0)(4,0)
\psline[linewidth=1pt,linestyle=dashed,dash=2pt 2pt]{-}(3.5,1)(4,1)
\psarc[linewidth=1pt,linestyle=dashed,dash=2pt 2pt](-0.5,1){0.5}{270}{360}
\psarc[linewidth=1pt,linestyle=dashed,dash=2pt 2pt](3.5,0){0.5}{90}{180}
\psline[linewidth=1pt]{-}(-0.5,0)(3.5,0)
\psline[linewidth=1pt]{-}(-0.5,1)(3.5,1)
\psset{linewidth=1pt}
\psarc(1.5,1){0.5}{0}{180}
\psarc[linecolor=gray](1.5,0){0.5}{0}{180}
\psbezier{-}(0,0)(0,0.5)(1,0.5)(1,1)
\psarc(2.5,1){0.5}{180}{360}
\psdots(0,0)(1,0)(2,0)(3,0)(0,1)(1,1)(2,1)(3,1)
\end{pspicture}\qquad
\begin{pspicture}(-1,0)(4,1.5)
\psline[linewidth=1pt,linestyle=dashed,dash=2pt 2pt]{-}(-1,0)(-0.5,0)
\psline[linewidth=1pt,linestyle=dashed,dash=2pt 2pt]{-}(-1,1)(-0.5,1)
\psline[linewidth=1pt,linestyle=dashed,dash=2pt 2pt]{-}(3.5,0)(4,0)
\psline[linewidth=1pt,linestyle=dashed,dash=2pt 2pt]{-}(3.5,1)(4,1)
\psarc[linewidth=1pt,linestyle=dashed,dash=2pt 2pt](-0.5,0){0.5}{0}{90}
\psarc[linewidth=1pt,linestyle=dashed,dash=2pt 2pt](3.5,0){0.5}{90}{180}
\psline[linewidth=1pt]{-}(-0.5,0)(3.5,0)
\psline[linewidth=1pt]{-}(-0.5,1)(3.5,1)
\psset{linewidth=1pt}
\psarc(1.5,1){0.5}{0}{180}
\psarc[linecolor=gray](1.5,0){0.5}{0}{180}
\psarc(0.5,1){0.5}{180}{360}
\psarc(2.5,1){0.5}{180}{360}
\psdots(0,0)(1,0)(2,0)(3,0)(0,1)(1,1)(2,1)(3,1)
\end{pspicture}
\qquad
\begin{pspicture}(-1,0)(4,1.5)
\psline[linewidth=1pt,linestyle=dashed,dash=2pt 2pt]{-}(-1,0)(-0.5,0)
\psline[linewidth=1pt,linestyle=dashed,dash=2pt 2pt]{-}(-1,1)(-0.5,1)
\psline[linewidth=1pt,linestyle=dashed,dash=2pt 2pt]{-}(3.5,0)(4,0)
\psline[linewidth=1pt,linestyle=dashed,dash=2pt 2pt]{-}(3.5,1)(4,1)
\psarc[linewidth=1pt,linestyle=dashed,dash=2pt 2pt](-0.5,0){0.5}{0}{90}
\psarc[linewidth=1pt,linestyle=dashed,dash=2pt 2pt](3.5,0){0.5}{90}{180}
\psarc[linewidth=1pt,linestyle=dashed,dash=2pt 2pt](-0.5,1){0.5}{270}{360}
\psarc[linewidth=1pt,linestyle=dashed,dash=2pt 2pt](3.5,1){0.5}{180}{270}
\psline[linewidth=1pt]{-}(-0.5,0)(3.5,0)
\psline[linewidth=1pt]{-}(-0.5,1)(3.5,1)
\psset{linewidth=1pt}
\psarc(1.5,1){0.5}{0}{180}
\psbezier[linecolor=gray]{-}(1,0)(1,0.5)(2,0.5)(2,0)
\psbezier{-}(1,1)(1,0.5)(2,0.5)(2,1)
\psdots(0,0)(1,0)(2,0)(3,0)(0,1)(1,1)(2,1)(3,1)
\end{pspicture}
\end{equation*}
For simplicity, any arcs between the gray and black curves were omitted. There could be some. By adding this curve on top of the diagram, one of the following situations occurs: a pair of points on the bottom are joined, one point on the bottom and one of the top are joined, two on the tops are joined, a loop is closed. Since the points connected by the new arc are neighbors, the extended curve can be deformed, without crossing the others in $c$, into the corresponding diagram below
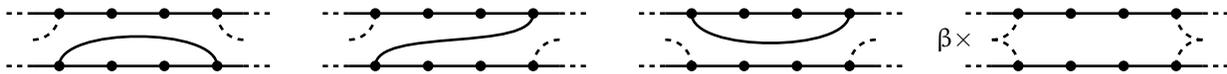
\begin{equation*}
\psset{unit=0.7}
\begin{pspicture}(-1,0)(4,1.5)
\psline[linewidth=1pt,linestyle=dashed,dash=2pt 2pt]{-}(-1,0)(-0.5,0)
\psline[linewidth=1pt,linestyle=dashed,dash=2pt 2pt]{-}(-1,1)(-0.5,1)
\psline[linewidth=1pt,linestyle=dashed,dash=2pt 2pt]{-}(3.5,0)(4,0)
\psline[linewidth=1pt,linestyle=dashed,dash=2pt 2pt]{-}(3.5,1)(4,1)
\psarc[linewidth=1pt,linestyle=dashed,dash=2pt 2pt](-0.5,1){0.5}{270}{360}
\psarc[linewidth=1pt,linestyle=dashed,dash=2pt 2pt](3.5,1){0.5}{180}{270}
\psline[linewidth=1pt]{-}(-0.5,0)(3.5,0)
\psline[linewidth=1pt]{-}(-0.5,1)(3.5,1)
\psset{linewidth=1pt}
\psbezier{-}(0,0)(0,0.75)(3,0.75)(3,0)
\psdots(0,0)(1,0)(2,0)(3,0)(0,1)(1,1)(2,1)(3,1)
\end{pspicture}\qquad
\begin{pspicture}(-1,0)(4,1.5)
\psline[linewidth=1pt,linestyle=dashed,dash=2pt 2pt]{-}(-1,0)(-0.5,0)
\psline[linewidth=1pt,linestyle=dashed,dash=2pt 2pt]{-}(-1,1)(-0.5,1)
\psline[linewidth=1pt,linestyle=dashed,dash=2pt 2pt]{-}(3.5,0)(4,0)
\psline[linewidth=1pt,linestyle=dashed,dash=2pt 2pt]{-}(3.5,1)(4,1)
\psarc[linewidth=1pt,linestyle=dashed,dash=2pt 2pt](-0.5,1){0.5}{270}{360}
\psarc[linewidth=1pt,linestyle=dashed,dash=2pt 2pt](3.5,0){0.5}{90}{180}
\psline[linewidth=1pt]{-}(-0.5,0)(3.5,0)
\psline[linewidth=1pt]{-}(-0.5,1)(3.5,1)
\psset{linewidth=1pt}
\psbezier{-}(0,0)(0,0.75)(3,0.25)(3,1)
\psdots(0,0)(1,0)(2,0)(3,0)(0,1)(1,1)(2,1)(3,1)
\end{pspicture}\qquad
\begin{pspicture}(-1,0)(4,1.5)
\psline[linewidth=1pt,linestyle=dashed,dash=2pt 2pt]{-}(-1,0)(-0.5,0)
\psline[linewidth=1pt,linestyle=dashed,dash=2pt 2pt]{-}(-1,1)(-0.5,1)
\psline[linewidth=1pt,linestyle=dashed,dash=2pt 2pt]{-}(3.5,0)(4,0)
\psline[linewidth=1pt,linestyle=dashed,dash=2pt 2pt]{-}(3.5,1)(4,1)
\psarc[linewidth=1pt,linestyle=dashed,dash=2pt 2pt](-0.5,0){0.5}{0}{90}
\psarc[linewidth=1pt,linestyle=dashed,dash=2pt 2pt](3.5,0){0.5}{90}{180}
\psline[linewidth=1pt]{-}(-0.5,0)(3.5,0)
\psline[linewidth=1pt]{-}(-0.5,1)(3.5,1)
\psset{linewidth=1pt}
\psbezier{-}(0,1)(0,0.25)(3,0.25)(3,1)
\psdots(0,0)(1,0)(2,0)(3,0)(0,1)(1,1)(2,1)(3,1)
\end{pspicture}
\qquad
\begin{pspicture}(-1.2,0)(4,1.5)
\rput(-1.2,0.5){$\beta\times$}
\psline[linewidth=1pt,linestyle=dashed,dash=2pt 2pt]{-}(-1,0)(-0.5,0)
\psline[linewidth=1pt,linestyle=dashed,dash=2pt 2pt]{-}(-1,1)(-0.5,1)
\psline[linewidth=1pt,linestyle=dashed,dash=2pt 2pt]{-}(3.5,0)(4,0)
\psline[linewidth=1pt,linestyle=dashed,dash=2pt 2pt]{-}(3.5,1)(4,1)
\psarc[linewidth=1pt,linestyle=dashed,dash=2pt 2pt](-0.5,0){0.5}{0}{90}
\psarc[linewidth=1pt,linestyle=dashed,dash=2pt 2pt](3.5,0){0.5}{90}{180}
\psarc[linewidth=1pt,linestyle=dashed,dash=2pt 2pt](-0.5,1){0.5}{270}{360}
\psarc[linewidth=1pt,linestyle=dashed,dash=2pt 2pt](3.5,1){0.5}{180}{270}
\psline[linewidth=1pt]{-}(-0.5,0)(3.5,0)
\psline[linewidth=1pt]{-}(-0.5,1)(3.5,1)
\psset{linewidth=1pt}
\psdots(0,0)(1,0)(2,0)(3,0)(0,1)(1,1)(2,1)(3,1)
\end{pspicture}
\end{equation*}
Note that the pairing between contiguous points on the bottom can be safely erased as long as we keep track that the link state under construction must have this arc.

The process has reduced by $2$ the number of points on the top and bottom sides of the rectangle. If the new diagram is not the unit, the process is repeated. Suppose then that a unit has been reached and that this unit connects $\Delta$ pairs of points, $0\le \Delta\le N$. Then the pairing of $(N-\Delta)$ points of the link has already been determined and all eigenstates of $\rho(c)$ in $B_N$ will share these pairings. The remaining $\Delta$ points can be tied to infinity or among each other since the restriction of $\rho(c)$ to these points acts as the unit.

To get the maximum number of defects in an eigenstate, we connect the $\Delta$ remaining points to infinity. Then $\delta(c)=\Delta$ and there is precisely one link eigenstate $u_c$ with this maximum number of defects. How many link eigenstates with $d$ defects are there? It is simply the number of ways to connect $(\delta(c)-d)$ of the remaining $\delta(c)$ points of $u_c$, that is 
$\dim V_{\delta(c)}^d=\left(\begin{smallmatrix}\delta(c)\\ (\delta(c)-d)/2\end{smallmatrix}\right)-
\left(\begin{smallmatrix}\delta(c)\\ (\delta(c)-d)/2-1\end{smallmatrix}\right)$.\hfill$\square$
\medskip

Let $C=\sum_c \gamma_c c$. It is useful to split the sum over connectivities into sums over connectivities with a given number $\delta$. Then 
$$C_d=\sum_{\text{\rm $c$ with $\delta(c)=d$}}\gamma_c c\qquad \text{\rm and}\qquad
C=\sum_{0\le d\le N}C_d$$
and
$$\tau(C_d)=\sum_{\text{\rm $c$ with $\delta(c)=d$}}\gamma_c \tau(c)=\sum_{\text{\rm $c$ with $\delta(c)=d$}}\gamma_c\beta^{\#(c)}\qquad \text{\rm and}\qquad
\tau(C)=\sum_{0\le d\le N}\tau(C_d).$$

Let $N$ be even and fix a number of defects $d$ with $0\le d\le N$ and denote by $\text{\rm tr}_d\rho(c)$ the trace of the diagonal block of $\rho(c)$ acting on the subspace $V_N^d$. All connectivities $c$ with $\delta(c)=d$ have a single diagonal element in $\rho(c)$ contributing to $\text{\rm tr}_d \rho(c)$. It comes from the eigenstate $u_c$ constructed in the proof above. The proof has also shown that the connectivity $c$ acts as the unit on $\delta(c)$ points. By definition $\tau(c)=\beta^{\#(c)}$ and therefore only $(\#(c)-\delta(c))$ loops are closed when the diagrams of $u_c$ and $c$ are joined. The eigenvalue of $u_c$ must be $\beta^{\#(c)-\delta(c)}$. Such connectivities $c$, with $\delta(c)=d$, also contribute to $\text{\rm tr}_{d'} \rho(C)$ with $d'<\delta(c)$. Equivalently $\text{\rm tr}_d \rho(C)$ gets contribution from connectivities $c$ that have their $\delta(c)>d$. If $d'\ge d$, then $\rho(c) $ with $\delta(c)=d'$ has precisely $(\dim V_{d'}^{d})$ link eigenstates with $d$ defects. The eigenvalue of these states can be obtained by an argument similar to the one above and is $\beta^{\#(c)-d'}$. Then let
\begin{align*}\text{\rm tr}_d \rho(C)&=\sum_{d'\ge d}\sum_{\text{\rm $c$ with $\delta(c)=d'$}}
\gamma_c\beta^{\#(c)-d'}\dim V_{d'}^{d}\\
&=\sum_{d'\ge d}\tau(C_{d'})\beta^{-d'}\dim V_{d'}^{d}.
\end{align*}
The numbers of defects $d$ and $d'$ have the same parity as $N$ and are thus even. They take $(\frac{N}2+1)$ possible values. Let $M$ be a $(\frac N2+1)\times(\frac N2+1)$ matrix whose elements are labeled by numbers of defects
\begin{equation}M_{dd'}=\begin{cases}\beta^{-d'}\dim V_{d'}^{d},& d'\ge d,\\
                       0,& d'<d.\end{cases}
\label{eq:matM}\end{equation}
Then $\text{\rm tr}_d\rho(C)=\sum_{d'}M_{dd'}\tau(C_{d'})$. The quantities of interest are the $\tau(C_d)$. The inverse of $M$ can be computed. (The expression is given at the end of this section.) Then
$$\tau(C_d)=\sum_{d'}(M^{-1})_{dd'}\text{\rm tr}_{d'}\rho(C)$$
and 
$$\tau(C)=\sum_{0\le d, d'\le N}(M^{-1})_{dd'}\text{\rm tr}_{d'}\rho(C).$$
The sum $\sum_d (M^{-1})_{dd'}$ will be done in another lemma below. This therefore proves the following relationship between $\tau(C)$ and the trace of $\rho(C)$. 
\begin{Proposition}Let $N$ be even and $W\in \text{\rm End}(V_N)$ be the linear transformation that acts as a multiple of the identity on each $V_N^d$ with \fbox{$\left.W\right|_{V_N^d}=  \frac{\sin (d+1)\lambda}{\sin\lambda}\cdot id$} where $\beta=2\cos\lambda$. Then $\tau(C)=\tr (\rho(C)W)$ for all $C\in TL_N(\beta)$.\label{thm:deuxTraces}
\end{Proposition}

Here are the proofs of the announced lemmas.

\begin{Lemme} Let $N$ be even. The inverse of the matrix $M$ introduced in \eqref{eq:matM} is
$$(M^{-1})_{dd'}=(-1)^{(d+d')/2}\beta^d\begin{pmatrix}(d+d')/2 \\ d\end{pmatrix}$$
if $d'\ge d$ and $0$ otherwise.
\end{Lemme}

\noindent{\scshape Proof\ } Recall that the indices $d$ and $d'$ are even integers in the interval $[0,N]$. It is easier to work with integers $i,j,\dots$ in the range $[0,\frac N2]$. With these indices we therefore have to prove that
$$\sum_{j=i}^l M_{ij}(M^{-1})_{jl}=\sum_{j=i}^l\frac1{\beta^{2j}}\left(\begin{pmatrix}2j\\ j-i\end{pmatrix}-
\begin{pmatrix}2j\\ j-i-1\end{pmatrix}\right)(-1)^{j+l}\beta^{2j}\begin{pmatrix}j+l\\
2j\end{pmatrix}=\delta_{il}$$
when $i\le l$. (We use the convention that the binomial coefficient $(\begin{smallmatrix} a\\b\end{smallmatrix})$ is zero if $b>a$ or if $a$ or $b$ are negative.) Using the two identities
$$\begin{pmatrix}r\\ m\end{pmatrix}\begin{pmatrix}m\\ k\end{pmatrix}=\begin{pmatrix}
r\\ k\end{pmatrix}\begin{pmatrix}r-k\\ m-k\end{pmatrix}\qquad
\text{\rm and}\qquad
\sum_{i=n-r}^{l-m}(-1)^i\begin{pmatrix}l\\ m+i\end{pmatrix}\begin{pmatrix}r+i\\ n\end{pmatrix}=
(-1)^{l+m}\begin{pmatrix}r-m\\ n-l\end{pmatrix}
$$
that hold for integer $l\ge 0$ and any integers $r,m,k,n$
(see eqs.~(5.21) and (5.24) of \cite{GKP} for proofs), we can compute the first sum 
$$\sum_{j=i}^l(-1)^{j+l}\begin{pmatrix}2j\\ j-i\end{pmatrix}\begin{pmatrix}j+l\\
2j\end{pmatrix}=\sum_{j=i}^l(-1)^{j+l}\begin{pmatrix}l+i\\ j+i\end{pmatrix}
\begin{pmatrix}j+l\\ l+i\end{pmatrix}=\begin{pmatrix}l-i\\0\end{pmatrix}.$$
The second sum is done similarly and we find
$$\sum_{j=i}^l M_{ij}(M^{-1})_{jl}=\begin{pmatrix}l-i\\0\end{pmatrix}-
\begin{pmatrix}l-i-1\\0\end{pmatrix}.$$
As long as both $l-i$ and $l-i-1$ are non-negative, they are both $1$ and $(MM^{-1})_{il}=0$ when $i<l$. 
But when $i=l$, the second binomial coefficient above is zero and $(MM^{-1})_{ii}=1$ as claimed.\hfill$\square$

\begin{Lemme} Again, let $N$ be even. With the above notation the number $w_d=\sum_{d'\le d}(M^{-1})_{d'd}$ is equal to ${\sin \lambda(d+1)}/{\sin\lambda}$ for $\beta=2\cos \lambda$.
\end{Lemme}

\noindent{\scshape Proof\ }Again we use integers $j,l,\dots$ in the range $[0,\frac N2]$ instead of the even numbers $d, d'$. We therefore compute
$$w_j\sin\lambda=\sum_{l=0}^j(-1)^{l+j}\begin{pmatrix}l+j\\ 2l\end{pmatrix}\sin\lambda\,(2\cos\lambda)^{2l}.$$
By expanding both trigonometric functions of the right-hand side into exponentials and regrouping into a sum of sine functions, we obtain
$$w_j\sin\lambda=\sum_{l=0}^j(-1)^{l+j}\begin{pmatrix}l+j\\ 2l\end{pmatrix}
\sum_{k=0}^l \left( \begin{pmatrix}2l\\ k+l\end{pmatrix}-\begin{pmatrix}2l\\ k+l+1\end{pmatrix}\right)\sin\lambda(2k+1).$$
Exchanging the order of the two sums we find again sums involving products of two binomial coefficients. These are very similar to those in the previous proof and, indeed, using again (5.21) and (5.24) of \cite{GKP}, we can bring these sums to the form
$$w_j\sin\lambda=\sum_{k=0}^j\left(\begin{pmatrix}j-k\\ 0\end{pmatrix}-\begin{pmatrix}j-k-1\\0\end{pmatrix}\right)\sin\lambda(2k+1).$$
As before both binomial coefficients are equal to one for all $k$, except for $k=j$ when the second one vanishes. The only term left is therefore $\sin\lambda(2j+1)$ as expected.\hfill$\square$

\subsection{$Q$-Potts models with cylindrical boundary conditions}
\label{sec:QPotts}

\begin{center}{\em For this section and the next one, $N$ is even.}\end{center}

The $Q$-Potts spin models, $Q\in \{1,2,3,4\}$, are closely related to the double row matrix $D_N(\lambda, u)$ with $\beta^2=(2\cos \lambda)^2=Q$. This section and the next detail this relationship.

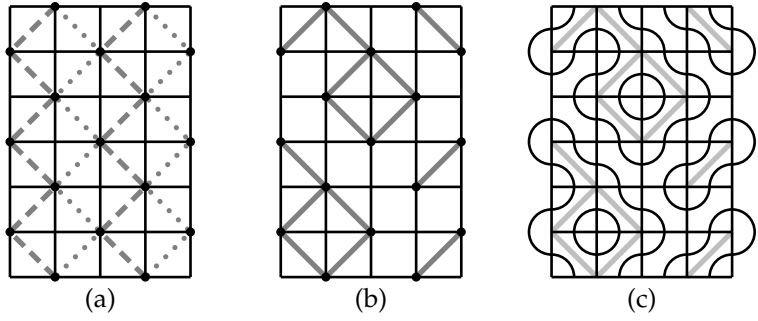
\begin{figure}[h!]
\begin{center}
\psset{unit=0.6}
\begin{pspicture}(0,-0.5)(17,6)
\psset{linewidth=1pt}
\psline[linecolor=gray,linestyle=dashed,linewidth=2pt]{-}(1,0)(0,1)(1,2)(0,3)(1,4)(0,5)(1,6)
\psline[linecolor=gray,linestyle=dashed,linewidth=2pt]{-}(3,0)(2,1)(3,2)(2,3)(3,4)(2,5)(3,6)
\psline[linecolor=gray,linestyle=dotted,linewidth=2pt]{-}(1,0)(2,1)(1,2)(2,3)(1,4)(2,5)(1,6)
\psline[linecolor=gray,linestyle=dotted,linewidth=2pt]{-}(3,0)(4,1)(3,2)(4,3)(3,4)(4,5)(3,6)
\psline{-}(0,6)(4,6)
\psline{-}(0,5)(4,5)
\psline{-}(0,4)(4,4)
\psline{-}(0,3)(4,3)
\psline{-}(0,2)(4,2)
\psline{-}(0,1)(4,1)
\psline{-}(0,0)(4,0)
\psline{-}(4,0)(4,6)
\psline{-}(3,0)(3,6)
\psline{-}(2,0)(2,6)
\psline{-}(1,0)(1,6)
\psline{-}(0,0)(0,6)
\pscircle*(1,0){0.1}\pscircle*(3,0){0.1}
\pscircle*(0,1){0.1}\pscircle*(2,1){0.1}\pscircle*(4,1){0.1}
\pscircle*(1,2){0.1}\pscircle*(3,2){0.1}
\pscircle*(0,3){0.1}\pscircle*(2,3){0.1}\pscircle*(4,3){0.1}
\pscircle*(1,4){0.1}\pscircle*(3,4){0.1}
\pscircle*(0,5){0.1}\pscircle*(2,5){0.1}\pscircle*(4,5){0.1}
\pscircle*(1,6){0.1}\pscircle*(3,6){0.1}
\rput(2,-0.5){(a)}
\psline[linecolor=gray,linewidth=2pt]{-}(6,1)(7,0)
\psline[linecolor=gray,linewidth=2pt]{-}(8,1)(7,0)
\psline[linecolor=gray,linewidth=2pt]{-}(10,1)(9,0)
\psline[linecolor=gray,linewidth=2pt]{-}(6,1)(7,2)
\psline[linecolor=gray,linewidth=2pt]{-}(8,1)(7,2)
\psline[linecolor=gray,linewidth=2pt]{-}(6,3)(7,2)
\psline[linecolor=gray,linewidth=2pt]{-}(9,2)(10,3)
\psline[linecolor=gray,linewidth=2pt]{-}(8,3)(7,4)
\psline[linecolor=gray,linewidth=2pt]{-}(8,3)(9,4)
\psline[linecolor=gray,linewidth=2pt]{-}(8,5)(7,4)
\psline[linecolor=gray,linewidth=2pt]{-}(8,5)(9,4)
\psline[linecolor=gray,linewidth=2pt]{-}(6,5)(7,6)
\psline[linecolor=gray,linewidth=2pt]{-}(7,6)(8,5)
\psline[linecolor=gray,linewidth=2pt]{-}(9,6)(10,5)
\psline{-}(6,6)(10,6)
\psline{-}(6,5)(10,5)
\psline{-}(6,4)(10,4)
\psline{-}(6,3)(10,3)
\psline{-}(6,2)(10,2)
\psline{-}(6,1)(10,1)
\psline{-}(6,0)(10,0)
\psline{-}(10,0)(10,6)
\psline{-}(9,0)(9,6)
\psline{-}(8,0)(8,6)
\psline{-}(7,0)(7,6)
\psline{-}(6,0)(6,6)
\pscircle*(7,0){0.1}\pscircle*(9,0){0.1}
\pscircle*(6,1){0.1}\pscircle*(8,1){0.1}\pscircle*(10,1){0.1}
\pscircle*(7,2){0.1}\pscircle*(9,2){0.1}
\pscircle*(6,3){0.1}\pscircle*(8,3){0.1}\pscircle*(10,3){0.1}
\pscircle*(7,4){0.1}\pscircle*(9,4){0.1}
\pscircle*(6,5){0.1}\pscircle*(8,5){0.1}\pscircle*(10,5){0.1}
\pscircle*(7,6){0.1}\pscircle*(9,6){0.1}
\rput(8,-0.5){(b)}
\psline[linecolor=lightgray,linewidth=2pt]{-}(12,1)(13,0) 
\psline[linecolor=lightgray,linewidth=2pt]{-}(14,1)(13,0)
\psline[linecolor=lightgray,linewidth=2pt]{-}(16,1)(15,0)
\psline[linecolor=lightgray,linewidth=2pt]{-}(12,1)(13,2)
\psline[linecolor=lightgray,linewidth=2pt]{-}(14,1)(13,2)
\psline[linecolor=lightgray,linewidth=2pt]{-}(12,3)(13,2)
\psline[linecolor=lightgray,linewidth=2pt]{-}(15,2)(16,3)
\psline[linecolor=lightgray,linewidth=2pt]{-}(14,3)(13,4)
\psline[linecolor=lightgray,linewidth=2pt]{-}(14,3)(15,4)
\psline[linecolor=lightgray,linewidth=2pt]{-}(14,5)(13,4)
\psline[linecolor=lightgray,linewidth=2pt]{-}(14,5)(15,4)
\psline[linecolor=lightgray,linewidth=2pt]{-}(12,5)(13,6)
\psline[linecolor=lightgray,linewidth=2pt]{-}(13,6)(14,5)
\psline[linecolor=lightgray,linewidth=2pt]{-}(15,6)(16,5)
\psline{-}(12,6)(16,6) 
\psline{-}(12,5)(16,5)
\psline{-}(12,4)(16,4)
\psline{-}(12,3)(16,3)
\psline{-}(12,2)(16,2)
\psline{-}(12,1)(16,1)
\psline{-}(12,0)(16,0)
\psline{-}(16,0)(16,6)
\psline{-}(15,0)(15,6)
\psline{-}(14,0)(14,6)
\psline{-}(13,0)(13,6)
\psline{-}(12,0)(12,6)
\psarc{-}(12,1){0.5}{90}{270} 
\psarc{-}(12,3){0.5}{90}{270}
\psarc{-}(12,5){0.5}{90}{270}
\psarc{-}(16,1){0.5}{270}{450}
\psarc{-}(16,3){0.5}{270}{450}
\psarc{-}(16,5){0.5}{270}{450}
\psarc{-}(12,0){0.5}{0}{90}\psarc{-}(13,1){0.5}{180}{270} 
\psarc{-}(13,1){0.5}{0}{90}\psarc{-}(14,2){0.5}{180}{270} 
\psarc{-}(14,1){0.5}{0}{90}\psarc{-}(15,2){0.5}{180}{270} 
\psarc{-}(12,2){0.5}{0}{90}\psarc{-}(13,3){0.5}{180}{270} 
\psarc{-}(13,2){0.5}{0}{90}\psarc{-}(14,3){0.5}{180}{270} 
\psarc{-}(12,3){0.5}{0}{90}\psarc{-}(13,4){0.5}{180}{270} 
\psarc{-}(13,3){0.5}{0}{90}\psarc{-}(14,4){0.5}{180}{270} 
\psarc{-}(14,4){0.5}{0}{90}\psarc{-}(15,5){0.5}{180}{270} 
\psarc{-}(15,4){0.5}{0}{90}\psarc{-}(16,5){0.5}{180}{270} 
\psarc{-}(13,5){0.5}{0}{90}\psarc{-}(14,6){0.5}{180}{270} 
\psarc{-}(14,5){0.5}{0}{90}\psarc{-}(15,6){0.5}{180}{270} 
\psarc{-}(15,5){0.5}{0}{90}\psarc{-}(16,6){0.5}{180}{270} 
\psarc{-}(13,1){0.5}{270}{360}\psarc{-}(14,0){0.5}{90}{180} 
\psarc{-}(14,1){0.5}{270}{360}\psarc{-}(15,0){0.5}{90}{180} 
\psarc{-}(15,1){0.5}{270}{360}\psarc{-}(16,0){0.5}{90}{180} 
\psarc{-}(12,2){0.5}{270}{360}\psarc{-}(13,1){0.5}{90}{180} 
\psarc{-}(15,2){0.5}{270}{360}\psarc{-}(16,1){0.5}{90}{180} 
\psarc{-}(14,3){0.5}{270}{360}\psarc{-}(15,2){0.5}{90}{180} 
\psarc{-}(15,3){0.5}{270}{360}\psarc{-}(16,2){0.5}{90}{180} 
\psarc{-}(14,4){0.5}{270}{360}\psarc{-}(15,3){0.5}{90}{180} 
\psarc{-}(15,4){0.5}{270}{360}\psarc{-}(16,3){0.5}{90}{180} 
\psarc{-}(12,5){0.5}{270}{360}\psarc{-}(13,4){0.5}{90}{180} 
\psarc{-}(13,5){0.5}{270}{360}\psarc{-}(14,4){0.5}{90}{180} 
\psarc{-}(12,6){0.5}{270}{360}\psarc{-}(13,5){0.5}{90}{180} 
\rput(14,-0.5){(c)}
\end{pspicture}
\caption{A $N\times(2M)$ grid with $N=4$ and $M=3$ with a spin lattice at $45^\circ$: (a) with all vertices and bonds drawn, (b) a graph $\mathcal G$ and (c) the corresponding loop configuration.}\label{fig:spin}
\end{center}
\end{figure}

Let a lattice of spins drawn on a $N\times (2M)$ rectangle, with $N$ even, as shown in Figure \ref{fig:spin} (a). The spins occupy the sites marked by dots. Nearest neighbors are indicated by gray bonds. The lattice is closed by periodic boundary conditions in the vertical direction, that is, the spins in the bottom line are identified to those on the top one. The spins on the leftmost and rightmost columns have only two nearest neighbors. As all other spins, they may take any of the $Q$ states available, so that free boundary conditions hold in the horizontal direction.  A configuration $\sigma$ is a choice, for each spin, of one of the $Q$ states. Allowing for anisotropy, we define the energy of that configuration to be $E_\sigma=-J\sum_{\langle i,j\rangle(J)} \delta_{\sigma_i\sigma_j} -K\sum_{\langle i,j\rangle(K)} \delta_{\sigma_i\sigma_j}$ where $\sum_{\langle i,j\rangle(J)}$ stands for a sum over all nearest-neighbors pairs $i,j$ with $i$ an odd column and $j$ in the next. These bonds are represented by dots in Figure \ref{fig:spin} (a). Dashed lines are used for the $K$ bonds. The Boltzmann weight of $\sigma$ is $e^{-E_\sigma/k_BT}/Z$ with the normalisation constant (the partition function) $Z=\sum_\sigma e^{-E_\sigma/k_BT}$. 

Let $\mu$ and $\nu$ be the restriction of $\sigma$ to two neighboring lines containing $\frac{N}2$ spins. These are then separated by a line with $\frac{N}2+1$ spins. Let $\rho$ be the states of the spins on this intermediary line. It is customary to define the (spin) transfer matrix $\tilde\Lambda$, a $Q^{N/2}\times Q^{N/2}$ real matrix, by
$$\tilde\Lambda_{\mu,\nu}=\sum_\rho \prod_{\langle i,j\rangle(J)}\exp(\gamma_J(\delta_{\mu_i\rho_j}+\delta_{\rho_j\nu_i}))  \prod_{\langle i,j\rangle(K)}\exp(\gamma_K(\delta_{\mu_i\rho_j}+\delta_{\rho_j\nu_i}))$$
where $\sum_\rho$ is the sum over all intermediary states and the pairs $\langle i,j\rangle$ are restricted to the bonds between the lines described by $\mu$ and $\nu$. 
The positive constant $\gamma_X$ is the ratio of the interaction constant $X\in\{J,K\}$ and the temperature in dimensionless unit. The spectrum of $\tilde\Lambda$ is known for $Q=2$ at critical temperature (see for example \cite{OPW}) and the partition function is $Z=Z_{N\times M}=\text{\rm tr\,} \tilde\Lambda^M$. 

The partition function $Z_{N\times M}$ can also be expressed in terms of a sum over Fortuin-Kasteleyn graphs. The exponential $e^{-E_\sigma/k_BT}$ that occurs in the Boltzmann weight of $\sigma$ can be written as $$\prod_{\langle i,j\rangle(J)}(1+v_J\delta_{\sigma_i\sigma_j})\prod_{\langle i,j\rangle(K)}(1+v_K\delta_{\sigma_i\sigma_j})$$ where $v_X=e^{\gamma_X}-1$. Therefore
$$
Z_{N\times M}=\sum_{\sigma}\prod_{\langle i,j\rangle(J)}(1+v_J\delta_{\sigma_i\sigma_j})\prod_{\langle i,j\rangle(K)}(1+v_K\delta_{\sigma_i\sigma_j})\\
= \sum_{\mathcal G}\sum_{\sigma|\mathcal G}v_J^{N_{bJ}}v_K^{N_{bK}}.
$$
In the last line, the sum $\sum_{\mathcal G}$ stands for the sum over possible graphs. The set of vertices is common to all these graphs and coincides with the set of spins. The sum $\sum_{\mathcal G}$ is therefore over the possible sets of bonds and $N_{bX}$ is the number of X-bonds in $\mathcal G$. The inner sum $\sum_{\sigma|\mathcal G}$ is over all possible configurations $\sigma$ compatible with the graph $\mathcal G$. To have the equality with the previous expression, we define a configuration $\sigma$ to be compatible with $\mathcal G$ if the states of two spins coincide when there is a bond in $\mathcal G$ between them. The sum $\sum_{\sigma|\mathcal G}$ amounts then to a factor of $Q$ for each connected component of $\mathcal G$. Then 
$$Z_{N\times M}=\sum_{\mathcal G} v_J^{N_{bJ}}v_K^{N_{bK}}Q^{N_c}$$
where $N_c$ is the number of connected components of $\mathcal G$. 
(The later expression has been known for many years \cite{Essam}.)

We represent a graph $\mathcal G$ on the lattice by the diagram constructed as follows. Two corners of each box of Figure \ref{fig:spin} (a) are occupied by spins, that is, vertices of $\mathcal G$. Each box is replaced by one where two quarter-circles are drawn with centers on opposite corners. If a bond exists in $\mathcal G$ between the two vertices of $\mathcal G$, the centers are chosen so that the two quarter-circles do not touch the bond. If there is no bond, the quarter-circles cross the diagonal where the bond would have been. The two states for quarter-circles are shown below. In the first the full diagonal indicates the existence of a bond in $\mathcal G$, in the second the dotted diagonal its absence.
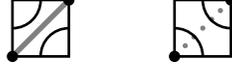
\begin{equation*}
\psset{unit=0.75}
\psset{linewidth=1pt}
\begin{pspicture}(0,0)(1,1)
\psline{-}(0,0)(0,1)(1,1)(1,0)(0,0)
\psline[linewidth=2pt,linecolor=gray]{-}(0,0)(1,1)
\psarc{-}(1,0){0.5}{90}{180}
\psarc{-}(0,1){0.5}{270}{360}
\pscircle*(0,0){0.1}\pscircle*(1,1){0.1}
\end{pspicture}
\qquad\qquad
\begin{pspicture}(0,0)(1,1)
\psline{-}(0,0)(0,1)(1,1)(1,0)(0,0)
\psline[linestyle=dotted, dotsep=3pt,linewidth=2pt,linecolor=gray]{-}(0,0)(1,1)
\psarc{-}(0,0){0.5}{0}{90}
\psarc{-}(1,1){0.5}{180}{270}
\pscircle*(0,0){0.1}\pscircle*(1,1){0.1}
\end{pspicture}
\end{equation*}
In Figure \ref{fig:spin} (b) a graph $\mathcal G$ is shown with its bonds drawn in gray. In (c) the same graph is represented with delimiting quarter-circles, instead of gray bonds. It is clear that representations with bonds (b) and with loops (c) are in one-to-one correspondence. If half-circles are added at the boundary of the grid as shown in (c), all loops formed by the quarter-circles are closed and it is possible to express the partition function in terms of data of the loop configurations only. (For the example of the Figure, $N_{bJ}=6,N_{bK}=8,N_b= N_{bJ} + N_{bK}=14$, $N_c=3$ and $\#(\mathcal G)=5$.) This can be done using the following lemma, a variation of Euler's formula for triangulations.
\begin{Lemme}\label{lem:euler}The relation $N_c=\frac12(\#(\mathcal G)+N_s-N_b)$ holds for any graph $\mathcal G$ on the strip with $N\times (2M)$ boxes. Here the number of spins $N_s=(N+1)\times M$ is independent of $\mathcal G$, but the number of bonds $N_b$, the number of closed loops $\#(\mathcal G)$ and the number of connected components $N_c$ depend on $\mathcal G$.
\end{Lemme}
\noindent{\scshape Proof\ \ }The formula holds when $\mathcal G$ is the graph with all possible bonds. Then there is a single connected component ($N_c=1$), the number of closed loops is $2+(N-1)M$ and the number of bonds $2NM$. Then $\frac12(\#(\mathcal G)+N_s-N_b)=1=N_c$ as claimed. It is then sufficient to prove that, if the relationship holds for a graph $\mathcal G$, it also holds for any $\mathcal G'$ obtained from $\mathcal G$ by removal of a single bond. Three cases have to be studied: {\em (i)} the bond being removed is in itself a whole connected subgraph, {\em (ii)} the subgraph containing the bond remains connected after its removal and {\em (iii)} the subgraph is broken into two connected components by its removal. 

In the case {\em (i)}, the removal of the bond creates two connected components out of the single one and two closed loops from the one circumscribing the bond:
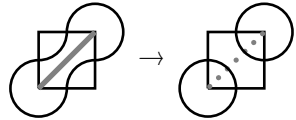
\begin{equation*}
\psset{unit=0.75}
\psset{linewidth=1pt}
\begin{pspicture}(0,-0.5)(4,1.5)
\psline{-}(0,0)(0,1)(1,1)(1,0)(0,0)
\psline[linewidth=2pt,linecolor=gray]{-}(0,0)(1,1)
\psarc{-}(1,0){0.5}{90}{180}
\psarc{-}(0,1){0.5}{270}{360}
\psarc{-}(1,1){0.5}{-90}{180}
\psarc{-}(0,0){0.5}{90}{360}
\rput(2,0.5){$\rightarrow$}
\psline{-}(3,0)(3,1)(4,1)(4,0)(3,0)
\psline[linestyle=dotted, dotsep=3pt,linewidth=2pt,linecolor=gray]{-}(3,0)(4,1)
\psarc{-}(4,1){0.5}{0}{360}
\psarc{-}(3,0){0.5}{0}{360}
\end{pspicture}
\end{equation*}
Then $N_c(\mathcal G')=N_c(\mathcal G)+1$, $\#(\mathcal G')=\#(\mathcal G)+1$ and $N_b(\mathcal G')=N_b(\mathcal G)-1$ and
\begin{equation*}N_c(\mathcal G')
=N_c(\mathcal G)+1=\frac12(\#(\mathcal G)+N_s-N_b(\mathcal G))+1
=\frac12(\#(\mathcal G')+N_s-N_b(\mathcal G'))
\end{equation*}
as before.

In case {\em (ii)}, the subgraph of $\mathcal G$ remains connected after the removal of the bond. This means that the two vertices tied by the bond are also tied to the component by other bonds. In other words, it is possible to go from one vertex to the other visiting bonds of the subgraph other than the one being removed. The subgraph thus contains a cycle.
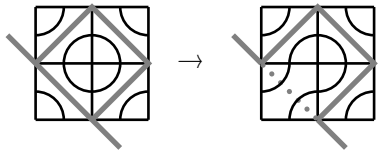
\begin{equation*}
\psset{unit=0.75}
\psset{linewidth=1pt}
\begin{pspicture}(-0.5,-0.5)(6,2.5)
\psline{-}(0,0)(0,2)
\psline{-}(1,0)(1,2)
\psline{-}(2,0)(2,2)
\psline{-}(0,0)(2,0)
\psline{-}(0,1)(2,1)
\psline{-}(0,2)(2,2)
\psline[linewidth=2pt,linecolor=gray]{-}(-0.5,1.5)(1.5,-0.5)
\psline[linewidth=2pt,linecolor=gray]{-}(0,1)(1,2)(2,1)(1,0)
\psarc{-}(1,1){0.5}{0}{360}
\psarc{-}(2,0){0.5}{90}{180}
\psarc{-}(0,2){0.5}{270}{360}
\psarc{-}(2,2){0.5}{180}{270}
\psarc{-}(0,0){0.5}{0}{90}
\rput(2.75,1){$\rightarrow$}
\psline{-}(4,0)(4,2)
\psline{-}(5,0)(5,2)
\psline{-}(6,0)(6,2)
\psline{-}(4,0)(6,0)
\psline{-}(4,1)(6,1)
\psline{-}(4,2)(6,2)
\psline[linewidth=2pt,linecolor=gray]{-}(3.5,1.5)(4,1)
\psline[linewidth=2pt,linecolor=gray]{-}(5,0)(5.5,-0.5)
\psline[linestyle=dotted, dotsep=3pt,linewidth=2pt,linecolor=gray]{-}(5,0)(4,1)
\psline[linewidth=2pt,linecolor=gray]{-}(4,1)(5,2)(6,1)(5,0)
\psarc{-}(5,1){0.5}{-90}{180}
\psarc{-}(6,0){0.5}{90}{180}
\psarc{-}(4,2){0.5}{270}{360}
\psarc{-}(6,2){0.5}{180}{270}
\psarc{-}(4,1){0.5}{270}{360}
\psarc{-}(5,0){0.5}{90}{180}
\end{pspicture}
\end{equation*}
The removal of the bond will therefore decrease the number of loops: $\#(\mathcal G')=\#(\mathcal G)-1$. Since $N_b(\mathcal G')=N_b(\mathcal G)-1$ and $N_c(\mathcal G')=N_c(\mathcal G)$, the identity is again easily verified. (Note that {\em (ii)} includes the case of a subgraph that winds non-trivially around the strip.)

In case {\em (iii)}, the two quarter-circles on each side of the bond must belong to the same closed loop if its removal breaks the subgraph in two.
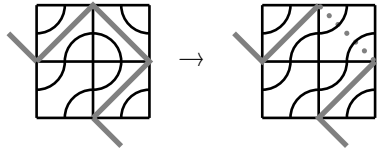
\begin{equation*}
\psset{unit=0.75}
\psset{linewidth=1pt}
\begin{pspicture}(-0.5,-0.5)(6,2.5)
\psline{-}(0,0)(0,2)
\psline{-}(1,0)(1,2)
\psline{-}(2,0)(2,2)
\psline{-}(0,0)(2,0)
\psline{-}(0,1)(2,1)
\psline{-}(0,2)(2,2)
\psline[linewidth=2pt,linecolor=gray]{-}(-0.5,1.5)(0,1)
\psline[linewidth=2pt,linecolor=gray]{-}(1,0)(1.5,-0.5)
\psline[linewidth=2pt,linecolor=gray]{-}(0,1)(1,2)(2,1)(1,0)
\psarc{-}(1,1){0.5}{-90}{180}
\psarc{-}(2,0){0.5}{90}{180}
\psarc{-}(0,2){0.5}{270}{360}
\psarc{-}(2,2){0.5}{180}{270}
\psarc{-}(0,1){0.5}{270}{360}
\psarc{-}(1,0){0.5}{90}{180}
\rput(2.75,1){$\rightarrow$}
\psline{-}(4,0)(4,2)
\psline{-}(5,0)(5,2)
\psline{-}(6,0)(6,2)
\psline{-}(4,0)(6,0)
\psline{-}(4,1)(6,1)
\psline{-}(4,2)(6,2)
\psline[linewidth=2pt,linecolor=gray]{-}(3.5,1.5)(4,1)(5,2)
\psline[linewidth=2pt,linecolor=gray]{-}(6,1)(5,0)(5.5,-0.5)
\psline[linestyle=dotted,dotsep=3pt,linewidth=2pt,linecolor=gray]{-}(6,1)(5,2)
\psarc{-}(5,1){0.5}{-90}{0}
\psarc{-}(5,1){0.5}{90}{180}
\psarc{-}(6,0){0.5}{90}{180}
\psarc{-}(4,2){0.5}{270}{360}
\psarc{-}(5,2){0.5}{270}{360}
\psarc{-}(6,1){0.5}{90}{180}
\psarc{-}(4,1){0.5}{270}{360}
\psarc{-}(5,0){0.5}{90}{180}
\end{pspicture}
\end{equation*}
The data associated to $\mathcal G'$ are $N_c(\mathcal G')=N_c(\mathcal G)+1$, $\#(\mathcal G')=\#(\mathcal G) + 1$ and $N_b(\mathcal G')=N_b(\mathcal G)-1$. Again an easy check shows that the identity is preserved. Note that the case {\em (i)} is in fact a particular case of {\em (iii)}.\hfill $\square$

\smallskip

The critical temperature of $Q$-Potts models is known to be such that $v_Jv_K=Q$ 
\cite{Baxter}. We will use the parametrization $\sqrt{Q} = 2 \cos \lambda$, $v_J/\sqrt{Q} =\sin(\lambda-u)/\sin(u)$, 
where $\lambda\in(0,\frac{\pi}2)$ and $u\in[0,\lambda]$ is the anisotropy parameter. The previous lemma allows to write the partition function at criticality as
\begin{align*}Z_{N\times M}
&=\sum_{\mathcal G} v_J^{N_{bJ}}v_K^{N_{bK}} Q^{N_c}=\sum_{\mathcal G}Q^{(\#(\mathcal G)+N_s)/2}\left(\frac{\sin(\lambda-u)}{\sin(u)}\right)^{N_{bJ}}\left(\frac{\sin(u)}{\sin(\lambda-u)}\right)^{N_{bK}} \\
&=\frac{Q^{N_s/2}}{(\sin(u)\sin(\lambda-u))^{NM}}\sum_{\mathcal G}Q^{\#(\mathcal G)/2}\sin^{N_{bJ}}(\lambda-u) \sin^{NM-N_{bJ}}(u) \sin^{N_{bK}}(u) \sin^{NM-N_{bK}}(\lambda-u)
\end{align*}
which is the form \eqref{eq:partitionFunction} up to the overall factor $\kappa^M$ where $\kappa=\left(Q^{(N+1)/2}/(\sin(u)\sin(\lambda-u))^{N}\right)$. Moreover, the weight given to every graph in $Z_{N\times M}$ is precisely the one in the definition of $D_N$. If we define $\Lambda=\tilde\Lambda/\kappa$, the Proposition \ref{thm:deuxTraces} leads to a relationship between the traces of $\Lambda$ and of $D_N$:
\begin{equation}\textrm{\fbox{$\text{\rm tr\,}\Lambda^M=\text{\rm tr\,}(\rho(D_N^M)W),\qquad \text{\rm for all\ }M\in\mathbb N.$}}\label{eq:spinLoop}\end{equation}

The matrix $\rho(D_N^M)$ is block-triangular, that is, the blocks acting from $V_N^d$ to $V_N^{d'}$ are zero if $d<d'$. The set of eigenvalues of $\rho(D_N^M)$ are then the union of those of the diagonal blocks of $\rho(D_N^M)$. Let $\delta$ be an eigenvalue of the diagonal block of $\rho(D_N)$ acting from $V_N^d$ to $V_N^d$. On $V_N^d$, the matrix $W$ acts as a multiple of the identity. Let $w$ be this factor. Then, because $W$ is diagonal, $\delta^M w$ is an eigenvalue of $\rho(D_N^M)W$. Let the eigenvalues of $\Lambda$ be in decreasing order of absolute values: $|\lambda_1|\ge |\lambda_2|\ge\dots$ Similarly, let the eigenvalues of $\rho(D_N)$ be ordered as $|\delta_1|\ge|\delta_2|\ge\dots$ The identity \eqref{eq:spinLoop} can be written as
\begin{equation}\sum_i\lambda_i^M=\sum_j\delta_j^M w_j, \qquad M\in\mathbb N.\label{eq:spinLoop2}\end{equation}
The factors $w_j$ are determined, for each eigenvalue $\delta_j$, as described above. The first sum contains $Q^{N/2}$ eigenvalues, the second $\left(\begin{smallmatrix}N\\ N/2\end{smallmatrix}\right)$. Note that, if an eigenvalue $\delta_j$ is degenerate, the various copies might have different weights $w_j$. Write $w(\delta)$ for the sum of the weights $w_j$ over all eigenvalues equal to $\delta$. 

Let $\Delta$ be $\max\{|\lambda_1|,|\delta_1|\}$. Suppose for the time being that no other eigenvalues of $\Lambda$ share the absolute value of $\lambda_1$; suppose, similarly, that the same statement holds for the spectrum of $\rho(D_N)$ and the absolute value of $\delta_1$. By dividing \eqref{eq:spinLoop2} by $\Delta^M$ and taking the limit $M\rightarrow \infty$, one can see that $|\lambda_1|>|\delta_1|$ is impossible. One can have $|\delta_1|>|\lambda_1|$ however, if $w(\delta_1)$ is zero. 
Finally one can have $|\delta_1|=|\lambda_1|$ and then the multiplicity $\text{\rm mult\,}\lambda_1$ of the eigenvalue of the spin matrix must be equal to $w(\delta_1)$. One can then subtract from both members of \eqref{eq:spinLoop2} the contribution $\delta_1w(\delta_1)$ of the maximum eigenvalue. The previous argument can then be repeated with the new sums. 

What happens when some of the eigenvalues in either spectrum have the same absolute values but are distinct? Suppose that all eigenvalues with absolute values larger than a given $r>0$ have been subtracted from the sum in \eqref{eq:spinLoop2}. And suppose that several $\lambda_i$ and $\delta_j$ have the same absolute value $r$ and let $S=\{\alpha_1, \alpha_2, \dots , \alpha_n\}$ 
be the set of distinct phases. The equality \eqref{eq:spinLoop2} has then the form
\begin{equation}\sum_{\alpha\in S}c_\alpha e^{i\alpha M}=v_M/r^M, \qquad M\in\mathbb N\label{eq:phases}\end{equation}
where $v_M$ is the sum over all eigenvalues with absolute value less than $r$ in either spectrum. For large $M$, the right-hand side is therefore small. The coefficient $c_\alpha$ in the left-hand side is 
\begin{equation}c_\alpha=\begin{cases}
\text{\rm mult }(re^{i\alpha}),& \text{\rm if $re^{i\alpha}$ is an eigenvalue of only $\Lambda$,}\\
-w(re^{i\alpha}),&\text{\rm if $re^{i\alpha}$ is an eigenvalue of only $\rho(D_M)$,}\\
\text{\rm mult }(re^{i\alpha})-w(re^{i\alpha}),&\text{\rm if $re^{i\alpha}$ is an eigenvalue of both.}
\end{cases}\label{eq:cCases}
\end{equation}
Let $n$ be the number of (distinct) elements of $S$ and consider $n$ consecutive equations \eqref{eq:phases} with index $M, M+1, \dots, M+n-1$. The system of equations has the form $A_Mc=w_M$ where $c$ is the vector containing the $c_\alpha$ and $w_M$ the vector of components $v_M/r^M, v_{M+1}/r^{M+1}, \dots$ The matrix $A_M$ is
$$\begin{pmatrix}e^{i\alpha_1 M} & e^{i\alpha_2 M} & \dots & e^{i\alpha_n M}\\
                 e^{i\alpha_1 (M+1)} & e^{i\alpha_2 (M+1)} & \dots & e^{i\alpha_n (M+1)}\\
                 \vdots & \vdots & \ddots & \vdots \\
                 e^{i\alpha_1 (M+n-1)} & e^{i\alpha_2 (M+n-1)} & \dots & e^{i\alpha_n (M+n-1)}\\
\end{pmatrix}$$
and, after extracting a global phase $\prod_{\alpha\in S}e^{i\alpha M}$, is a Vandermonde matrix. The determinant of $A$ is therefore the product of a phase that depends on $M$ with the Vandermonde determinant that is independent of it. So the vector $c$ is given by $A_M^{-1}w_M$ and this should hold independently of $M$. When $M$ is taken to infinity, $A_M^{-1}v_M$ goes to zero and the vector $c$ is therefore zero. The fact that $c_\alpha$ is zero for all elements of $S$ has different consequences depending on the case in \eqref{eq:cCases}. The first case cannot happen: if $re^{i\alpha}$ is in the spectrum of $\Lambda$ only, then $\text{\rm mult }re^{i\alpha}>0$. Consequently, all eigenvalues of $\Lambda$ must be eigenvalues of $\rho(D_M)$. The second case may happen. It then forces the corresponding $\delta=re^{i\alpha}$ to have a $w(\delta)=0$. And, in the third case, some eigenvalues $\lambda$ and $\delta$ are equal and $\text{\rm mult }\lambda=w(\delta)$. We have then proved the following result.

\begin{Proposition}\label{propo:pottsLoops}Define the $Q$-Potts model ($Q=\beta^2\in\{1,2,3\}$) on a lattice $N\times (2M)$ as in Figure \ref{fig:spin} with free boundary conditions on the left and right sides, and periodic boundary conditions in the vertical direction. 
\begin{itemize}
\item[{\em (i)}] The partition function of the model is given by $Z_{N \times M} = \kappa^M \text{\rm tr\,}(\rho(D_N^M)W),\text{\rm\ for all\ }M\in\mathbb N$. $W$ has been defined in Proposition \ref{thm:deuxTraces} 
\item[{\em (ii)}] The eigenvalues $\lambda$ of the spin transfer matrix $\Lambda$ of the $Q$-Potts model belong to the spectrum of the link representation $\rho(D_N)$ of the double-row transfer matrix and $\text{\rm mult\,} \lambda$ in $\Lambda$ is equal to the weight $w(\lambda)$, that is the sum of factors $w_j$ of $\lambda$ in $\rho(D_N)$.
\item[{\em (iii)}] If an eigenvalue $\delta$ of $\rho(D_N)$ does not belong to the spectrum of $\Lambda_N$, the sum $w(\delta)$ of its weights is zero.
\end{itemize}\end{Proposition}

\subsection{$Q$-Potts models with fixed, free and mixed boundary conditions}
\label{sec:otherboundary}

The trace of the spin transfer matrix  $\textrm{tr}\ \tilde\Lambda$ is the partition function for the lattice with periodic boundary condition in the vertical direction and free on the left and right sides. Of course there is more information in $\tilde\Lambda$ than just this particular partition function. Other partition functions, for spins at the top and bottom of the lattice (see Figure \ref{fig:spin2} (a)) in fixed, free or mixed boundary conditions while those on the vertical edges remain free,
are also of interest. 
In the limit $N\rightarrow\infty$, they were computed in \cite{Dubailetal}, using Coulomb gas arguments. (The lattice used there is slightly different than ours.) This section shows how these partition functions can be extracted from $\rho(D_N(\lambda,u))$ for any $N$.

\begin{Definition} The Gram product $\langle \cdot | \cdot \rangle_G$ is a bilinear form on $V_N$. It is defined by its value on pairs of vectors in $B_N$. For $u_1$, $u_2 \in B_N$, it is computed by reflecting $u_2$ in a horizontal mirror and then connecting the end points of $u_2$ to those of $u_1$. Then $\langle u_1|u_2 \rangle_G$ is non-zero only if every defect of $u_1$ ends up into one of $u_2$ and vice versa; when this occurs, it is equal to $\beta^n$ where $n$ is the number of loops closed by connecting $u_1$ and $u_2$. 
Finally, we denote by $G$ the matrix representing the bilinear form in the link basis:  $G_{vw}=\langle v|w \rangle_G$.
\label{sec:Gram}\end{Definition}
For example, $\langle 
%
\begin{pspicture}(0.05,0)(2.0,0.5)
\psset{unit=0.3}
\psset{dotsize=2.6pt 0}
\psdots(0.6,0)(1.2,0)(1.8,0)(2.4,0)(3,0)(3.6,0)(4.2,0)(4.8,0)(5.4,0)(6,0)
\psline{-}(0.6,0)(0.6,1)
\psline{-}(2.4,0)(2.4,1)
\psarc{-}(1.5,0){0.3}{0}{180}
\psarc{-}(3.9,0){0.3}{0}{180}
\psarc{-}(5.1,0){0.3}{0}{180}
\psbezier{-}(3,0)(3,1)(6,1)(6,0)
\end{pspicture}
|
%
\begin{pspicture}(0.05,0)(2.0,0.5)
\psset{unit=0.3}
\psset{dotsize=2.6pt 0}
\psdots(0.6,0)(1.2,0)(1.8,0)(2.4,0)(3,0)(3.6,0)(4.2,0)(4.8,0)(5.4,0)(6,0)
\psline{-}(0.6,0)(0.6,1)
\psline{-}(1.2,0)(1.2,1)
\psline{-}(3.0,0)(3.0,1)
\psline{-}(6.0,0)(6.0,1)
\psarc{-}(2.1,0){0.3}{0}{180}
\psarc{-}(4.5,0){0.3}{0}{180}
\psbezier{-}(3.6,0)(3.6,0.8)(5.4,0.8)(5.4,0)
\end{pspicture}
\rangle_G
= 0$ as vectors with different numbers of defects have vanishing Gram product, but the following product is not:
\begin{equation*}\big\langle 
%
\psset{unit=0.5}
\begin{pspicture}(0.05,0)(6.5,1)
\psset{dotsize=2.6pt 0}
\psdots(0.6,0)(1.2,0)(1.8,0)(2.4,0)(3,0)(3.6,0)(4.2,0)(4.8,0)(5.4,0)(6,0)
\psline{-}(0.6,0)(0.6,1)
\psline{-}(2.4,0)(2.4,1)
\psarc{-}(1.5,0){0.3}{0}{180}
\psarc{-}(3.9,0){0.3}{0}{180}
\psarc{-}(5.1,0){0.3}{0}{180}
\psbezier{-}(3,0)(3,1)(6,1)(6,0)
\end{pspicture}
\big|
%
\begin{pspicture}(0.05,0)(6.5,1)
\psset{dotsize=2.6pt 0}
\psdots(0.6,0)(1.2,0)(1.8,0)(2.4,0)(3,0)(3.6,0)(4.2,0)(4.8,0)(5.4,0)(6,0)
\psline{-}(0.6,0)(0.6,1)
\psline{-}(3.6,0)(3.6,1)
\psarc{-}(1.5,0){0.3}{0}{180}
\psarc{-}(2.7,0){0.3}{0}{180}
\psarc{-}(5.1,0){0.3}{0}{180}
\psbezier{-}(4.2,0)(4.2,0.8)(6,0.8)(6,0)
\end{pspicture}
\big\rangle_G
\rightarrow 
\begin{pspicture}(0.05,0)(6.5,1)
\psset{dotsize=2.6pt 0}
\psdots(0.6,0)(1.2,0)(1.8,0)(2.4,0)(3,0)(3.6,0)(4.2,0)(4.8,0)(5.4,0)(6,0)
\psline{-}(0.6,0)(0.6,1)
\psline{-}(2.4,0)(2.4,1)
\psarc{-}(1.5,0){0.3}{0}{180}
\psarc{-}(3.9,0){0.3}{0}{180}
\psarc{-}(5.1,0){0.3}{0}{180}
\psbezier{-}(3,0)(3,1)(6,1)(6,0)
\psline{-}(0.6,0)(0.6,-1)
\psarc{-}(1.5,0){0.3}{180}{360}
\psarc{-}(2.7,0){0.3}{180}{360}
\psarc{-}(5.1,0){0.3}{180}{360}
\psline{-}(3.6,0)(3.6,-1)
\psbezier{-}(4.2,0)(4.2,-0.8)(6,-0.8)(6,0)
\end{pspicture}
\end{equation*} 
Because two loops are closed, 
$\langle 
%
\begin{pspicture}(0.05,0)(2.0,0.5)
\psset{unit=0.3}
\psset{dotsize=2.6pt 0}
\psdots(0.6,0)(1.2,0)(1.8,0)(2.4,0)(3,0)(3.6,0)(4.2,0)(4.8,0)(5.4,0)(6,0)
\psline{-}(0.6,0)(0.6,1)
\psline{-}(2.4,0)(2.4,1)
\psarc{-}(1.5,0){0.3}{0}{180}
\psarc{-}(3.9,0){0.3}{0}{180}
\psarc{-}(5.1,0){0.3}{0}{180}
\psbezier{-}(3,0)(3,1)(6,1)(6,0)
\end{pspicture}
|
%
\begin{pspicture}(0.05,0)(2.0,0.5)
\psset{unit=0.3}
\psset{dotsize=2.6pt 0}
\psdots(0.6,0)(1.2,0)(1.8,0)(2.4,0)(3,0)(3.6,0)(4.2,0)(4.8,0)(5.4,0)(6,0)
\psline{-}(0.6,0)(0.6,1)
\psline{-}(3.6,0)(3.6,1)
\psarc{-}(1.5,0){0.3}{0}{180}
\psarc{-}(2.7,0){0.3}{0}{180}
\psarc{-}(5.1,0){0.3}{0}{180}
\psbezier{-}(4.2,0)(4.2,0.8)(6,0.8)(6,0)
\end{pspicture}
\rangle_G
= \beta^2$. For more, see \cite{Westbury}, where this product is actually defined on quotient spaces with fixed number of defects.
\begin{figure}[h!]
\begin{center}
\psset{unit=0.6}
\begin{pspicture}(0,-2.5)(8,8)
\psset{linewidth=1pt}
\psline[linecolor=gray,linestyle=dashed,linewidth=2pt]{-}(1,0)(0,1)(1,2)(0,3)(1,4)(0,5)(1,6)
\psline[linecolor=gray,linestyle=dashed,linewidth=2pt]{-}(3,0)(2,1)(3,2)(2,3)(3,4)(2,5)(3,6)
\psline[linecolor=gray,linestyle=dashed,linewidth=2pt]{-}(5,0)(4,1)(5,2)(4,3)(5,4)(4,5)(5,6)
\psline[linecolor=gray,linestyle=dashed,linewidth=2pt]{-}(7,0)(6,1)(7,2)(6,3)(7,4)(6,5)(7,6)
\psline[linecolor=gray,linestyle=dotted,linewidth=2pt]{-}(1,0)(2,1)(1,2)(2,3)(1,4)(2,5)(1,6)
\psline[linecolor=gray,linestyle=dotted,linewidth=2pt]{-}(3,0)(4,1)(3,2)(4,3)(3,4)(4,5)(3,6)
\psline[linecolor=gray,linestyle=dotted,linewidth=2pt]{-}(5,0)(6,1)(5,2)(6,3)(5,4)(6,5)(5,6)
\psline[linecolor=gray,linestyle=dotted,linewidth=2pt]{-}(7,0)(8,1)(7,2)(8,3)(7,4)(8,5)(7,6)
\psline{-}(0,6)(8,6)
\psline{-}(0,5)(8,5)
\psline{-}(0,4)(8,4)
\psline{-}(0,3)(8,3)
\psline{-}(0,2)(8,2)
\psline{-}(0,1)(8,1)
\psline{-}(0,0)(8,0)
\psline{-}(8,0)(8,6)
\psline{-}(7,0)(7,6)
\psline{-}(6,0)(6,6)
\psline{-}(5,0)(5,6)
\psline{-}(4,0)(4,6)
\psline{-}(3,0)(3,6)
\psline{-}(2,0)(2,6)
\psline{-}(1,0)(1,6)
\psline{-}(0,0)(0,6)
\pscircle*[linecolor=lightgray](1,0){0.3}\psarc(1,0){0.3}{0}{360}
\pscircle*[linecolor=lightgray](3,0){0.3}\psarc(3,0){0.3}{0}{360}
\pscircle*[linecolor=lightgray](5,0){0.3}\psarc(5,0){0.3}{0}{360}
\pscircle*[linecolor=lightgray](7,0){0.3}\psarc(7,0){0.3}{0}{360}
\pscircle*[linecolor=white](1,6){0.3}\psarc(1,6){0.3}{0}{360}
\pscircle*[linecolor=white](3,6){0.3}\psarc(3,6){0.3}{0}{360}
\pscircle*[linecolor=white](5,6){0.3}\psarc(5,6){0.3}{0}{360}
\pscircle*[linecolor=white](7,6){0.3}\psarc(7,6){0.3}{0}{360}
\pscircle*(1,0){0.1}\pscircle*(3,0){0.1}\pscircle*(5,0){0.1}\pscircle*(7,0){0.1}
\pscircle*(0,1){0.1}\pscircle*(2,1){0.1}\pscircle*(4,1){0.1}\pscircle*(6,1){0.1}\pscircle*(8,1){0.1}
\pscircle*(1,2){0.1}\pscircle*(3,2){0.1}\pscircle*(5,2){0.1}\pscircle*(7,2){0.1}
\pscircle*(0,3){0.1}\pscircle*(2,3){0.1}\pscircle*(4,3){0.1}\pscircle*(6,3){0.1}\pscircle*(8,3){0.1}
\pscircle*(1,4){0.1}\pscircle*(3,4){0.1}\pscircle*(5,4){0.1}\pscircle*(7,4){0.1}
\pscircle*(0,5){0.1}\pscircle*(2,5){0.1}\pscircle*(4,5){0.1}\pscircle*(6,5){0.1}\pscircle*(8,5){0.1}
\pscircle*(1,6){0.1}\pscircle*(3,6){0.1}\pscircle*(5,6){0.1}\pscircle*(7,6){0.1}
\rput(4,-2.5){(a)}
\end{pspicture}\qquad
\begin{pspicture}(0,-2.5)(8,8)
\psset{linewidth=1pt}
\psline[linecolor=gray,linestyle=dashed,linewidth=2pt]{-}(1,0)(0,1)(1,2)(0,3)(1,4)(0,5)(1,6)
\psline[linecolor=gray,linestyle=dashed,linewidth=2pt]{-}(3,0)(2,1)(3,2)(2,3)(3,4)(2,5)(3,6)
\psline[linecolor=gray,linestyle=dashed,linewidth=2pt]{-}(5,0)(4,1)(5,2)(4,3)(5,4)(4,5)(5,6)
\psline[linecolor=gray,linestyle=dashed,linewidth=2pt]{-}(7,0)(6,1)(7,2)(6,3)(7,4)(6,5)(7,6)
\psline[linecolor=gray,linestyle=dotted,linewidth=2pt]{-}(1,0)(2,1)(1,2)(2,3)(1,4)(2,5)(1,6)
\psline[linecolor=gray,linestyle=dotted,linewidth=2pt]{-}(3,0)(4,1)(3,2)(4,3)(3,4)(4,5)(3,6)
\psline[linecolor=gray,linestyle=dotted,linewidth=2pt]{-}(5,0)(6,1)(5,2)(6,3)(5,4)(6,5)(5,6)
\psline[linecolor=gray,linestyle=dotted,linewidth=2pt]{-}(7,0)(8,1)(7,2)(8,3)(7,4)(8,5)(7,6)
\psline{-}(0,6)(8,6)
\psline{-}(0,5)(8,5)
\psline{-}(0,4)(8,4)
\psline{-}(0,3)(8,3)
\psline{-}(0,2)(8,2)
\psline{-}(0,1)(8,1)
\psline{-}(0,0)(8,0)
\psline{-}(8,0)(8,6)
\psline{-}(7,0)(7,6)
\psline{-}(6,0)(6,6)
\psline{-}(5,0)(5,6)
\psline{-}(4,0)(4,6)
\psline{-}(3,0)(3,6)
\psline{-}(2,0)(2,6)
\psline{-}(1,0)(1,6)
\psline{-}(0,0)(0,6)
\psbezier[linestyle=dashed,dash=.30 .11 .08 .11,linewidth=2pt,linecolor=gray](1,0)(1,-1)(3,-1.5)(4,-1.5)
\psbezier[linestyle=dashed,dash=.30 .11 .08 .11,linewidth=2pt,linecolor=gray](7,0)(7,-1)(5,-1.5)(4,-1.5)
\psbezier[linestyle=dashed,dash=.30 .11 .08 .11,linewidth=2pt,linecolor=gray](3,0)(3,-1)(3.66,-1.5)(4,-1.5)
\psbezier[linestyle=dashed,dash=.30 .11 .08 .11,linewidth=2pt,linecolor=gray](5,0)(5,-1)(4.33,-1.5)(4,-1.5)
\psbezier[linestyle=dashed,dash=.30 .11 .08 .11,linewidth=2pt,linecolor=gray](1,6)(1,7)(3,7.5)(4,7.5)
\psbezier[linestyle=dashed,dash=.30 .11 .08 .11,linewidth=2pt,linecolor=gray](7,6)(7,7)(5,7.5)(4,7.5)
\psbezier[linestyle=dashed,dash=.30 .11 .08 .11,linewidth=2pt,linecolor=gray](3,6)(3,7)(3.66,7.5)(4,7.5)
\psbezier[linestyle=dashed,dash=.30 .11 .08 .11,linewidth=2pt,linecolor=gray](5,6)(5,7)(4.33,7.5)(4,7.5)
\pscircle*[linecolor=white](4,7.5){0.3}\psarc(4,7.5){0.3}{0}{360}\pscircle*(4,7.5){0.1}
\pscircle*[linecolor=lightgray](4,-1.5){0.3}\psarc(4,-1.5){0.3}{0}{360}\pscircle*(4,-1.5){0.1}
\pscircle*(1,0){0.1}\pscircle*(3,0){0.1}\pscircle*(5,0){0.1}\pscircle*(7,0){0.1}
\pscircle*(0,1){0.1}\pscircle*(2,1){0.1}\pscircle*(4,1){0.1}\pscircle*(6,1){0.1}\pscircle*(8,1){0.1}
\pscircle*(1,2){0.1}\pscircle*(3,2){0.1}\pscircle*(5,2){0.1}\pscircle*(7,2){0.1}
\pscircle*(0,3){0.1}\pscircle*(2,3){0.1}\pscircle*(4,3){0.1}\pscircle*(6,3){0.1}\pscircle*(8,3){0.1}
\pscircle*(1,4){0.1}\pscircle*(3,4){0.1}\pscircle*(5,4){0.1}\pscircle*(7,4){0.1}
\pscircle*(0,5){0.1}\pscircle*(2,5){0.1}\pscircle*(4,5){0.1}\pscircle*(6,5){0.1}\pscircle*(8,5){0.1}
\pscircle*(1,6){0.1}\pscircle*(3,6){0.1}\pscircle*(5,6){0.1}\pscircle*(7,6){0.1}
\rput(4,-2.5){(b)}
\end{pspicture} \qquad
\begin{pspicture}(0,-2.5)(8,8)
\psset{linewidth=1pt}
\psline{-}(0,6)(8,6)
\psline{-}(0,5)(8,5)
\psline{-}(0,4)(8,4)
\psline{-}(0,3)(8,3)
\psline{-}(0,2)(8,2)
\psline{-}(0,1)(8,1)
\psline{-}(0,0)(8,0)
\psline{-}(8,0)(8,6)
\psline{-}(7,0)(7,6)
\psline{-}(6,0)(6,6)
\psline{-}(5,0)(5,6)
\psline{-}(4,0)(4,6)
\psline{-}(3,0)(3,6)
\psline{-}(2,0)(2,6)
\psline{-}(1,0)(1,6)
\psline{-}(0,0)(0,6)
\psbezier[linewidth=2pt,linecolor=gray](1,0)(1,-1)(3,-1.5)(4,-1.5)
\psbezier[linewidth=2pt,linecolor=gray](7,0)(7,-1)(5,-1.5)(4,-1.5)
\psbezier[linewidth=2pt,linecolor=gray](3,0)(3,-1)(3.66,-1.5)(4,-1.5)
\psbezier[linewidth=2pt,linecolor=gray](5,0)(5,-1)(4.33,-1.5)(4,-1.5)
\psbezier[linewidth=2pt,linecolor=gray](1,6)(1,7)(3,7.5)(4,7.5)
\psbezier[linewidth=2pt,linecolor=gray](7,6)(7,7)(5,7.5)(4,7.5)
\psbezier[linewidth=2pt,linecolor=gray](3,6)(3,7)(3.66,7.5)(4,7.5)
\psbezier[linewidth=2pt,linecolor=gray](5,6)(5,7)(4.33,7.5)(4,7.5)
\psline[linecolor=gray,linewidth=2pt]{-}(0,1)(1,0)(2,1)(3,2)(2,3)(1,2)
\psline[linecolor=gray,linewidth=2pt]{-}(2,3)(3,4)(4,5)(5,6)
\psline[linecolor=gray,linewidth=2pt]{-}(0,5)(1,4)(2,5)(3,6)
\psline[linecolor=gray,linewidth=2pt]{-}(4,1)(5,0)(6,1)(7,0)
\psline[linecolor=gray,linewidth=2pt]{-}(8,1)(7,2)(6,3)
\psline[linecolor=gray,linewidth=2pt]{-}(8,3)(7,4)
\psline[linecolor=gray,linewidth=2pt]{-}(6,5)(5,4)
\psarc{-}(2,0){0.5}{180}{360}\psarc{-}(4,0){0.5}{180}{360}\psarc{-}(6,0){0.5}{180}{360}
\psarc{-}(2,6){0.5}{0}{180}\psarc{-}(4,6){0.5}{0}{180}\psarc{-}(6,6){0.5}{0}{180}
\psarc{-}(0,1){0.5}{90}{270}\psarc{-}(0,3){0.5}{90}{270}\psarc{-}(0,5){0.5}{90}{270}
\psarc{-}(8,1){0.5}{-90}{90}\psarc{-}(8,3){0.5}{-90}{90}\psarc{-}(8,5){0.5}{-90}{90}
\psbezier{-}(0.5,0)(0.5,-2.5)(7.5,-2.5)(7.5,0)
\psbezier{-}(0.5,6)(0.5,8.5)(7.5,8.5)(7.5,6)
\pscircle*(1,0){0.1}\pscircle*(3,0){0.1}\pscircle*(5,0){0.1}\pscircle*(7,0){0.1}
\pscircle*(0,1){0.1}\pscircle*(2,1){0.1}\pscircle*(4,1){0.1}\pscircle*(6,1){0.1}\pscircle*(8,1){0.1}
\pscircle*(1,2){0.1}\pscircle*(3,2){0.1}\pscircle*(5,2){0.1}\pscircle*(7,2){0.1}
\pscircle*(0,3){0.1}\pscircle*(2,3){0.1}\pscircle*(4,3){0.1}\pscircle*(6,3){0.1}\pscircle*(8,3){0.1}
\pscircle*(1,4){0.1}\pscircle*(3,4){0.1}\pscircle*(5,4){0.1}\pscircle*(7,4){0.1}
\pscircle*(0,5){0.1}\pscircle*(2,5){0.1}\pscircle*(4,5){0.1}\pscircle*(6,5){0.1}\pscircle*(8,5){0.1}
\pscircle*(1,6){0.1}\pscircle*(3,6){0.1}\pscircle*(5,6){0.1}\pscircle*(7,6){0.1}
\pscircle*(4,-1.5){0.1}\pscircle*(4,7.5){0.1}
\psarc{-}(0,0){0.5}{0}{90}\psarc{-}(1,1){0.5}{180}{270}
\psarc{-}(2,0){0.5}{90}{180}\psarc{-}(1,1){0.5}{270}{360}
\psarc{-}(3,0){0.5}{90}{180}\psarc{-}(2,1){0.5}{270}{360}
\psarc{-}(3,0){0.5}{0}{90}\psarc{-}(4,1){0.5}{180}{270}
\psarc{-}(4,0){0.5}{0}{90}\psarc{-}(5,1){0.5}{180}{270}
\psarc{-}(6,0){0.5}{90}{180}\psarc{-}(5,1){0.5}{270}{360}
\psarc{-}(6,0){0.5}{0}{90}\psarc{-}(7,1){0.5}{180}{270}
\psarc{-}(7,0){0.5}{0}{90}\psarc{-}(8,1){0.5}{180}{270}
\psarc{-}(0,1){0.5}{0}{90}\psarc{-}(1,2){0.5}{180}{270}
\psarc{-}(2,1){0.5}{90}{180}\psarc{-}(1,2){0.5}{270}{360}
\psarc{-}(3,1){0.5}{90}{180}\psarc{-}(2,2){0.5}{270}{360}
\psarc{-}(4,1){0.5}{90}{180}\psarc{-}(3,2){0.5}{270}{360}
\psarc{-}(4,1){0.5}{0}{90}\psarc{-}(5,2){0.5}{180}{270}
\psarc{-}(6,1){0.5}{90}{180}\psarc{-}(5,2){0.5}{270}{360}
\psarc{-}(6,1){0.5}{0}{90}\psarc{-}(7,2){0.5}{180}{270}
\psarc{-}(7,1){0.5}{0}{90}\psarc{-}(8,2){0.5}{180}{270}
\psarc{-}(1,2){0.5}{90}{180}\psarc{-}(0,3){0.5}{270}{360}
\psarc{-}(2,2){0.5}{90}{180}\psarc{-}(1,3){0.5}{270}{360}
\psarc{-}(2,2){0.5}{0}{90}\psarc{-}(3,3){0.5}{180}{270}
\psarc{-}(3,2){0.5}{0}{90}\psarc{-}(4,3){0.5}{180}{270}
\psarc{-}(5,2){0.5}{90}{180}\psarc{-}(4,3){0.5}{270}{360}
\psarc{-}(5,2){0.5}{0}{90}\psarc{-}(6,3){0.5}{180}{270}
\psarc{-}(6,2){0.5}{0}{90}\psarc{-}(7,3){0.5}{180}{270}
\psarc{-}(7,2){0.5}{0}{90}\psarc{-}(8,3){0.5}{180}{270}
\psarc{-}(0,3){0.5}{0}{90}\psarc{-}(1,4){0.5}{180}{270}
\psarc{-}(2,3){0.5}{90}{180}\psarc{-}(1,4){0.5}{270}{360}
\psarc{-}(3,3){0.5}{90}{180}\psarc{-}(2,4){0.5}{270}{360}
\psarc{-}(4,3){0.5}{90}{180}\psarc{-}(3,4){0.5}{270}{360}
\psarc{-}(4,3){0.5}{0}{90}\psarc{-}(5,4){0.5}{180}{270}
\psarc{-}(6,3){0.5}{90}{180}\psarc{-}(5,4){0.5}{270}{360}
\psarc{-}(6,3){0.5}{0}{90}\psarc{-}(7,4){0.5}{180}{270}
\psarc{-}(7,3){0.5}{0}{90}\psarc{-}(8,4){0.5}{180}{270}
\psarc{-}(0,4){0.5}{0}{90}\psarc{-}(1,5){0.5}{180}{270}
\psarc{-}(2,4){0.5}{90}{180}\psarc{-}(1,5){0.5}{270}{360}
\psarc{-}(3,4){0.5}{90}{180}\psarc{-}(2,5){0.5}{270}{360}
\psarc{-}(4,4){0.5}{90}{180}\psarc{-}(3,5){0.5}{270}{360}
\psarc{-}(5,4){0.5}{90}{180}\psarc{-}(4,5){0.5}{270}{360}
\psarc{-}(6,4){0.5}{90}{180}\psarc{-}(5,5){0.5}{270}{360}
\psarc{-}(7,4){0.5}{90}{180}\psarc{-}(6,5){0.5}{270}{360}
\psarc{-}(7,4){0.5}{0}{90}\psarc{-}(8,5){0.5}{180}{270}
\psarc{-}(0,5){0.5}{0}{90}\psarc{-}(1,6){0.5}{180}{270}
\psarc{-}(2,5){0.5}{90}{180}\psarc{-}(1,6){0.5}{270}{360}
\psarc{-}(3,5){0.5}{90}{180}\psarc{-}(2,6){0.5}{270}{360}
\psarc{-}(4,5){0.5}{90}{180}\psarc{-}(3,6){0.5}{270}{360}
\psarc{-}(5,5){0.5}{90}{180}\psarc{-}(4,6){0.5}{270}{360}
\psarc{-}(6,5){0.5}{90}{180}\psarc{-}(5,6){0.5}{270}{360}
\psarc{-}(6,5){0.5}{0}{90}\psarc{-}(7,6){0.5}{180}{270}
\psarc{-}(8,5){0.5}{90}{180}\psarc{-}(7,6){0.5}{270}{360}
\rput(4,-2.5){(c)}
\end{pspicture}
\caption{The states of spins marked by white or gray circles
in (a) are specified by boundary conditions. To obtain fixed boundary conditions, the spins are tied to a spurious spin in the desired state like in (b) and the interaction between the new spin and those at the boundary is sent to infinity. 
Again $K$ bonds are represented by dashed lines, $J$ bonds by dotted lines and $L$ bonds by alternating dots and dashes.}
In (c) a FK graph (with $N_c^* = 7$) with the fixed boundary conditions on the top and bottom edges of the lattice. Note that this configuration contributes to the partition function $Z^{(b)}$, but is excluded from $Z^{(a)}$. \label{fig:spin2}
\end{center}
\end{figure}
\begin{Lemme} The partition functions on a lattice $N\times (2M)$ as in Figures \ref{fig:spin} and \ref{fig:spin2} with free boundary conditions on the left and right edges depend on the boundary conditions on the top and bottom edges as follows.
\begin{itemize}
\item[{\em (i)}] 
\begin{pspicture}(-0.2,-0.1)(0.2,0.3)\psset{unit=0.6}
\pscircle*[linecolor=white](0,0){0.3}\psarc(0,0){0.3}{0}{360}\pscircle*(0,0){0.1}
\end{pspicture}
and 
\begin{pspicture}(-0.2,-0.1)(0.2,0.3)\psset{unit=0.6}
\pscircle*[linecolor=lightgray](0,0){0.3}\psarc(0,0){0.3}{0}{360}\pscircle*(0,0){0.1}
\end{pspicture}
are fixed and have distinct values:
\begin{align}Z_{N\times M}^{(a)} = K(\beta,u)\beta^{-N-2} \Big( &\langle 
\begin{pspicture}(0.05,0)(2.0,0.5)
\psset{unit=0.3}
\psset{dotsize=2.6pt 0}
\psdots(0.6,0)(1.2,0)(1.8,0)(2.4,0)(3,0)(5,0)(5.6,0)(6.2,0)
\psarc{-}(1.5,0){0.3}{0}{180}
\psarc{-}(2.7,0){0.3}{0}{180}
\psarc{-}(5.3,0){0.3}{0}{180}
\rput(4,0){$...$}
\psbezier{-}(0.6,0)(0.6,1)(6.2,1)(6.2,0)
\end{pspicture}
|\, D^M_N(\lambda, u)
\begin{pspicture}(0.05,0)(2.0,0.5)
\psset{unit=0.3}
\psset{dotsize=2.6pt 0}
\psdots(0.6,0)(1.2,0)(1.8,0)(2.4,0)(3,0)(5,0)(5.6,0)(6.2,0)
\psarc{-}(1.5,0){0.3}{0}{180}
\psarc{-}(2.7,0){0.3}{0}{180}
\psarc{-}(5.3,0){0.3}{0}{180}
\rput(4,0){$...$}
\psbezier{-}(0.6,0)(0.6,1)(6.2,1)(6.2,0)
\end{pspicture}
\rangle_G\notag \\
& - \beta \langle 
\begin{pspicture}(0.05,0)(2.0,0.5)
\psset{unit=0.3}
\psset{dotsize=2.6pt 0}
\psdots(0.6,0)(1.2,0)(1.8,0)(2.4,0)(3,0)(5,0)(5.6,0)(6.2,0)
\psline{-}(0.6,0)(0.6,1)
\psarc{-}(1.5,0){0.3}{0}{180}
\psarc{-}(2.7,0){0.3}{0}{180}
\psarc{-}(5.3,0){0.3}{0}{180}
\psline{-}(6.2,0)(6.2,1)
\rput(4,0){$...$}
\end{pspicture} 
 | \, D^M_N(\lambda, u)
\begin{pspicture}(0.05,0)(2.0,0.5)
\psset{unit=0.3}
\psset{dotsize=2.6pt 0}
\psdots(0.6,0)(1.2,0)(1.8,0)(2.4,0)(3,0)(5,0)(5.6,0)(6.2,0)
\psline{-}(0.6,0)(0.6,1)
\psarc{-}(1.5,0){0.3}{0}{180}
\psarc{-}(2.7,0){0.3}{0}{180}
\psarc{-}(5.3,0){0.3}{0}{180}
\psline{-}(6.2,0)(6.2,1)
\rput(4,0){$...$}
\end{pspicture} 
 \rangle_G \Big);\label{eq:autreZa}\end{align}
\item[{\em (ii)}] 
\begin{pspicture}(-0.2,-0.1)(0.2,0.3)\psset{unit=0.6}
\pscircle*[linecolor=white](0,0){0.3}\psarc(0,0){0.3}{0}{360}\pscircle*(0,0){0.1}
\end{pspicture}
and 
\begin{pspicture}(-0.2,-0.1)(0.2,0.3)\psset{unit=0.6}
\pscircle*[linecolor=lightgray](0,0){0.3}\psarc(0,0){0.3}{0}{360}\pscircle*(0,0){0.1}
\end{pspicture}
are fixed and have the same value:
\begin{align}Z_{N\times M}^{(b)} = K(\beta,u)\beta^{-N-2} \Big( &\langle 
\begin{pspicture}(0.05,0)(2.0,0.5)
\psset{unit=0.3}
\psset{dotsize=2.6pt 0}
\psdots(0.6,0)(1.2,0)(1.8,0)(2.4,0)(3,0)(5,0)(5.6,0)(6.2,0)
\psarc{-}(1.5,0){0.3}{0}{180}
\psarc{-}(2.7,0){0.3}{0}{180}
\psarc{-}(5.3,0){0.3}{0}{180}
\rput(4,0){$...$}
\psbezier{-}(0.6,0)(0.6,1)(6.2,1)(6.2,0)
\end{pspicture}
| \, D^M_N(\lambda, u) 
\begin{pspicture}(0.05,0)(2.0,0.5)
\psset{unit=0.3}
\psset{dotsize=2.6pt 0}
\psdots(0.6,0)(1.2,0)(1.8,0)(2.4,0)(3,0)(5,0)(5.6,0)(6.2,0)
\psarc{-}(1.5,0){0.3}{0}{180}
\psarc{-}(2.7,0){0.3}{0}{180}
\psarc{-}(5.3,0){0.3}{0}{180}
\rput(4,0){$...$}
\psbezier{-}(0.6,0)(0.6,1)(6.2,1)(6.2,0)
\end{pspicture}
\rangle_G\notag\\
& + \beta(\beta^2- 1) \langle 
\begin{pspicture}(0.05,0)(2.0,0.5)
\psset{unit=0.3}
\psset{dotsize=2.6pt 0}
\psdots(0.6,0)(1.2,0)(1.8,0)(2.4,0)(3,0)(5,0)(5.6,0)(6.2,0)
\psline{-}(0.6,0)(0.6,1)
\psarc{-}(1.5,0){0.3}{0}{180}
\psarc{-}(2.7,0){0.3}{0}{180}
\psarc{-}(5.3,0){0.3}{0}{180}
\psline{-}(6.2,0)(6.2,1)
\rput(4,0){$...$}
\end{pspicture} 
| \, D^M_N(\lambda, u) 
\begin{pspicture}(0.05,0)(2.0,0.5)
\psset{unit=0.3}
\psset{dotsize=2.6pt 0}
\psdots(0.6,0)(1.2,0)(1.8,0)(2.4,0)(3,0)(5,0)(5.6,0)(6.2,0)
\psline{-}(0.6,0)(0.6,1)
\psarc{-}(1.5,0){0.3}{0}{180}
\psarc{-}(2.7,0){0.3}{0}{180}
\psarc{-}(5.3,0){0.3}{0}{180}
\psline{-}(6.2,0)(6.2,1)
\rput(4,0){$...$}
\end{pspicture} 
\rangle_G \Big);\label{eq:autreZb}\end{align}
\item[{\em (iii)}] 
\begin{pspicture}(-0.2,-0.1)(0.2,0.3)\psset{unit=0.6}
\pscircle*[linecolor=white](0,0){0.3}\psarc(0,0){0.3}{0}{360}\pscircle*(0,0){0.1}
\end{pspicture}
and 
\begin{pspicture}(-0.2,-0.1)(0.2,0.3)\psset{unit=0.6}
\pscircle*[linecolor=lightgray](0,0){0.3}\psarc(0,0){0.3}{0}{360}\pscircle*(0,0){0.1}
\end{pspicture}
are free:
\begin{equation}Z_{N\times M}^{(c)} = K(\beta,u) \big\langle 
\begin{pspicture}(0.05,0)(2.0,0.5)
\psset{unit=0.3}
\psset{dotsize=2.6pt 0}
\psdots(0.6,0)(1.2,0)(1.8,0)(2.4,0)(3,0)(3.6,0)(5.6,0)(6.2,0)
\psarc{-}(0.9,0){0.3}{0}{180}
\psarc{-}(2.1,0){0.3}{0}{180}
\psarc{-}(3.3,0){0.3}{0}{180}
\psarc{-}(5.9,0){0.3}{0}{180}
\rput(4.6,0){$...$}
\end{pspicture}
| \, D^M_N(\lambda, u) 
\begin{pspicture}(0.05,0)(2.0,0.5)
\psset{unit=0.3}
\psset{dotsize=2.6pt 0}
\psdots(0.6,0)(1.2,0)(1.8,0)(2.4,0)(3,0)(3.6,0)(5.6,0)(6.2,0)
\psarc{-}(0.9,0){0.3}{0}{180}
\psarc{-}(2.1,0){0.3}{0}{180}
\psarc{-}(3.3,0){0.3}{0}{180}
\psarc{-}(5.9,0){0.3}{0}{180}
\rput(4.6,0){$...$}
\end{pspicture}
 \big\rangle_G ;\label{eq:autreZc}\end{equation}
%
%
\item[{\em (iv)}] 
\begin{pspicture}(-0.2,-0.1)(0.2,0.3)\psset{unit=0.6}
\pscircle*[linecolor=white](0,0){0.3}\psarc(0,0){0.3}{0}{360}\pscircle*(0,0){0.1}
\end{pspicture}
is free,  
\begin{pspicture}(-0.2,-0.1)(0.2,0.3)\psset{unit=0.6}
\pscircle*[linecolor=lightgray](0,0){0.3}\psarc(0,0){0.3}{0}{360}\pscircle*(0,0){0.1}
\end{pspicture}
is fixed:
\begin{equation}Z_{N\times M}^{(d)} = K(\beta,u) \beta^{-N/2-1}\big\langle 
\begin{pspicture}(0.05,0)(2.0,0.5)
\psset{unit=0.3}
\psset{dotsize=2.6pt 0}
\psdots(0.6,0)(1.2,0)(1.8,0)(2.4,0)(3,0)(5,0)(5.6,0)(6.2,0)
\psarc{-}(1.5,0){0.3}{0}{180}
\psarc{-}(2.7,0){0.3}{0}{180}
\psarc{-}(5.3,0){0.3}{0}{180}
\rput(4,0){$...$}
\psbezier{-}(0.6,0)(0.6,1)(6.2,1)(6.2,0)
\end{pspicture}
| \, D^M_N(\lambda, u)
\begin{pspicture}(0.05,0)(2.0,0.5)
\psset{unit=0.3}
\psset{dotsize=2.6pt 0}
\psdots(0.6,0)(1.2,0)(1.8,0)(2.4,0)(3,0)(3.6,0)(5.6,0)(6.2,0)
\psarc{-}(0.9,0){0.3}{0}{180}
\psarc{-}(2.1,0){0.3}{0}{180}
\psarc{-}(3.3,0){0.3}{0}{180}
\psarc{-}(5.9,0){0.3}{0}{180}
\rput(4.6,0){$...$}
\end{pspicture}
\big\rangle_G \label{eq:autreZd}\end{equation}
\end{itemize}
where $$K(\beta,u)=\frac{\beta^{N_s}}{\big(\sin(u)\sin(\lambda-u)\big)^{NM}} \qquad \textrm{and} \qquad N_s = (N+1)M+\textstyle{\frac12}N.$$
\end{Lemme}
\noindent{\scshape Proof\ \ } 
The condition (i) that boundary spins be all fixed to a single value is implemented as in Figure \ref{fig:spin2} (b): all spins from the top are connected to a spurious spin with fixed value. The same is done for those on the bottom. By putting a factor $e^{-N\gamma_L}$ and sending the interaction $L$ of these bonds to infinity, we get the desired partition function:
\begin{equation*}Z_{N\times M}^{(a)} = \displaystyle\lim_{L\to \infty} e^{-N\gamma_L} \sum_{\bar\sigma } e^{-E_{\bar\sigma}/k_BT},\qquad  E_{\bar\sigma} =-J\sum_{\langle i,j\rangle(J)} \delta_{\bar\sigma_i\bar\sigma_j} -K\sum_{\langle i,j\rangle(K)} \delta_{\bar\sigma_i\bar\sigma_j} -L\sum_{\langle i,j\rangle(L)} \delta_{\bar\sigma_i\bar\sigma_j} \end{equation*}
where $\bar{\sigma}$ is the set of all spin configurations in Figure \ref{fig:spin2} (b), with 
\begin{pspicture}(-0.2,-0.1)(0.2,0.3)\psset{unit=0.6}
\pscircle*[linecolor=white](0,0){0.3}\psarc(0,0){0.3}{0}{360}\pscircle*(0,0){0.1}
\end{pspicture}
and 
\begin{pspicture}(-0.2,-0.1)(0.2,0.3)\psset{unit=0.6}
\pscircle*[linecolor=lightgray](0,0){0.3}\psarc(0,0){0.3}{0}{360}\pscircle*(0,0){0.1}
\end{pspicture}
fixed to distinct states.
\begin{align*}
Z_{N\times M}^{(a)} &= \displaystyle\lim_{L\to \infty} e^{-N\gamma_L} \sum_{\bar{\sigma}} e^{-E_{\bar\sigma}/k_BT} \\
& =  \displaystyle\lim_{L\to \infty} e^{-N\gamma_L} \sum_{\bar\sigma} \prod_{\langle i,j\rangle(J)}(1+v_J\delta_{\bar\sigma_i\bar\sigma_j})\prod_{\langle i,j\rangle(K)}(1+v_K\delta_{\bar\sigma_i\bar\sigma_j}) \prod_{\langle i,j\rangle(L)}(1+v_L\delta_{\bar\sigma_i\bar\sigma_j}) \\
& = \sum_{\sigma} \prod_{\langle i,j\rangle(J)}(1+v_J\delta_{\sigma_i\sigma_j})\prod_{\langle i,j\rangle(K)}(1+v_K\delta_{\sigma_i\sigma_j}) \\
& = \sum_{\sigma} \sum_{\mathcal G | \sigma} v_J^{Nb_J}v_K^{Nb_K}.
\end{align*}
The first sum is now on spin configurations $\sigma$ with 
\begin{pspicture}(-0.2,-0.1)(0.2,0.3)\psset{unit=0.6}
\pscircle*[linecolor=white](0,0){0.3}\psarc(0,0){0.3}{0}{360}\pscircle*(0,0){0.1} \end{pspicture}
and 
\begin{pspicture}(-0.2,-0.1)(0.2,0.3)\psset{unit=0.6}
\pscircle*[linecolor=lightgray](0,0){0.3}\psarc(0,0){0.3}{0}{360}\pscircle*(0,0){0.1}
\end{pspicture}
of Figure \ref{fig:spin2} (a) fixed, and $\mathcal G | \sigma$ represents the Fortuin-Kasteleyn loop configurations that are in the restricted set compatible with $\sigma$. Inverting the sums, the restriction on $\sigma$ is moved to the set of FK configurations; it becomes the set $\mathcal G^{a}$ of all graphs that do not contain a subgraph connecting the top and bottom edges of the lattice. The partition function is then
\begin{equation}
Z_{N\times M}^{(a)} = \sum_{\mathcal G_a} \sum_{\sigma | \mathcal G_a} v_J^{Nb_J}v_K^{Nb_K} = \sum_{\mathcal G_a}  v_J^{Nb_J}v_K^{Nb_K} Q^{N_c^*}.
\label{eq:intermedZ}\end{equation}
At the last step, we summed over all $\sigma$ compatible with a given graph in $\mathcal G_a$. Since the clusters touching the boundary have fixed spin, the resulting weight is $Q^{N_c^*}$ where $N_c^*$ is the number of closed components not touching the top and lower boundaries, that is, $N_c^* = N_c -2$. Lemma \ref{lem:euler} was proved for a lattice drawn on the geometry of a ribbon. In the present case, the lattice is drawn on a rectangle, but the relationship remains $N_c=\frac12(\#(\mathcal G)+N_s^*-N_b^*)$ when $N_s^*$ and $N_b^*$ include the spurious sites and the bonds joining them to the lattice. This relation is proved using the same arguments. In terms of $N_s$ and $N_b$, the numbers without the spurious sites and bonds, one gets $N_c^*=\frac12(\#(\mathcal G)+N_s-N_b-N-2)$. At the critical temperature and for the parametrization of the interactions used before, the partition function is now
\begin{equation}
Z_{N\times M}^{(a)} = \frac{\beta^{N_s - N-2}}{\big(\sin(u) \sin(\lambda-u)\big)^{NM}} \sum_{\mathcal G_a}\underbrace{\beta^{\#(\mathcal G)}\sin^{N_{bJ}}(\lambda-u) \sin^{NM-N_{bJ}}(u) \sin^{N_{bK}}(u) \sin^{NM-N_{bK}}(\lambda-u)}_{X(\mathcal G)}.
\end{equation}
To perform the final sum, we write $\mathcal{G}^a \cup \mathcal{G}_{cr}= \mathcal{G}$ where the subscript $ \mathcal{G}_{cr}$ is the set of crossing graphs. The two sets $\mathcal G_a$ and $\mathcal G_{\text{\rm cr}}$ are disjoint. The only way to close the loops arriving at the horizontal edges that is compatible with the bonds ending at one of the spurious sites is by adding the set of arcs $\begin{pspicture}(0.05,0)(2.0,0.3)
\psset{unit=0.3}
\psset{dotsize=2.6pt 0}
\psdots(0.6,0)(1.2,0)(1.8,0)(2.4,0)(3,0)(5,0)(5.6,0)(6.2,0)
\psarc{-}(1.5,0){0.3}{0}{180}
\psarc{-}(2.7,0){0.3}{0}{180}
\psarc{-}(5.3,0){0.3}{0}{180}
\rput(4,0){$...$}
\psbezier{-}(0.6,0)(0.6,1)(6.2,1)(6.2,0)
\end{pspicture}$ on the top and the bottom of the lattice. Therefore 
$\sum_{\mathcal G} X(\mathcal G)=\langle 
%
\begin{pspicture}(0.05,0)(2.0,0.3)
\psset{unit=0.3}
\psset{dotsize=2.6pt 0}
\psdots(0.6,0)(1.2,0)(1.8,0)(2.4,0)(3,0)(5,0)(5.6,0)(6.2,0)
\psarc{-}(1.5,0){0.3}{0}{180}
\psarc{-}(2.7,0){0.3}{0}{180}
\psarc{-}(5.3,0){0.3}{0}{180}
\rput(4,0){$...$}
\psbezier{-}(0.6,0)(0.6,1)(6.2,1)(6.2,0)
\end{pspicture} 
|D_N^M(\lambda,u)
\begin{pspicture}(0.05,0)(2.0,0.3)
\psset{unit=0.3}
\psset{dotsize=2.6pt 0}
\psdots(0.6,0)(1.2,0)(1.8,0)(2.4,0)(3,0)(5,0)(5.6,0)(6.2,0)
\psarc{-}(1.5,0){0.3}{0}{180}
\psarc{-}(2.7,0){0.3}{0}{180}
\psarc{-}(5.3,0){0.3}{0}{180}
\rput(4,0){$...$}
\psbezier{-}(0.6,0)(0.6,1)(6.2,1)(6.2,0)
\end{pspicture} 
%
 \rangle_G$. Note that the Gram product adds the proper power of $\beta$ to account for the loops that are closed by adding the arcs at the lower boundary. The argument for $\sum_{\mathcal G_{cr}} X(\mathcal G)$ is similar. This time, the Gram product of
$ D_N^M(\lambda,u) 
 %
\begin{pspicture}(0.05,0)(2.0,0.3)
\psset{unit=0.3}
\psset{dotsize=2.6pt 0}
\psdots(0.6,0)(1.2,0)(1.8,0)(2.4,0)(3,0)(5,0)(5.6,0)(6.2,0)
\psline{-}(0.6,0)(0.6,1)
\psarc{-}(1.5,0){0.3}{0}{180}
\psarc{-}(2.7,0){0.3}{0}{180}
\psarc{-}(5.3,0){0.3}{0}{180}
\psline{-}(6.2,0)(6.2,1)
\rput(4,0){$...$}
\end{pspicture} 
$
with $ 
 %
\begin{pspicture}(0.05,0)(2.0,0.3)
\psset{unit=0.3}
\psset{dotsize=2.6pt 0}
\psdots(0.6,0)(1.2,0)(1.8,0)(2.4,0)(3,0)(5,0)(5.6,0)(6.2,0)
\psline{-}(0.6,0)(0.6,1)
\psarc{-}(1.5,0){0.3}{0}{180}
\psarc{-}(2.7,0){0.3}{0}{180}
\psarc{-}(5.3,0){0.3}{0}{180}
\psline{-}(6.2,0)(6.2,1)
\rput(4,0){$...$}
\end{pspicture} 
$ excludes any non-crossing graph. Moreover each graph contributing to the latter Gram product also appears in
$\langle 
%
\begin{pspicture}(0.05,0)(2.0,0.3)
\psset{unit=0.3}
\psset{dotsize=2.6pt 0}
\psdots(0.6,0)(1.2,0)(1.8,0)(2.4,0)(3,0)(5,0)(5.6,0)(6.2,0)
\psarc{-}(1.5,0){0.3}{0}{180}
\psarc{-}(2.7,0){0.3}{0}{180}
\psarc{-}(5.3,0){0.3}{0}{180}
\rput(4,0){$...$}
\psbezier{-}(0.6,0)(0.6,1)(6.2,1)(6.2,0)
\end{pspicture} 
|D_N^M(\lambda,u)
\begin{pspicture}(0.05,0)(2.0,0.3)
\psset{unit=0.3}
\psset{dotsize=2.6pt 0}
\psdots(0.6,0)(1.2,0)(1.8,0)(2.4,0)(3,0)(5,0)(5.6,0)(6.2,0)
\psarc{-}(1.5,0){0.3}{0}{180}
\psarc{-}(2.7,0){0.3}{0}{180}
\psarc{-}(5.3,0){0.3}{0}{180}
\rput(4,0){$...$}
\psbezier{-}(0.6,0)(0.6,1)(6.2,1)(6.2,0)
\end{pspicture} 
%
 \rangle_G$, but with a weight that has an additional factor $\beta$ for the outer loop that is left open here. Adding by hand a factor of $\beta$ to the second sum yields \eqref{eq:autreZa}.

The steps leading to (i) can be used for (ii) and carry through up to (\ref{eq:intermedZ}), with $\mathcal{G}_a$ replaced with $\mathcal{G}_b = \mathcal{G}$. Here, however, $N_c^*$ is $N_c-1$ when a crossing occurs and $N_c-2$ when none exists:
\begin{align*}
Z_{N\times M}^{(b)} &= \left(\sum_{\mathcal G_a} + \sum_{\mathcal G_{cr}}\right)  v_J^{Nb_J}v_K^{Nb_K} Q^{N_c^*} = Z_{N\times M}^{(a)} + \frac{\beta^{N_s - N}}{(\sin(u) \sin(\lambda-u))^{NM}} \sum_{\mathcal G_{cr}} X(\mathcal{G}) \\
&= Z_{N\times M}^{(a)} + K(\beta, u) \beta^{1-N}  \langle 
%
\begin{pspicture}(0.05,0)(2.0,0.5)
\psset{unit=0.3}
\psset{dotsize=2.6pt 0}
\psdots(0.6,0)(1.2,0)(1.8,0)(2.4,0)(3,0)(5,0)(5.6,0)(6.2,0)
\psline{-}(0.6,0)(0.6,1)
\psarc{-}(1.5,0){0.3}{0}{180}
\psarc{-}(2.7,0){0.3}{0}{180}
\psarc{-}(5.3,0){0.3}{0}{180}
\psline{-}(6.2,0)(6.2,1)
\rput(4,0){$...$}
\end{pspicture} 
|D_N^M(\lambda,u) 
%
\begin{pspicture}(0.05,0)(2.0,0.5)
\psset{unit=0.3}
\psset{dotsize=2.6pt 0}
\psdots(0.6,0)(1.2,0)(1.8,0)(2.4,0)(3,0)(5,0)(5.6,0)(6.2,0)
\psline{-}(0.6,0)(0.6,1)
\psarc{-}(1.5,0){0.3}{0}{180}
\psarc{-}(2.7,0){0.3}{0}{180}
\psarc{-}(5.3,0){0.3}{0}{180}
\psline{-}(6.2,0)(6.2,1)
\rput(4,0){$...$}
\end{pspicture} 
\rangle_G
\end{align*}
and the result follows. 

Cases (iii) and (iv) are simpler. Their proofs do not require any new idea and are omitted.
\hfill $\square$

\medskip

By the last remark of Definition \ref{sec:Gram}, any element of the form $\langle v | D^M_N w \rangle_G$ can be expressed as $\sum_y G_{vy} \rho(D^M_N)_{yw}$: the information on $Z_{N \times M}$ is contained in $\rho(D^M_N)$ as suggested earlier.

%
%

\section{A central element of $TL_N$}
\label{sec:fn}

Let $\lambda$ be fixed. The element $D_N(\lambda, u)\in TL_N$ is a homogeneous polynomial in $\sin u$ and $\sin(\lambda-u)$ of order $2N$. Its coefficients are elements of $TL_N$. Moreover, since $D_N(\lambda,u)=D_N(\lambda,u+\pi)$ and $D_N(\lambda,u) = D_N(\lambda,\lambda -u)$ (see \cite{Behrend}), the double-row matrix $D_N(\lambda,u)$ can be written as a Fourier series in $v=u-\lambda/2$ ($v=0$ is then the isotropic case):
\begin{equation}\label{eq:fourier}D_N(\lambda,v+\lambda/2) = \textstyle{\frac12}C_0+\sum_{i=1}^{N} C_{2i}(\lambda) \cos(2i v)\end{equation}
where the Fourier coefficients $C_{2i}(\lambda)$ are again elements of $TL_N$. The commutation relation $[D_N(\lambda,u),D_N(\lambda,v)]=0$ implies that $[D_N(\lambda,u),C_{2i}(\lambda)] = 0$ for all $i$ or simply $[C_{2i}(\lambda), C_{2j}(\lambda)]=0$ for all $i,j$. One of these Fourier coefficients is easier than the others to study since it is a central element of $TL_N$. Even though it is central, we shall see in the present and following sections that it carries much of the information about the Jordan structure. We start our analysis by this coefficient.

\subsection{The last Fourier coefficient of $D_N(\lambda, u)$}
\label{sec:cmax}

\begin{Definition} The braid 2-box (see for example \cite{PRZ}) is defined as:
\begin{alignat*}{2}
\psset{linewidth=1pt}
\begin{pspicture}(-0.5,-0.1)(0.5,0.5)\lw
\psline{-}(-0.5,-0.5)(0.5,-0.5)(0.5,0.5)(-0.5,0.5)(-0.5,-0.5)
\psset{linecolor=myc}\unlw
\psline{-}(-0.5,0)(0.5,0)
\psline{-}(0,-0.5)(0,-0.15)
\psline{-}(0,0.15)(0,0.5)
\end{pspicture}\ & = &\ -i e^{-i\lambda/2}	\displaystyle\lim_{u\to -i \infty} \frac1{\sin(\lambda-u)} \ \
\begin{pspicture}(-0.5,-0.1)(0.5,0.5)\lw
\psline{-}(-0.5,-0.5)(0.5,-0.5)(0.5,0.5)(-0.5,0.5)(-0.5,-0.5)
\psarc[linewidth=0.5pt]{-}(-0.5,-0.5){0.25}{0}{90}
\rput(0,0){$u$}
\end{pspicture}\ \ &=  \ \ i e^{i\lambda/2}\ \ 
\begin{pspicture}(-0.5,-0.1)(0.5,0.5)\lw
\psline{-}(-0.5,-0.5)(0.5,-0.5)(0.5,0.5)(-0.5,0.5)(-0.5,-0.5)
\psset{linecolor=myc}\unlw
\psarc{-}(-0.5,-0.5){0.5}{0}{90}
\psarc{-}(0.5,0.5){0.5}{180}{270}
\end{pspicture}\ \ -\ i e^{-i\lambda/2}\ \ 
\begin{pspicture}(-0.5,-0.1)(0.5,0.5)\lw
\psline{-}(-0.5,-0.5)(0.5,-0.5)(0.5,0.5)(-0.5,0.5)(-0.5,-0.5)
\psset{linecolor=myc}\unlw
\psarc{-}(0.5,-0.5){0.5}{90}{180}
\psarc{-}(-0.5,0.5){0.5}{270}{360}
\end{pspicture}\\
&&\ &\\
\psset{linewidth=1pt}
\begin{pspicture}(-0.5,-0.1)(0.5,0.5)\lw
\psline{-}(-0.5,-0.5)(0.5,-0.5)(0.5,0.5)(-0.5,0.5)(-0.5,-0.5)
\psset{linecolor=myc}\unlw
\psline{-}(0,-0.5)(0,0.5)
\psline{-}(-0.5,0)(-0.15,0)
\psline{-}(0.15,0)(0.5,0)
\end{pspicture}\ & = & i e^{i\lambda/2}	\displaystyle\lim_{u\to +i \infty} \frac1{\sin(\lambda-u)} \ \
\begin{pspicture}(-0.5,-0.1)(0.5,0.5)\lw
\psline{-}(-0.5,-0.5)(0.5,-0.5)(0.5,0.5)(-0.5,0.5)(-0.5,-0.5)
\psarc[linewidth=0.5pt]{-}(-0.5,-0.5){0.25}{0}{90}
\rput(0,0){$u$}
\end{pspicture}\ \ &=  \ \ -i e^{-i\lambda/2}\ \ 
\begin{pspicture}(-0.5,-0.1)(0.5,0.5)\lw
\psline{-}(-0.5,-0.5)(0.5,-0.5)(0.5,0.5)(-0.5,0.5)(-0.5,-0.5)
\psset{linecolor=myc}\unlw
\psarc{-}(-0.5,-0.5){0.5}{0}{90}
\psarc{-}(0.5,0.5){0.5}{180}{270}
\end{pspicture}\ \ +\ i e^{i\lambda/2}\ \ 
\begin{pspicture}(-0.5,-0.1)(0.5,0.5)\lw
\psline{-}(-0.5,-0.5)(0.5,-0.5)(0.5,0.5)(-0.5,0.5)(-0.5,-0.5)
\psset{linecolor=myc}\unlw
\psarc{-}(0.5,-0.5){0.5}{90}{180}
\psarc{-}(-0.5,0.5){0.5}{270}{360}
\end{pspicture}
\end{alignat*}
We define the double-row braid matrix:
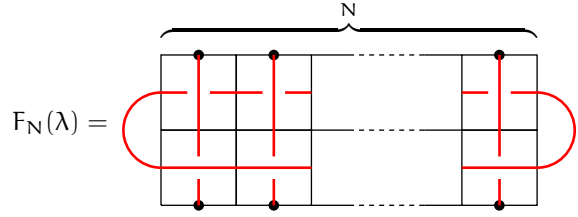
\begin{equation*}
F_N(\lambda) = 
\ \ \ \ \ \ \
\psset{linewidth=1pt}\overbrace{
\begin{pspicture}(0,0)(5,1.2)
\psdots(0.5,-1)(1.5,-1)(4.5,-1)
\psdots(0.5,1)(1.5,1)(4.5,1)
\lw
\psline{-}(0,-1)(1,-1)(1,0)(0,0)(0,-1)
\psline{-}(1,-1)(2,-1)(2,0)(1,0)(1,-1)
\psline{-}(4,-1)(5,-1)(5,0)(4,0)(4,-1)
\psline{-}(0,0)(1,0)(1,1)(0,1)(0,0)
\psline{-}(1,0)(2,0)(2,1)(1,1)(1,0)
\psline{-}(4,0)(5,0)(5,1)(4,1)(4,0)
\psline{-}(2,-1)(2.5,-1)\psline[linestyle=dashed,dash=2pt 2pt]{-}(2.5,0)(3.5,0)\psline{-}(3.5,-1)(4,-1)
\psline{-}(2,0)(2.5,0)\psline[linestyle=dashed,dash=2pt 2pt]{-}(2.5,1)(3.5,1)\psline{-}(3.5,0)(4,0)
\psline{-}(2,1)(2.5,1)\psline[linestyle=dashed,dash=2pt 2pt]{-}(2.5,-1)(3.5,-1)\psline{-}(3.5,1)(4,1)
\psset{linecolor=myc}\unlw
\psarc{-}(0,0){0.5}{90}{270}
\psarc{-}(5,0){0.5}{270}{450}
\psline{-}(0,-0.5)(2,-0.5)
\psline{-}(4,-0.5)(5,-0.5)
\psline{-}(0.5,0)(0.5,1)
\psline{-}(1.5,0)(1.5,1)
\psline{-}(4.5,0)(4.5,1)
\psline{-}(0.5,-1)(0.5,-0.65)
\psline{-}(0.5,-0.35)(0.5,0)
\psline{-}(1.5,-1)(1.5,-0.65)
\psline{-}(1.5,-0.35)(1.5,0)
\psline{-}(4.5,-1)(4.5,-0.65)
\psline{-}(4.5,-0.35)(4.5,0)
\psline{-}(0,0.5)(0.35,0.5)
\psline{-}(0.65,0.5)(1.35,0.5)
\psline{-}(1.65,0.5)(2,0.5)
\psline{-}(4.0,0.5)(4.35,0.5)
\psline{-}(4.65,0.5)(5,0.5)
\end{pspicture}}^N
\end{equation*}
\vskip 0.75cm
\end{Definition}
Note that\begin{pspicture}(-0.35,-0.1)(0.35,0.5)
\psset{unit=0.3}
\psline[linewidth=0.3pt]{-}(-0.5,-0.5)(0.5,-0.5)(0.5,0.5)(-0.5,0.5)(-0.5,-0.5)
\psset{linecolor=myc}
\psline[linewidth=0.3pt]{-}(-0.5,0)(0.5,0)
\psline[linewidth=0.3pt]{-}(0,-0.5)(0,-0.15)
\psline[linewidth=0.3pt]{-}(0,0.15)(0,0.5)
\end{pspicture}and\begin{pspicture}(-0.35,-0.1)(0.35,0.5)
\psset{unit=0.3}
\psline[linewidth=0.3pt]{-}(-0.5,-0.5)(0.5,-0.5)(0.5,0.5)(-0.5,0.5)(-0.5,-0.5)
\psset{linecolor=myc}
\psline[linewidth=0.3pt]{-}(0,-0.5)(0,0.5)
\psline[linewidth=0.3pt]{-}(-0.5,0)(-0.15,0)
\psline[linewidth=0.3pt]{-}(0.15,0)(0.5,0)
\end{pspicture}can be obtained from one another by a simple rotation by $\frac\pi2$ of the connectivity boxes. Consequently one gets rid of the small arc giving the orientation of the box. Note also that the weights of one are the complex conjugate of those of the other.

Let $c$ be an element of $TL_N$. We shall denote by $\rho(c)_{d',d}$ the submatrix of its link representation containing lines associated with basis elements in $B_N^{d'}$ and columns with those in $B_N^{d}$. Because $\rho(c)$ never increases the number of defects, $\rho(c)_{d',d}=0$ whenever $d'\ge d$.

\begin{Lemme}\label{lem:fn}
\begin{itemize}
\item[(i)]
The last Fourier coefficient of $D_N(\lambda,u)$ is given by
\begin{equation*}
C_{2N}(\lambda)= 2^{-2N+1} F_N(\lambda).
\end{equation*}
\item[(ii)] The diagonal submatrices $\rho(F_N)_{d,d}$ are 
\begin{equation} \rho(F_{N}(\lambda))_{d,d} = 2(-1)^d \, \cos(\lambda(d+1)) \ id_{ \textrm{dim} V_N^d}.\label{eq:cmaxdiag}\end{equation}
\item[(iii)] $F_N(\lambda)$ is in the center of $TL_N(\beta)$ with $\beta=2\cos\lambda$.
\end{itemize}
\end{Lemme}
\
\noindent{\scshape Proof\ \ } To prove {\em (i)}, the limit of 
$$D_N(\lambda,v+\lambda/2)=\textstyle{\frac12}C_0+\sum_{j=1}^NC_{2j}(\lambda)\cos (2jv)=\frac12\sum_{-N\le j\le N}C_{|2j|}e^{2ijv}$$
is taken with the appropriate power of the factor defining the braid box. The limit of the left-hand side gives
\begin{equation*}
\psset{unit=0.5}
\psset{linewidth=1pt}
\begin{pspicture}(-16.75,0)(5.5,10.5)
\psdots(1.5,0.5)(2.5,1.5)(4.5,3.5)
\psdots(1.5,9.5)(2.5,8.5)(4.5,6.5)
\lw
\rput(-9.75,5){$\displaystyle{\lim_{u\rightarrow -i\infty} \frac{(-ie^{-i\lambda/2})^{2N}D_N(\lambda,u)}{\sin^{2N}(\lambda-u)}=\lim_{u\rightarrow -i\infty} \frac{(-ie^{-i\lambda/2})^{2N}}{\sin^{2N}(\lambda-u)}}$}
\psline{-}(1,0)(3,2)(2,3)(0,1)(1,0)
\psline{-}(4,3)(5,4)(4,5)(3,4)(4,3)
\psline{-}(4,5)(5,6)(4,7)(3,6)(4,5)
\psline{-}(2,7)(3,8)(1,10)(0,9)(2,7)
\psline{-}(2,1)(1,2)\psline{-}(2,9)(1,8)
\psline[linestyle=dashed,dash=2pt 2pt]{-}(3,2)(4,3)
\psline[linestyle=dashed,dash=2pt 2pt]{-}(2,3)(3,4)
\psline[linestyle=dashed,dash=2pt 2pt]{-}(3,8)(4,7)
\psline[linestyle=dashed,dash=2pt 2pt]{-}(2,7)(3,6)
\psarc[linewidth=0.5pt]{-}(1,0){0.3535}{45}{135}
\psarc[linewidth=0.5pt]{-}(2,1){0.3535}{45}{135}
\psarc[linewidth=0.5pt]{-}(4,3){0.3535}{45}{135}
\psarc[linewidth=0.5pt]{-}(4,5){0.3535}{45}{135}
\psarc[linewidth=0.5pt]{-}(2,7){0.3535}{45}{135}
\psarc[linewidth=0.5pt]{-}(1,8){0.3535}{45}{135}
\psset{linecolor=myc}\unlw
\psarc{-}(4,5){0.7071}{-45}{45}
\psarc{-}(0,1){0.7071}{180}{315}
\psarc{-}(0,9){0.7071}{45}{180}
\psline{-}(0.5,1.5)(0.5,8.5)\psline{-}(1.5,2.5)(1.5,7.5)\psline{-}(2.5,3.5)(2.5,6.5)\psline{-}(3.5,4.5)(3.5,5.5)
\psline{-}(-.7071,1)(-.7071,9)
\rput(1,1){$u$}
\rput(2,2){$u$}
\rput(4,4){$u$}
\rput(4,6){$u$}
\rput(2,8){$u$}
\rput(1,9){$u$}
\end{pspicture}
\psset{linewidth=1pt}
\begin{pspicture}(-2.5,0)(7.5,10.5)
\psdots(1.5,0.5)(2.5,1.5)(4.5,3.5)
\psdots(1.5,9.5)(2.5,8.5)(4.5,6.5)
\lw
\rput(-2.25,5){$=$}
\psline{-}(1,0)(3,2)(2,3)(0,1)(1,0)
\psline{-}(4,3)(5,4)(4,5)(3,4)(4,3)
\psline{-}(4,5)(5,6)(4,7)(3,6)(4,5)
\psline{-}(2,7)(3,8)(1,10)(0,9)(2,7)
\psline{-}(2,1)(1,2)\psline{-}(2,9)(1,8)
\psline[linestyle=dashed,dash=2pt 2pt]{-}(3,2)(4,3)
\psline[linestyle=dashed,dash=2pt 2pt]{-}(2,3)(3,4)
\psline[linestyle=dashed,dash=2pt 2pt]{-}(3,8)(4,7)
\psline[linestyle=dashed,dash=2pt 2pt]{-}(2,7)(3,6)
\psset{linecolor=myc}\unlw
\psarc{-}(4,5){0.7071}{-45}{45}
\psarc{-}(0,1){0.7071}{180}{315}
\psarc{-}(0,9){0.7071}{45}{180}
\psline{-}(0.5,1.5)(0.5,8.5)\psline{-}(1.5,2.5)(1.5,7.5)\psline{-}(2.5,3.5)(2.5,6.5)\psline{-}(3.5,4.5)(3.5,5.5)
\psline{-}(-.7071,1)(-.7071,9)
\lw
\psline[linewidth=1pt]{-}(3.5,5.5)(4.5,6.5)
\psline[linewidth=1pt]{-}(3.5,6.5)(3.85,6.15)
\psline[linewidth=1pt]{-}(4.15,5.85)(4.5,5.5)
\psline[linewidth=1pt]{-}(1.5,7.5)(2.5,8.5)
\psline[linewidth=1pt]{-}(1.5,8.5)(1.85,8.15)
\psline[linewidth=1pt]{-}(2.15,7.85)(2.5,7.5)
\psline[linewidth=1pt]{-}(0.5,8.5)(1.5,9.5)
\psline[linewidth=1pt]{-}(0.5,9.5)(0.85,9.15)
\psline[linewidth=1pt]{-}(1.15,8.85)(1.5,8.5)

\psline[linewidth=1pt]{-}(3.5,3.5)(4.5,4.5)
\psline[linewidth=1pt]{-}(3.5,4.5)(3.85,4.15)
\psline[linewidth=1pt]{-}(4.15,3.85)(4.5,3.5)

\psline[linewidth=1pt]{-}(1.5,2.5)(1.85,2.15)
\psline[linewidth=1pt]{-}(2.15,1.85)(2.5,1.5)

\psline[linewidth=1pt]{-}(0.5,1.5)(0.85,1.15)
\psline[linewidth=1pt]{-}(1.15,0.85)(1.5,0.5)

\psline[linewidth=1pt]{-}(0.5,0.5)(2.5,2.5)
\rput(7,5.){$=F_N(\lambda).$}
\end{pspicture}
\end{equation*}
That of the right-hand side is
$$(-ie^{-i\lambda/2})^{2N}\lim_{u\rightarrow -i\infty}\frac12\sum_{-N\le j\le N} 
C_{|2j|}\frac{e^{2ij(u-\lambda/2)}}{\sin^{2N}(\lambda-u)}=2^{2N-1}C_{2N}.$$

To prove {\em (ii)} and do other computations, several identities will be used. The following identities are well-known and proved for example in \cite{PRZ}: 
\\
\begin{equation*}
\psset{unit=0.6}
\psset{linewidth=1pt}
\begin{pspicture}(0,-0.1)(4.1,1.0)\lw
\psline{-}(0,0)(1,1)(2,0)(1,-1)(0,0)
\psline{-}(2,0)(3,1)(4,0)(3,-1)(2,0)
\psset{linecolor=myc}\unlw
\psarc{-}(2,0){0.71}{45}{135}
\psarc{-}(2,0){0.71}{225}{315}
\psline{-}(3.5,0.5)(2.5,-0.5)
\psline{-}(0.5,0.5)(1.5,-0.5)
\psline{-}(0.5,-0.5)(0.85,-0.15)\psline{-}(1.5,0.5)(1.15,0.15)
\psline{-}(2.5,0.5)(2.85,0.15)\psline{-}(3.5,-0.5)(3.15,-0.15)
\end{pspicture} =
\begin{pspicture}(-0.1,-0.1)(2.2,1.0)\lw
\psline{-}(0,0)(1,1)(2,0)(1,-1)(0,0)
\psset{linecolor=myc}\unlw
\psarc{-}(1,-1){0.71}{45}{135}
\psarc{-}(1,1){0.71}{225}{315}
\rput(-1.5,-1.8){(a) Inversion relation}
\end{pspicture}
\qquad
\psset{linewidth=1pt}
\begin{pspicture}(0,-0.1)(4.1,1.0)\lw
\psline{-}(0,0)(1,1)(2,0)(1,-1)(0,0)
\psline{-}(2,0)(3,1)(4,0)(3,-1)(2,0)
\psset{linecolor=myc}\unlw
\psarc{-}(2,0){0.71}{45}{135}
\psarc{-}(2,0){0.71}{225}{315}
\psarc{-}(2,0){0.71}{315}{45}
\psarc{-}(4,0){0.71}{135}{225}
\psline{-}(0.5,0.5)(1.5,-0.5)
\psline{-}(0.5,-0.5)(0.85,-0.15)\psline{-}(1.5,0.5)(1.15,0.15)
\end{pspicture} = i e^{i\frac{3\lambda}{2}}
\begin{pspicture}(-0.1,-0.1)(2.2,1.0)\lw
\psline{-}(0,0)(1,1)(2,0)(1,-1)(0,0)
\psset{linecolor=myc}\unlw
\psarc{-}(0,0){0.71}{315}{45}
\psarc{-}(2,0){0.71}{135}{225}
\rput(-2,-1.8){(b) Twist relation}
\end{pspicture}\qquad
\begin{pspicture}(-0.5,-0.1)(3.58421,1.0)\lw
\psset{unit=1.41421}
\psline(0,-1)(1,-1)(1,1)(0,1)(0,-1)
\psline(0,0)(1,0)(1.707107,0.707107)(2.41421,0)(1.707107,-0.707107)(1,0)
\psset{linecolor=myc}\unlw
\psarc{-}(1,0){0.5}{45}{90}
\psarc{-}(1,0){0.5}{270}{315}
\psline{-}(0,-0.5)(1,-0.5)
\psline{-}(0,0.5)(0.35,0.5)\psline{-}(0.65,0.5)(1,0.5)
\psline{-}(0.5,1)(0.5,0)
\psline{-}(0.5,-1)(0.5,-0.65)\psline{-}(0.5,0)(0.5,-0.35)
\psline{-}(1.353553,-0.353553)(2.06066,0.353553)
\psline{-}(1.353553,0.353553)(1.60104,0.106066)\psline{-}(2.06066,-0.353553)(1.81317,-0.106066)
\end{pspicture} =
\begin{pspicture}(0.675786,-0.1)(3.2,1.0)\lw
\psset{unit=1.41421}
\psline(2,-1)(3,-1,)(3,1,)(2,1,)(2,-1)
\psline(3,0)(2,0)(1.29289,0.707107)(0.585786,0)(1.29289,-0.707107)(2,0)
\psset{linecolor=myc}\unlw
\psarc{-}(2,0){0.5}{90}{135}
\psarc{-}(2,0){0.5}{225}{270}
\psline{-}(2,0.5)(3,0.5)
\psline{-}(2,-0.5)(2.35,-0.5)\psline{-}(2.65,-0.5)(3,-0.5)
\psline{-}(2.5,-1)(2.5,0)
\psline{-}(2.5,1)(2.5,0.65)\psline{-}(2.5,0)(2.5,0.35)
\psline{-}(0.93934,-0.353553)(1.64645,0.353553)
\psline{-}(0.93934,0.353553)(1.18683,0.106066)\psline{-}(1.64645,-0.353553)(1.39896,-0.106066)
\rput(-0.0,-1.4){(c) Yang-Baxter equation}
\end{pspicture}
\end{equation*}
\\
\vspace{4mm}

\noindent The next one characterizes the action of two braid boxes on an arc:
\begin{equation}
\label{eq:semicircle}
\psset{unit=1}
\psset{linewidth=1pt}
\begin{pspicture}(-0.5,0)(2.5,1.0)\lw
\psline{-}(0,-0.5)(2,-0.5)(2,0.5)(0,0.5)(0,-0.5)
\psline{-}(1,-0.5)(1,0.5)
\psset{linecolor=myc}\unlw
\psline{-}(0,0)(2,0)
\psline{-}(0.5,-0.5)(0.5,-0.15)\psline{-}(0.5,0.5)(0.5,0.15)
\psline{-}(1.5,-0.5)(1.5,-0.15)\psline{-}(1.5,0.5)(1.5,0.15)
\psset{linecolor=myc2}
\psarc{-}(1,0.5){0.5}{0}{180}
\end{pspicture} 
=
\begin{pspicture}(-0.5,0)(2.5,1.0)\lw
\psline{-}(0,-0.5)(2,-0.5)(2,0.5)(0,0.5)(0,-0.5)
\psline{-}(1,-0.5)(1,0.5)
\psset{linecolor=myc}\unlw
\psarc{-}(1,-0.5){0.5}{0}{180}
\psarc{-}(0,0.5){0.5}{270}{360}
\psarc{-}(2,0.5){0.5}{180}{270}
\psset{linecolor=myc2}
\psarc{-}(1,0.5){0.5}{0}{180}
\end{pspicture}
=
\begin{pspicture}(-0.5,0)(2.5,1.0)\lw
\psline{-}(0,-0.5)(2,-0.5)(2,0.5)(0,0.5)(0,-0.5)
\psline{-}(1,-0.5)(1,0.5)
\psset{linecolor=myc}\unlw
\psline{-}(0,0)(0.35,0)\psline{-}(1.35,0)(0.65,0)\psline{-}(1.65,0)(2,0)
\psline{-}(0.5,-0.5)(0.5,0.5)
\psline{-}(1.5,-0.5)(1.5,0.5)
\psset{linecolor=myc2}
\psarc{-}(1,0.5){0.5}{0}{180}
\end{pspicture}
\end{equation}
\\
It can be verified by direct computation. Therefore, when $\rho(F_N(\lambda))$ acts on a link state, the arcs of the state propagate down the diagrammatic representation of $F_N$. For instance:
\\
\\
\begin{align}
\label{eq:semicexemple}
\psset{unit=0.8}\begin{pspicture}(-0.5,-0.2)(8.5,2.0)
\psdots(0.5,1)(1.5,1)(2.5,1)(3.5,1)(4.5,1)(5.5,1)(6.5,1)(7.5,1)
\psdots(0.5,-1)(1.5,-1)(2.5,-1)(3.5,-1)(4.5,-1)(5.5,-1)(6.5,-1)(7.5,-1)
\lw
\psline{-}(0,-1)(0,1)(8,1)(8,-1)(0,-1)
\psline{-}(0,0)(8,0)
\psline{-}(1,-1)(1,1)\psline{-}(2,-1)(2,1)\psline{-}(3,-1)(3,1)\psline{-}(4,-1)(4,1)\psline{-}(5,-1)(5,1)\psline{-}(6,-1)(6,1)\psline{-}(7,-1)(7,1)
\psset{linecolor=myc}\unlw
\psarc{-}(0,0){0.5}{90}{270}
\psarc{-}(8,0){0.5}{270}{90}
\psline{-}(0,-0.5)(8,-0.5)
\psline{-}(0,0.5)(0.35,0.5)
\psline{-}(0.65,0.5)(1.35,0.5)\psline{-}(1.65,0.5)(2.35,0.5)\psline{-}(2.65,0.5)(3.35,0.5)\psline{-}(3.65,0.5)(4.35,0.5)\psline{-}(4.65,0.5)(5.35,0.5)\psline{-}(5.65,0.5)(6.35,0.5)\psline{-}(6.65,0.5)(7.35,0.5)\psline{-}(7.65,0.5)(8,0.5)
\psline{-}(0.5,0)(0.5,1)\psline{-}(1.5,0)(1.5,1)\psline{-}(2.5,0)(2.5,1)\psline{-}(3.5,0)(3.5,1)\psline{-}(4.5,0)(4.5,1)\psline{-}(5.5,0)(5.5,1)\psline{-}(6.5,0)(6.5,1)\psline{-}(7.5,0)(7.5,1)
\psline{-}(0.5,0)(0.5,-0.35)\psline{-}(0.5,-0.65)(0.5,-1)
\psline{-}(1.5,0)(1.5,-0.35)\psline{-}(1.5,-0.65)(1.5,-1)
\psline{-}(2.5,0)(2.5,-0.35)\psline{-}(2.5,-0.65)(2.5,-1)
\psline{-}(3.5,0)(3.5,-0.35)\psline{-}(3.5,-0.65)(3.5,-1)
\psline{-}(4.5,0)(4.5,-0.35)\psline{-}(4.5,-0.65)(4.5,-1)
\psline{-}(5.5,0)(5.5,-0.35)\psline{-}(5.5,-0.65)(5.5,-1)
\psline{-}(6.5,0)(6.5,-0.35)\psline{-}(6.5,-0.65)(6.5,-1)
\psline{-}(7.5,0)(7.5,-0.35)\psline{-}(7.5,-0.65)(7.5,-1)
\psset{linecolor=myc2}
\psline{-}(0.5,1)(0.5,2)
\psarc{-}(3,1){0.5}{0}{180}
\psbezier{-}(1.5,1)(1.5,2)(4.5,2)(4.5,1)
\psarc{-}(7,1){0.5}{0}{180}
\psline{-}(5.5,1)(5.5,2)
\end{pspicture}
\
\ &=
\
\psset{unit=0.8}\begin{pspicture}(-0.5,-0.2)(8.5,2.0)
\psdots(0.5,1)(1.5,1)(2.5,1)(3.5,1)(4.5,1)(5.5,1)(6.5,1)(7.5,1)
\psdots(0.5,-1)(1.5,-1)(2.5,-1)(3.5,-1)(4.5,-1)(5.5,-1)(6.5,-1)(7.5,-1)
\lw
\psline{-}(0,-1)(0,1)(8,1)(8,-1)(0,-1)
\psline{-}(0,0)(8,0)
\psline{-}(1,-1)(1,1)\psline{-}(2,-1)(2,1)\psline{-}(3,-1)(3,1)\psline{-}(4,-1)(4,1)
\psline{-}(5,-1)(5,1)\psline{-}(6,-1)(6,1)\psline{-}(7,-1)(7,1)
\psset{linecolor=myc}\unlw
\psarc{-}(0,0){0.5}{90}{270}
\psarc{-}(8,0){0.5}{270}{90}
\psline{-}(0,-0.5)(2,-0.5)\psline{-}(4,-0.5)(6,-0.5)
\psline{-}(0,0.5)(0.35,0.5)\psline{-}(0.65,0.5)(1.35,0.5)\psline{-}(1.65,0.5)(2.0,0.5)\psline{-}(4.0,0.5)(4.35,0.5)\psline{-}(4.65,0.5)(5.35,0.5)\psline{-}(5.65,0.5)(6.0,0.5)
\psline{-}(0.5,0)(0.5,1)\psline{-}(1.5,0)(1.5,1)\psline{-}(4.5,0)(4.5,1)\psline{-}(5.5,0)(5.5,1)\psline{-}(0.5,0)(0.5,-0.35)\psline{-}(0.5,-0.65)(0.5,-1)
\psline{-}(1.5,0)(1.5,-0.35)\psline{-}(1.5,-0.65)(1.5,-1)
\psline{-}(4.5,0)(4.5,-0.35)\psline{-}(4.5,-0.65)(4.5,-1)
\psline{-}(5.5,0)(5.5,-0.35)\psline{-}(5.5,-0.65)(5.5,-1)
\psarc{-}(3,-1){0.5}{0}{180}\psarc{-}(3,0){0.5}{0}{180}
\psarc{-}(2,1){0.5}{270}{360}\psarc{-}(4,1){0.5}{180}{270}
\psarc{-}(2,0){0.5}{270}{360}\psarc{-}(4,0){0.5}{180}{270}
\psarc{-}(7,-1){0.5}{0}{180}\psarc{-}(7,0){0.5}{0}{180}
\psarc{-}(6,1){0.5}{270}{360}\psarc{-}(8,1){0.5}{180}{270}
\psarc{-}(6,0){0.5}{270}{360}\psarc{-}(8,0){0.5}{180}{270}
\psset{linecolor=myc2}
\psline{-}(0.5,1)(0.5,2)
\psarc{-}(3,1){0.5}{0}{180}
\psbezier{-}(1.5,1)(1.5,2)(4.5,2)(4.5,1)
\psarc{-}(7,1){0.5}{0}{180}
\psline{-}(5.5,1)(5.5,2)
\end{pspicture}
\\
\nonumber \\
\nonumber \\
\ &=
\psset{unit=0.8}\begin{pspicture}(-0.5,-0.2)(8.5,2.0)
\psdots(0.5,1)(1.5,1)(2.5,1)(3.5,1)(4.5,1)(5.5,1)(6.5,1)(7.5,1)
\psdots(0.5,-1)(1.5,-1)(2.5,-1)(3.5,-1)(4.5,-1)(5.5,-1)(6.5,-1)(7.5,-1)\lw
\psline{-}(0,-1)(0,1)(8,1)(8,-1)(0,-1)
\psline{-}(0,0)(8,0)
\psline{-}(1,-1)(1,1)\psline{-}(2,-1)(2,1)\psline{-}(3,-1)(3,1)\psline{-}(4,-1)(4,1)
\psline{-}(5,-1)(5,1)\psline{-}(6,-1)(6,1)\psline{-}(7,-1)(7,1)
\psset{linecolor=myc}\unlw
\psarc{-}(0,0){0.5}{90}{270}
\psarc{-}(8,0){0.5}{270}{90}
\psline{-}(0,-0.5)(1,-0.5)
\psline{-}(0,0.5)(0.35,0.5)
\psline{-}(0.65,0.5)(1,0.5)
\psline{-}(5,0.5)(5.35,0.5)
\psline{-}(5,-0.5)(6,-0.5)
\psline{-}(5.65,0.5)(6.0,0.5)
\psline{-}(0.5,0)(0.5,1)
\psline{-}(5.5,0)(5.5,1)
\psline{-}(0.5,0)(0.5,-0.35)\psline{-}(0.5,-0.65)(0.5,-1)
\psarc{-}(1,1){0.5}{270}{360}\psarc{-}(2,0){0.5}{90}{180}
\psarc{-}(1,0){0.5}{270}{360}\psarc{-}(2,-1){0.5}{90}{180}
\psarc{-}(4,0){0.5}{0}{90}\psarc{-}(5,1){0.5}{180}{270}
\psarc{-}(4,-1){0.5}{0}{90}\psarc{-}(5,0){0.5}{180}{270}
\psline{-}(5.5,0)(5.5,-0.35)\psline{-}(5.5,-0.65)(5.5,-1)
\psarc{-}(3,-1){0.5}{0}{180}\psarc{-}(3,0){0.5}{0}{180}
\psarc{-}(2,1){0.5}{270}{360}\psarc{-}(4,1){0.5}{180}{270}
\psarc{-}(2,0){0.5}{270}{360}\psarc{-}(4,0){0.5}{180}{270}
\psarc{-}(7,-1){0.5}{0}{180}\psarc{-}(7,0){0.5}{0}{180}
\psarc{-}(6,1){0.5}{270}{360}\psarc{-}(8,1){0.5}{180}{270}
\psarc{-}(6,0){0.5}{270}{360}\psarc{-}(8,0){0.5}{180}{270}
\psset{linecolor=myc2}
\psline{-}(0.5,1)(0.5,2)
\psarc{-}(3,1){0.5}{0}{180}
\psbezier{-}(1.5,1)(1.5,2)(4.5,2)(4.5,1)
\psarc{-}(7,1){0.5}{0}{180}
\psline{-}(5.5,1)(5.5,2)
\end{pspicture}
\nonumber
\end{align}
\\
\\
This simple observation proves that $\rho(F_{N}(\lambda))_{d,d}$ will be a constant times the identity: the arcs in the incoming link state are carried through and are part of the outgoing one, and  if any pair of entering defects annihilates, the image is not in $V_N^d$. When an arc goes through the transfer matrix, it forces one of the two states for each box it visits, and it creates equal numbers of both box states so that the factors $ie^{i\lambda/2}$ and $-ie^{-i\lambda/2}$ cancel. All the boxes whose state has been forced by the arc can then be removed since they contribute globally by a factor of $1$. The matrix block $\rho(F_{N}(\lambda))_{d,d}$ is therefore a multiple of the identity matrix and the factor is obtained by calculating the action of $F_d(\lambda)$ on the link state with $d$ defects. The sum in the top left box is first done explicitly.
\begin{equation*}
\psset{unit=0.8}\begin{pspicture}(-0.6,-0.2)(5.6,1.8)
\psdots(0.5,1)(1.5,1)(2.5,1)(4.5,1)
\psdots(0.5,-1)(1.5,-1)(2.5,-1)(4.5,-1)
\lw
\psline{-}(0,-1)(0,1)(3,1)(3,-1)(0,-1)
\psline{-}(4,-1)(4,1)(5,1)(5,-1)(4,-1)
\psline{-}(0,0)(3,0)\psline{-}(4,0)(5,0)
\psline{-}(1,-1)(1,1)\psline{-}(2,-1)(2,1)
\psline{-}(4,-1)(4,1)\psline{-}(5,-1)(5,1)
\psline[linestyle=dashed,dash=2pt 2pt]{-}(3,-1)(4,-1)
\psline[linestyle=dashed,dash=2pt 2pt]{-}(3,0)(4,0)
\psline[linestyle=dashed,dash=2pt 2pt]{-}(3,1)(4,1)
\psset{linecolor=myc}\unlw
\psarc{-}(0,0){0.5}{90}{270}
\psarc{-}(5,0){0.5}{270}{90}
\psline{-}(0,-0.5)(3,-0.5)
\psline{-}(4,-0.5)(5,-0.5)
\psline{-}(0,0.5)(0.35,0.5)\psline{-}(0.65,0.5)(1,0.5)
\psline{-}(1.0,0.5)(1.35,0.5)\psline{-}(1.65,0.5)(2,0.5)
\psline{-}(2.0,0.5)(2.35,0.5)\psline{-}(2.65,0.5)(3,0.5)
\psline{-}(4.0,0.5)(4.35,0.5)\psline{-}(4.65,0.5)(5,0.5)
\psline{-}(0.5,0)(0.5,1)\psline{-}(1.5,0)(1.5,1)
\psline{-}(2.5,0)(2.5,1)\psline{-}(4.5,0)(4.5,1)
\psline{-}(0.5,0)(0.5,-0.35)\psline{-}(0.5,-0.65)(0.5,-1)
\psline{-}(1.5,0)(1.5,-0.35)\psline{-}(1.5,-0.65)(1.5,-1)
\psline{-}(2.5,0)(2.5,-0.35)\psline{-}(2.5,-0.65)(2.5,-1)
\psline{-}(4.5,0)(4.5,-0.35)\psline{-}(4.5,-0.65)(4.5,-1)
\psset{linecolor=myc2}
\psline{-}(0.5,1)(0.5,1.8)
\psline{-}(1.5,1)(1.5,1.8)
\psline{-}(2.5,1)(2.5,1.8)
\psline{-}(4.5,1)(4.5,1.8)
\end{pspicture}
\ =- i e^{-i \lambda/2}
\begin{pspicture}(-0.6,-0.2)(5.6,1.8)
\psdots(0.5,1)(1.5,1)(2.5,1)(4.5,1)
\psdots(0.5,-1)(1.5,-1)(2.5,-1)(4.5,-1)\lw
\psline{-}(0,-1)(0,1)(3,1)(3,-1)(0,-1)
\psline{-}(4,-1)(4,1)(5,1)(5,-1)(4,-1)
\psline{-}(0,0)(3,0)
\psline{-}(4,0)(5,0)
\psline{-}(1,-1)(1,1)\psline{-}(2,-1)(2,1)\psline{-}(3,-1)(3,1)\psline{-}(5,-1)(5,1)
\psline[linestyle=dashed,dash=2pt 2pt]{-}(3,-1)(4,-1)
\psline[linestyle=dashed,dash=2pt 2pt]{-}(3,0)(4,0)
\psline[linestyle=dashed,dash=2pt 2pt]{-}(3,1)(4,1)
\psset{linecolor=myc}\unlw
\psarc{-}(0,0){0.5}{90}{270}
\psarc{-}(5,0){0.5}{270}{90}
\psline{-}(0,-0.5)(3,-0.5)\psline{-}(4,-0.5)(5,-0.5)
\psline{-}(1.0,0.5)(1.35,0.5)\psline{-}(1.65,0.5)(2,0.5)
\psline{-}(2.0,0.5)(2.35,0.5)\psline{-}(2.65,0.5)(3,0.5)
\psline{-}(4.0,0.5)(4.35,0.5)\psline{-}(4.65,0.5)(5,0.5)
\psline{-}(1.5,0)(1.5,1)\psline{-}(2.5,0)(2.5,1)\psline{-}(4.5,0)(4.5,1)
\psline{-}(0.5,0)(0.5,-0.35)\psline{-}(0.5,-0.65)(0.5,-1)
\psline{-}(1.5,0)(1.5,-0.35)\psline{-}(1.5,-0.65)(1.5,-1)
\psline{-}(2.5,0)(2.5,-0.35)\psline{-}(2.5,-0.65)(2.5,-1)
\psline{-}(4.5,0)(4.5,-0.35)\psline{-}(4.5,-0.65)(4.5,-1)
\psarc{-}(0,0){0.5}{0}{90}
\psarc{-}(1,1){0.5}{180}{270}
\psset{linecolor=myc2}
\psline{-}(0.5,1)(0.5,1.8)
\psline{-}(1.5,1)(1.5,1.8)
\psline{-}(2.5,1)(2.5,1.8)
\psline{-}(4.5,1)(4.5,1.8)
\end{pspicture}
\ + i e^{i \lambda/2}
\begin{pspicture}(-0.6,-0.2)(5.6,1.8)
\psdots(0.5,1)(1.5,1)(2.5,1)(4.5,1)
\psdots(0.5,-1)(1.5,-1)(2.5,-1)(4.5,-1)\lw
\psline{-}(0,-1)(0,1)(3,1)(3,-1)(0,-1)
\psline{-}(4,-1)(4,1)(5,1)(5,-1)(4,-1)
\psline{-}(0,0)(3,0)\psline{-}(4,0)(5,0)
\psline{-}(1,-1)(1,1)\psline{-}(2,-1)(2,1)
\psline{-}(3,-1)(3,1)\psline{-}(4,-1)(4,1)
\psline[linestyle=dashed,dash=2pt 2pt]{-}(3,-1)(4,-1)
\psline[linestyle=dashed,dash=2pt 2pt]{-}(3,0)(4,0)
\psline[linestyle=dashed,dash=2pt 2pt]{-}(3,1)(4,1)
\psset{linecolor=myc}\unlw
\psarc{-}(0,0){0.5}{90}{270}
\psarc{-}(5,0){0.5}{270}{90}
\psline{-}(0,-0.5)(3,-0.5)\psline{-}(4,-0.5)(5,-0.5)
\psline{-}(1.0,0.5)(1.35,0.5)\psline{-}(1.65,0.5)(2,0.5)
\psline{-}(2.0,0.5)(2.35,0.5)\psline{-}(2.65,0.5)(3,0.5)
\psline{-}(4.0,0.5)(4.35,0.5)\psline{-}(4.65,0.5)(5,0.5)
\psline{-}(1.5,0)(1.5,1)\psline{-}(2.5,0)(2.5,1)\psline{-}(4.5,0)(4.5,1)
\psline{-}(0.5,0)(0.5,-0.35)\psline{-}(0.5,-0.65)(0.5,-1)
\psline{-}(1.5,0)(1.5,-0.35)\psline{-}(1.5,-0.65)(1.5,-1)
\psline{-}(2.5,0)(2.5,-0.35)\psline{-}(2.5,-0.65)(2.5,-1)
\psline{-}(4.5,0)(4.5,-0.35)\psline{-}(4.5,-0.65)(4.5,-1)
\psarc{-}(0,1){0.5}{270}{360}
\psarc{-}(1,0){0.5}{90}{180}
\psset{linecolor=myc2}
\psline{-}(0.5,1)(0.5,1.8)
\psline{-}(1.5,1)(1.5,1.8)
\psline{-}(2.5,1)(2.5,1.8)
\psline{-}(4.5,1)(4.5,1.8)
\end{pspicture}
\end{equation*}
\\ \\
Because only the diagonal element is of interest, states with a smaller number of defects are not considered. In the following equalities, the ``$d$'' over the equal sign stresses the fact that equalities are up to states whose number of defects is smaller than $d$. In many braid boxes, this constraint chooses one of the two states. 
The first term above is
\begin{align*}(-ie^{-i\lambda/2})\psset{unit=0.8}
\begin{pspicture}(-0.6,0)(5.6,1.85)
\psdots(0.5,1)(1.5,1)(2.5,1)(4.5,1)
\psdots(0.5,-1)(1.5,-1)(2.5,-1)(4.5,-1)\lw
\psline{-}(0,-1)(0,1)(3,1)(3,-1)(0,-1)
\psline{-}(4,-1)(4,1)(5,1)(5,-1)(4,-1)
\psline{-}(0,0)(3,0)
\psline{-}(4,0)(5,0)
\psline{-}(1,-1)(1,1)\psline{-}(2,-1)(2,1)\psline{-}(3,-1)(3,1)\psline{-}(5,-1)(5,1)
\psline[linestyle=dashed,dash=2pt 2pt]{-}(3,-1)(4,-1)
\psline[linestyle=dashed,dash=2pt 2pt]{-}(3,0)(4,0)
\psline[linestyle=dashed,dash=2pt 2pt]{-}(3,1)(4,1)
\psset{linecolor=myc}\unlw
\psarc{-}(0,0){0.5}{90}{270}
\psarc{-}(5,0){0.5}{270}{90}
\psline{-}(0,-0.5)(3,-0.5)\psline{-}(4,-0.5)(5,-0.5)
\psline{-}(1.0,0.5)(1.35,0.5)\psline{-}(1.65,0.5)(2,0.5)
\psline{-}(2.0,0.5)(2.35,0.5)\psline{-}(2.65,0.5)(3,0.5)
\psline{-}(4.0,0.5)(4.35,0.5)\psline{-}(4.65,0.5)(5,0.5)
\psline{-}(1.5,0)(1.5,1)\psline{-}(2.5,0)(2.5,1)\psline{-}(4.5,0)(4.5,1)
\psline{-}(0.5,0)(0.5,-0.35)\psline{-}(0.5,-0.65)(0.5,-1)
\psline{-}(1.5,0)(1.5,-0.35)\psline{-}(1.5,-0.65)(1.5,-1)
\psline{-}(2.5,0)(2.5,-0.35)\psline{-}(2.5,-0.65)(2.5,-1)
\psline{-}(4.5,0)(4.5,-0.35)\psline{-}(4.5,-0.65)(4.5,-1)
\psarc{-}(0,0){0.5}{0}{90}
\psarc{-}(1,1){0.5}{180}{270}
\psset{linecolor=myc2}
\psline{-}(0.5,1)(0.5,1.8)
\psline{-}(1.5,1)(1.5,1.8)
\psline{-}(2.5,1)(2.5,1.8)
\psline{-}(4.5,1)(4.5,1.8)
\end{pspicture}
&
\overset{d}{=}
(-ie^{-i\lambda/2})^2
\psset{unit=0.8}\begin{pspicture}(-0.6,0.)(5.6,1.85)
\psdots(0.5,1)(1.5,1)(2.5,1)(4.5,1)
\psdots(0.5,-1)(1.5,-1)(2.5,-1)(4.5,-1)\lw
\psline{-}(0,-1)(0,1)(3,1)(3,-1)(0,-1)
\psline{-}(5,-1)(5,1)(4,1)(4,-1)(5,-1)
\psline{-}(0,0)(3,0)\psline{-}(5,0)(4,0)
\psline{-}(1,-1)(1,1)\psline{-}(2,-1)(2,1)\
\psline{-}(3,-1)(3,1)\psline{-}(4,-1)(4,1)
\psline[linestyle=dashed,dash=2pt 2pt]{-}(3,-1)(4,-1)
\psline[linestyle=dashed,dash=2pt 2pt]{-}(3,0)(4,0)
\psline[linestyle=dashed,dash=2pt 2pt]{-}(3,1)(4,1)
\psset{linecolor=myc}\unlw
\psarc{-}(0,0){0.5}{90}{270}
\psarc{-}(5,0){0.5}{270}{90}
\psline{-}(0,-0.5)(3,-0.5)\psline{-}(5,-0.5)(4,-0.5)
\psline{-}(2.0,0.5)(2.35,0.5)\psline{-}(2.65,0.5)(3,0.5)
\psline{-}(4.0,0.5)(4.35,0.5)\psline{-}(4.65,0.5)(5,0.5)
\psline{-}(2.5,0)(2.5,1)\psline{-}(4.5,0)(4.5,1)
\psline{-}(0.5,0)(0.5,-0.35)\psline{-}(0.5,-0.65)(0.5,-1)
\psline{-}(1.5,0)(1.5,-0.35)\psline{-}(1.5,-0.65)(1.5,-1)
\psline{-}(2.5,0)(2.5,-0.35)\psline{-}(2.5,-0.65)(2.5,-1)
\psline{-}(4.5,0)(4.5,-0.35)\psline{-}(4.5,-0.65)(4.5,-1)
\psarc{-}(0,0){0.5}{0}{90}
\psarc{-}(1,0){0.5}{0}{90}
\psarc{-}(1,1){0.5}{180}{270}
\psarc{-}(2,1){0.5}{180}{270}
\psset{linecolor=myc2}
\psline{-}(0.5,1)(0.5,1.8)
\psline{-}(1.5,1)(1.5,1.8)
\psline{-}(2.5,1)(2.5,1.8)
\psline{-}(4.5,1)(4.5,1.8)
\end{pspicture}\\
\\
&
\overset{d}{=}
(-ie^{-i\lambda/2})^{(2d-1)}
\psset{unit=0.8}\begin{pspicture}(-0.6,0.)(5.6,2.45)
\psdots(0.5,1)(1.5,1)(2.5,1)(4.5,1)
\psdots(0.5,-1)(1.5,-1)(2.5,-1)(4.5,-1)\lw
\psline{-}(0,-1)(0,1)(3,1)(3,-1)(0,-1)
\psline{-}(5,-1)(5,1)(4,1)(4,-1)(5,-1)
\psline{-}(0,0)(3,0)\psline{-}(5,0)(4,0)
\psline{-}(1,-1)(1,1)\psline{-}(2,-1)(2,1)
\psline{-}(3,-1)(3,1)\psline{-}(4,-1)(4,1)
\psline[linestyle=dashed,dash=2pt 2pt]{-}(3,-1)(4,-1)
\psline[linestyle=dashed,dash=2pt 2pt]{-}(3,0)(4,0)
\psline[linestyle=dashed,dash=2pt 2pt]{-}(3,1)(4,1)
\psset{linecolor=myc}\unlw
\psarc{-}(0,0){0.5}{90}{270}
\psarc{-}(5,0){0.5}{270}{90}
\psline{-}(0,-0.5)(1,-0.5)
\psline{-}(0.5,0)(0.5,-0.35)\psline{-}(0.5,-0.65)(0.5,-1)
\psarc{-}(0,0){0.5}{0}{90}
\psarc{-}(1,0){0.5}{270}{90}
\psarc{-}(2,0){0.5}{270}{90}
\psarc{-}(4,0){0.5}{270}{90}
\psarc{-}(1,1){0.5}{180}{270}
\psarc{-}(2,1){0.5}{180}{270}
\psarc{-}(3,1){0.5}{180}{270}
\psarc{-}(5,1){0.5}{180}{270}
\psarc{-}(2,-1){0.5}{90}{180}
\psarc{-}(3,-1){0.5}{90}{180}
\psarc{-}(5,-1){0.5}{90}{180}
\psset{linecolor=myc2}
\psline{-}(0.5,1)(0.5,1.8)
\psline{-}(1.5,1)(1.5,1.8)
\psline{-}(2.5,1)(2.5,1.8)
\psline{-}(4.5,1)(4.5,1.8)
\end{pspicture}\\
\\
&
=(-1)^d  e^{-i \lambda(d+1)}
\psset{unit=0.8}\begin{pspicture}(-0.6,0.25)(5.6,2.45)
\psdots(0.5,1)(1.5,1)(2.5,1)(4.5,1)
\psdots(0.5,-1)(1.5,-1)(2.5,-1)(4.5,-1)\lw
\psline{-}(0,-1)(0,1)(3,1)(3,-1)(0,-1)
\psline{-}(5,-1)(5,1)(4,1)(4,-1)(5,-1)
\psline{-}(0,0)(3,0)\psline{-}(5,0)(4,0)
\psline{-}(1,-1)(1,1)\psline{-}(2,-1)(2,1)
\psline{-}(3,-1)(3,1)\psline{-}(4,-1)(4,1)
\psline[linestyle=dashed,dash=2pt 2pt]{-}(3,-1)(4,-1)
\psline[linestyle=dashed,dash=2pt 2pt]{-}(3,0)(4,0)
\psline[linestyle=dashed,dash=2pt 2pt]{-}(3,1)(4,1)
\psset{linecolor=myc}\unlw
\psarc{-}(5,0){0.5}{270}{90}
\psarc{-}(1,0){0.5}{270}{90}
\psarc{-}(2,0){0.5}{270}{90}
\psarc{-}(4,0){0.5}{270}{90}
\psarc{-}(1,1){0.5}{180}{270}
\psarc{-}(2,1){0.5}{180}{270}
\psarc{-}(3,1){0.5}{180}{270}
\psarc{-}(5,1){0.5}{180}{270}
\psarc{-}(1,-1){0.5}{90}{180}
\psarc{-}(2,-1){0.5}{90}{180}
\psarc{-}(3,-1){0.5}{90}{180}
\psarc{-}(5,-1){0.5}{90}{180}
\psset{linecolor=myc2}
\psline{-}(0.5,1)(0.5,1.8)
\psline{-}(1.5,1)(1.5,1.8)
\psline{-}(2.5,1)(2.5,1.8)
\psline{-}(4.5,1)(4.5,1.8)
\end{pspicture}
\end{align*}
\\
\\
\\
We have used the twist relation to obtain the last equality. The second term is computed as follows.
\begin{align*}(ie^{i\lambda/2})
\psset{unit=0.8}\begin{pspicture}(-0.6,0.2)(5.6,1.9)
\psdots(0.5,1)(1.5,1)(2.5,1)(4.5,1)
\psdots(0.5,-1)(1.5,-1)(2.5,-1)(4.5,-1)\lw
\psline{-}(0,-1)(0,1)(3,1)(3,-1)(0,-1)
\psline{-}(0,0)(3,0)
\psline{-}(5,-1)(5,1)(4,1)(4,-1)(5,-1)
\psline{-}(5,0)(4,0)
\psline{-}(1,-1)(1,1)\psline{-}(2,-1)(2,1)\psline{-}(3,-1)(3,1)
\psline{-}(4,-1)(4,1)\unlw
\psarc[linecolor=myc]{-}(0,0){0.5}{90}{270}
\psarc[linecolor=myc]{-}(5,0){0.5}{270}{90}
\psline[linestyle=dashed,dash=2pt 2pt]{-}(3,-1)(4,-1)
\psline[linestyle=dashed,dash=2pt 2pt]{-}(3,0)(4,0)
\psline[linestyle=dashed,dash=2pt 2pt]{-}(3,1)(4,1)
\psline[linecolor=myc]{-}(0,-0.5)(3,-0.5)\psline[linecolor=myc]{-}(5,-0.5)(4,-0.5)
\psline[linecolor=myc]{-}(1.0,0.5)(1.35,0.5)\psline[linecolor=myc]{-}(1.65,0.5)(2,0.5)
\psline[linecolor=myc]{-}(2.0,0.5)(2.35,0.5)\psline[linecolor=myc]{-}(2.65,0.5)(3,0.5)
\psline[linecolor=myc]{-}(4.0,0.5)(4.35,0.5)\psline[linecolor=myc]{-}(4.65,0.5)(5,0.5)
\psline[linecolor=myc]{-}(1.5,0)(1.5,1)\psline[linecolor=myc]{-}(2.5,0)(2.5,1)
\psline[linecolor=myc]{-}(4.5,0)(4.5,1)
\psline[linecolor=myc]{-}(0.5,0)(0.5,-0.35)\psline[linecolor=myc]{-}(0.5,-0.65)(0.5,-1)
\psline[linecolor=myc]{-}(1.5,0)(1.5,-0.35)\psline[linecolor=myc]{-}(1.5,-0.65)(1.5,-1)
\psline[linecolor=myc]{-}(2.5,0)(2.5,-0.35)\psline[linecolor=myc]{-}(2.5,-0.65)(2.5,-1)
\psline[linecolor=myc]{-}(4.5,0)(4.5,-0.35)\psline[linecolor=myc]{-}(4.5,-0.65)(4.5,-1)
\psarc[linecolor=myc]{-}(1,0){0.5}{90}{180}
\psarc[linecolor=myc]{-}(0,1){0.5}{270}{360}
\psset{linecolor=myc2}
\psline{-}(0.5,1)(0.5,1.8)
\psline{-}(1.5,1)(1.5,1.8)
\psline{-}(2.5,1)(2.5,1.8)
\psline{-}(4.5,1)(4.5,1.8)
\end{pspicture}
&\overset{d}{=} (ie^{i\lambda/2})^2
\psset{unit=0.8}\begin{pspicture}(-0.6,0.2)(5.6,1.9)
\psdots(0.5,1)(1.5,1)(2.5,1)(4.5,1)
\psdots(0.5,-1)(1.5,-1)(2.5,-1)(4.5,-1)\lw
\psline{-}(0,-1)(0,1)(3,1)(3,-1)(0,-1)
\psline{-}(0,0)(3,0)
\psline{-}(5,-1)(5,1)(4,1)(4,-1)(5,-1)
\psline{-}(5,0)(4,0)
\psline{-}(1,-1)(1,1)\psline{-}(2,-1)(2,1)
\psline{-}(3,-1)(3,1)\psline{-}(4,-1)(4,1)
\psline[linestyle=dashed,dash=2pt 2pt]{-}(3,-1)(4,-1)
\psline[linestyle=dashed,dash=2pt 2pt]{-}(3,0)(4,0)
\psline[linestyle=dashed,dash=2pt 2pt]{-}(3,1)(4,1)\unlw
\psset{linecolor=myc}
\psarc{-}(0,0){0.5}{90}{270}
\psarc{-}(5,0){0.5}{270}{90}
\psline{-}(0,-0.5)(3,-0.5)\psline{-}(5,-0.5)(4,-0.5)
\psline{-}(2.0,0.5)(2.35,0.5)\psline{-}(2.65,0.5)(3,0.5)
\psline{-}(4.0,0.5)(4.35,0.5)\psline{-}(4.65,0.5)(5,0.5)
\psline{-}(2.5,0)(2.5,1)\psline{-}(4.5,0)(4.5,1)
\psline{-}(0.5,0)(0.5,-0.35)\psline{-}(0.5,-0.65)(0.5,-1)
\psline{-}(1.5,0)(1.5,-0.35)\psline{-}(1.5,-0.65)(1.5,-1)
\psline{-}(2.5,0)(2.5,-0.35)\psline{-}(2.5,-0.65)(2.5,-1)
\psline{-}(4.5,0)(4.5,-0.35)\psline{-}(4.5,-0.65)(4.5,-1)
\psarc{-}(1,0){0.5}{90}{180}
\psarc{-}(0,1){0.5}{270}{360}
\psarc{-}(2,0){0.5}{90}{180}
\psarc{-}(1,1){0.5}{270}{360}
\psset{linecolor=myc2}
\psline{-}(0.5,1)(0.5,1.8)
\psline{-}(1.5,1)(1.5,1.8)
\psline{-}(2.5,1)(2.5,1.8)
\psline{-}(4.5,1)(4.5,1.8)
\end{pspicture}
+
\begin{pspicture}(-0.6,0.2)(5.6,1.9)
\psdots(0.5,1)(1.5,1)(2.5,1)(4.5,1)
\psdots(0.5,-1)(1.5,-1)(2.5,-1)(4.5,-1)\lw
\psline{-}(0,-1)(0,1)(3,1)(3,-1)(0,-1)
\psline{-}(0,0)(3,0)
\psline{-}(5,-1)(5,1)(4,1)(4,-1)(5,-1)
\psline{-}(5,0)(4,0)
\psline{-}(1,-1)(1,1)\psline{-}(2,-1)(2,1)
\psline{-}(3,-1)(3,1)\psline{-}(4,-1)(4,1)
\psline[linestyle=dashed,dash=2pt 2pt]{-}(3,-1)(4,-1)
\psline[linestyle=dashed,dash=2pt 2pt]{-}(3,0)(4,0)
\psline[linestyle=dashed,dash=2pt 2pt]{-}(3,1)(4,1)
\psset{linecolor=myc}\unlw
\psarc{-}(0,0){0.5}{90}{270}
\psarc{-}(5,0){0.5}{270}{90}
\psline{-}(0,-0.5)(3,-0.5)\psline{-}(5,-0.5)(4,-0.5)
\psline{-}(2.0,0.5)(2.35,0.5)\psline{-}(2.65,0.5)(3,0.5)
\psline{-}(4.0,0.5)(4.35,0.5)\psline{-}(4.65,0.5)(5,0.5)
\psline{-}(2.5,0)(2.5,1)\psline{-}(4.5,0)(4.5,1)
\psline{-}(0.5,0)(0.5,-0.35)\psline{-}(0.5,-0.65)(0.5,-1)
\psline{-}(1.5,0)(1.5,-0.35)\psline{-}(1.5,-0.65)(1.5,-1)
\psline{-}(2.5,0)(2.5,-0.35)\psline{-}(2.5,-0.65)(2.5,-1)
\psline{-}(4.5,0)(4.5,-0.35)\psline{-}(4.5,-0.65)(4.5,-1)
\psarc{-}(1,0){0.5}{0}{180}
\psarc{-}(0,1){0.5}{270}{360}
\psarc{-}(2,1){0.5}{180}{270}
\psset{linecolor=myc2}
\psline{-}(0.5,1)(0.5,1.8)
\psline{-}(1.5,1)(1.5,1.8)
\psline{-}(2.5,1)(2.5,1.8)
\psline{-}(4.5,1)(4.5,1.8)
\end{pspicture}
\\
\\
&\overset{d}{=}  (ie^{i\lambda/2})^d
\psset{unit=0.8}\begin{pspicture}(-0.6,0.3)(5.6,2.45)
\psdots(0.5,1)(1.5,1)(2.5,1)(4.5,1)
\psdots(0.5,-1)(1.5,-1)(2.5,-1)(4.5,-1)\lw
\psline{-}(0,-1)(0,1)(3,1)(3,-1)(0,-1)
\psline{-}(0,0)(3,0)
\psline{-}(5,-1)(5,1)(4,1)(4,-1)(5,-1)
\psline{-}(5,0)(4,0)
\psline{-}(1,-1)(1,1)\psline{-}(2,-1)(2,1)
\psline{-}(3,-1)(3,1)\psline{-}(4,-1)(4,1)
\psline[linestyle=dashed,dash=2pt 2pt]{-}(3,-1)(4,-1)
\psline[linestyle=dashed,dash=2pt 2pt]{-}(3,0)(4,0)
\psline[linestyle=dashed,dash=2pt 2pt]{-}(3,1)(4,1)
\psset{linecolor=myc}\unlw
\psarc{-}(0,0){0.5}{90}{270}
\psarc{-}(5,0){0.5}{270}{90}
\psline{-}(0,-0.5)(3,-0.5)\psline{-}(5,-0.5)(4,-0.5)
\psline{-}(0.5,0)(0.5,-0.35)\psline{-}(0.5,-0.65)(0.5,-1)
\psline{-}(1.5,0)(1.5,-0.35)\psline{-}(1.5,-0.65)(1.5,-1)
\psline{-}(2.5,0)(2.5,-0.35)\psline{-}(2.5,-0.65)(2.5,-1)
\psline{-}(4.5,0)(4.5,-0.35)\psline{-}(4.5,-0.65)(4.5,-1)
\psarc{-}(3,0){0.5}{90}{180}
\psarc{-}(1,0){0.5}{90}{180}
\psarc{-}(2,0){0.5}{90}{180}
\psarc{-}(5,0){0.5}{90}{180}
\psarc{-}(0,1){0.5}{270}{360}
\psarc{-}(1,1){0.5}{270}{360}
\psarc{-}(2,1){0.5}{270}{360}
\psarc{-}(4,1){0.5}{270}{360}
\psset{linecolor=myc2}
\psline{-}(0.5,1)(0.5,1.8)
\psline{-}(1.5,1)(1.5,1.8)
\psline{-}(2.5,1)(2.5,1.8)
\psline{-}(4.5,1)(4.5,1.8)
\end{pspicture}
\\
\\
& = (ie^{i\lambda/2})^{(d+1)}e^{i\lambda}
\psset{unit=0.8}\begin{pspicture}(-0.6,0.3)(5.6,2.45)
\psdots(0.5,1)(1.5,1)(2.5,1)(4.5,1)
\psdots(0.5,-1)(1.5,-1)(2.5,-1)(4.5,-1)\lw
\psline{-}(0,-1)(0,1)(3,1)(3,-1)(0,-1)
\psline{-}(0,0)(3,0)
\psline{-}(5,-1)(5,1)(4,1)(4,-1)(5,-1)
\psline{-}(5,0)(4,0)
\psline{-}(1,-1)(1,1)\psline{-}(2,-1)(2,1)
\psline{-}(3,-1)(3,1)\psline{-}(4,-1)(4,1)
\psline[linestyle=dashed,dash=2pt 2pt]{-}(3,-1)(4,-1)
\psline[linestyle=dashed,dash=2pt 2pt]{-}(3,0)(4,0)
\psline[linestyle=dashed,dash=2pt 2pt]{-}(3,1)(4,1)
\psset{linecolor=myc}\unlw
\psarc{-}(0,0){0.5}{90}{270}
\psline{-}(0,-0.5)(3,-0.5)
\psline{-}(0.5,0)(0.5,-0.35)\psline{-}(0.5,-0.65)(0.5,-1)
\psline{-}(1.5,0)(1.5,-0.35)\psline{-}(1.5,-0.65)(1.5,-1)
\psline{-}(2.5,0)(2.5,-0.35)\psline{-}(2.5,-0.65)(2.5,-1)
\psarc{-}(4,-1){0.5}{0}{90}
\psarc{-}(3,0){0.5}{90}{180}
\psarc{-}(1,0){0.5}{90}{180}
\psarc{-}(2,0){0.5}{90}{180}
\psarc{-}(0,1){0.5}{270}{360}
\psarc{-}(1,1){0.5}{270}{360}
\psarc{-}(2,1){0.5}{270}{360}
\psarc{-}(4,1){0.5}{270}{360}
\psset{linecolor=myc2}
\psline{-}(0.5,1)(0.5,1.8)
\psline{-}(1.5,1)(1.5,1.8)
\psline{-}(2.5,1)(2.5,1.8)
\psline{-}(4.5,1)(4.5,1.8)
\end{pspicture}
\\
\\
& \overset{d}{=} (-1)^d e^{i\lambda(d+1)}
\psset{unit=0.8}\begin{pspicture}(-0.6,-0.3)(5.6,2.45)
\psdots(0.5,1)(1.5,1)(2.5,1)(4.5,1)
\psdots(0.5,-1)(1.5,-1)(2.5,-1)(4.5,-1)\lw
\psline{-}(0,-1)(0,1)(3,1)(3,-1)(0,-1)
\psline{-}(0,0)(3,0)
\psline{-}(5,-1)(5,1)(4,1)(4,-1)(5,-1)
\psline{-}(5,0)(4,0)
\psline{-}(1,-1)(1,1)\psline{-}(2,-1)(2,1)
\psline{-}(3,-1)(3,1)\psline{-}(4,-1)(4,1)
\psline[linestyle=dashed,dash=2pt 2pt]{-}(3,-1)(4,-1)
\psline[linestyle=dashed,dash=2pt 2pt]{-}(3,0)(4,0)
\psline[linestyle=dashed,dash=2pt 2pt]{-}(3,1)(4,1)
\psset{linecolor=myc}\unlw
\psarc{-}(0,0){0.5}{90}{270}
\psarc{-}(0,1){0.5}{270}{360}
\psarc{-}(1,1){0.5}{270}{360}
\psarc{-}(2,1){0.5}{270}{360}
\psarc{-}(4,1){0.5}{270}{360}
\psarc{-}(1,0){0.5}{90}{270}
\psarc{-}(2,0){0.5}{90}{270}
\psarc{-}(3,0){0.5}{90}{270}
\psarc{-}(0,-1){0.5}{0}{90}
\psarc{-}(1,-1){0.5}{0}{90}
\psarc{-}(2,-1){0.5}{0}{90}
\psarc{-}(4,-1){0.5}{0}{90}
\psset{linecolor=myc2}
\psline{-}(0.5,1)(0.5,1.8)
\psline{-}(1.5,1)(1.5,1.8)
\psline{-}(2.5,1)(2.5,1.8)
\psline{-}(4.5,1)(4.5,1.8)
\end{pspicture}
\end{align*}
\\
\\
The last term of the first line does not contribute to $\rho(F_N(\lambda))_{d,d}$ because (\ref{eq:semicircle}) causes the arc to propagate downwards to the bottom, decreasing the number of defects by $2$. The sum of the two contributions is the constant in (\ref{eq:cmaxdiag}). (Another proof of this identity is given in Appendix \ref{app:b}.)

To prove statement {\em (iii)}, that $F_{N}$ is a central element in $TL_N$, it is sufficient to verify that $F_N$ commutes with the generators $e_i$. This is a direct consequence of the identity \eqref{eq:semicircle} and an easy computation. For example
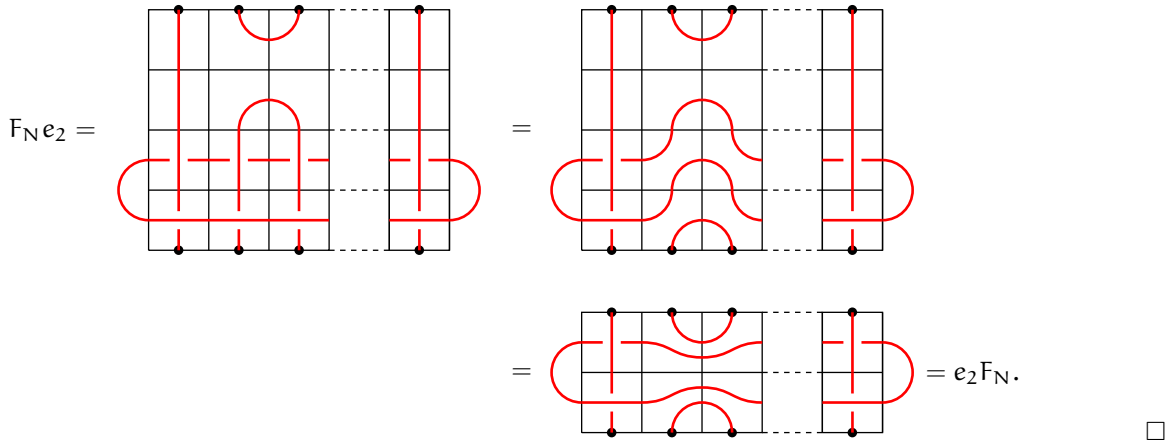
\begin{align*}
\psset{unit=0.8}\begin{pspicture}(-1.6,-0.2)(5.6,3.)
\rput(-1.6,1.){$F_Ne_2=$}
\psdots(0.5,3)(1.5,3)(2.5,3)(4.5,3)
\psdots(0.5,-1)(1.5,-1)(2.5,-1)(4.5,-1)
\lw
\psline{-}(0,-1)(0,3)(3,3)(3,-1)(0,-1)
\psline{-}(4,-1)(4,3)(5,3)(5,-1)(4,-1)
\psline{-}(0,1)(3,1)\psline(4,1)(5,1)
\psline{-}(0,2)(3,2)\psline(4,2)(5,2)
\psline{-}(0,0)(3,0)\psline{-}(4,0)(5,0)
\psline{-}(1,-1)(1,3)\psline{-}(2,-1)(2,3)
\psline{-}(4,-1)(4,3)\psline{-}(5,-1)(5,3)
\psline[linestyle=dashed,dash=2pt 2pt]{-}(3,-1)(4,-1)
\psline[linestyle=dashed,dash=2pt 2pt]{-}(3,0)(4,0)
\psline[linestyle=dashed,dash=2pt 2pt]{-}(3,1)(4,1)
\psline[linestyle=dashed,dash=2pt 2pt]{-}(3,2)(4,2)
\psline[linestyle=dashed,dash=2pt 2pt]{-}(3,3)(4,3)
\psset{linecolor=myc}\unlw
\psarc{-}(0,0){0.5}{90}{270}
\psarc{-}(5,0){0.5}{270}{90}
\psline{-}(0,-0.5)(3,-0.5)
\psline{-}(4,-0.5)(5,-0.5)
\psline{-}(0,0.5)(0.35,0.5)\psline{-}(0.65,0.5)(1,0.5)
\psline{-}(1.0,0.5)(1.35,0.5)\psline{-}(1.65,0.5)(2,0.5)
\psline{-}(2.0,0.5)(2.35,0.5)\psline{-}(2.65,0.5)(3,0.5)
\psline{-}(4.0,0.5)(4.35,0.5)\psline{-}(4.65,0.5)(5,0.5)
\psline{-}(0.5,0)(0.5,3)\psline{-}(1.5,0)(1.5,1)
\psline{-}(2.5,0)(2.5,1)\psline{-}(4.5,0)(4.5,3)
\psarc{-}(2,1){0.5}{0}{180}
\psarc{-}(2,3){0.5}{180}{360}
\psline{-}(0.5,0)(0.5,-0.35)\psline{-}(0.5,-0.65)(0.5,-1)
\psline{-}(1.5,0)(1.5,-0.35)\psline{-}(1.5,-0.65)(1.5,-1)
\psline{-}(2.5,0)(2.5,-0.35)\psline{-}(2.5,-0.65)(2.5,-1)
\psline{-}(4.5,0)(4.5,-0.35)\psline{-}(4.5,-0.65)(4.5,-1)
\end{pspicture}
&\psset{unit=0.8}\begin{pspicture}(-1.6,-0.2)(5.6,3.)
\rput(-1.0,1.){$=$}
\psdots(0.5,3)(1.5,3)(2.5,3)(4.5,3)
\psdots(0.5,-1)(1.5,-1)(2.5,-1)(4.5,-1)
\lw
\psline{-}(0,-1)(0,3)(3,3)(3,-1)(0,-1)
\psline{-}(4,-1)(4,3)(5,3)(5,-1)(4,-1)
\psline{-}(0,1)(3,1)\psline(4,1)(5,1)
\psline{-}(0,2)(3,2)\psline(4,2)(5,2)
\psline{-}(0,0)(3,0)\psline{-}(4,0)(5,0)
\psline{-}(1,-1)(1,3)\psline{-}(2,-1)(2,3)
\psline{-}(4,-1)(4,3)\psline{-}(5,-1)(5,3)
\psline[linestyle=dashed,dash=2pt 2pt]{-}(3,-1)(4,-1)
\psline[linestyle=dashed,dash=2pt 2pt]{-}(3,0)(4,0)
\psline[linestyle=dashed,dash=2pt 2pt]{-}(3,1)(4,1)
\psline[linestyle=dashed,dash=2pt 2pt]{-}(3,2)(4,2)
\psline[linestyle=dashed,dash=2pt 2pt]{-}(3,3)(4,3)
\psset{linecolor=myc}\unlw
\psarc{-}(0,0){0.5}{90}{270}
\psarc{-}(5,0){0.5}{270}{90}
\psline{-}(0,-0.5)(1,-0.5)
\psarc{-}(2,0){0.5}{0}{180}
\psarc{-}(2,-1){0.5}{0}{180}
\psarc{-}(1,0){0.5}{-90}{0}
\psarc{-}(1,1){0.5}{-90}{0}
\psarc{-}(3,0){0.5}{180}{270}
\psarc{-}(3,1){0.5}{180}{270}
\psline{-}(4,-0.5)(5,-0.5)
\psline{-}(0,0.5)(0.35,0.5)\psline{-}(0.65,0.5)(1,0.5)
\psline{-}(4.0,0.5)(4.35,0.5)\psline{-}(4.65,0.5)(5,0.5)
\psline{-}(0.5,0)(0.5,3)
\psline{-}(4.5,0)(4.5,3)
\psarc{-}(2,1){0.5}{0}{180}
\psarc{-}(2,3){0.5}{180}{360}
\psline{-}(0.5,0)(0.5,-0.35)\psline{-}(0.5,-0.65)(0.5,-1)
\psline{-}(4.5,0)(4.5,-0.35)\psline{-}(4.5,-0.65)(4.5,-1)
\end{pspicture}\\
&
\psset{unit=0.8}\begin{pspicture}(-1.6,-0.2)(7.6,2.5)
\rput(-1,0.){$=$}
\psdots(0.5,1)(1.5,1)(2.5,1)(4.5,1)
\psdots(0.5,-1)(1.5,-1)(2.5,-1)(4.5,-1)
\lw
\psline{-}(0,-1)(0,1)(3,1)(3,-1)(0,-1)
\psline{-}(4,-1)(4,1)(5,1)(5,-1)(4,-1)
\psline{-}(0,0)(3,0)\psline{-}(4,0)(5,0)
\psline{-}(1,-1)(1,1)\psline{-}(2,-1)(2,1)
\psline{-}(4,-1)(4,1)\psline{-}(5,-1)(5,1)
\psline[linestyle=dashed,dash=2pt 2pt]{-}(3,-1)(4,-1)
\psline[linestyle=dashed,dash=2pt 2pt]{-}(3,0)(4,0)
\psline[linestyle=dashed,dash=2pt 2pt]{-}(3,1)(4,1)
\psset{linecolor=myc}\unlw
\psarc{-}(0,0){0.5}{90}{270}
\psarc{-}(5,0){0.5}{270}{90}
\psline{-}(0,-0.5)(1,-0.5)
\psline{-}(4,-0.5)(5,-0.5)
\psline{-}(0,0.5)(0.35,0.5)\psline{-}(0.65,0.5)(1,0.5)
\psline{-}(4.0,0.5)(4.35,0.5)\psline{-}(4.65,0.5)(5,0.5)
\psline{-}(0.5,0)(0.5,1)\psline{-}(4.5,0)(4.5,1)
\psline{-}(0.5,0)(0.5,-0.35)\psline{-}(0.5,-0.65)(0.5,-1)
\psline{-}(4.5,0)(4.5,-0.35)\psline{-}(4.5,-0.65)(4.5,-1)
\psarc{-}(2,-1){0.5}{0}{180}
\psarc{-}(2,1){0.5}{180}{360}
\psbezier{-}(1,-0.5)(1.5,-0.5)(1.5,-0.25)(2.,-0.25)
\psbezier{-}(3,-0.5)(2.5,-0.5)(2.5,-0.25)(2.,-0.25)
\psbezier{-}(1,0.5)(1.5,0.5)(1.5,0.25)(2.,0.25)
\psbezier{-}(3,0.5)(2.5,0.5)(2.5,0.25)(2.,0.25)
\rput(6.5,0){$=e_2F_N.$}
\end{pspicture}
\end{align*}
\hfill $\square$

\medskip

Even though equation \eqref{eq:semicircle} was introduced as a tool to prove the previous lemma, it implies that $\rho(F_N(\lambda))$ can be computed recursively with minimal cost. In fact, the knowledge of $\rho(F_{n}(\lambda))$, $n<N$, determines with no further computation all columns of $\rho(F_N(\lambda))$ except the last one. This can be seen easily. Let $v$ be any vector in $B_N$ that has at least one arc, that is, any basis vector except the last one that has $N$ defects. Then, when $\rho(F_N(\lambda))$ acts on $v$, all boxes under the arcs of $v$ are fixed by \eqref{eq:semicircle} and only the boxes of $\rho(F_N(\lambda))$ under defects of $v$ remain to be summed. Moreover these boxes are connected horizontally by the arcs of $v$ that have moved downward into $\rho(F_N(\lambda))$. All boxes that have been fixed by the arcs can be removed and the remaining column of $\rho(F_N(\lambda))$ glued together. If $v\in B_N^d$, then the action of $\rho(F_N(\lambda))$ on $v$ is obtained by the action of $\rho(F_d)$ on the vector with $d$ defects. Since $d<N$, this information is contained in the last column of $\rho(F_d(\lambda))$ and, to obtain the components of the vector $\rho(F_N(\lambda))v$, one has simply to insert back the arcs into the basis vectors that correspond to non-zero elements of this last column. This observation that allows for a recursive computation of $\rho(F_N(\lambda))$ can be completed by an efficient algorithm to compute the last column of $\rho(F_N(\lambda))$. This algorithm is described in Appendix \ref{app:b}.

\subsection{The Wenzl-Jones Projector}

A family of linear transformations $P^d$ will play a central role in the rest of this paper. They are constructed from the Wenzl-Jones projectors which we first introduce.

\begin{Definition}\label{def:fN} For each $N\ge 1$, define $WJ_N\in TL_N(\beta)$ to be $WJ_1=id$ and, for $N>1$,
\begin{equation*}
\psset{unit=0.565685}\begin{pspicture}(0,-6)(7,6)
\rput(-1.4,0){$ WJ_N =$}
\lw
\psline{-}(2,2)(0,0)(2,-2)
\psline{-}(3,3)(6,6)(7,5)(6,4)(7,3)(6,2)(7,1)(6,0)(7,-1)(6,-2)(7,-3)(6,-4)(7,-5)(6,-6)(4,-4)(5,-3)(4,-2)(5,-1)(4,0)(5,1)(4,2)(5,3)(4,4)
\psline{-}(3,-3)(4,-4)
\psline{-}(5,5)(6,4)(5,3)(6,2)(5,1)(6,0)(5,-1)(6,-2)(5,-3)(6,-4)(5,-5)
\psline{-}(1,1)(2,0)(1,-1)
\psline[linestyle=dashed,dash=2pt 2pt]{-}(2,2)(3,3)
\psline[linestyle=dashed,dash=2pt 2pt]{-}(2,-2)(3,-3)
\psline[linestyle=dashed,dash=2pt 2pt]{-}(3,-3)(4,-2)(3,-1)(4,0)(3,1)(4,2)(3,3)
\psline[linestyle=dashed,dash=2pt 2pt]{-}(2,2)(3,1)(2,0)(3,-1)(2,-2)
\rput(6,5){$_1$}
\rput(6,3){$_1$}
\rput(6,1){$_1$}
\rput(6,-1){$_1$}
\rput(6,-3){$_1$}
\rput(6,-5){$_1$}
\rput(5,4){$_2$}
\rput(5,2){$_2$}
\rput(5,0){$_2$}
\rput(5,-2){$_2$}
\rput(5,-4){$_2$}
\rput(4,3){$_3$}
\rput(4,1){$_3$}
\rput(4,-1){$_3$}
\rput(4,-3){$_3$}
\rput(2,1){$_{N-2}$}
\rput(2,-1){$_{N-2}$}
\rput(1,0){$_{N-1}$}\unlw
\psdots(0.5,0.5)(1.5,1.5)(2.5,2.5)(3.5,3.5)(4.5,4.5)(5.5,5.5)(6.5,5.5)
\psdots(0.5,-0.5)(1.5,-1.5)(2.5,-2.5)(3.5,-3.5)(4.5,-4.5)(5.5,-5.5)(6.5,-5.5)

\psset{linecolor=myc}
\psarc{-}(6,-4){0.7071}{-45}{45}
\psarc{-}(6,-2){0.7071}{-45}{45}
\psarc{-}(6,0){0.7071}{-45}{45}
\psarc{-}(6,2){0.7071}{-45}{45}
\psarc{-}(6,4){0.7071}{-45}{45} 
\end{pspicture}
\end{equation*}
and each box stands for
\begin{equation}
\psset{linewidth=1pt}
\begin{pspicture}(-0.5,-0.5)(0.4,0.5)
\lw
\psline{-}(-0.5,0)(0,-0.5)(0.5,0)(0,0.5)(-0.5,0)
\rput(0,0){$_{k}$}
\end{pspicture}\ \ 
\begin{pspicture}(-0.2,-0.5)(0.2,0.5)
\rput(0,-0.03){$_{=}$}\lw
\end{pspicture}
\underbrace{
\begin{pspicture}(-0.5,-0.5)(0.5,0.5)\lw
\psline{-}(-0.5,0)(0,-0.5)(0.5,0)(0,0.5)(-0.5,0)
\psset{linecolor=myc}\unlw
\psarc{-}(-0.5,0){0.3536}{-45}{45}
\psarc{-}(0.5,0){0.3536}{135}{225}
\end{pspicture}}_{id}
\begin{pspicture}(-0.6,-0.5)(0.6,0.5)
\rput(0,-0.03){$_{+\ \frac{S_k}{S_{k+1}} }$}
\end{pspicture}
\underbrace{\begin{pspicture}(-0.5,-0.5)(0.5,0.5)\lw
\psline{-}(-0.5,0)(0,-0.5)(0.5,0)(0,0.5)(-0.5,0)
\psset{linecolor=myc}\unlw
\psarc{-}(0,0.5){0.3536}{225}{315}
\psarc{-}(0,-0.5){0.3536}{45}{135}
\end{pspicture}}_{e_{N-k}}
\begin{pspicture}(-0.6,-0.5)(0.6,0.5)
\rput(0,-0.03){$_{=\  \frac1{S_{k+1}} }$}\lw
\end{pspicture}
\begin{pspicture}(-0.5,-0.5)(0.5,0.5)\lw
\psline{-}(-0.5,0)(0,-0.5)(0.5,0)(0,0.5)(-0.5,0)
\rput(0,0){$_{-k \lambda}$}
\psarc[linewidth=0.5pt]{-}(0,-0.5){0.176777}{45}{135}
\end{pspicture}
\label{eq:kbox}
\end{equation}
\\
where the following compact notation is used from now on:
\begin{equation*}S_k = \sin(k \Lambda), \qquad C_k = \cos(k \Lambda),  \qquad \Lambda = \pi-\lambda.\end{equation*}
\end{Definition}

\noindent Note that with this notation, $\beta = -2 C_1 = -S_2/S_1$ and $\Lambda\in(\frac\pi2,\pi)$. Thoughout the next sections, we will relax this last constraint and take $\Lambda\in \mathbb{R}$ even though the corresponding loop model is unphysical, having negative Boltzmann weights.

This diagrammatic definition appears in \cite{Zhou} and, more recently, in \cite{PRZ}. But the object itself has been known to mathematicians for a long time (see \cite{Jones}, \cite{Wenzl}, \cite{KauffmanLins}). Note that, due to the factors $S_{k+1}=\sin(\Lambda(k+1))$ in the denominator, the $WJ_N$ may fail to exist in certain of the algebras $TL_N(\beta)$. A closer look will be given to this difficulty starting in the following subsections.

\begin{Proposition}
\label{sec:WJcarac} 
The $WJ_N$ are non-zero elements and satisfy the following properties:
\begin{itemize}
\item[{(i)}]  $WJ_N e_i = e_i WJ_N =0$ for $N \ge 2$ and $i=1,2,...,N-1$;
\item[{(ii)}]  $(WJ_N)^2 = WJ_N$.
\end{itemize}
\end{Proposition}
\noindent{\scshape Proof\ \ }
To prove statement {\em (i)}, the following two identities are needed: 
\begin{alignat}{3}
\psset{linewidth=1pt}
\begin{pspicture}(-0.5,0.1)(1,1)\lw
\psline{-}(-0.5,0)(0,-0.5)(0.5,0)(0,0.5)(-0.5,0)
\psline{-}(0.5,0)(0,0.5)(0.5,1)(1,0.5)(0.5,0)
\rput(0.5,0.5){$_{k}$}
\rput(0,0){$_{k+1}$}\unlw
\psset{linecolor=myc}
\psarc{-}(0,0.5){0.3536}{45}{225}
\end{pspicture}\ \ & = \ \frac{S_k}{S_{k+1}}\ \ 
\begin{pspicture}(-0.5,0.1)(1,1)\lw
\psline{-}(-0.5,0)(0,-0.5)(0.5,0)(0,0.5)(-0.5,0)
\psline{-}(0.5,0)(0,0.5)(0.5,1)(1,0.5)(0.5,0)\unlw
\psset{linecolor=myc}
\psarc{-}(0,0.5){0.3536}{45}{225}
\psarc{-}(0.5,0){0.3536}{45}{225}
\psarc{-}(-0.5,0){0.3536}{-45}{45}
\psarc{-}(0.5,1){0.3536}{225}{315}
\end{pspicture}\ +\ \ 
\begin{pspicture}(-0.5,0.1)(1,1)\lw
\psline{-}(-0.5,0)(0,-0.5)(0.5,0)(0,0.5)(-0.5,0)
\psline{-}(0.5,0)(0,0.5)(0.5,1)(1,0.5)(0.5,0)\unlw
\psset{linecolor=myc}
\psarc{-}(0,0.5){0.3536}{-45}{225}
\psarc{-}(0.5,0){0.3536}{135}{225}
\psarc{-}(-0.5,0){0.3536}{-45}{45}
\psarc{-}(1,0.5){0.3536}{135}{225}
\end{pspicture}\ +\ \frac{S_k}{S_{k+2}}\ \ 
\begin{pspicture}(-0.5,0.1)(1,1)\lw
\psline{-}(-0.5,0)(0,-0.5)(0.5,0)(0,0.5)(-0.5,0)
\psline{-}(0.5,0)(0,0.5)(0.5,1)(1,0.5)(0.5,0)\unlw
\psset{linecolor=myc}
\psarc{-}(0,0.5){0.3536}{45}{225}
\psarc{-}(0.5,0){0.3536}{45}{135}
\psarc{-}(0,0.5){0.3536}{-135}{-45}
\psarc{-}(0,-0.5){0.3536}{45}{135}
\psarc{-}(0.5,1){0.3536}{225}{315}
\end{pspicture}\ +\ \frac{S_{k+1}}{S_{k+2}}\ \ 
\begin{pspicture}(-0.5,0.1)(1,1)\lw
\psline{-}(-0.5,0)(0,-0.5)(0.5,0)(0,0.5)(-0.5,0)
\psline{-}(0.5,0)(0,0.5)(0.5,1)(1,0.5)(0.5,0)\unlw
\psset{linecolor=myc}
\psarc{-}(0,0.5){0.3536}{0}{360}
\psarc{-}(0,-0.5){0.3536}{45}{135}
\psarc{-}(1,0.5){0.3536}{135}{225}
\end{pspicture}\notag
\\ & \notag
\\ & =  \ \frac{S_k}{S_{k+1}}\ \ 
\begin{pspicture}(-0.5,0.1)(1,1)\lw
\psline{-}(-0.5,0)(0,-0.5)(0.5,0)(0,0.5)(-0.5,0)
\psline{-}(0.5,0)(0,0.5)(0.5,1)(1,0.5)(0.5,0)\unlw
\psset{linecolor=myc}
\psarc{-}(0,0.5){0.3536}{45}{225}
\psarc{-}(0.5,0){0.3536}{45}{225}
\psarc{-}(-0.5,0){0.3536}{-45}{45}
\psarc{-}(0.5,1){0.3536}{225}{315}
\end{pspicture}\ +\  \underbrace{\left( 1+\frac{S_k}{S_{k+2}}-\frac{S_2}{S_1}\frac{S_{k+1}}{S_{k+2}}\right)}_0\ \ 
\begin{pspicture}(-0.5,0.1)(1,1)\lw
\psline{-}(-0.5,0)(0,-0.5)(0.5,0)(0,0.5)(-0.5,0)
\psline{-}(0.5,0)(0,0.5)(0.5,1)(1,0.5)(0.5,0)\unlw
\psset{linecolor=myc}
\psarc{-}(0,-0.5){0.3536}{45}{135}
\psarc{-}(1,0.5){0.3536}{135}{225}
\end{pspicture}  
\ =  \ \frac{S_k}{S_{k+1}}\ \ 
\begin{pspicture}(-0.5,0.1)(1,1)\lw
\psline{-}(-0.5,0)(0,-0.5)(0.5,0)(0,0.5)(-0.5,0)
\psline{-}(0.5,0)(0,0.5)(0.5,1)(1,0.5)(0.5,0)\unlw
\psset{linecolor=myc}
\psarc{-}(0,0.5){0.3536}{45}{225}
\psarc{-}(0.5,0){0.3536}{45}{225}
\psarc{-}(-0.5,0){0.3536}{-45}{45}
\psarc{-}(0.5,1){0.3536}{225}{315}
\end{pspicture}
\notag \\
\notag \\
\psset{linewidth=1pt}
\begin{pspicture}(-0.5,0)(0.5,0.5)\lw
\psline{-}(-0.5,0)(0,-0.5)(0.5,0)(0,0.5)(-0.5,0)
\rput(0,0){$_{1}$}\unlw
\psset{linecolor=myc}
\psarc{-}(0,0.5){0.3536}{-45}{225}
\end{pspicture}\ \ & = \ \
\begin{pspicture}(-0.5,0)(0.5,0.5)\lw
\psline{-}(-0.5,0)(0,-0.5)(0.5,0)(0,0.5)(-0.5,0)\unlw
\psset{linecolor=myc}
\psarc{-}(0,0.5){0.3536}{-45}{225}
\psarc{-}(0.5,0){0.3536}{135}{225}
\psarc{-}(-0.5,0){0.3536}{-45}{45}
\end{pspicture}\ +\ \frac{S_1}{S_{2}}\ \ 
\begin{pspicture}(-0.5,0)(0.5,0.5)\lw
\psline{-}(-0.5,0)(0,-0.5)(0.5,0)(0,0.5)(-0.5,0)\unlw
\psset{linecolor=myc}
\psarc{-}(0,0.5){0.3536}{-45}{225}
\psarc{-}(0,0.5){0.3536}{225}{315}
\psarc{-}(0,-0.5){0.3536}{45}{135}
\end{pspicture}\  = \ \left(1- \frac{S_2}{S_1}\frac{S_1}{S_2} \right)\
\begin{pspicture}(-0.5,0)(0.5,0.5)\lw
\psline{-}(-0.5,0)(0,-0.5)(0.5,0)(0,0.5)(-0.5,0)\unlw
\psset{linecolor=myc}
\psarc{-}(0,-0.5){0.3536}{45}{135}
\end{pspicture} \ = \ 0\label{eq:diamant}
\end{alignat}

\smallskip

\noindent Note that the second identity is just $WJ_2 e_1 = 0$. For $N>2$, 
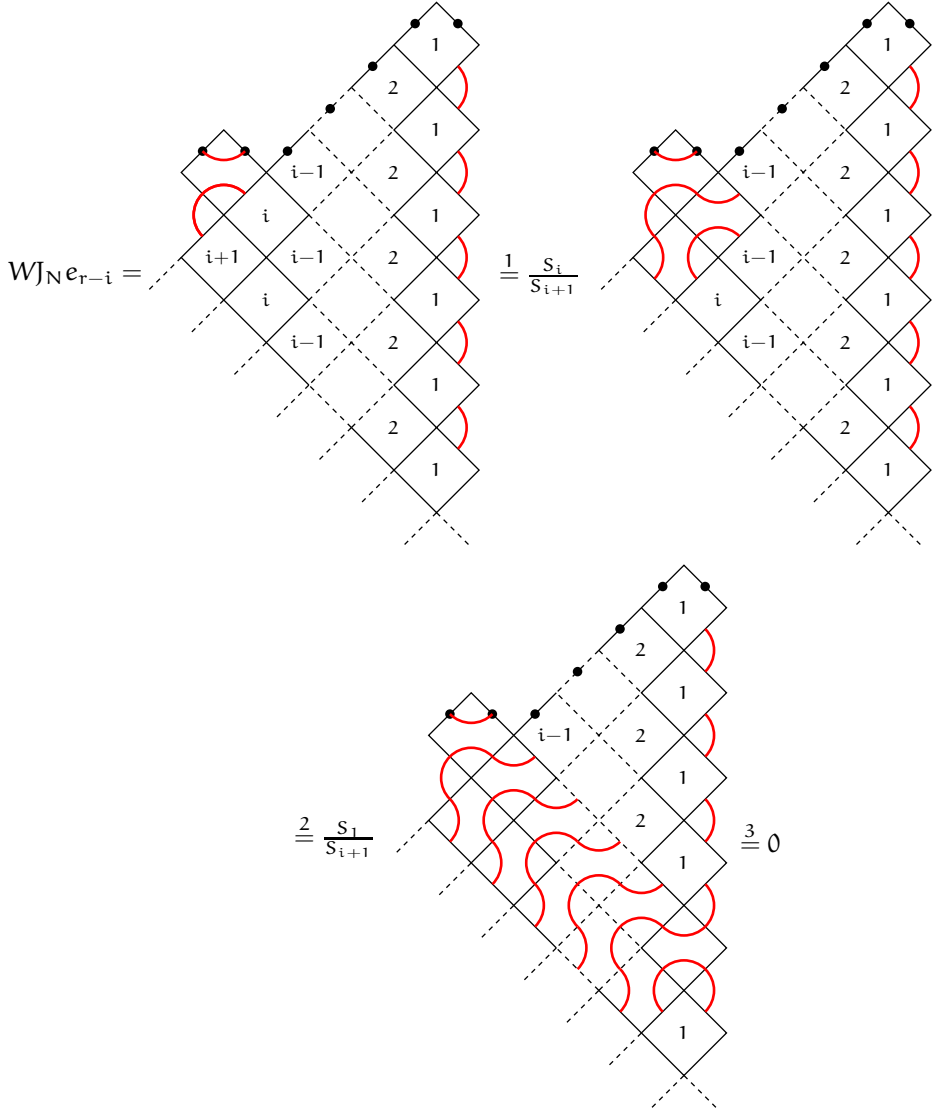
\begin{equation*}
\psset{unit=0.565685}\begin{pspicture}(0,-6.75)(7,6)
\psdots(0.5,2.5)(1.5,2.5)(2.5,2.5)(3.5,3.5)(4.5,4.5)(5.5,5.5)(6.5,5.5)
\lw
\rput(-2.5,-0.38){$WJ_N e_{r-i} =$}
\psline{-}(1,1)(0,2)(1,3)(2,2)
\psline{-}(2,2)(0,0)(2,-2)
\psline{-}(4,4)(6,6)(7,5)(6,4)(7,3)(6,2)(7,1)(6,0)(7,-1)(6,-2)(7,-3)(6,-4)(7,-5)(6,-6)(4,-4)
\psline[linestyle=dashed,dash=2pt 2pt]{-}(4,-4)(5,-3)(4,-2)(5,-1)(4,0)(5,1)(4,2)(5,3)(4,4)
\psline{-}(5,5)(6,4)(5,3)(6,2)(5,1)(6,0)(5,-1)(6,-2)(5,-3)(6,-4)(5,-5)
\psline{-}(1,1)(2,0)(1,-1)
\psline{-}(2,2)(3,3)
\psline{-}(2,-2)(3,-3)
\psline[linestyle=dashed,dash=2pt 2pt]{-}(4,-4)(3,-3)(4,-2)(3,-1)(4,0)(3,1)(4,2)(3,3)(4,4)
\psline{-}(2,2)(3,1)(2,0)(3,-1)(2,-2)
\psline[linestyle=dashed,dash=2pt 2pt](0,0)(-0.75,-0.75)
\psline[linestyle=dashed,dash=2pt 2pt](1,-1)(0.25,-1.75)
\psline[linestyle=dashed,dash=2pt 2pt](2,-2)(1.25,-2.75)
\psline[linestyle=dashed,dash=2pt 2pt](3,-3)(2.25,-3.75)
\psline[linestyle=dashed,dash=2pt 2pt](4,-4)(3.25,-4.75)
\psline[linestyle=dashed,dash=2pt 2pt](5,-5)(4.25,-5.75)
\psline[linestyle=dashed,dash=2pt 2pt](6,-6)(5.25,-6.75)
\psline[linestyle=dashed,dash=2pt 2pt](6,-6)(6.75,-6.75)
\rput(6,5){$_1$}
\rput(6,3){$_1$}
\rput(6,1){$_1$}
\rput(6,-1){$_1$}
\rput(6,-3){$_1$}
\rput(6,-5){$_1$}
\rput(5,4){$_2$}
\rput(5,2){$_2$}
\rput(5,0){$_2$}
\rput(5,-2){$_2$}
\rput(5,-4){$_2$}
\rput(3,-2){$_{i-1}$}
\rput(3,0){$_{i-1}$}
\rput(3,2){$_{i-1}$}
\rput(2,1){$_{i}$}
\rput(2,-1){$_{i}$}
\rput(1,0){$_{i+1}$}
\psset{linecolor=myc}\unlw
\psarc{-}(1,3){0.7071}{-135}{-45}
\psarc{-}(6,-4){0.7071}{-45}{45}
\psarc{-}(6,-2){0.7071}{-45}{45}
\psarc{-}(6,0){0.7071}{-45}{45}
\psarc{-}(6,2){0.7071}{-45}{45}
\psarc{-}(6,4){0.7071}{-45}{45} 
\psarc{-}(1,1){0.7071}{45}{225} 
\psarc{-}(1,1){0.7071}{45}{225} 
\end{pspicture}
\ \  \ \
\begin{pspicture}(-3,-6.75)(5,6)
\psdots(0.5,2.5)(1.5,2.5)(2.5,2.5)(3.5,3.5)(4.5,4.5)(5.5,5.5)(6.5,5.5)
\lw
\rput(-2.2,-0.38){$ \overset{1}{=} \frac{S_i}{S_{i+1}}$}
\psline{-}(1,1)(0,2)(1,3)(2,2)
\psline{-}(2,2)(0,0)(2,-2)
\psline{-}(4,4)(6,6)(7,5)(6,4)(7,3)(6,2)(7,1)(6,0)(7,-1)(6,-2)(7,-3)(6,-4)(7,-5)(6,-6)(4,-4)
\psline[linestyle=dashed,dash=2pt 2pt]{-}(4,-4)(5,-3)(4,-2)(5,-1)(4,0)(5,1)(4,2)(5,3)(4,4)
\psline{-}(5,5)(6,4)(5,3)(6,2)(5,1)(6,0)(5,-1)(6,-2)(5,-3)(6,-4)(5,-5)
\psline{-}(1,1)(2,0)(1,-1)
\psline{-}(2,2)(3,3)
\psline{-}(2,-2)(3,-3)
\psline[linestyle=dashed,dash=2pt 2pt]{-}(4,-4)(3,-3)(4,-2)(3,-1)(4,0)(3,1)(4,2)(3,3)(4,4)
\psline{-}(2,2)(3,1)(2,0)(3,-1)(2,-2)
\psline[linestyle=dashed,dash=2pt 2pt](0,0)(-0.75,-0.75)
\psline[linestyle=dashed,dash=2pt 2pt](1,-1)(0.25,-1.75)
\psline[linestyle=dashed,dash=2pt 2pt](2,-2)(1.25,-2.75)
\psline[linestyle=dashed,dash=2pt 2pt](3,-3)(2.25,-3.75)
\psline[linestyle=dashed,dash=2pt 2pt](4,-4)(3.25,-4.75)
\psline[linestyle=dashed,dash=2pt 2pt](5,-5)(4.25,-5.75)
\psline[linestyle=dashed,dash=2pt 2pt](6,-6)(5.25,-6.75)
\psline[linestyle=dashed,dash=2pt 2pt](6,-6)(6.75,-6.75)
\rput(6,5){$_1$}
\rput(6,3){$_1$}
\rput(6,1){$_1$}
\rput(6,-1){$_1$}
\rput(6,-3){$_1$}
\rput(6,-5){$_1$}
\rput(5,4){$_2$}
\rput(5,2){$_2$}
\rput(5,0){$_2$}
\rput(5,-2){$_2$}
\rput(5,-4){$_2$}
\rput(3,-2){$_{i-1}$}
\rput(3,0){$_{i-1}$}
\rput(3,2){$_{i-1}$}
\rput(2,-1){$_{i}$}
\psset{linecolor=myc}\unlw
\psarc{-}(1,3){0.7071}{-135}{-45}
\psarc{-}(6,-4){0.7071}{-45}{45}
\psarc{-}(6,-2){0.7071}{-45}{45}
\psarc{-}(6,0){0.7071}{-45}{45}
\psarc{-}(6,2){0.7071}{-45}{45}
\psarc{-}(6,4){0.7071}{-45}{45} 
\psarc{-}(1,1){0.7071}{45}{225} 
\psarc{-}(2,0){0.7071}{45}{225}
\psarc{-}(0,0){0.7071}{-45}{45}
\psarc{-}(2,2){0.7071}{-135}{-45} 
\end{pspicture}
\end{equation*}
\begin{equation*}
\psset{unit=0.565685}\begin{pspicture}(-3,-6.75)(7,6)
\psdots(0.5,2.5)(1.5,2.5)(2.5,2.5)(3.5,3.5)(4.5,4.5)(5.5,5.5)(6.5,5.5)\lw
\rput(-2.2,-0.38){$ \overset{2}{=} \frac{S_1}{S_{i+1}}$}
\rput(7.8,-0.38){$ \overset{3}{=}0$}
\psline{-}(1,1)(0,2)(1,3)(2,2)
\psline{-}(2,2)(0,0)(2,-2)
\psline{-}(4,4)(6,6)(7,5)(6,4)(7,3)(6,2)(7,1)(6,0)(7,-1)(6,-2)(7,-3)(6,-4)(7,-5)(6,-6)(4,-4)
\psline[linestyle=dashed,dash=2pt 2pt]{-}(4,-4)(5,-3)(4,-2)(5,-1)(4,0)(5,1)(4,2)(5,3)(4,4)
\psline{-}(5,5)(6,4)(5,3)(6,2)(5,1)(6,0)(5,-1)(6,-2)(5,-3)(6,-4)(5,-5)
\psline{-}(1,1)(2,0)(1,-1)
\psline{-}(2,2)(3,3)
\psline{-}(2,-2)(3,-3)
\psline[linestyle=dashed,dash=2pt 2pt]{-}(4,-4)(3,-3)(4,-2)(3,-1)(4,0)(3,1)(4,2)(3,3)(4,4)
\psline{-}(2,2)(3,1)(2,0)(3,-1)(2,-2)
\psline[linestyle=dashed,dash=2pt 2pt](0,0)(-0.75,-0.75)
\psline[linestyle=dashed,dash=2pt 2pt](1,-1)(0.25,-1.75)
\psline[linestyle=dashed,dash=2pt 2pt](2,-2)(1.25,-2.75)
\psline[linestyle=dashed,dash=2pt 2pt](3,-3)(2.25,-3.75)
\psline[linestyle=dashed,dash=2pt 2pt](4,-4)(3.25,-4.75)
\psline[linestyle=dashed,dash=2pt 2pt](5,-5)(4.25,-5.75)
\psline[linestyle=dashed,dash=2pt 2pt](6,-6)(5.25,-6.75)
\psline[linestyle=dashed,dash=2pt 2pt](6,-6)(6.75,-6.75)
\rput(6,5){$_1$}
\rput(6,3){$_1$}
\rput(6,1){$_1$}
\rput(6,-1){$_1$}
\rput(6,-5){$_1$}
\rput(5,4){$_2$}
\rput(5,2){$_2$}
\rput(5,0){$_2$}
\rput(3,2){$_{i-1}$}
\psset{linecolor=myc}\unlw
\psarc{-}(1,3){0.7071}{-135}{-45}
\psarc{-}(6,-4){0.7071}{-45}{45}
\psarc{-}(6,-2){0.7071}{-45}{45}
\psarc{-}(6,0){0.7071}{-45}{45}
\psarc{-}(6,2){0.7071}{-45}{45}
\psarc{-}(6,4){0.7071}{-45}{45} 
\psarc{-}(1,1){0.7071}{45}{225} 
\psarc{-}(2,0){0.7071}{45}{225}
\psarc{-}(3,-1){0.7071}{45}{225}
\psarc{-}(4,-2){0.7071}{45}{225}
\psarc{-}(5,-3){0.7071}{45}{225}
\psarc{-}(6,-4){0.7071}{45}{225}
\psarc{-}(0,0){0.7071}{-45}{45}
\psarc{-}(1,-1){0.7071}{-45}{45}
\psarc{-}(2,-2){0.7071}{-45}{45}
\psarc{-}(3,-3){0.7071}{-45}{45}
\psarc{-}(4,-4){0.7071}{-45}{45}
\psarc{-}(2,2){0.7071}{-135}{-45} 
\psarc{-}(3,1){0.7071}{-135}{-45} 
\psarc{-}(4,0){0.7071}{-135}{-45} 
\psarc{-}(5,-1){0.7071}{-135}{-45} 
\psarc{-}(6,-2){0.7071}{-135}{-45} 
\end{pspicture}
\end{equation*}
Here, the first identity has been used once after equality $1$ and $i-1$ more times after equality $2$, allowing bubbles to propagate in the lower right direction. Finally, the second identity has been used after equality $3$ and the bubble vanishes. To obtain the relation $e_{r-i} WJ_N = 0$, the generator $e_{r-i}$ is applied from below, and the bubbles propagate in the upper right direction. 

Statement {\em (ii)} is trivially true for $WJ_1$. Now let $G \in TL_N(\beta)$. It can be written as $G = \alpha_{id}(G)\, id + \sum_{\{c\}^*} \alpha_c(G) c $ where $\{c\}^*$ is the set of all connectivities in $TL_N(\beta)$  (as described in section \ref{sec:TLconnect}), excluding the identity. Since every connectivity in $\{c^*\}$ is a finite product of the $e_i$'s, $\left(\sum_{\{c*\}} \alpha_c c \right)  \times WJ_N = 0$ and $G WJ_N = \alpha_{id}(G) WJ_N$. Choosing $G=WJ_N$, it only remains to show that $\alpha_{id}(WJ_N) = 1$. This is indeed true, as the only configuration giving the identity is the one where every box is chosen to be
\psset{linewidth=1pt}
\begin{pspicture}(-0.2,-0.1)(0.15,0.2)\lw
\psline{-}(-0.2,0)(0,0.2)(0.2,0)(0,-0.2)(-0.2,0)
\psset{linecolor=myc}
\psarc{-}(0.2,0){0.14}{135}{225}
\psarc{-}(-0.2,0){0.14}{-45}{45}
\end{pspicture}
. The normalization of the boxes in eq.~(\ref{eq:kbox}) ensures that the total weight $\alpha_{id}(WJ_N)$ is $1$. \hfill$\square$

\medskip

Kauffman and Lins \cite{KauffmanLins} introduce the family of $WJ_N\in TL_N$ recursively and they show that there is a unique non-zero element in $TL_N$ that has the two properties stated in the above proposition. The two definitions, theirs and the present, must therefore coincide. The element of $TL_N$ obtained by reflecting the diagram defining $WJ_N$ along a vertical axis also has the properties 
of Proposition \ref{sec:WJcarac}; by unicity it must also coincide with $WJ_N$.

To introduce the family of linear transformations $P^d$, the following subspaces of the link space will be needed. Let $UpTo_d$ be the span of $\cup_{e\le d}B_N^e$ where $B_N^e$ is the link basis of the subspace of vectors with precisely $e$ defects. Therefore $UpTo_d$ is the subspaces of all vectors whose components have up to $d$ defects. The subspaces $UpTo_d$ depend also on $N$; we chose not to add the label $N$ to the notation. The natural filtration of the link space and the fact that connectivities $c \in TL_N$ do not increase the number of defects are formulated easily in terms of these subspaces: if $N$ is even
$$UpTo_0\subset UpTo_2\subset \dots\subset  UpTo_N=V_N\qquad \text{\rm and}\qquad
\rho(c)UpTo_d\subset UpTo_d,\quad \text{\rm for all } d$$
and similarly for $N$ odd. We also define $UpTo_{-1}=UpTo_{-2}=\{0\}$.
\begin{Definition}\label{def:Pd}The linear transformation $P^d:UpTo_d\rightarrow UpTo_d$, $0\le d\le N$ with $N$ and $d$ of the same parity, is defined by $P^0=id$ and $P^1=id$ and, for $d\ge 2$, by its action on the basis $\cup_{e\le d}B_N^e$:
\begin{itemize}
\item[(i)] $P^dv=0$ if $v\in B^e_N$ with $e<d$;
\item[(ii)] for $v\in B_N^d$, the following procedure is followed: first, all arcs are removed from $v$, leaving its $d$ defects; second, the action of $\rho(WJ_d)$ on these $d$ defects is computed; and finally, the arcs of $v$ are then reinserted, unchanged and in their original positions, into each vectors of the linear combination obtained from $\rho(WJ_d)v^d$.
\end{itemize}
\end{Definition}
The definition of $P^d$ on the elements of $B_N^d$ is somewhat awkward. But it has a diagrammatic representation that is easy to work with, namely the action of $P^d$ on $v\in B_N^d$ is drawn from top to bottom as first the projector $WJ_d$ acting on $d$ defects followed by the insertion of the arcs between the defects coming out $WJ_d$. Here is the action of $P^2$ on two vectors of $V_4^2$, given first in the form $P^2v$, second diagrammatically, and last, as a linear combination of basis elements:
\begin{align*}
P^2\big(\psset{unit=1}
\psset{linewidth=1pt}
\begin{pspicture}(3.8,0.1)(5.4,1)
\psset{linecolor=black}
\psdots(4.0,0)(4.4,0)(4.8,0)(5.2,0)
\unlw
\psset{linecolor=myc2}
\psline{-}(4,0)(4,0.5)
\psline{-}(4.4,0)(4.4,0.5)
\psarc{-}(5,0){0.2}{0}{180}
\psset{linecolor=black}
\end{pspicture}\big) &=
\psset{unit=1}
\psset{linewidth=1pt}
\begin{pspicture}(3.8,0.1)(5.4,1.2)
\psset{linecolor=black}
\psdots(4.0,0)(4.4,0)(4.8,0)(5.2,0)
\lw
\psline{-}(4.2,0.3)(4.6,0.7)(4.2,1.1)(3.8,0.7)(4.2,0.3)\unlw
\rput(4.2,0.7){$_1$}
\psset{linecolor=myc2}
\psline{-}(4,0)(4,0.5)
\psline{-}(4.4,0)(4.4,0.5)
\psline{-}(4,0.9)(4,1.2)
\psline{-}(4.4,0.9)(4.4,1.2)
\psarc{-}(5,0){0.2}{0}{180}
\psset{linecolor=black}
\end{pspicture}= 
\psset{unit=1}
\psset{linewidth=1pt}
\begin{pspicture}(3.8,0.1)(5.4,1)
\psset{linecolor=black}
\psdots(4.0,0)(4.4,0)(4.8,0)(5.2,0)
\unlw
\psset{linecolor=myc2}
\psline{-}(4,0)(4,0.5)
\psline{-}(4.4,0)(4.4,0.5)
\psarc{-}(5,0){0.2}{0}{180}
\psset{linecolor=black}
\end{pspicture}+\frac{S_1}{S_2}
\psset{unit=1}
\psset{linewidth=1pt}
\begin{pspicture}(3.8,0.1)(5.4,1)
\psset{linecolor=black}
\psdots(4.0,0)(4.4,0)(4.8,0)(5.2,0)
\unlw
\psset{linecolor=myc2}
\psarc{-}(4.2,0){0.2}{0}{180}
\psarc{-}(5,0){0.2}{0}{180}
\end{pspicture} \\
P^2\big(\psset{unit=1}
\psset{linewidth=1pt}
\begin{pspicture}(3.8,0.1)(5.4,1)
\psset{linecolor=black}
\psdots(4.0,0)(4.4,0)(4.8,0)(5.2,0)
\unlw
\psset{linecolor=myc2}
\psline{-}(4,0)(4,0.5)
\psline{-}(5.2,0)(5.2,0.5)
\psarc{-}(4.6,0){0.2}{0}{180}
\psset{linecolor=black}
\end{pspicture}\big) &=
\psset{unit=1}
\psset{linewidth=1pt}
\begin{pspicture}(6,0.1)(7.4,1.2)
\psset{linecolor=black}
\psdots(6,0)(7.2,0)(6.4,0)(6.8,0)\lw
\psline{-}(6.6,0.3)(7.0,0.7)(6.6,1.1)(6.2,0.7)(6.6,0.3)\unlw
\rput(6.6,0.7){$_1$}
\psset{linecolor=myc2}
\psbezier{-}(6.0,0)(6.0,0.4)(6.4,0.5)(6.4,0.5)
\psbezier{-}(7.2,0)(7.2,0.4)(6.8,0.5)(6.8,0.5)
\psarc{-}(6.6,0){0.2}{0}{180}
\psline{-}(6.4,0.9)(6.4,1.2)
\psline{-}(6.8,0.9)(6.8,1.2)
\end{pspicture}=\psset{unit=1}
\psset{linewidth=1pt}
\begin{pspicture}(3.8,0.1)(5.4,1)
\psset{linecolor=black}
\psdots(4.0,0)(4.4,0)(4.8,0)(5.2,0)
\unlw
\psset{linecolor=myc2}
\psline{-}(4,0)(4,0.5)
\psline{-}(5.2,0)(5.2,0.5)
\psarc{-}(4.6,0){0.2}{0}{180}
\psset{linecolor=black}
\end{pspicture}+\frac{S_1}{S_2}
\psset{unit=1}
\psset{linewidth=1pt}
\begin{pspicture}(3.8,0.1)(5.4,1)
\psset{linecolor=black}
\psdots(4.0,0)(4.4,0)(4.8,0)(5.2,0)
\unlw
\psset{linecolor=myc2}
\psarc{-}(4.6,0){0.2}{0}{180}
\psbezier{-}(4,0)(4,0.4)(5.2,0.4)(5.2,0)
\psset{linecolor=black}
\end{pspicture}
\end{align*}
Even though the $P^d$'s are similar to the $WJ_N$, they are different by one property, namely statement {\em (i)} of Lemma \ref{sec:WJcarac} does not make sense for $P^d$. The product $e_iWJ_N$ amounts to closing two neighboring defects coming out of $WJ_N$. But, for vectors in $B_N^d$, defects are not necessarily contiguous since they may be separated by arcs. Still one sees, in the second example above, that a generalization of this property can be proposed. Indeed, if $c$ is any connectivity that ties positions $1$ and $4$, then $cP^2
(\psset{unit=0.4}
\psset{linewidth=1pt}
\begin{pspicture}(3.8,0.0)(5.4,0.4)
\psset{linecolor=black}
\psset{dotsize=2.0pt 0}
\psdots(4.0,0)(4.4,0)(4.8,0)(5.2,0)
\unlw
\psset{linecolor=myc2}
\psline{-}(4,0)(4,0.5)
\psline{-}(5.2,0)(5.2,0.5)
\psarc{-}(4.6,0){0.2}{0}{180}
\psset{linecolor=black}
\end{pspicture})$ will be zero:
$$\psset{unit=1}
\psset{linewidth=1pt}
\begin{pspicture}(3.4,-0.5)(6.2,0.5)
\rput(3.5,0){$cP^2\big($}
\psset{linecolor=black}
\psdots(4.0,0)(4.4,0)(4.8,0)(5.2,0)
\unlw
\psset{linecolor=myc2}
\psline{-}(4,0)(4,0.5)
\psline{-}(5.2,0)(5.2,0.5)
\psarc{-}(4.6,0){0.2}{0}{180}
\psset{linecolor=black}
\rput(5.6,0){$\big) =$}
\end{pspicture}\psset{unit=1}
\psset{linewidth=1pt}
\begin{pspicture}(3.8,-1)(6.6,0.5)
\psset{linecolor=black}
\psdots(4.0,0)(4.4,0)(4.8,0)(5.2,0)
\psdots(4.0,-0.8)(4.4,-0.8)(4.8,-0.8)(5.2,-0.8)
\lw
\psline{-}(3.8,0)(5.4,0)(5.4,-0.8)(3.8,-0.8)(3.8,0)
\unlw
\psset{linecolor=myc2}
\psline{-}(4,0)(4,0.5)
\psline{-}(5.2,0)(5.2,0.5)
\psarc{-}(4.6,0){0.2}{0}{360}
\psbezier{-}(4,0)(4,-0.4)(5.2,-0.4)(5.2,0)
\psset{linecolor=black}
\rput(6.,-0.5){${\displaystyle{-\frac{1}{\beta}}}$}
\end{pspicture}
\psset{unit=1}
\psset{linewidth=1pt}
\begin{pspicture}(3.8,-1)(5.4,0.5)
\psset{linecolor=black}
\psdots(4.0,0)(4.4,0)(4.8,0)(5.2,0)
\psdots(4.0,-0.8)(4.4,-0.8)(4.8,-0.8)(5.2,-0.8)
\lw
\psline{-}(3.8,0)(5.4,0)(5.4,-0.8)(3.8,-0.8)(3.8,0)\unlw
\psset{linecolor=myc2}
\psarc{-}(4.6,0){0.2}{0}{360}
\psbezier{-}(4,0)(4,-0.4)(5.2,-0.4)(5.2,0)
\psbezier{-}(4,0)(4,0.4)(5.2,0.4)(5.2,0)
\psset{linecolor=black}
\rput(6,-0.5){$=0.$}
\end{pspicture}
$$
This observation can be generalized.
\begin{Proposition}\label{propo:Pd} The linear transformations $P^d$ have the following properties.
\begin{itemize}
\item[(i)] $P^d$ is a projector.
\item[(ii)] Let $v\in B_N^d$ and $i,j$ be positions of two contiguous defects of $v$. If $c$ is a connectivity in $TL_N$ that join $i$ and $j$, then $cP^dv=0$.
\item[(iii)] The restriction $\rho(F_N(\lambda))|_{UpTo_d}$ commutes with $P^d$ and, on $UpTo_d$,
$$\rho(F_N(\lambda))P^d=-2C_{d+1}P^d.$$
\end{itemize}
\end{Proposition}
\noindent{\scshape Proof\ \ } If $v\in UpTo_e$ with $e<d$, then $P^dv=0$ and therefore $(P^d)^2v=P^dv$. If $v\in B_N^d$, then $P^dv=v+w$ for some $w\in UpTo_{d-2}$ and therefore $P^dw=0$. Then $(P^d)^2v=P^d(v+w)=P^dv$.

The second statement is the equivalent of the property $e_iWJ_N=0$ proved in Proposition \ref{sec:WJcarac} and is actually an immediate consequence of it. The action of $P^d$ on $V_N^d$ cannot be written as the action of a given element of $TL_N$ as it was for $WJ_N$. However the proof of {\em (i)} in the previous proposition is purely in terms of connectivities. For example, the two first identities opening the proof  do not state on which defects they act; they are statements about connectivities only. If a connectivity $c$ closes to adjacent defects of $v$, then $cP^dv$ will contain an arc, from $i$ to $j$, encircling whichever arcs that lie between positions $i$ and $j$. These arcs will have to be closed by $c$ and, for the purpose of studying connectivities, the loops closed that way can be forgotten. The tie between $i$ and $j$ will introduce therefore the same connectivity as that introduced by $e_i$ in the proof of Proposition \ref{sec:WJcarac}, leading to $cP^dv=0$.

For {\em (iii)}, note first that, if $v\in B_N^e$, $e<d$, then $\rho(F_N)P^dv=0$. But, because $\rho(F_N) UpTo_e\subset UpTo_e$ for all $e$, $\rho(F_N)v\in UpTo_e$ and $P^d(\rho(F_N)v)=0$. If $v\in B_N^d$, then statement {\em (ii)} of Lemma \ref{lem:fn} shows that $\rho(F_N)v=-2C_{d+1}v+w$ for some $w\in UpTo_{d-2}$. Therefore $P^d(\rho(F_N)v)=-2C_{d+1}P^dv$. The fact that $\rho(F_N)(P^dv)=-2C_{d+1}P^dv$ follows from \eqref{eq:semicircle}. The latter identity allows to push $\rho(F_N)$ upward through all arcs in $v$ as the following example shows.
\begin{align*}
\rho(F_8)(P^2v)&=
\psset{unit=0.7}\begin{pspicture}(-0.7,-0.2)(8.4,3)
\psdots(0.5,1)(1.5,1)(2.5,1)(3.5,1)(4.5,1)(5.5,1)(6.5,1)(7.5,1)
\psdots(0.5,-1)(1.5,-1)(2.5,-1)(3.5,-1)(4.5,-1)(5.5,-1)(6.5,-1)(7.5,-1)
\lw
\psline{-}(3,2)(2.5,2.5)(3,3)(3.5,2.5)(3,2)
\psline{-}(0,-1)(0,1)(8,1)(8,-1)(0,-1)
\psline{-}(0,0)(8,0)
\psline{-}(1,-1)(1,1)\psline{-}(2,-1)(2,1)\psline{-}(3,-1)(3,1)\psline{-}(4,-1)(4,1)\psline{-}(5,-1)(5,1)\psline{-}(6,-1)(6,1)\psline{-}(7,-1)(7,1)
\rput(3,2.5){$1$}
\psset{linecolor=myc}\unlw
\psarc{-}(0,0){0.5}{90}{270}
\psarc{-}(8,0){0.5}{270}{90}
\psline{-}(0,-0.5)(8,-0.5)
\psline{-}(0,0.5)(0.35,0.5)
\psline{-}(0.65,0.5)(1.35,0.5)\psline{-}(1.65,0.5)(2.35,0.5)\psline{-}(2.65,0.5)(3.35,0.5)\psline{-}(3.65,0.5)(4.35,0.5)\psline{-}(4.65,0.5)(5.35,0.5)\psline{-}(5.65,0.5)(6.35,0.5)\psline{-}(6.65,0.5)(7.35,0.5)\psline{-}(7.65,0.5)(8,0.5)
\psline{-}(0.5,0)(0.5,1)\psline{-}(1.5,0)(1.5,1)\psline{-}(2.5,0)(2.5,1)\psline{-}(3.5,0)(3.5,1)\psline{-}(4.5,0)(4.5,1)\psline{-}(5.5,0)(5.5,1)\psline{-}(6.5,0)(6.5,1)\psline{-}(7.5,0)(7.5,1)
\psline{-}(0.5,0)(0.5,-0.35)\psline{-}(0.5,-0.65)(0.5,-1)
\psline{-}(1.5,0)(1.5,-0.35)\psline{-}(1.5,-0.65)(1.5,-1)
\psline{-}(2.5,0)(2.5,-0.35)\psline{-}(2.5,-0.65)(2.5,-1)
\psline{-}(3.5,0)(3.5,-0.35)\psline{-}(3.5,-0.65)(3.5,-1)
\psline{-}(4.5,0)(4.5,-0.35)\psline{-}(4.5,-0.65)(4.5,-1)
\psline{-}(5.5,0)(5.5,-0.35)\psline{-}(5.5,-0.65)(5.5,-1)
\psline{-}(6.5,0)(6.5,-0.35)\psline{-}(6.5,-0.65)(6.5,-1)
\psline{-}(7.5,0)(7.5,-0.35)\psline{-}(7.5,-0.65)(7.5,-1)
\psset{linecolor=myc2}
\psarc{-}(3,1){0.5}{0}{180}
\psbezier{-}(1.5,1)(1.5,2)(4.5,2)(4.5,1)
\psbezier{-}(0.5,1)(0.5,2)(2.25,2.25)(2.75,2.25)
\psbezier{-}(5.5,1)(5.5,2)(3.75,2.25)(3.25,2.25)
\psarc{-}(7,1){0.5}{0}{180}
\psline{-}(2.75,2.75)(2.75,3.2)
\psline{-}(3.25,2.75)(3.25,3.2)
\end{pspicture}
\
\ =
\begin{pspicture}(-0.6,-0.2)(8.5,2.0)
\psdots(0.5,1)(1.5,1)(2.5,1)(3.5,1)(4.5,1)(5.5,1)(6.5,1)(7.5,1)
\psdots(0.5,-1)(1.5,-1)(2.5,-1)(3.5,-1)(4.5,-1)(5.5,-1)(6.5,-1)(7.5,-1)\lw
\psline{-}(0,-1)(0,1)(8,1)(8,-1)(0,-1)
\psline{-}(0,0)(8,0)
\psline{-}(1,-1)(1,1)\psline{-}(2,-1)(2,1)\psline{-}(3,-1)(3,1)\psline{-}(4,-1)(4,1)
\psline{-}(5,-1)(5,1)\psline{-}(6,-1)(6,1)\psline{-}(7,-1)(7,1)
\rput(3,2.5){$1$}
\psline{-}(3,2)(2.5,2.5)(3,3)(3.5,2.5)(3,2)
\psset{linecolor=myc}\unlw
\psarc{-}(0,0){0.5}{90}{270}
\psarc{-}(8,0){0.5}{270}{90}
\psline{-}(0,-0.5)(1,-0.5)
\psline{-}(0,0.5)(0.35,0.5)
\psline{-}(0.65,0.5)(1,0.5)
\psline{-}(5,0.5)(5.35,0.5)
\psline{-}(5,-0.5)(6,-0.5)
\psline{-}(5.65,0.5)(6.0,0.5)
\psline{-}(0.5,0)(0.5,1)
\psline{-}(5.5,0)(5.5,1)
\psline{-}(0.5,0)(0.5,-0.35)\psline{-}(0.5,-0.65)(0.5,-1)
\psarc{-}(1,1){0.5}{270}{360}\psarc{-}(2,0){0.5}{90}{180}
\psarc{-}(1,0){0.5}{270}{360}\psarc{-}(2,-1){0.5}{90}{180}
\psarc{-}(4,0){0.5}{0}{90}\psarc{-}(5,1){0.5}{180}{270}
\psarc{-}(4,-1){0.5}{0}{90}\psarc{-}(5,0){0.5}{180}{270}
\psline{-}(5.5,0)(5.5,-0.35)\psline{-}(5.5,-0.65)(5.5,-1)
\psarc{-}(3,-1){0.5}{0}{180}\psarc{-}(3,0){0.5}{0}{180}
\psarc{-}(2,1){0.5}{270}{360}\psarc{-}(4,1){0.5}{180}{270}
\psarc{-}(2,0){0.5}{270}{360}\psarc{-}(4,0){0.5}{180}{270}
\psarc{-}(7,-1){0.5}{0}{180}\psarc{-}(7,0){0.5}{0}{180}
\psarc{-}(6,1){0.5}{270}{360}\psarc{-}(8,1){0.5}{180}{270}
\psarc{-}(6,0){0.5}{270}{360}\psarc{-}(8,0){0.5}{180}{270}
\psset{linecolor=myc2}
\psarc{-}(3,1){0.5}{0}{180}
\psbezier{-}(1.5,1)(1.5,2)(4.5,2)(4.5,1)
\psbezier{-}(0.5,1)(0.5,2)(2.25,2.25)(2.75,2.25)
\psbezier{-}(5.5,1)(5.5,2)(3.75,2.25)(3.25,2.25)
\psarc{-}(7,1){0.5}{0}{180}
\psline{-}(2.75,2.75)(2.75,3.2)
\psline{-}(3.25,2.75)(3.25,3.2)
\end{pspicture}\\
& =
\psset{unit=0.7}\begin{pspicture}(-0.5,-0.9)(8.5,4.5)
\psdots(0.5,-1)(1.5,-1)(2.5,-1)(3.5,-1)(4.5,-1)(5.5,-1)(6.5,-1)(7.5,-1)
\psdots(2.5,0)(3.5,0)(2.5,2)(3.5,2)
\lw
\psline{-}(2,0.)(2,2.)(4,2.)(4,0.)(2,0.)
\psline{-}(2,1.)(4,1.)
\psline{-}(3,0.)(3,2.)
\psline{-}(3,2.25)(2.5,2.75)(3,3.25)(3.5,2.75)(3,2.25)
\rput(3,2.75){$1$}
\psset{linecolor=myc}\unlw
\psarc{-}(2.,1){0.5}{90}{270}
\psarc{-}(4.,1){0.5}{-90}{90}
\psline{-}(2.5,0)(2.5,0.35)
\psline{-}(3.5,0)(3.5,0.35)
\psline{-}(2.5,0.65)(2.5,2)
\psline{-}(3.5,0.65)(3.5,2)
\psline{-}(2,0.5)(4,0.5)
\psline{-}(2,1.5)(2.35,1.5)
\psline{-}(2.65,1.5)(3.35,1.5)
\psline{-}(3.65,1.5)(4,1.5)
\psset{linecolor=myc2}
\psarc{-}(3,-1){0.5}{0}{180}
\psbezier{-}(1.5,-1)(1.5,0)(4.5,0)(4.5,-1)
\psbezier{-}(0.5,-1)(0.5,-0.5)(2.5,-0.25)(2.5,0.)
\psbezier{-}(5.5,-1)(5.5,-0.5)(3.5,-0.25)(3.5,0.)
\psbezier{-}(2.5,2)(2.5,2.25)(2.65,2.5)(2.75,2.5)
\psbezier{-}(3.5,2)(3.5,2.25)(3.35,2.5)(3.25,2.5)
\psline{-}(2.75,3)(2.75,3.45)
\psline{-}(3.25,3)(3.25,3.45)
\psarc{-}(7,-1){0.5}{0}{180}
\end{pspicture}
=-2C_{3}P^2v.
\end{align*}
To understand the third equality, note that the only boxes of $\rho(F_N)$ remaining to be summed over are those directly under the defects of $v$ and that these boxes are connected as if the columns containing them were glued to one another.  The arcs between them either connect these columns or they reproduce on the bottom line the original arcs of $v$. One can therefore draw the remaining boxes together, without the connecting arcs, but one then has to add the original arcs below. The last equality follows from the fact that $F_d(\lambda)$ commutes with $WJ_d$ and that $\rho(F_d(\lambda))_{d,d}=-2C_{d+1}\ id$. This closes both the proof of the commutativity of $P^d$ and of $\rho(F_N)|_{UpTo_d}$ and that of the eigenvalue of $\rho(F_N)$ on the subspace onto which $P^d$ projects. \hfill$\square$

\subsection{Eigenvectors of $\rho(F_N(\lambda))$ for non-critical $\lambda$'s}

The previous proposition has shown that $P^dv$, $v\in B_N^d$, is an eigenvector of $\rho(F_N(\lambda))$. Therefore one may hope that the projectors $P^d$, whenever they exist, can diagonalize $\rho(F_N(\lambda))$.

\begin{Definition} The number $\lambda$ (or $\Lambda)$ is said to be critical for $N$ if $e^{i\Lambda}=e^{i(\pi-\lambda)}$ is a $(2l)$-th root of unity, for some $l$ in the range $2\le l\le N$. Otherwise, $\lambda$ is said non-critical or generic (for $N$).
\end{Definition}

Let $N$ be fixed and $\lambda$ non-critical for this $N$. Let $PB_N^d$ be the set
$PB_N^d=\{P^dv\ |\ v\in B_N^d\}$.
Because, for any such $v\in B_N^d$, $P^dv=v+w$ for some $w\in UpTo_{d-2}$, the set $\cup_{e\le d}PB_N^e$ is a basis for $UpTo_d$ and $PB_N=\cup_{0\le d\le N}PB_N^d$ a basis for $V_N$. We shall keep on writing, somewhat abusively, that $PB_N^d$, like $B_N^d$, spans the sector with $d$ defects. Here is $PB_4$ as an example:
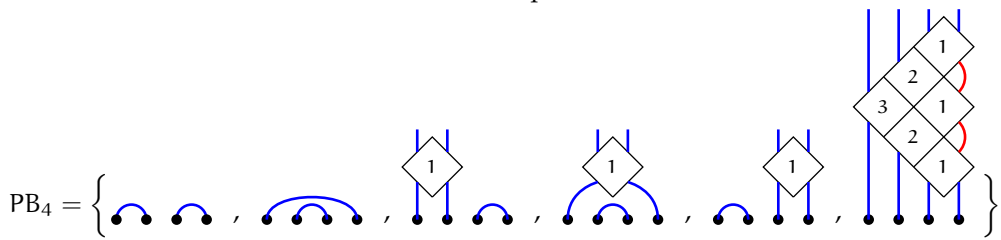
\begin{equation*}
PB_4=\bigg\{
\psset{unit=1}
\psset{linewidth=1pt}
\begin{pspicture}(-0.1,0.1)(11.5,2.5)
\psset{linecolor=myc2}
\psarc{-}(0.2,0){0.2}{0}{180}
\psarc{-}(1,0){0.2}{0}{180}
\psset{linecolor=black}
\psdots(0.0,0)(0.4,0)(0.8,0)(1.2,0)
\psdots(2.0,0)(2.4,0)(2.8,0)(3.2,0)
\rput(1.6,0){,}
\psset{linecolor=myc2}
\psarc{-}(2.6,0){0.2}{0}{180}
\psbezier{-}(2,0)(2,0.4)(3.2,0.4)(3.2,0)
\psset{linecolor=black}
\psdots(4.0,0)(4.4,0)(4.8,0)(5.2,0)
\rput(3.6,0){,}\lw
\psline{-}(4.2,0.3)(4.6,0.7)(4.2,1.1)(3.8,0.7)(4.2,0.3)\unlw
\rput(4.2,0.7){$_1$}
\psset{linecolor=myc2}
\psline{-}(4,0)(4,0.5)
\psline{-}(4.4,0)(4.4,0.5)
\psline{-}(4,0.9)(4,1.2)
\psline{-}(4.4,0.9)(4.4,1.2)
\psarc{-}(5,0){0.2}{0}{180}
\psset{linecolor=black}
\rput(5.6,0){,}
\psdots(6,0)(7.2,0)(6.4,0)(6.8,0)\lw
\psline{-}(6.6,0.3)(7.0,0.7)(6.6,1.1)(6.2,0.7)(6.6,0.3)\unlw
\rput(6.6,0.7){$_1$}
\psset{linecolor=myc2}
\psbezier{-}(6.0,0)(6.0,0.4)(6.4,0.5)(6.4,0.5)
\psbezier{-}(7.2,0)(7.2,0.4)(6.8,0.5)(6.8,0.5)
\psarc{-}(6.6,0){0.2}{0}{180}
\psline{-}(6.4,0.9)(6.4,1.2)
\psline{-}(6.8,0.9)(6.8,1.2)
\psset{linecolor=black}
\rput(7.6,0){,}
\psdots(8.0,0)(8.4,0)(8.8,0)(9.2,0)\lw
\psline{-}(9,0.3)(9.4,0.7)(9,1.1)(8.6,0.7)(9,0.3)\unlw
\rput(9,0.7){$_1$}
\psset{linecolor=myc2}
\psarc{-}(8.2,0){0.2}{0}{180}
\psline{-}(8.8,0)(8.8,0.5)
\psline{-}(9.2,0)(9.2,0.5)
\psline{-}(8.8,0.9)(8.8,1.2)
\psline{-}(9.2,0.9)(9.2,1.2)
\psset{linecolor=black}
\rput(9.6,0){,}
\psdots(10,0)(10.4,0)(10.8,0)(11.2,0)\lw
\psline{-}(11,0.3)(9.8,1.5)(11.0,2.7)(11.4,2.3)(11.0,1.9)(11.4,1.5)(11.0,1.1)(11.4,0.7)(11.0,0.3)
\psline{-}(10.6,0.7)(11,1.1)(10.6,1.5)(11,1.9)(10.6,2.3)
\psline{-}(10.2,1.1)(10.6,1.5)(10.2,1.9)\unlw
\psset{linecolor=myc2}
\psline{-}(10,0)(10,1.3)
\psline{-}(10.4,0)(10.4,0.9)
\psline{-}(10.8,0)(10.8,0.5)
\psline{-}(11.2,0)(11.2,0.5)
\psline{-}(10,1.7)(10,2.8)
\psline{-}(10.4,2.1)(10.4,2.8)
\psline{-}(10.8,2.5)(10.8,2.8)
\psline{-}(11.2,2.5)(11.2,2.8)
\psset{linecolor=myc}
\psarc{-}(11,1.1){0.28284}{-45}{45}
\psarc{-}(11,1.9){0.28284}{-45}{45}
\psset{linecolor=black}
\rput(11,0.7){$_1$}\rput(11,1.5){$_1$}\rput(11,2.3){$_1$}
\rput(10.6,1.1){$_2$}\rput(10.6,1.9){$_2$}
\rput(10.2,1.5){$_3$}\
\end{pspicture}
\bigg\}.
\label{eq:basePB4}
\end{equation*}
As an immediate consequence of Proposition \ref{propo:Pd} {\em (iii)}, we get the following result.
\begin{Proposition} For $\lambda$ generic,
$PB_N$ is a basis of eigenvectors of $F_N(\lambda)$.
\end{Proposition}

The first few $WJ_N$ are $$WJ_1=id, \qquad WJ_2 = id + \frac{S_1}{S_2} e_1, \qquad WJ_3 =id + \frac{S_2}{S_3} (e_1 + e_2) + \frac{S_1}{S_3} (e_1e_2 + e_2e_1).$$ 
Even though these and other
$WJ_N$'s may be singular at some critical $\lambda_c$, the limit $\lim_{\lambda\rightarrow\lambda_c}P^dv$ might exist. For instance, the projector $WJ_3$ is singular at $\Lambda_c=2\pi/3$ in the range $\Lambda\in(\frac{\pi}2,\pi)$. However, $\lim_{\Lambda\to\Lambda_c}P^3(\psset{unit=0.4}
\psset{linewidth=1pt}
\begin{pspicture}(-0.1,0.0)(0.8,0.4)
\psset{linecolor=black}
\psset{dotsize=2.0pt 0}
\psdots(0.0,0)(0.4,0)(0.8,0)
\unlw
\psset{linecolor=myc2}
\psline{-}(0,0)(0,0.5)
\psline{-}(0.4,0)(0.4,0.5)
\psline{-}(0.8,0)(0.8,0.5)
\psset{linecolor=black}
\end{pspicture}) = 
\psset{linewidth=1pt}
\begin{pspicture}(-0.1,0.0)(0.8,0.4)
\psset{linecolor=black}
\psset{dotsize=2.0pt 0}
\psdots(0.0,0)(0.4,0)(0.8,0)
\unlw
\psset{linecolor=myc2}
\psline{-}(0,0)(0,0.5)
\psline{-}(0.4,0)(0.4,0.5)
\psline{-}(0.8,0)(0.8,0.5)
\psset{linecolor=black}
\end{pspicture} - \frac12 (
\psset{linewidth=1pt}
\begin{pspicture}(-0.1,0.0)(0.8,0.4)
\psset{linecolor=black}
\psset{dotsize=2.0pt 0}
\psdots(0.0,0)(0.4,0)(0.8,0)
\unlw
\psset{linecolor=myc2}
\psline{-}(0,0)(0,0.5)
\psarc{-}(0.6,0){0.2}{0}{180}
\psset{linecolor=black}
\end{pspicture} 
+ 
\psset{linewidth=1pt}
\begin{pspicture}(-0.1,0.0)(0.9,0.4)
\psset{linecolor=black}
\psset{dotsize=2.0pt 0}
\psdots(0.0,0)(0.4,0)(0.8,0)
\unlw
\psset{linecolor=myc2}
\psline{-}(0.8,0)(0.8,0.5)
\psarc{-}(0.2,0){0.2}{0}{180}
\psset{linecolor=black}
\end{pspicture}
)$. The value $\Lambda_c = 2\pi/3$ is the one for percolation. It is clear that the condition of genericity on $\lambda$ is somewhat too restrictive. We shall come back to this in the next section.

\begin{Proposition} Let $\lambda$ be generic for $N$. Then
\begin{itemize}
\item[(i)] the subspace spanned by $PB_N^d$ is stable under $TL_N$;
\item[(ii)] in the basis $PB_N$, the matrix $\rho(D_N(\lambda,u))$ is block diagonal, each block corresponding to a sector spanned by $PB_N^d$, $0\le d\le N$.
\end{itemize}
\end{Proposition}
\noindent{\scshape Proof\ \ } The second statement follows from the first. For the latter, it is sufficient to study the action of the generators $e_i$'s on each element $v\in PB_N^d$ of the basis. (The examples drawn below may help.) Three cases occur: 
\begin{itemize}
\item[{\em (a)}] the arc of $e_i$ connects two defects leaving $P^d$ and the result is $0$ by Proposition \ref{propo:Pd};
\item[{\em (b)}] it connects one defect leaving $P^d$ and one arc of $v$. The resulting link state is different from the original one, but is still an element of $PB_N^d$;
\item[{\em (c)}] it connects no defects leaving $P^d$. The resulting link state is also an element of $PB_N^d$, up to a possible factor $\beta$. \hfill$\square$
\end{itemize} 

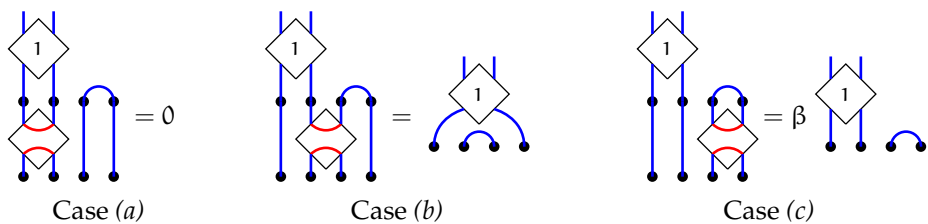
\begin{equation*}
\begin{pspicture}(0,-0.3)(1.2,1)
\psdots(0.0,0)(0.4,0)(0.8,0)(1.2,0)
\psdots(0.0,-1)(0.4,-1)(0.8,-1)(1.2,-1)\lw
\psline{-}(0.2,0.3)(0.6,0.7)(0.2,1.1)(-0.2,0.7)(0.2,0.3)
\psline{-}(0.2,-0.9)(0.6,-0.5)(0.2,-0.1)(-0.2,-0.5)(0.2,-0.9)\unlw
\rput(0.2,0.7){$_1$}
\psset{linecolor=myc}
\psarc{-}(0.2,-0.1){0.282843}{225}{315}
\psarc{-}(0.2,-0.9){0.282843}{45}{135}
\psset{linecolor=myc2}
\psline{-}(0,0)(0,0.5)
\psline{-}(0.4,0)(0.4,0.5)
\psline{-}(0,0.9)(0,1.2)
\psline{-}(0.4,0.9)(0.4,1.2)
\psarc{-}(1,0){0.2}{0}{180}
\psline{-}(0,0)(0,-0.3)
\psline{-}(0.4,0)(0.4,-0.3)
\psline{-}(0,-0.7)(0,-1)
\psline{-}(0.4,-0.7)(0.4,-1)
\psline{-}(0.8,0)(0.8,-1)
\psline{-}(1.2,0)(1.2,-1)
\rput(1.0,-1.45){Case {\em (a)}}
\end{pspicture} \ \ = 0 \qquad \qquad
\begin{pspicture}(0,-0.3)(1.2,1)
\psdots(0.0,0)(0.4,0)(0.8,0)(1.2,0)
\psdots(0.0,-1)(0.4,-1)(0.8,-1)(1.2,-1)\lw
\psline{-}(0.2,0.3)(0.6,0.7)(0.2,1.1)(-0.2,0.7)(0.2,0.3)
\psline{-}(0.6,-0.9)(1,-0.5)(0.6,-0.1)(0.2,-0.5)(0.6,-0.9)\unlw
\rput(0.2,0.7){$_1$}
\psset{linecolor=myc}
\psarc{-}(0.6,-0.1){0.282843}{225}{315}
\psarc{-}(0.6,-0.9){0.282843}{45}{135}
\psset{linecolor=myc2}
\psline{-}(0,0)(0,0.5)
\psline{-}(0.4,0)(0.4,0.5)
\psline{-}(0,0.9)(0,1.2)
\psline{-}(0.4,0.9)(0.4,1.2)
\psarc{-}(1,0){0.2}{0}{180}
\psline{-}(0.8,0)(0.8,-0.3)
\psline{-}(0.4,0)(0.4,-0.3)
\psline{-}(0.8,-0.7)(0.8,-1)
\psline{-}(0.4,-0.7)(0.4,-1)
\psline{-}(0.0,0)(0.0,-1)
\psline{-}(1.2,0)(1.2,-1)
\rput(1.55,-1.45){Case {\em (b)}}
\end{pspicture}  \ \ = 
\begin{pspicture}(-0.2,0.3)(1.5,2)
\psdots(0,0)(1.2,0)(0.4,0)(0.8,0)\lw
\psline{-}(0.6,0.3)(1.0,0.7)(0.6,1.1)(0.2,0.7)(0.6,0.3)\unlw
\rput(0.6,0.7){$_1$}
\psset{linecolor=myc2}
\psbezier{-}(0.0,0)(0.0,0.4)(0.4,0.5)(0.4,0.5)
\psbezier{-}(1.2,0)(1.2,0.4)(0.8,0.5)(0.8,0.5)
\psarc{-}(0.6,0){0.2}{0}{180}
\psline{-}(0.4,0.9)(0.4,1.2)
\psline{-}(0.8,0.9)(0.8,1.2)
\end{pspicture}\qquad \qquad
\begin{pspicture}(0,-0.3)(1.2,1)
\psdots(0.0,0)(0.4,0)(0.8,0)(1.2,0)
\psdots(0.0,-1)(0.4,-1)(0.8,-1)(1.2,-1)\lw
\psline{-}(0.2,0.3)(0.6,0.7)(0.2,1.1)(-0.2,0.7)(0.2,0.3)
\psline{-}(1,-0.9)(1.4,-0.5)(1,-0.1)(0.6,-0.5)(1,-0.9)\unlw
\rput(0.2,0.7){$_1$}
\psset{linecolor=myc}
\psarc{-}(1,-0.1){0.282843}{225}{315}
\psarc{-}(1,-0.9){0.282843}{45}{135}
\psset{linecolor=myc2}
\psline{-}(0,0)(0,0.5)
\psline{-}(0.4,0)(0.4,0.5)
\psline{-}(0,0.9)(0,1.2)
\psline{-}(0.4,0.9)(0.4,1.2)
\psarc{-}(1,0){0.2}{0}{180}
\psline{-}(0.8,0)(0.8,-0.3)
\psline{-}(1.2,0)(1.2,-0.3)
\psline{-}(0.8,-0.7)(0.8,-1)
\psline{-}(1.2,-0.7)(1.2,-1)
\psline{-}(0.0,0)(0.0,-1)
\psline{-}(0.4,0)(0.4,-1)
\rput(1.55,-1.45){Case {\em (c)}}
\end{pspicture}  \ \ = \beta
\begin{pspicture}(-0.3,0.3)(1.5,2)
\psdots(0.0,0)(0.4,0)(0.8,0)(1.2,0)\lw
\psline{-}(0.2,0.3)(0.6,0.7)(0.2,1.1)(-0.2,0.7)(0.2,0.3)\unlw
\rput(0.2,0.7){$_1$}
\psset{linecolor=myc2}
\psline{-}(0,0)(0,0.5)
\psline{-}(0.4,0)(0.4,0.5)
\psline{-}(0,0.9)(0,1.2)
\psline{-}(0.4,0.9)(0.4,1.2)
\psarc{-}(1,0){0.2}{0}{180}
\end{pspicture} 
\end{equation*}
\\
\\

Note that, in view of the example given just before the proposition, the hypothesis that $\lambda$ be generic may be replaced by the less stringent hypothesis that all vectors in $PB_N$ exist or be obtained by the limit process described earlier.

%
%

\section{The Jordan structure of $\rho(D_N(\lambda,u))$}
\label{sec:jordan}

\subsection{Jordan blocks and families of linear transformations}\label{sub:jordan}

The singularity of $WJ_N$ and of the projectors $P^d$ at critical values of $\lambda$ may appear as a weakness of these tools. However it is exactly this singular behavior that allows to probe the Jordan structure of $\rho(F_N(\lambda))$ and, eventually, of $\rho(D_N(\lambda,u))$. 

The study of the Jordan structure of $\rho(F_N(\lambda))$ as a function of $\lambda$ is an example of the study of linear operators depending on parameters. In the early chapters of his book \cite{Kato}, Kato gives examples of each possible singular behaviors of such families. The one that interests us here is given by linear transformations that, in a basis $\{e_1,e_2\}$, have the form of the simple $2\times 2$ matrix
$$m(\lambda)=\begin{pmatrix} \lambda & 1 \\ 0 & 0\end{pmatrix}.$$
Its eigenvectors, when $\lambda\neq 0$, are
$$v_1=e_1\qquad \text{\rm and}\qquad v_2=e_2-\frac{e_1}{\lambda}.$$
Note that the change of basis $\{e_1,e_2\}\rightarrow \{v_1,v_2\}$ is singular at the critical value $\lambda_c=0$. The similarity with $\rho(F_N(\lambda))$ and the change of basis $B_N\rightarrow PB_N$ is revealing. Let us draw the parallel as follows:
$$\begin{matrix}
\text{\rm span }e_1 & \leftrightarrow & UpTo_e& \\
\text{\rm span }\{e_1,e_2\} & \leftrightarrow & UpTo_d, & \text{\rm with }e<d\\
\{e_1,e_2\}\rightarrow \{v_1,v_2\} & \leftrightarrow & B_N\rightarrow PB_N
\end{matrix}$$
One notices that the component of $v_2$ in the subspace $\mathbb Ce_2$ is constant and equal to one, exactly like the vector $P^dv$, for $v\in UpTo_d$, whose component in $V_N^d$ remains $v$ for all $\lambda$. The singularity, in both $v_2$ and $P^dv$, occurs in the subspaces $\mathbb Ce_1\leftrightarrow UpTo_e$. Finally, because $v_2=v_2(\lambda)$ has a simple pole, it can be written as a Laurent polynomial around its singular point $\lambda_c=0$:
$$v_2(\lambda)=r+\frac{s}{\lambda-\lambda_c}$$
and this form is unique if $r,s\in\mathbb C^2$ are chosen to be constant vectors. Then, the matrix $m(\lambda)$ at $\lambda=\lambda_c$ has a $2\times 2$ Jordan block with eigenvalue $\mu=0$ and
\begin{align*}
m(\lambda)r&=\mu r+\text{\rm constant}\times s,\\
m(\lambda)s&=\mu s.
\end{align*}
(A rescaling of $s$ might be necessary to change the constant to the usual $1$ in the Jordan block.) The fact that the regular and singular parts of $v_2$ are proportional to the vectors that form the Jordan block is not a coincidence. This will be the case for $F_N(\lambda)$ and it is the goal of this section to prove it.

We end this subsection by recalling basic properties of the Jordan structure of block triangular matrices. Let $A\in\mathbb C^{n\times n}$ and let $\spec A$ be the set of distinct eigenvalues of $A$. Let $T$ be the matrix that puts $A$ into its Jordan form. (The Jordan form is unique only up to a permutation of its Jordan block. We shall refer to it nonetheless as {\em the} Jordan form of $A$.) Then $T^{-1}AT$ is block diagonal with each of its blocks being of the form 
$$\begin{pmatrix} \mu & 1 & & &  \\
                  0 & \mu & 1 & & \\
                    & & \ddots & \ddots & \\ 
                    & & & \mu & 1\\
                    & & & & \mu
\end{pmatrix}$$
for some $\mu\in \spec A$. For each $\mu\in \text{\rm spec }A$, define the subspace
$$V_A(\mu)=\{v\in\mathbb C^n\ |\ (A-\mu\, \text{\rm id})^n v=0\}.$$
Clearly $\mathbb C^{n}=\oplus_{\mu\in\text{\rm spec }A} V_A(\mu)$. Moreover $A|_{V_A(\mu)}=\mu\times id + n$ where $n$ is a nilpotent matrix.

\begin{Lemme}\label{lem:jordanAB}\begin{itemize}
\item[{(i)}] Let $A,B\in \mathbb C^{n\times n}$ such that $[A,B]=0$. Then $BV_A(\mu)\subset V_A(\mu)$ for all $\mu\in \spec A$.
\item[{(ii)}] Let $\{A_i,1\le i\le m\}$ be a set of $n\times n$ commuting matrices. Let $\mu_i\in\text{\rm spec\ }A_i$ be a choice of $m$ eigenvalues, one for each $A_i$. Then $W(\{\mu_1,\mu_2,\dots,\mu_m\})=\cap_{1\le i\le m}V_{A_i}(\mu_i)$ is stable under all $A_j$. Moreover, if $W_1, W_2, \dots, W_l$ is a list of all non-trivial subspaces obtained by such intersections, then $\mathbb C^n=\oplus_{1\le j\le l}W_j$.
\end{itemize}
\end{Lemme}

\noindent{\scshape Proof\ \ } For {\em (i)}, let $v\in V_A(\mu)$. Then $(A-\mu I)^n (Bv)=B(A-\mu I)^n v =0$
and $Bv\in V_A(\mu)$. 

For the stability of the $W(\{\mu_i\})$, note simply that $A_jW(\{\mu_i\})=\cap_{1\le i\le m}A_jV_{A_i}(\mu_i)\subset W(\{\mu_i\})$ by {\em (i)}. Clearly, if the two ordered sets $\{\mu_i\}$ and $\{\nu_i\}$ are distinct, then $W(\{\mu_i\})\cap W(\{\nu_i\})=\{0\}$ since, for at least one $j$, the eigenvalues $\mu_j$ and $\nu_j$ are distinct and $V_{A_j}(\mu_j)\cap V_{A_j}(\nu_j)=\{0\}$. Therefore, if $W_1, W_2, \dots, W_l$ correspond to different choices of eigenvalues, their pairwise intersections are trivial. Because $\mathbb C^n=\oplus_{\mu\in\text{\rm spec\ }A_i}V_{A_i}(\mu)$ and $V_{A_i}(\mu) = \sum'_{\mu_j \in \text{\rm spec\ }A_j} W(\{\mu_1,\mu_2,...\})$, where $\sum'$ means that $\mu_i$ is fixed to $\mu$, the sum of $W_1, W_2, \dots, W_l$ must be $\mathbb C^n$.
\hfill $\square$

\medskip

Let $M$ be a matrix $\in\mathbb C^{n\times n}$ and let $n_i, 1<i<m$, be positive numbers with $\sum_{1\le i\le m} n_i=n$. We say that $M$ is a {\em block-triangular matrix} if it is partitioned in block $m_{ij}$, $1\le i,j\le n$ where $m_{ij}$ is a $n_i\times n_j$ submatrix and all $m_{ij}=0$ when $i>j$. We now make three observations that will be useful for the study of block triangular matrices.

The most obvious one is $\text{\rm spec\ }M=\cup_{i=1}^m \text{\rm spec\ }m_{ii}$. This has already been used in section \ref{sec:QPotts}.

The second property allows for the identification of non-trivial Jordan blocks using submatrices of a block triangular matrix. Let $\mu$ be an eigenvalue of $M$ whose degeneracy is larger than 1. Suppose that $\mu$ is an eigenvalue of more than a single diagonal block $m_{kk}$ and let $i$ and $j$ be the indices of the first and last diagonal blocks that have $\mu$ as an eigenvalue and suppose $i<j$. Let $M_{\text{\rm top}}$, $M_{\text{\rm mid}}$ and $M_{\text{\rm bot}}$ the submatrices of $M$ along its diagonal that gather all blocks with indices smaller than $i$ for $M_{\text{\rm top}}$, between $i$ and $j$ (included) for 
$M_{\text{\rm mid}}$ and larger than $j$ for $M_{\text{\rm bot}}$. In other words
$$M=\begin{pmatrix} M_{\text{\rm top}} & X & Y \\
                     0 & M_{\text{\rm mid}} & Z\\
                     0 & 0 & M_{\text{\rm bot}}
\end{pmatrix}$$
for certain blocks $X, Y, Z$. Let $v$ be a vector in $V(\mu)$ and let $v=(v_1, v_2, ..., v_m)$, each $v_i$ having $n_i$ components. This vector $v$ solves the equation $(M-\mu)^p v=0$ for some $p$, and we start the study of this equation with its bottom part, namely the equation
$$(M_{\text{\rm bot}}-\mu \ id)^p \begin{pmatrix}v_{j+1}\\ \vdots \\ v_{m}\end{pmatrix}=0.$$
Since $\mu$ is not an eigenvalue of any of the diagonal blocks of $M_{\text{\rm bot}}$ by hypothesis, then $M_{\text{\rm bot}}-\mu \ id$ is non-singular, so are its powers, and all the $v_k$, $k>j$ must be zero. Recall now that the number of eigenvectors associated to $\mu$ is given by $\dim \ker(M-\mu \ id)$. We have just seen that solving $Mv=\mu v$ is equivalent to solving
$$\begin{pmatrix} M_{\text{\rm top}} & X\\
                  0 & M_{\text{\rm mid}}\end{pmatrix}
\begin{pmatrix} v_1\\ \vdots \\ v_j
\end{pmatrix}=\mu\begin{pmatrix} v_1\\ \vdots \\ v_j
\end{pmatrix}.$$
The number of eigenvectors associated to $\mu$ is therefore $\sum_{1\le k\le j}n_k-\text{\rm rank\ }\left(
\left(\begin{smallmatrix} M_{\text{\rm top}} & X\\
0 & M_{\text{\rm mid}}\end{smallmatrix}\right)-\mu \ id\right)$. But $M_{\text{\rm top}}$ does not have $\mu$ as eigenvalue by hypothesis and $M_{\text{\rm top}}-\mu \ id$ is non-singular. So the first $\sum_{1\le k< i} n_k$ columns are of maximal rank. Therefore the number of eigenvectors of $M$ associated to $\mu$ is $\sum_{i\le k\le j}n_k-\text{\rm rank\ }\left( M_{\text{\rm mid}}-\mu \ id\right)=\dim \ker (M_{\text{\rm mid}}-\mu \ id)$. We therefore conclude that {\em the number of eigenvectors of $M$ associated to $\mu$ is the number of eigenvectors of $M_{\text{\rm mid}}$ associated to $\mu$}.
\begin{Definition}\begin{itemize}
\item[(i)] Let $V_i,1\le i\le N$ be a filtration of subspaces $V_1\subset V_2\subset\dots\subset V_N=\mathbb C^n$. Set $V_n=\{0\}$ for $n\le 0$. A vector $v$ is said to belong strictly to $V_i$ if $v\in V_i$ but $v\not\in V_{i-1}$. We then write $v\overset{\circ}{\in} V_i$.
\item[(ii)] For such a filtration let $A:\mathbb C^n\rightarrow \mathbb C^n$ be a linear transformation such that $AV_i\subset V_i$ for all $i$. The linear transformation $A$ is said to have a Jordan block between $V_i$ and $V_j$, $i>j$, if there exists $\mu\in\text{\rm spec\ }A$ and two vectors $v$ and $w$ in $V_A(\mu)$ such that $v\overset{\circ}{\in} V_i$, $w\overset{\circ}{\in}V_j$ and $(A-\mu\, \text{\rm id})v=w$.
\end{itemize}
\end{Definition}
By this definition, a vector $v$ strictly in $V_i$ is always non-zero. The second part of the definition is fairly intuitive, but it helps in formulating the third property. Suppose as before that $M$ is a block triangular matrix. Set $V_1$ to be the subspace of vectors whose components after the $n_1$ first ones are all zero. Similarly the vectors in $V_j$ have their components zero if they are after the first $(n_1+n_2+\dots+n_j)$ ones. Because $M$ is block triangular, it satisfies $MV_i\subset V_i$. Suppose now that the diagonal blocks $A_{ii}$ and $A_{jj}$, $i<j$, both share the eigenvalues $\mu$ and let $M_{\text{\rm mid}}$ as before. Then, an argument similar to the previous one shows that $M$ has a Jordan block between $V_i$ and $V_j$ if and only if $M_{\text{\rm mid}}$ does.

The following lemma is an easy consequence of the definitions and properties just introduced.

\begin{Lemme}\label{lem:lePetit} The matrix
$\rho(D_N(\lambda,u))$ has no Jordan blocks between two subspaces $UpTo_d$ and $UpTo_{d'}$ with defects $d$ and $d'$ if $\cos (\lambda(d+1)) \neq \cos (\lambda(d'+1)) $.
\end{Lemme}
\noindent{\scshape Proof\ \ } The block $\rho(F_N(\lambda))_{d,d}$ has a unique, degenerate eigenvalue  $\mu_d=2 (-1)^d \cos(\lambda(d+1)) $. Its subspaces $V_{F_N}(\mu_d)$ are therefore spanned by vectors strictly in some $UpTo_{d'}$'s whose $d'$ satisfies $\cos(\lambda(d+1))=\cos(\lambda(d'+1))$. Since $F_N$ and $D_N$ commute, $\rho(D_N(\lambda,u))$ must share the same property by Lemma \ref{lem:jordanAB}.\hfill$\square$

\smallskip

As a consequence, for $\lambda/\pi$ irrational, $\rho(D_N(\lambda,u))$ will never have any Jordan blocks.

%
%

\subsection{The singularities of $P^d$}

To carry through the program outlined in the previous paragraph, the first task is to identify the singularities of all $P^dv$, $v\in B_N^d$. This is somewhat simplified by the fact that, for $v\in B_N^d$, the computation of $P^dv$ starts by that of the action of $\rho(WJ_d)$ on the vector with $d$ defects. This vector will be denoted $v^d\in B_d^d$. Any singularities of $P^dv$ for $v\in B_N^d$ are therefore readable from those of $\rho(WJ_d)(v^d)$. This simplification is welcome as the explicit computation of $\rho(WJ_d)(v^d)$ is itself difficult. The aim could be to give all components of this vector explicitly. We tried that for a while, but realized that a more modest goal is sufficient. It is enough to calculate a few well-chosen coefficients and relate all others to these. The following two lemmas give all the information needed to proceed with our goal. Their proof is highly technical and done in Appendix \ref{app:a}.

Since only the coefficients of $\rho(WJ_d)(v^d)$ are needed, we shall denote them as $P^d_w$ where $w$ is some link basis vectors. Therefore $P^d_w$ is the matrix element of $P^d$ at position $(w,v^d)$. Let $\{m^m\}$ denote the basis vector with $m$ concentric arcs joining together the first $2m$ points, all other being joined to defects. 
\begin{Lemme} Concentric bubbles in leftmost position.
\begin{equation*}
P^d_{\{m^m\}} = \frac1{ \left( 2 S_{1/2} \right)^m} \prod_{i=0}^{m-1} \frac{S_{(d-m-i)/2}}{C_{(d-i)/2}},
\end{equation*}
where, again, the following compact notation has been used:
$S_k = \sin(k \Lambda)$, $C_k = \cos(k \Lambda)$, $\Lambda = \pi-\lambda$.
\end{Lemme}
These basis vectors $w=\{m^m\}$ are well-chosen as their components contain all possible singularities of $P^d$.
\begin{Lemme}\label{sec:anylinkstate}
Let $w$ be any link state with $k$ arcs. Then 
$$P^d_w=\sum_{i=1}^k\alpha_iP^d_{\{i^i\}}$$
for some functions $\alpha_i$ analytic in $\Lambda$. Therefore
\begin{equation}
\{\Lambda \in \mathbb{R}\, |\, P^d_{w} \ \textrm{diverges at }\Lambda\}  \subset  \cup_{i=1}^k \{\Lambda \in  \mathbb{R}\, |\, P^d_{\{i^i\}} \ \textrm{diverges at } \Lambda\}  \subset \cup_{i=0}^{k-1} \{\Lambda \in  \mathbb{R}\, |\, C_{(d-i)/2}=0 \}.
\label{eq:inclusion}
\end{equation}
\end{Lemme}

%
%
\subsection{Jordan blocks of $\rho(F_N(\lambda))$}

We first study the coefficient $P_{\{m^m\}}^d$ as a function of $q=e^{i\lambda}$. 

\begin{Lemme}Let $\Lambda_c=\pi a/b$ with $a,b$ coprime integers and $b$ non-zero and $q_c=e^{i(\pi-\Lambda_c)}$. Then
\begin{itemize}
\item for $m$ in the range $1\le m< b$, the coefficient $\pmm$ has a pole at $q_c$ if $\left.C_{(d-m+1)/2}\right|_{q=q_c}=0$;
\item if $m=m_1b+m_2$ with $m_1>0$ and $b>m_2\ge 0$, then one of $\pmm\pm P^d_{\{m_2^{m_2}\}}/(2S_{1/2})^{m_1b}$ is regular at $q_c$;
\item all poles of $\pmm(q), m>0,$ are simple.
\end{itemize}
\end{Lemme}

\noindent Note that the lemma does not preclude singularities of $\pmm$ other than the zeroes of $C_{(d-m+1)/2}$. In fact, $\pmm$ will have in general poles at zeroes of some of the $C_{(d-i)/2}$ with $0\le i<m-1$.

\medskip

\noindent{\scshape Proof:} The following properties characterize the number $C_{n/2}$ and $S_{n/2}$, $n\in\mathbb Z$, at $\Lambda=\Lambda_c$. First, if $a$ is even, then $C_{n/2}$, $n\in\mathbb Z$, is never zero. Indeed the argument of $C_{n/2}=\cos (\pi an/2b)$ is $\pi$ times a fraction whose denominator is odd. This argument can never be of the form $(2i+1)\pi/2$. From now on, we assume that $a$ is odd. Second, at $\Lambda=\Lambda_c$, $C_{(n+2b)/2}=-C_{n/2}$ and the values taken by $|C_{n/2}|, n\in\mathbb Z,$ belong to any set of $2b$ consecutive $C_i$'s, namely to any set $\{|C_{j/2}|, |C_{(j+1)/2}|,\dots, |C_{(j+2b-1)/2}|\}$ for any $j$. The values $|S_{n/2}|$ have the same property. Third, for all $n$, $|C_{(n+b)/2}|=|S_{n/2}|$. And finally, for any $j\in\mathbb Z$, the sets $\{C_{(j+i)/2},0\le i<2b\}$ (and therefore $\{S_{(j+i)/2},0\le i<2b\}$) contain precisely one zero. To prove this, clearly $C_{b/2}=\cos \pi a/2=0$ as $a$ is odd. Moreover, if $C_{i/2}=C_{j/2}=0$ with $|j-i|<2b$, then $\Lambda_c i/2=(2k-1)\pi/2$ and $\Lambda_c j/2=(2l-1)\pi/2$ for some $k,l\in\mathbb Z$ and $a(i-j)=2(l-k)b$. Since $a$ is odd and $a, b$ are coprimes, $(i-j)$ must be a multiple of $2b$ and the constraint $|j-i|<2b$ implies that $i=j$.

We now study $C_{n/2}=C_{n/2}(q)$ in a neighborhood of $q_c$. If $\cos \Lambda_c n/2=0$, then
\begin{align*}
\cos \Lambda n/2 &=\cos (\Lambda-\Lambda_c+\Lambda_c)n/2=\pm \sin (\Lambda-\Lambda_c)n/2\\
&=\pm \frac i2\frac{(q-q_c)}{(qq_c)^{n/2}}\sum_{j=0}^{n-1}q^jq_c^{n-1-j}
\end{align*}
where $q=e^{i(\pi-\Lambda)}$ and $C_{n/2}(q)$ goes to zero linearly in $(q-q_c)$ as $q$ goes to $q_c$.

The formula
$$\pmm = \frac1{(2S_{1/2})^m}\prod_{i=0}^{m-1}\frac{S_{(d-m-i)/2}}{C_{(d-i)/2}}$$
reveals that the only possible poles of $\pmm$ are at the zeroes of the cosine functions $C_{(d-i)/2}$. There will be a singularity of $\pmm$ at a zero of a given cosine of the denominator if and only if this zero is not cancelled by a zero of one of the sine functions of the numerator. Therefore $\pmm$ will have a pole at zeroes of $C_{(d-m+1)/2}$ if no sine functions of the numerator are equal, up to a sign, to $S_{(d-(m+b-1))/2}$. There is no such cancellation because the hypothesis $m<b$ implies that $d-(m+b-1)<d-(2m-1)$ and therefore that $d-(m+b-1)$ is not in the range $[d-(2m-1), d-m]$ of the index of the sine functions.

Suppose now that $m=m_1b+m_2$ with $m_1\ge 1$ and $b>m_2\ge 0$. If $m_1$ is even, the last $m_1b$ terms of the product (corresponding to the values of the index $m_2\le i\le m_1b+m_2-1$) contain $m_1/2$ complete sets of the possible absolute values of both the sine and cosine functions. The product of these $m_1b$ terms is therefore $\pm 1$. Using the periodicity of the sine, one gets
$$\pmm = \pm\frac1{(2S_{1/2})^m}\prod_{i=0}^{m_2-1}\frac{S_{(d-m_1b-m_2-i)/2}}{C_{(d-i)/2}}+\mathcal O((q-q_c)^0)=\pm\frac1{(2S_{1/2})^{m_1b}}P^d_{\{m_2^{m_2}\}}+\mathcal O((q-q_c)^0).$$
If $m_1$ is odd, the terms of the product are split as follows. The $m_2$ first cosines are gathered with the last $m_2$ sines in a first product, and all the remaining sines and cosines in a second:
$$\pmm=\frac1{(2S_{1/2})^m}\prod_{i=0}^{m_2-1}\frac{S_{(d-2m_1 b-m_2-i)/2}}{C_{(d-i)/2}}
\prod_{i=0}^{m_1b-1}\frac{S_{(d-m_1b-m_2-i)/2}}{C_{(d-m_2-i)/2}}.$$
Because $m_1$ is odd, $S_{(d-m_1b-m_2-i)/2}=\pm C_{(d-m_2-i)/2}$ at $q=q_c$ and the last product is $\pm 1$. Again the periodicity of trigonometric functions gives the same result as for $m_1$ even.

Thus the coefficient $P^b_{\{n^n\}}$ with $n\ge b$ has the same singularities as $\pmm$ with $m\equiv n\mod b$, $0\le m<b$. But the denominator of $\pmm$, $1\le m<b-1$, has less than $b$ cosine functions among $\{C_{d/2}, C_{(d-1)/2}, \dots,\allowbreak C_{(d-(b-2))/2}\}$ and at most one of these can vanish for a given $\Lambda_c$. Hence poles of $\pmm$, $m\ge 1$, are simple.\hfill$\square$

\begin{Lemme} Let $\Lambda_c=\pi-\lambda_c=\pi a/b$ with $a,b$ coprime integers and $b$ non-zero and set $q_c=e^{i(\pi-\Lambda_c)}$. Then
\begin{itemize}
\item if $a$ is even: $P^d(q)(v^d)$ is not singular at $\Lambda_c=\pi a/b$;
\item if $a$ is odd: 
$P^d(q)(v^d)$ has a singularity at $\Lambda_c=\pi a/b$ if and only if there exists $d'$ satisfying
\begin{equation}\label{eq:laCondition}
\textrm{\fbox{$d-d'<2b \qquad$ and $\qquad {\textstyle{\frac12}}(d+d')\equiv b-1\mod 2b.$}}
\end{equation}
$d'$ is the largest integer for which there exists a $w\in B^{d'}_N$ 
such that the coefficient $P^d_w$ is singular. (By definition of $P^d$, $d'$ is strictly smaller than $d$.) In the following, $d'$ will retain this definition.
\end{itemize}
\end{Lemme}

\noindent{\scshape Proof:} The only poles of $P^d(q)(v^d)$ are at zeroes of $C_{(d-i)/2}$, $1\le i<b-1$. But if $a$ is even, and $b$ is therefore odd,
then none of these cosines vanish at $\Lambda=\Lambda_c$.

Let then $a$ be odd. In order for $d'$ to be the largest defect number with some of the $P^d_w$, $w\in B^{d'}_N$,  
singular, one must have that $P^d_v$ is regular for any $v\in B^e_N$, 
 $e>d'$. This means that all $P^d_{\{n^n\}}$ with $n<m=\frac12 (d-d')$ are regular. Because of the inclusion of sets of singularities of Lemma \ref{sec:anylinkstate}, the coefficient $\pmm$ must be singular. Hence the definition of the number $d'$ is equivalent to the requirement that all $P^d_{\{n^n\}}$ with $n<m=\frac12 (d-d')$ be regular and $\pmm$ be singular at $\Lambda_c$. This $\Lambda_c$ must therefore be a zero of $C_{(d-m+1)/2}$, that is $d-m+1=(2k+1)b$ for some $k\in\mathbb Z$ or, equivalently,
$\frac{(d+d')}2\equiv b-1\mod 2b.$\hfill $\square$

\medskip

Let $\Lambda_c=\pi a/b$, with $a,b$ coprimes and $b$ non-zero, $q_c=e^{i(\pi-\Lambda_c)}$, the pair $(d,d')$ solving \eqref{eq:laCondition} and $v\in B_N^d$. 
 Because the poles of $P^d(q)(v^d)$ are simple, the vector $P^d(q)v$ can be written as 
$$P^d(q)v=r_{q_c}(q,v)+\frac{s_{q_c}(v)}{q-q_c}$$
where $s_{q_c}(v)$ is the residue of $P^d(q)v$ at $q_c$. As a function of $q$, the vector $r_{q_c}(q,v)$ is then analytic in a neighborhood of $q_c$.

Recall that $\upto_d=\bigoplus_{e\le d}V^{e}_N$ is the subspace spanned by basis vectors with at most $d$ defects. Our next goal is to study the action of $F_N(\lambda)$ on the projections of $r_{q_c}$ and $s_{q_c}$ on the quotient $\upto_d/\upto_{d'-2}$. Let $v\in \upto_d$ and denote by $v_{[d',d]}$ the subset of its components along the link basis vectors with defect number in the range $[d',d]$. Clearly the coordinate vectors $v_{[d',d]}$, $v\in B_N^d$, and the projected vectors $\pi(v)\in \upto_d/\upto_{d'-2}$ are in one-to-one correspondence. (Here $\pi$ denotes the canonical projection $\upto_d\rightarrow \upto_d/\upto_{d'-2}$.) Let $M:V_N\rightarrow V_N$ be a linear transformation and denote by $M_{[d',d]}$ the diagonal block of its matrix form in the link basis that includes lines and columns in sectors $d'$ to $d$. Because $F_N(\lambda)$ is upper triangular, $(F_N(\lambda)v)_{[d',d]}=F_N(\lambda)_{[d',d]}v_{[d',d]}$ for $v\in \upto_d$. (In other words, there is a natural action of $F_N(\lambda)$ on the quotient $\upto_d/\upto_{d'-2}$.)

\begin{Lemme} Let $(d',d)$ be a solution of \eqref{eq:laCondition}. With the notation above, if $\hat r=(r_{q_c}(q_c,v))_{[d',d]}$ and $\hat s=(s_{q_c}(v))_{[d',d]}$, then 
\begin{align*}
(F_N(\lambda_c))_{[d',d]}\hat r & =\mu_d\hat r+\alpha \hat s,\\
(F_N(\lambda_c))_{[d',d]}\hat s & =\mu_d\hat s
\end{align*}
where $\alpha\in\mathbb C^\times$ and $\mu_d=(-1)^d(q_c^{(d+1)}+q_c^{-(d+1)})$.
\end{Lemme}

\noindent In other words $\hat r$ and $\hat s$ are part of a Jordan block under the natural action of $F_N(\lambda)$ on $\upto_d/\upto_{d'-2}$.

\medskip

\noindent{\scshape Proof:} We write $F_N(q)$ instead of $F_N(\lambda)$, with $q=e^{i\lambda}$, to stress the dependency upon $q$ and compute the action of $F_N(q)$ for $q$ in a neighborhood of $q_c$, but distinct of it:
\begin{align*}
F_N(q)_{[d',d]}(r_{q_c}(q,v))_{[d',d]} & = F_N(q)_{[d',d]}\left((P^d(q)v)_{[d',d]}-
\frac{s_{q_c}(v)_{[d',d]}}{q-q_c}\right)\\
&= \mu_d(q)(P^d(q)v)_{[d',d]}-\mu_{d'}(q)\frac{\hat s}{q-q_c}\\
\intertext{because $P^d(q)v$ is an eigenvector of $F_N(q)$ for generic $q$ with eigenvalue $\mu_d(q)=2(-1)^d\cos (\lambda(d+1))=(-1)^d(q^{(d+1)}+q^{-(d+1)})$ and $s_{q_c}(v)_{[d',d]}=\hat s$ has components only in $V^{d'}_N$ and is therefore an eigenvector with eigenvalue $\mu_{d'}(q)$. Then }
&= \mu_q(r_{q_c}(q,v))_{[d',d]}+\left(\frac{\mu_d(q)-\mu_{d'}(q)}{q-q_c}\right)\hat s.
\end{align*}
The limit $q\rightarrow q_c$ of the first term is simply $\mu_d(q_c)\hat r$, since both factors are smooth at $q_c$. Since $d$ and $d'$ have the same parity, the difference $(\mu_d(q)-\mu_{d'}(q))$ can be written up to a sign as
$$\cos\lambda(d+1)-\cos\lambda(d'+1)=-2\sin \frac\lambda2(d-d')\sin\frac\lambda2(d+d'+2).$$
If $(d,d')$ is a solution of \eqref{eq:laCondition}, then $d+d'+2=2b(2k+1)$ for some integer $k$ and $\sin\frac\lambda2(d+d'+2)$ has a simple zero at $\lambda=\lambda_c$.
Could $\sin\frac\lambda2(d-d')$ also vanish at $\lambda=\lambda_c$? This would mean that both $\lambda_c(d+1)$ and $\lambda_c(d'+1)$ are integer multiples of $\pi$. Because $a$ and $b$ are coprimes, this implies that $2b$ divides both $(d+1)$ and $(d'+1)$. The requirement that $|d-d'|<2b$ implies that there are no such pair $(d,d')$ with distinct $d$ and $d'$. The limit $\alpha=\lim_{q\rightarrow q_c}(\mu_d(q)-\mu_{d'}(q))/(q-q_c)$ therefore exists and is non-zero. The second equation of the statement follows from the previous observation that $\hat s$ is an eigenvector of $F_N(\lambda)_{[d',d]}$ and its eigenvalue coincides with $\mu_d(q_c)$ when $q\rightarrow q_c$.\hfill$\square$

\medskip

Because the constant $\alpha$ is non-zero, it can be absorbed in the definition of $\hat s$. By the discussion in section \ref{sub:jordan}, the previous lemma implies that $\rho(F_N(\lambda_c))$ has a Jordan block between the subspace $UpTo_d$ and $UpTo_{d'}$. The next result provides the linear condition determining the size of the Jordan block.

\begin{Lemme} Let $\lambda_c$ be fixed and note $F_N$ for $F_N(\lambda_c)$. Let $\hat x_0$ and $\hat x_1$ be two vectors in $\upto_d/\upto_{d'-2}$ that are respectively projections of some vectors strictly in $\upto_d$ and $\upto_{d'}$.
The two following statements are equivalent. 
\begin{itemize}
\item $\hat x_0$ and $\hat x_1$ satisfy
\begin{align*}
(F_N)_{[d',d]}\hat x_0 & = \mu \hat x_0+\hat x_1\\
(F_N)_{[d',d]}\hat x_1 &= \mu \hat x_1;
\end{align*}
\item there exists $\{x_0,x_1,\dots, x_i\}\subset V_N, i\ge 1,$ such that
\begin{align*}
F_Nx_j&=\mu x_j+x_{j+1}, \qquad 0\le j\le i-1\\
F_Nx_i&=\mu x_i
\end{align*}
where $x_j\overset{\circ}{\in}UpTo_{d_j}$ with $d_{j+1}<d_j$ and $(x_0)_{[d',d]}=\hat x_0$ and $(x_1)_{[d',d]}=\hat x_1$.
\end{itemize}
\end{Lemme}

\noindent{\scshape Proof:} Due to the block-triangular structure of $F_N$, the statement $\Leftarrow$ is immediate. To prove the statement $\Rightarrow$, we solve recursively on the number of defects $e< d'$. Suppose that a partial solution has been obtained for the components in sectors $e+2$ to $d$ of $l+1$ vectors:
\begin{align*}
(F_Nx_j)_{[e+2,d]}&=\mu (x_j)_{[e+2,d]}+(x_{j+1})_{[e+2,d]}, \qquad 0\le j<l\\
(F_Nx_l)_{[e+2,d]}&=\mu (x_l)_{[e+2,d]}.
\end{align*}
The next step is to determine the components of these vectors along the basis vectors in $B_N^e$. The block matrix $(F_N)_{[e,e]}$ is a multiple of the identity, say by the factor $\sigma$. If $a=\sigma-\mu$ is non-zero, then the system of equations for the components in $B_N^e$ is
\begin{align*}
a(x_j)_{[e,e]}+X_e(x_j)_{[e+2,d]}&=(x_{j+1})_{[e,e]},\qquad 0\le j<l\\
a(x_l)_{[e,e]}+X_e(x_l)_{[e+2,d]}&=0
\end{align*}
where the rectangular matrix $X_e$ contains the parts of the lines of $F_N$ to the right of the block $(F_N)_{[e,e]}$. The solution to this system is unique and can be found by solving first the last equation, labeled by $l$, then the previous, and so on.

If $a=\sigma-\mu=0$, then the system is different and might not have a solution. It is
\begin{align}
X_e(x_j)_{[e+2,d]}&=(x_{j+1})_{[e,e]},\qquad 0\le j<l\\
X_e(x_l)_{[e+2,d]}&=0\label{eq:JordanSize}.
\end{align}
The last equation is actually a constraint on the components of $x_l$ that have already been fixed. If it is not satisfied, a new vector $x_{l+1}$ has to be introduced in the Jordan block. Its components in $B_N^e$ are $(x_{l+1})_{[e,e]}=X_e(x_l)_{[e+2,d]}$. Then the $j$-th equation can be used to determine $(x_{j+1})_{[e,e]}$ for $1<j\le l$. Finally one can set $(x_0)_{[e,e]}$ to zero. This determines the components in $\cup_{e\le f\le d}B_N^f$ 
of all the vectors in this extended Jordan block. Repeating the process until $e=0$ is reached leads to the second statement.\hfill$\square$

\medskip

The constraint \eqref{eq:JordanSize} will not be satisfied in general if $\hat x_0$ is chosen from the regular part of $P^d(v)$ for some $v\in B_N^d$. 
But, of course, a linear combination of vectors in $PB_N^d$ 
might. Therefore the number of Jordan blocks connecting the sectors $d$ and $d'$ is likely to be smaller than the dimension of $V_N^d$. 

The following proposition determines between which sectors the central element $F_N$ has Jordan blocks. It is an immediate consequence of the three previous lemmas.

\begin{Proposition}\label{thm:lePetit}
Let $\Lambda_c=\pi-\lambda_c=\pi a/b$, with $a,b$ coprimes and $b$ non-zero. Then
\begin{itemize}
\item if $a$ is even, $F_N(\lambda_c)$ has no Jordan blocks;
\item let $a$ be odd and $|d-d'|<2b$. Then $F_N(\lambda_c)$ has a Jordan block between the sectors $d$ and $d'$ if and only if 
$${\textstyle{\frac12}}(d+d')\equiv b-1\mod 2b.$$
\end{itemize}
\end{Proposition}

%
%
\subsection{Jordan blocks of $\rho(D_N(\lambda,u))$}

\begin{Proposition}\label{thm:laGrande}
\begin{itemize}
\item If $P^d(q=e^{i\lambda})(v^d)$ is regular at $\lambda$, then, for all $u$ and $N$, the matrix $\rho(D_N(\lambda, u))$ has no Jordan blocks between the sectors $d$ and $d'$, for any $d'<d$.
\item If $P^d(q=e^{i\lambda})(v^d)$ is singular at some $\lambda=\lambda_c$ and the pair $(d,d')$ satisfies \eqref{eq:laCondition}, then, for all $u$, but a finite number of values, and all $N(\ge d)$, there are some vectors $\overset{\circ}{\in}\upto_d$ that form Jordan blocks of $\rho(D_N(\lambda_c,u))$ with vectors in $\upto_d$. Moreover, if the diagonal block $\rho(D_N)_{d,d}$ is diagonalizable, then $\rho(D_N(\lambda_c,u))$ has a Jordan block between sectors $d$ and $d'$, for all $u$ except a finite number of values.
\end{itemize}
\end{Proposition}

\noindent{\scshape Proof:} If $P^d(e^{i\lambda})(v^d)$ is regular at $\lambda$, then all vectors in $PB_N^d$ are strictly in $\upto_d$ and their projections on $\upto_d/\upto_{d-2}$ form a basis of this quotient. The subspace spanned by $PB_N^d$ in $\upto_d$ is stable under the action of the generators of the Temperley-Lieb algebra and, therefore, of $\rho(D_N(\lambda,u))$. This set $PB_N^d$ can be completed into a basis of $\upto_d$ by a set $B$ of vectors strictly in $\upto_e$, for some $e<d$. In this basis $D_N(\lambda,u)$ has no matrix elements between the two subspaces spanned by $PB_N^d$ and $B$ respectively.

Expand $\rho(D_N(\lambda_c,u))$ as a trigonometric polynomial of $u$ as in \eqref{eq:fourier}. There are at most $N+1$ linearly independent coefficients in this polynomial, the last one is $\rho(F_N)$ up to a non-zero constant, and they all commute by \eqref{eq:commu}. Denote by $d_i,0\le i\le N$, their restrictions to $\upto_d$. These $d_i$ are therefore linear transformations of $\upto_d$. As shown in Lemma \ref{sub:jordan}, the vector space $UpTo_d$ can be written as a direct sum of subspaces $W(\{\mu_i\})=\cap_{0\le i\le N}V_{d_i}(\mu_i)$ where $\{\mu_i\}$ represents a choice of one eigenvalue $\mu_i$ for each $d_i$. The subspaces $W(\{\mu_i\})$ are stable under all $d_j$'s.

Suppose now that one of them, say $d_N$, has a Jordan block between the sectors $d$ and $d'$. Let $\mu$ be the eigenvalue of this block. Since the Jordan form of $d_N$ contains a Jordan block of size at least $2$, there exists a choice $\{\mu_i\}$ with $\mu_N=\mu$ such that the restriction of $d_N$ to $W_0=W(\{\mu_i\})$ has a non-trivial Jordan block. The restrictions of the $d_i$'s to this subspace $W_0$ are of the form $\mu_i \cdot id_{W_0}+n_i$ where $n_i$ is nilpotent. Because the $d_i$'s commute, so do the $n_i$'s. Hence, on this subspace $W_0$, the matrix $\rho(D_N(\lambda_c,u))|_{W_0}=f(u)\cdot id_{\textrm{dim}W_0}+n(u)$ where $f(u)$ is a trigonometric polynomial and $n(u)$ is a nilpotent matrix. (Recall that the sum of two commuting nilpotent matrices is nilpotent.) Because $n_N$ is non-zero, the matrix $n(u)$ is itself a non-zero trigonometric polynomial. It can vanish only at a finite number of values of $u$ and, generically, $\rho(D_N)|_{UpTo_d}$ has Jordan blocks.

To see if $\rho(D_N(\lambda_c,u))$ has a Jordan block between $UpTo_d$ and $UpTo_{d'}$, it is sufficient, by the last property of section \ref{sub:jordan}, to see if its action on the quotient $UpTo_d/UpTo_{d'-2}$ has such a block. Suppose now that the diagonal block $\rho(D_N(\lambda_c,u))_{d,d}$ is diagonalizable and pick $W_0$ as above. Then $W_0$ is stable  under $\rho(D_N)$ and $\rho(D_N)$ preserves the filtration $\dots\subset W_0\cap UpTo_{d'}\subset\dots\subset W_0\cap UpTo_d\subset \dots$ However the action of $\rho(D_N)$ on $(W_0\cap UpTo_d)/(W_0\cap UpTo_{d-2})$ cannot have a Jordan block since $\rho(D_N)_{d,d}$ is diagonalizable. But $\rho(D_N)$ can only have Jordan blocks between sectors upon which $F_N$ has the same eigenvalues by Lemma \ref{lem:lePetit}. Because $|d'-d|<2b$, the sectors $d$ and $d'$ are two such consecutive sectors. Therefore the actions of both $\rho(F_N)$ and $\rho(D_N)$ on $(W_0\cap UpTo_d)/(W_0 \cap UpTo_{d'-2})$ must have a Jordan block between the projections of $W_0\cap UpTo_d$ and $W_0\cap UpTo_{d'}$. The last statement follows, again by the last property of section \ref{sub:jordan}.\hfill$\square$

\medskip

Both the diagonalizability of the diagonal blocks $\rho(D_N)_{d,d}$ and the reality of their spectrum are delicate questions. A natural way to settle both questions is to find a scalar product on $V_N^d$ with respect to which $\rho(D_N)_{d,d}$ is self-adjoint. We found such a scalar product, for all $\beta$ and $N\le 12$, using a computer. In the basis $B_N$, the elements of the matrix representing the scalar products are polynomials in $\beta$. But we could not guess its form for general $N$. The situation is to be paralleled to the same two questions for the $XXZ$ Hamiltonian with the specific boundary conditions introduced in \cite{Alcaraz}. With these boundary conditions, $H_{XXZ}$ is invariant under $U_q(sl(2))$ \cite{PasquierSaleur}. Because $U_q(sl(2))$ has indecomposable representations when $q$ is a root of unity, it is not surprising that, at these values, this Hamiltonian is not diagonalizable. Korff and Weston \cite{Korff} identified a subspace to which the restriction of $H_{XXZ}$ is self-adjoint with respect to some scalar product. Due to the link between loop transfer matrices and the Hamiltonian $H_{XXZ}$, we had hoped their efforts would answer rigorously our questions for $\rho(D_N)$. However the subspace constructed in \cite{Korff} is too small to contain the image of the diagonal blocks of $\rho(D_N)$, even for the $N$ for which our computation answers affirmatively the questions of diagonalizability and of the reality of the spectrum.

\section{Conclusion}

The results of this paper show the existence of Jordan blocks between sectors $d$ and $d'$ for certain $\lambda$'s. For example, for the transfer matrices $\rho(D_N)$ related to critical polymers, the Ising and the $3$-Potts models by Proposition \ref{propo:pottsLoops}, the sectors tied by a line in the following diagrams are joined by Jordan blocks. 
\begin{equation*}
\begin{matrix}
\text{\rm Critical polymers ($\lambda=\pi/2$)} &
\text{\rm Ising model ($\lambda=\pi/4$)} &
\text{\rm 3-Potts model ($\lambda=\pi/6$)}\\
& \\
\text{\rm for $N$ even}\hfill &\\
\psset{unit=0.5}
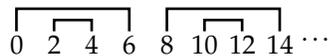
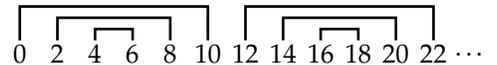
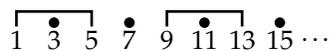
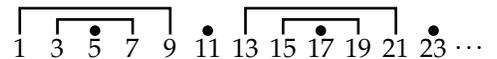
\begin{pspicture}(0,0)(6.3,0.8)
\rput(0,-0.15){0}
\rput(1,-0.15){2}
\rput(2,-0.15){4}
\rput(3,-0.15){6}
\rput(4,-0.15){8}
\rput(5,-0.15){10}
\rput(6,0){\dots}
\psline{-}(0,0.2)(0,0.5)(1,0.5)(1,0.2)
\psline{-}(2,0.2)(2,0.5)(3,0.5)(3,0.2)
\psline{-}(4,0.2)(4,0.5)(5,0.5)(5,0.2)
\end{pspicture}&
\ \qquad
\psset{unit=0.5}
\begin{pspicture}(0,0)(8.3,0.8)
\rput(0,-0.15){0}
\rput(1,-0.15){2}
\rput(2,-0.15){4}
\rput(3,-0.15){6}
\rput(4,-0.15){8}
\rput(5,-0.15){10}
\rput(6,-0.15){12}
\rput(7,-0.15){14}
\rput(8,0){\dots}
\psline{-}(0,0.2)(0,0.8)(3,0.8)(3,0.2)
\psline{-}(4,0.2)(4,0.8)(7,0.8)(7,0.2)
\psline{-}(1,0.2)(1,0.5)(2,0.5)(2,0.2)
\psline{-}(5,0.2)(5,0.5)(6,0.5)(6,0.2)
\end{pspicture}&
\ \qquad\psset{unit=0.5}
\begin{pspicture}(0,0)(12.3,1.1)
\rput(0,-0.15){0}
\rput(1,-0.15){2}
\rput(2,-0.15){4}
\rput(3,-0.15){6}
\rput(4,-0.15){8}
\rput(5,-0.15){10}
\rput(6,-0.15){12}
\rput(7,-0.15){14}
\rput(8,-0.15){16}
\rput(9,-0.15){18}
\rput(10,-0.15){20}
\rput(11,-0.15){22}
\rput(12,-0.15){\dots}
\psline{-}(0,0.2)(0,1.1)(5,1.1)(5,0.2)
\psline{-}(6,0.2)(6,1.1)(11,1.1)(11,0.2)
\psline{-}(1,0.2)(1,0.8)(4,0.8)(4,0.2)
\psline{-}(7,0.2)(7,0.8)(10,0.8)(10,0.2)
\psline{-}(2,0.2)(2,0.5)(3,0.5)(3,0.2)
\psline{-}(8,0.2)(8,0.5)(9,0.5)(9,0.2)
\end{pspicture}\\
& \\
\text{\rm for $N$ odd}\hfill &\\
\psset{unit=0.5}
\begin{pspicture}(0,0)(6.3,0.8)
\rput(0,-0.15){1}
\rput(1,-0.15){3}
\rput(2,-0.15){5}
\rput(3,-0.15){7}
\rput(4,-0.15){9}
\rput(5,-0.15){11}
\rput(6,-0.15){\dots}
\psdots(0,0.3)(1,0.3)(2,0.3)(3,0.3)(4,0.3)(5,0.3)
\end{pspicture}&
\ \qquad\psset{unit=0.5}
\begin{pspicture}(0,0)(8.3,0.8)
\rput(0,-0.15){1}
\rput(1,-0.15){3}
\rput(2,-0.15){5}
\rput(3,-0.15){7}
\rput(4,-0.15){9}
\rput(5,-0.15){11}
\rput(6,-0.15){13}
\rput(7,-0.15){15}
\rput(8,-0.15){\dots}
\psline{-}(0,0.2)(0,0.6)(2,0.6)(2,0.2)
\psline{-}(4,0.2)(4,0.6)(6,0.6)(6,0.2)
\psdots(1,0.3)(3,0.3)(5,0.3)(7,0.3)
\end{pspicture}&
\ \qquad\psset{unit=0.5}
\begin{pspicture}(0,0)(12.3,1.1)
\rput(0,-0.15){1}
\rput(1,-0.15){3}
\rput(2,-0.15){5}
\rput(3,-0.15){7}
\rput(4,-0.15){9}
\rput(5,-0.15){11}
\rput(6,-0.15){13}
\rput(7,-0.15){15}
\rput(8,-0.15){17}
\rput(9,-0.15){19}
\rput(10,-0.15){21}
\rput(11,-0.15){23}
\rput(12,-0.15){\dots}
\psdots(2,0.3)(5,0.3)(8,0.3)(11,0.3)
\psline{-}(0,0.2)(0,0.9)(4,0.9)(4,0.2)
\psline{-}(6,0.2)(6,0.9)(10,0.9)(10,0.2)
\psline{-}(1,0.2)(1,0.6)(3,0.6)(3,0.2)
\psline{-}(7,0.2)(7,0.6)(9,0.6)(9,0.2)
\end{pspicture}\\
\end{matrix}
\end{equation*}
\noindent However the transfer matrix $\rho(D_N(\lambda=\pi /3,u))$ that is tied with percolation has no Jordan blocks for any $N$. This particular result for $\lambda=\pi/3$ was suggested by Dubail, Jacobsen and Saleur \cite{Saleur}, based on their exploration for small $N$'s.

A few questions remain unanswered. The first concerns the order of the Jordan cells: our numerics suggest that all Jordan blocks are of size $2$, though a proof remains unknown, even for $F_N$. In the case of critical polymers ($\lambda = \pi /2$), an inversion equation for $D_N$ is known \cite{PRZ} and yields $F_N^2 = 0$ for $N$ even (and Jordan blocks all have size $2$), but $(F_N-2\ id)(F_N+2\ id) = 0$ for $N$ odd (no Jordan blocks). The size of the Jordan cells for $\rho(D_N)$ might be an even harder problem and we have not been able to prove that they are of size $2$ for any non-trivial cases. Another related question is that of possible Jordan blocks connecting sectors $d$ and $d'$ when $|d-d'|\ge 2b$. These have not been ruled out by Propositions \ref{thm:lePetit} and \ref{thm:laGrande}. Indeed, consider a non-zero linear combination of vectors in $UpTo_{d'}$ that are partners in Jordan blocks of vectors strictly in $UpTo_d$. Such a linear combination could fail to be strictly in $UpTo_{d'}$. It will then be strictly in $UpTo_{d''}$, with $d''<d'<d$. Of course the eigenvalues of $F_N$ in the sectors $d, d'$ and $d''$ should all be equal. For example, a Jordan block for $\lambda=\frac\pi4$ could exist between sectors $d=12$ and $d''=2$. Though limited to small $N$, our numerics have ruled out any such Jordan blocks. 

The goal of the computation was to decide whether, for simple boundary conditions corresponding to open ones on spin lattices, representations of the Virasoro algebra other than irreducible highest weight ones could appear in the thermodynamical limit. Our result is that this is indeed the case. One would probably like to count, for each eigenvalue of the limit operator $L_0$, how many Jordan blocks appear. Such information cannot be retrieved from our method, at least in the present state. 

One may still wonder what type of representations occur in the limit. Let $\mathcal H_N\propto\frac12\frac{\partial\ }{\partial u}D_N(\lambda,u)|_{u=0}$ where the proportionality factor is a power of $\sin \lambda$. Once the finite size corrections are properly handled, the large $N$ limit of the matrix $\rho(\mathcal H_N)$ is linked to $L_0$. It was argued in \cite{PRZ}, with strong numerical support, that the spectrum of the diagonal block of $\rho(\mathcal H_N)$ in the sector with $d$ defects reproduces the character of some highest weight modules associated with the central charge $c_{a,b}$ (see \eqref{eq:centralcharge}) and highest weight $\Delta_{1,d+1}=(ad^2-2d(b-a))/4b$. (Their numerical analysis allows to fix precisely the first few eigenvalues of $L_0$, but does not go deep enough to suggest whether the highest weight representation is irreducible.) If this is the case, then a pair $(d,d')$ solving $\frac12(d+d')\equiv b-1\mod 2b$ will lead to a pair of highest weights such that $\Delta_{1,d'+1}-\Delta_{1,d+1}\in \mathbb Z$. Therefore staggered modules, a large family of which was recently classified in \cite{Ridout}, are likely candidates for the representations of the Virasoro algebra appearing in the thermodynamical limit of the models we have studied.

We hope that the methods introduced here to probe the Jordan structure can be used for other boundary conditions as those introduced in \cite{PRZ} 
and in 
\cite{JacobsenSaleur2}. Both these sets are stated algebraically, in a way therefore that might help to extend the present results, and they might actually match one another. Moreover the latter were given a direct interpretation in terms of classical spin models. Indeed these conditions amount, for the $Q$-Potts models, to restrict spins on the boundaries to take values only in a subset $\{Q_1, Q_2,\dots, Q_s\}$ of the $Q$ possible values. Of course, for these new boundary conditions, the Jordan cell structure could be very different from the one found here (see, for example, \cite{PRR}).

%
%

\section*{Acknowledgements} 
AMD holds a scholarship and YSA a grant of the Canadian Natural Sciences and Engineering Research Council. This support is gratefully acknowledged. We have benefitted greatly from discussions with Paul Pearce and J\o rgen Rasmussen and from David Ridout's many remarks and explanations. Discussions with Jesper Jacobsen and Philippe di Francesco are also gratefully acknowledged.

%
%

\bigskip

\bigskip

\noindent{\LARGE\bfseries Appendices}

%
%

\appendix\section{The main lemmas}\label{app:a}

\subsection{Labeling link states and matrix elements}\label{sub:notation}

In this section, we introduce a new notation for link states in $B_N$ and for matrix elements of linear transformations on $V_N$, both of which will prove useful in later sections.

To describe $w \in B_N^{N-2k}$ a link state with $N-2k$ defects and $k$ arcs, we label by $k$ integers $n_i$, ranging from $1$ to $N-1$, the halfway points between points which $w$ connects: $w\rightarrow v^r_{n_1,n_2,\dots, n_k}$. We usually write the $n_i$'s, from left to right in the order the arcs need to be closed, going from the inner to the outer ones. Even with this ordering, there may be many $v^r_{n_1,n_2,\dots, n_k}$ for a given $w$, but they all determine $w$ completely. For example,
\begin{alignat*}{3}
v^{10}_{3,7} & = 
\begin{pspicture}(0.2,-0.1)(5.1,0.5)
\psset{unit=0.5}
\psdots(1,0)(2,0)(3,0)(4,0)(5,0)(6,0)(7,0)(8,0)(9,0)(10,0)
\psset{linecolor=myc2}
\psline{-}(1,0)(1,1)
\psline{-}(2,0)(2,1)
\psline{-}(5,0)(5,1)
\psline{-}(6,0)(6,1)
\psline{-}(9,0)(9,1)
\psline{-}(10,0)(10,1)
\psarc(3.5,0){0.5}{0}{180}
\psarc(7.5,0){0.5}{0}{180}
\rput(1.5,-0.25){$_1$}
\rput(2.5,-0.25){$_2$}
\rput(3.5,-0.25){$_3$}
\rput(4.5,-0.25){$_4$}
\rput(5.5,-0.25){$_5$}
\rput(6.5,-0.25){$_6$}
\rput(7.5,-0.25){$_7$}
\rput(8.5,-0.25){$_8$}
\rput(9.5,-0.25){$_9$}
\end{pspicture} 
, \qquad & v^{10}_{2,2,7,7}  & =\begin{pspicture}(0.2,-0.1)(5.1,0.5)
\psset{unit=0.5}
\psdots(1,0)(2,0)(3,0)(4,0)(5,0)(6,0)(7,0)(8,0)(9,0)(10,0)
\psset{linecolor=myc2}
\psline{-}(5,0)(5,1)\psline{-}(10,0)(10,1)
\psarc(2.5,0){0.5}{0}{180}
\psarc(7.5,0){0.5}{0}{180}
\psbezier{-}(1,0)(1,1)(4,1)(4,0)
\psbezier{-}(6,0)(6,1)(9,1)(9,0)
\rput(1.5,-0.25){$_1$}
\rput(2.5,-0.25){$_2$}
\rput(3.5,-0.25){$_3$}
\rput(4.5,-0.25){$_4$}
\rput(5.5,-0.25){$_5$}
\rput(6.5,-0.25){$_6$}
\rput(7.5,-0.25){$_7$}
\rput(8.5,-0.25){$_8$}
\rput(9.5,-0.25){$_9$}
\end{pspicture}, \\
&&& \\
v^{10}_{2,6,8,7} & = 
\begin{pspicture}(0.2,-0.1)(5.1,0.5)
\psset{unit=0.5}
\psdots(1,0)(2,0)(3,0)(4,0)(5,0)(6,0)(7,0)(8,0)(9,0)(10,0)
\psset{linecolor=myc2}
\psline{-}(1,0)(1,1)\psline{-}(4,0)(4,1)
\psarc(2.5,0){0.5}{0}{180}
\psarc(6.5,0){0.5}{0}{180}
\psarc(8.5,0){0.5}{0}{180}
\psbezier{-}(5,0)(5,1)(10,1)(10,0)
\rput(1.5,-0.25){$_1$}
\rput(2.5,-0.25){$_2$}
\rput(3.5,-0.25){$_3$}
\rput(4.5,-0.25){$_4$}
\rput(5.5,-0.25){$_5$}
\rput(6.5,-0.25){$_6$}
\rput(7.5,-0.25){$_7$}
\rput(8.5,-0.25){$_8$}
\rput(9.5,-0.25){$_9$}
\end{pspicture} 
, \qquad & v^{10}_{3,5,4,8,5}  & =\begin{pspicture}(0.2,-0.1)(5.1,0.5)
\psset{unit=0.5}
\psdots(1,0)(2,0)(3,0)(4,0)(5,0)(6,0)(7,0)(8,0)(9,0)(10,0)
\psset{linecolor=myc2}
\psarc(8.5,0){0.5}{0}{180}
\psarc(3.5,0){0.5}{0}{180}
\psarc(5.5,0){0.5}{0}{180}
\psbezier{-}(1,0)(1,1.4)(10,1.4)(10,0)
\psbezier{-}(2,0)(2,1.0)(7,1.0)(7,0)
\rput(1.5,-0.25){$_1$}
\rput(2.5,-0.25){$_2$}
\rput(3.5,-0.25){$_3$}
\rput(4.5,-0.25){$_4$}
\rput(5.5,-0.25){$_5$}
\rput(6.5,-0.25){$_6$}
\rput(7.5,-0.25){$_7$}
\rput(8.5,-0.25){$_8$}
\rput(9.5,-0.25){$_9$}
\end{pspicture}.
\end{alignat*}
As in section \ref{sec:jordan} the link state with $N$ defects and no arcs is labelled $v^N$.  We shall also shorten the notation whenever there is several arcs sharing the same halfpoints by writing $n_i^m$ for a group of $m$ successive identical $n_i$'s. For example $v^r_{2,2,7,7}$ could also be written as $v^r_{2^2, 7^2}$. Finally, we shall call a $1$-bubble a half-arc that encloses no other half-arc, a $2$-bubble a half-arc that encloses $1$-bubbles and an $n$-bubble a half arc that contains at least one $(n-1)$-bubble. So $v^{10}_{3,7}$ has $1$-bubbles only, $v^{10}_{2,2,7,7}$ drawn above has $1$-bubbles and $2$-bubbles and $v^{10}_{3,5,4,8,5}$ has a $3$-bubble.

The linear transformations on $V_N$, like $\rho(F_N)$ and $P^d$, will often be expressed in the basis $B_N$. To refer to a given matrix element of the linear transformation $A:V_N\rightarrow V_N$ (or on one of the subspaces $UpTo_d$), we shall use the bra-ket notation. Therefore the matrix element $A_{v,w}$, for $v,w\in B_N$, will be written as $\langle v\,|\,Aw\rangle$. (This is equivalent to introducing a scalar product on $V_N$ for which $B_N$ is an orthonormal basis.) As most of the computations will be done graphically, it is useful to develop tools to quickly select a given matrix element. To depict $\langle v\,|\,Aw\rangle$, we draw the ``incoming'' state $w$ with full line and usually on the top of the diagram. The ``outgoing'' state $v$ is depicted with dashed ones and usually on the bottom. The matrix elements of the identity are all zero unless $v=w$. For instance,
\begin{equation*}
\begin{pspicture}(-1.9,-0.3)(5.3,0.4)
\rput(-1.5,0){$\langle v^{10}_{2,6,8,7}\ |\ id\, v^{10}_{3,7,7} \rangle=$}
\psset{unit=0.5}
\psdots(1,0)(2,0)(3,0)(4,0)(5,0)(6,0)(7,0)(8,0)(9,0)(10,0)
\psset{linecolor=myc2}
\psline{-}(1,0)(1,1)\psline{-}(4,0)(4,1)
\psarc(2.5,0){0.5}{0}{180}
\psarc(6.5,0){0.5}{0}{180}
\psarc(8.5,0){0.5}{0}{180}
\psbezier{-}(5,0)(5,1)(10,1)(10,0)
\psset{linecolor=myc3,linestyle=dashed,dash=2pt 2pt}
\psarc(3.5,0){0.5}{180}{360}
\psline{-}(1,0)(1,-1)
\psline{-}(2,0)(2,-1)
\psline{-}(10,0)(10,-1)
\psline{-}(5,0)(5,-1)
\psarc(7.5,0){0.5}{180}{360}
\psbezier{-}(6,0)(6,-1)(9,-1)(9,0)
\rput(11.35,0.05){$=0$}
\end{pspicture}
\end{equation*}

\noindent In some computations the arcs or defects of the incoming state will cross the diagram representing the linear transformation $A$. When this occurs, a partial comparison of the the incoming and outgoing states is possible and the patterns can be replaced by either $1$ or $0$. In the latter case, this means that this particular contribution to $A$ is vanishing and can be ignored. Here are the patterns:
\begin{equation}
\begin{pspicture}(0.2,-0.1)(0.7,0.5)
\psset{unit=0.5}
\psdots(0,0)(1,0)
\psset{linecolor=myc2}
\psarc(0.5,0){0.5}{0}{180}
\psset{linecolor=myc3,linestyle=dashed,dash=2pt 2pt}
\psarc(0.5,0){0.5}{180}{360}
\end{pspicture} =
\begin{pspicture}(-0.2,-0.1)(0.2,0.5)
\psset{unit=0.5}
\psdots(0,0)
\psset{linecolor=myc2}
\psline{-}(0,0)(0,1)
\psset{linecolor=myc3,linestyle=dashed,dash=2pt 2pt}
\psline{-}(0,0)(0,-1)
\end{pspicture} =1
\begin{pspicture}(-2.2,-0.1)(0.7,0.5)
\psset{unit=0.5}
\psdots(0,0)(1,0)
\psset{linecolor=myc2}
\psarc(0.5,0){0.5}{0}{180}
\psset{linecolor=myc3,linestyle=dashed,dash=2pt 2pt}
\psline{-}(0,0)(0,-1)
\psline{-}(1,0)(1,-1)
\end{pspicture} =
\begin{pspicture}(-0.2,-0.1)(0.7,0.5)
\psset{unit=0.5}
\psdots(0,0)(1,0)
\psset{linecolor=myc2}
\psline{-}(0,0)(0,1)
\psset{linecolor=myc3,linestyle=dashed,dash=2pt 2pt}
\psarc(0.5,0){0.5}{180}{360}
\end{pspicture} =
\begin{pspicture}(-0.2,-0.1)(1.2,0.5)
\psset{unit=0.5}
\psdots(0,0)(1,0)(2,0)
\psset{linecolor=myc2}
\psarc(1.5,0){0.5}{0}{180}
\psset{linecolor=myc3,linestyle=dashed,dash=2pt 2pt}
\psarc(0.5,0){0.5}{180}{360}
\psline{-}(2,0)(2,-1)
\end{pspicture} =
\begin{pspicture}(-0.2,-0.1)(1.7,0.5)
\psset{unit=0.5}
\psdots(0,0)(1,0)(2,0)(3,0)
\psset{linecolor=myc2}
\psarc(1.5,0){0.5}{0}{180}
\psset{linecolor=myc3,linestyle=dashed,dash=2pt 2pt}
\psarc(0.5,0){0.5}{180}{360}
\psarc(2.5,0){0.5}{180}{360}
\end{pspicture} =0
\label{eq:dotprod}
\end{equation}
\\
\noindent and their images through a vertical mirror. 

By definition, the computation of $P^d:UpTo_d\rightarrow UpTo_d$, for a vector in $B_N^d\subset UpTo_d\subset V_N$, starts by the removal of all arcs, leaving only its $d$ defects. The singularities of any matrix elements $\langle v | P^d w\rangle$ can therefore be deduced from those of $\langle v'| P^d v^d\rangle$ for some $v'$ and for $v^d\in B_d^d$, the vector with all defects in $V_d$. We found useful to use another letter for the number of defects to underline the fact that $v^d\subset V_d$ belongs to a vector space in general distinct from $V_N$. We shall therefore compute matrix elements of $P^r$ in the column corresponding to $v^r$. These are the only matrix elements that are really needed and they will be labelled in any of the two following forms, the second being defined by the first that was just introduced: 
\begin{equation*} \langle v^r_{n_1,..., n_k} | P^r v^r \rangle=P^r_{\{n_1,...,n_k\}} \end{equation*}
The extreme case $P^r_{\{\}}$ stands for $\langle v^r|P^rv^r\rangle$ which is equal to $1$. 
We are now ready to give explicit expressions for some matrix elements $P^r_w$.

\subsection{$P^r_{\{m^m\}}$ and its singularities}

The singularities of all the matrix elements of $P^r$ will be understood from those of the $P^r_{\{m^m\}}$ for some $m$. We start by computing these key elements.

\begin{Lemme} Single bubble in first position:
\begin{equation*}P^r_{\{1\}} = \frac{S_{(r-1)/2}}{2 S_{1/2} C_{r/2}}. \end{equation*}
\label{sec:Pr1}\end{Lemme}
\noindent{\scshape Proof\ \ } The strategy is to start with the first lower diagonal row of boxes, applying eqs. (\ref{eq:kbox}) and (\ref{eq:dotprod}). To understand the first step below, note that the sum represented by the lowest box, labelled by ``$1$'', gives two terms. The first draws an arc between the two outgoing defects and is therefore zero by \eqref{eq:dotprod}. Therefore the only non-vanishing contribution comes from the second which is drawn. The other steps are similar.
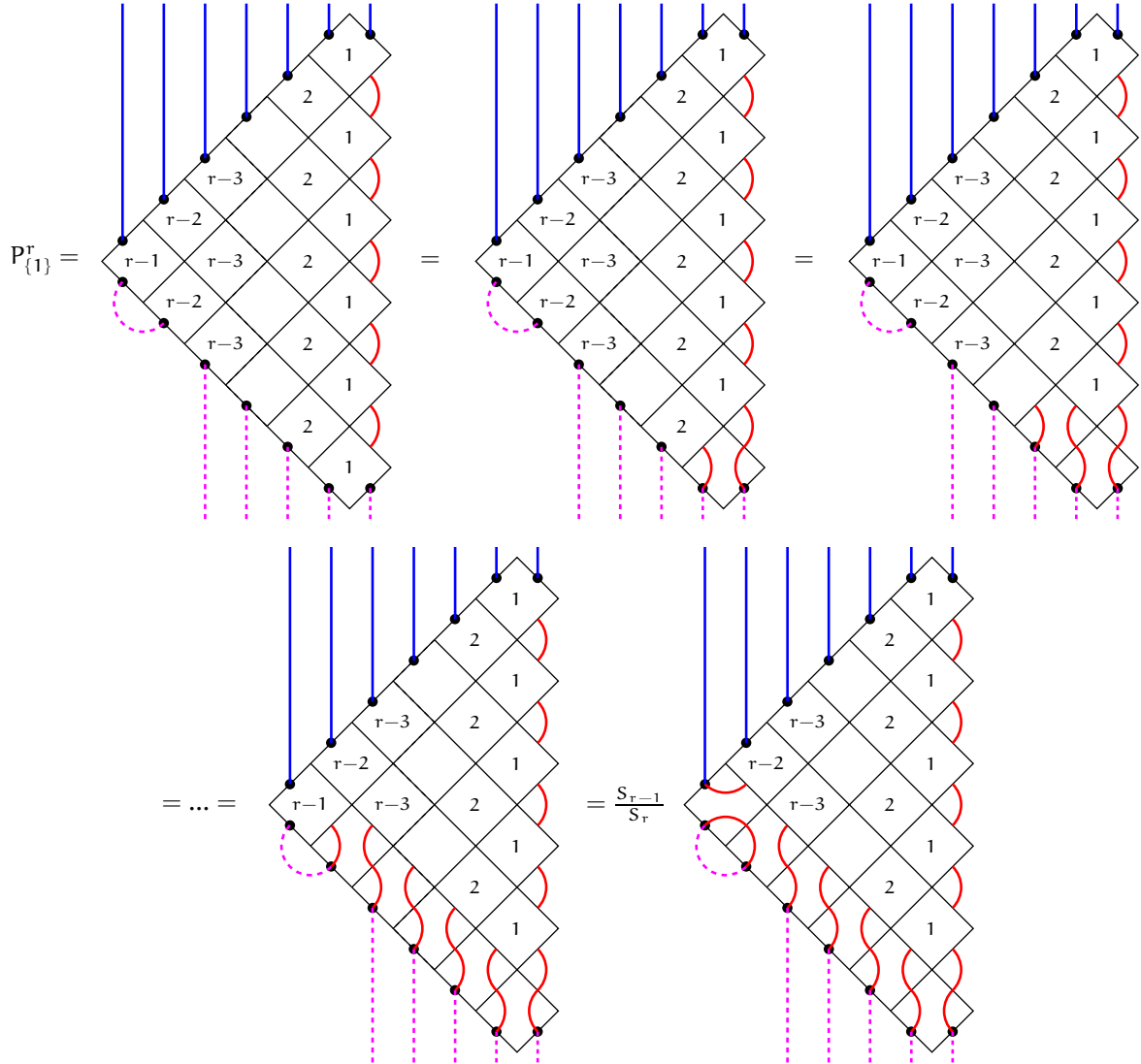
\begin{equation*}
\psset{unit=0.565685}\begin{pspicture}(-2.0,-6.0)(7,6.0)
\psdots(0.5,0.5)(1.5,1.5)(2.5,2.5)(4.5,4.5)(5.5,5.5)(6.5,5.5)
\psdots(0.5,-0.5)(1.5,-1.5)(2.5,-2.5)(4.5,-4.5)(5.5,-5.5)(6.5,-5.5)
\psdots(3.5,3.5)(3.5,-3.5)
\lw
\rput(-1.4,-0.0){$ P^r_{\{1\}}=$}
\psline{-}(2,2)(0,0)(2,-2)
\psline{-}(4,4)(6,6)(7,5)(6,4)(7,3)(6,2)(7,1)(6,0)(7,-1)(6,-2)(7,-3)(6,-4)(7,-5)(6,-6)(4,-4)
\psline{-}(4,-4)(5,-3)(4,-2)(5,-1)(4,0)(5,1)(4,2)(5,3)(4,4)
\psline{-}(5,5)(6,4)(5,3)(6,2)(5,1)(6,0)(5,-1)(6,-2)(5,-3)(6,-4)(5,-5)
\psline{-}(1,1)(2,0)(1,-1)
\psline{-}(2,2)(3,3)
\psline{-}(2,-2)(3,-3)
\psline{-}(4,-4)(3,-3)(4,-2)(3,-1)(4,0)(3,1)(4,2)(3,3)(4,4)
\psline[linestyle=dashed,dash=2pt 2pt]{-}(4,-4)(3,-3)(4,-2)(3,-1)(4,0)(3,1)(4,2)(3,3)(4,4)
\psline{-}(2,2)(3,1)(2,0)(3,-1)(2,-2)
\rput(6,5){$_1$}
\rput(6,3){$_1$}
\rput(6,1){$_1$}
\rput(6,-1){$_1$}
\rput(6,-3){$_1$}
\rput(6,-5){$_1$}
\rput(5,4){$_2$}
\rput(5,2){$_2$}
\rput(5,0){$_2$}
\rput(5,-2){$_2$}
\rput(5,-4){$_2$}
\rput(3,-2){$_{r-3}$}
\rput(3,0){$_{r-3}$}
\rput(3,2){$_{r-3}$}
\rput(2,1){$_{r-2}$}
\rput(2,-1){$_{r-2}$}
\rput(1,0){$_{r-1}$}
\psset{linecolor=myc}\unlw
\psarc{-}(6,-4){0.7071}{-45}{45}
\psarc{-}(6,-2){0.7071}{-45}{45}
\psarc{-}(6,0){0.7071}{-45}{45}
\psarc{-}(6,2){0.7071}{-45}{45}
\psarc{-}(6,4){0.7071}{-45}{45} 
\psset{linecolor=myc2}\unlw
\psline{-}(0.5,0.5)(0.5,6.25)
\psline{-}(1.5,1.5)(1.5,6.25)
\psline{-}(2.5,2.5)(2.5,6.25)
\psline{-}(3.5,3.5)(3.5,6.25)
\psline{-}(4.5,4.5)(4.5,6.25)
\psline{-}(5.5,5.5)(5.5,6.25)
\psline{-}(6.5,5.5)(6.5,6.25)
\psset{linecolor=myc3,linestyle=dashed,dash=2pt 2pt}
\psarc{-}(1,-1){0.7071}{135}{315} 
\psline{-}(2.5,-2.5)(2.5,-6.25)
\psline{-}(3.5,-3.5)(3.5,-6.25)
\psline{-}(4.5,-4.5)(4.5,-6.25)
\psline{-}(5.5,-5.5)(5.5,-6.25)
\psline{-}(6.5,-5.5)(6.5,-6.25)
\end{pspicture}
\begin{pspicture}(-2.05,-6.0)(7,6.0)
\psdots(0.5,0.5)(1.5,1.5)(2.5,2.5)(4.5,4.5)(5.5,5.5)(6.5,5.5)
\psdots(0.5,-0.5)(1.5,-1.5)(2.5,-2.5)(4.5,-4.5)(5.5,-5.5)(6.5,-5.5)
\psdots(3.5,3.5)(3.5,-3.5)
\lw
\rput(-1.1,-0.0){$ =$}
\psline{-}(2,2)(0,0)(2,-2)
\psline{-}(4,4)(6,6)(7,5)(6,4)(7,3)(6,2)(7,1)(6,0)(7,-1)(6,-2)(7,-3)(6,-4)(7,-5)(6,-6)(4,-4)
\psline{-}(4,-4)(5,-3)(4,-2)(5,-1)(4,0)(5,1)(4,2)(5,3)(4,4)
\psline{-}(5,5)(6,4)(5,3)(6,2)(5,1)(6,0)(5,-1)(6,-2)(5,-3)(6,-4)(5,-5)
\psline{-}(1,1)(2,0)(1,-1)
\psline{-}(2,2)(3,3)
\psline{-}(2,-2)(3,-3)
\psline{-}(4,-4)(3,-3)(4,-2)(3,-1)(4,0)(3,1)(4,2)(3,3)(4,4)
\psline[linestyle=dashed,dash=2pt 2pt]{-}(4,-4)(3,-3)(4,-2)(3,-1)(4,0)(3,1)(4,2)(3,3)(4,4)
\psline{-}(2,2)(3,1)(2,0)(3,-1)(2,-2)
\rput(6,5){$_1$}
\rput(6,3){$_1$}
\rput(6,1){$_1$}
\rput(6,-1){$_1$}
\rput(6,-3){$_1$}
\rput(5,4){$_2$}
\rput(5,2){$_2$}
\rput(5,0){$_2$}
\rput(5,-2){$_2$}
\rput(5,-4){$_2$}
\rput(3,-2){$_{r-3}$}
\rput(3,0){$_{r-3}$}
\rput(3,2){$_{r-3}$}
\rput(2,1){$_{r-2}$}
\rput(2,-1){$_{r-2}$}
\rput(1,0){$_{r-1}$}
\psset{linecolor=myc}\unlw
\psarc{-}(6,-4){0.7071}{-45}{45}
\psarc{-}(6,-2){0.7071}{-45}{45}
\psarc{-}(6,0){0.7071}{-45}{45}
\psarc{-}(6,2){0.7071}{-45}{45}
\psarc{-}(6,4){0.7071}{-45}{45}
\psarc{-}(7,-5){0.7071}{135}{225}
\psarc{-}(5,-5){0.7071}{-45}{45} 
\psset{linecolor=myc2}\unlw
\psline{-}(0.5,0.5)(0.5,6.25)
\psline{-}(1.5,1.5)(1.5,6.25)
\psline{-}(2.5,2.5)(2.5,6.25)
\psline{-}(3.5,3.5)(3.5,6.25)
\psline{-}(4.5,4.5)(4.5,6.25)
\psline{-}(5.5,5.5)(5.5,6.25)
\psline{-}(6.5,5.5)(6.5,6.25)
\psset{linecolor=myc3,linestyle=dashed,dash=2pt 2pt}
\psarc{-}(1,-1){0.7071}{135}{315} 
\psline{-}(2.5,-2.5)(2.5,-6.25)
\psline{-}(3.5,-3.5)(3.5,-6.25)
\psline{-}(4.5,-4.5)(4.5,-6.25)
\psline{-}(5.5,-5.5)(5.5,-6.25)
\psline{-}(6.5,-5.5)(6.5,-6.25)
\end{pspicture}
\begin{pspicture}(-2.05,-6.0)(7,6.0)
\psdots(0.5,0.5)(1.5,1.5)(2.5,2.5)(4.5,4.5)(5.5,5.5)(6.5,5.5)
\psdots(0.5,-0.5)(1.5,-1.5)(2.5,-2.5)(4.5,-4.5)(5.5,-5.5)(6.5,-5.5)
\psdots(3.5,3.5)(3.5,-3.5)
\lw
\rput(-1.1,-0.0){$ =$}
\psline{-}(2,2)(0,0)(2,-2)
\psline{-}(4,4)(6,6)(7,5)(6,4)(7,3)(6,2)(7,1)(6,0)(7,-1)(6,-2)(7,-3)(6,-4)(7,-5)(6,-6)(4,-4)
\psline{-}(4,-4)(5,-3)(4,-2)(5,-1)(4,0)(5,1)(4,2)(5,3)(4,4)
\psline{-}(5,5)(6,4)(5,3)(6,2)(5,1)(6,0)(5,-1)(6,-2)(5,-3)(6,-4)(5,-5)
\psline{-}(1,1)(2,0)(1,-1)
\psline{-}(2,2)(3,3)
\psline{-}(2,-2)(3,-3)
\psline{-}(4,-4)(3,-3)(4,-2)(3,-1)(4,0)(3,1)(4,2)(3,3)(4,4)
\psline{-}(2,2)(3,1)(2,0)(3,-1)(2,-2)
\rput(6,5){$_1$}
\rput(6,3){$_1$}
\rput(6,1){$_1$}
\rput(6,-1){$_1$}
\rput(6,-3){$_1$}
\rput(5,4){$_2$}
\rput(5,2){$_2$}
\rput(5,0){$_2$}
\rput(5,-2){$_2$}
\rput(3,-2){$_{r-3}$}
\rput(3,0){$_{r-3}$}
\rput(3,2){$_{r-3}$}
\rput(2,1){$_{r-2}$}
\rput(2,-1){$_{r-2}$}
\rput(1,0){$_{r-1}$}
\psset{linecolor=myc}\unlw
\psarc{-}(6,-4){0.7071}{-45}{45}
\psarc{-}(6,-2){0.7071}{-45}{45}
\psarc{-}(6,0){0.7071}{-45}{45}
\psarc{-}(6,2){0.7071}{-45}{45}
\psarc{-}(6,4){0.7071}{-45}{45}
\psarc{-}(6,-4){0.7071}{135}{225}
\psarc{-}(4,-4){0.7071}{-45}{45} 
\psarc{-}(7,-5){0.7071}{135}{225}
\psarc{-}(5,-5){0.7071}{-45}{45} 
\psset{linecolor=myc2}\unlw
\psline{-}(0.5,0.5)(0.5,6.25)
\psline{-}(1.5,1.5)(1.5,6.25)
\psline{-}(2.5,2.5)(2.5,6.25)
\psline{-}(3.5,3.5)(3.5,6.25)
\psline{-}(4.5,4.5)(4.5,6.25)
\psline{-}(5.5,5.5)(5.5,6.25)
\psline{-}(6.5,5.5)(6.5,6.25)
\psset{linecolor=myc3,linestyle=dashed,dash=2pt 2pt}
\psarc{-}(1,-1){0.7071}{135}{315} 
\psline{-}(2.5,-2.5)(2.5,-6.25)
\psline{-}(3.5,-3.5)(3.5,-6.25)
\psline{-}(4.5,-4.5)(4.5,-6.25)
\psline{-}(5.5,-5.5)(5.5,-6.25)
\psline{-}(6.5,-5.5)(6.5,-6.25)
\end{pspicture}
\end{equation*}
\begin{equation*}
\psset{unit=0.565685}\begin{pspicture}(-2.05,-6.0)(7,6.5)
\psdots(0.5,0.5)(1.5,1.5)(2.5,2.5)(4.5,4.5)(5.5,5.5)(6.5,5.5)
\psdots(0.5,-0.5)(1.5,-1.5)(2.5,-2.5)(4.5,-4.5)(5.5,-5.5)(6.5,-5.5)
\psdots(3.5,3.5)(3.5,-3.5)
\lw
\rput(-1.7,-0.0){$ = ... =$}
\psline{-}(2,2)(0,0)(2,-2)
\psline{-}(4,4)(6,6)(7,5)(6,4)(7,3)(6,2)(7,1)(6,0)(7,-1)(6,-2)(7,-3)(6,-4)(7,-5)(6,-6)(4,-4)
\psline{-}(4,-4)(5,-3)(4,-2)(5,-1)(4,0)(5,1)(4,2)(5,3)(4,4)
\psline{-}(5,5)(6,4)(5,3)(6,2)(5,1)(6,0)(5,-1)(6,-2)(5,-3)(6,-4)(5,-5)
\psline{-}(1,1)(2,0)(1,-1)
\psline{-}(2,2)(3,3)
\psline{-}(2,-2)(3,-3)
\psline[linestyle=dashed,dash=2pt 2pt]{-}(4,-4)(3,-3)(4,-2)(3,-1)(4,0)(3,1)(4,2)(3,3)(4,4)
\psline{-}(4,-4)(3,-3)(4,-2)(3,-1)(4,0)(3,1)(4,2)(3,3)(4,4)
\psline{-}(2,2)(3,1)(2,0)(3,-1)(2,-2)
\rput(6,5){$_1$}
\rput(6,3){$_1$}
\rput(6,1){$_1$}
\rput(6,-1){$_1$}
\rput(6,-3){$_1$}
\rput(5,4){$_2$}
\rput(5,2){$_2$}
\rput(5,0){$_2$}
\rput(5,-2){$_2$}
\rput(3,0){$_{r-3}$}
\rput(3,2){$_{r-3}$}
\rput(2,1){$_{r-2}$}
\rput(1,0){$_{r-1}$}
\psset{linecolor=myc}\unlw
\psarc{-}(6,-4){0.7071}{-45}{45}
\psarc{-}(6,-2){0.7071}{-45}{45}
\psarc{-}(6,0){0.7071}{-45}{45}
\psarc{-}(6,2){0.7071}{-45}{45}
\psarc{-}(6,4){0.7071}{-45}{45}
\psarc{-}(3,-1){0.7071}{135}{225}
\psarc{-}(1,-1){0.7071}{-45}{45} 
\psarc{-}(4,-2){0.7071}{135}{225}
\psarc{-}(2,-2){0.7071}{-45}{45} 
\psarc{-}(5,-3){0.7071}{135}{225}
\psarc{-}(3,-3){0.7071}{-45}{45} 
\psarc{-}(6,-4){0.7071}{135}{225}
\psarc{-}(4,-4){0.7071}{-45}{45} 
\psarc{-}(7,-5){0.7071}{135}{225}
\psarc{-}(5,-5){0.7071}{-45}{45} 
\psset{linecolor=myc2}\unlw
\psline{-}(0.5,0.5)(0.5,6.25)
\psline{-}(1.5,1.5)(1.5,6.25)
\psline{-}(2.5,2.5)(2.5,6.25)
\psline{-}(3.5,3.5)(3.5,6.25)
\psline{-}(4.5,4.5)(4.5,6.25)
\psline{-}(5.5,5.5)(5.5,6.25)
\psline{-}(6.5,5.5)(6.5,6.25)
\psset{linecolor=myc3,linestyle=dashed,dash=2pt 2pt}
\psarc{-}(1,-1){0.7071}{135}{315} 
\psline{-}(2.5,-2.5)(2.5,-6.25)
\psline{-}(3.5,-3.5)(3.5,-6.25)
\psline{-}(4.5,-4.5)(4.5,-6.25)
\psline{-}(5.5,-5.5)(5.5,-6.25)
\psline{-}(6.5,-5.5)(6.5,-6.25)
\end{pspicture}
\begin{pspicture}(-3.05,-6.0)(7,6.0)
\psdots(0.5,0.5)(1.5,1.5)(2.5,2.5)(4.5,4.5)(5.5,5.5)(6.5,5.5)
\psdots(0.5,-0.5)(1.5,-1.5)(2.5,-2.5)(4.5,-4.5)(5.5,-5.5)(6.5,-5.5)
\psdots(3.5,3.5)(3.5,-3.5)
\lw
\rput(-1.4,-0.0){$ = \frac{S_{r-1}}{S_r}$}
\psline{-}(2,2)(0,0)(2,-2)
\psline{-}(4,4)(6,6)(7,5)(6,4)(7,3)(6,2)(7,1)(6,0)(7,-1)(6,-2)(7,-3)(6,-4)(7,-5)(6,-6)(4,-4)
\psline{-}(4,-4)(5,-3)(4,-2)(5,-1)(4,0)(5,1)(4,2)(5,3)(4,4)
\psline{-}(5,5)(6,4)(5,3)(6,2)(5,1)(6,0)(5,-1)(6,-2)(5,-3)(6,-4)(5,-5)
\psline{-}(1,1)(2,0)(1,-1)
\psline{-}(2,2)(3,3)
\psline{-}(2,-2)(3,-3)
\psline{-}(4,-4)(3,-3)(4,-2)(3,-1)(4,0)(3,1)(4,2)(3,3)(4,4)
\psline{-}(2,2)(3,1)(2,0)(3,-1)(2,-2)
\rput(6,5){$_1$}
\rput(6,3){$_1$}
\rput(6,1){$_1$}
\rput(6,-1){$_1$}
\rput(6,-3){$_1$}
\rput(5,4){$_2$}
\rput(5,2){$_2$}
\rput(5,0){$_2$}
\rput(5,-2){$_2$}
\rput(3,0){$_{r-3}$}
\rput(3,2){$_{r-3}$}
\rput(2,1){$_{r-2}$}
\psset{linecolor=myc}\unlw
\psarc{-}(6,-4){0.7071}{-45}{45}
\psarc{-}(6,-2){0.7071}{-45}{45}
\psarc{-}(6,0){0.7071}{-45}{45}
\psarc{-}(6,2){0.7071}{-45}{45}
\psarc{-}(6,4){0.7071}{-45}{45}
\psarc{-}(3,-1){0.7071}{135}{225}
\psarc{-}(1,-1){0.7071}{-45}{135} 
\psarc{-}(1,1){0.7071}{-135}{-45} 
\psarc{-}(4,-2){0.7071}{135}{225}
\psarc{-}(2,-2){0.7071}{-45}{45} 
\psarc{-}(5,-3){0.7071}{135}{225}
\psarc{-}(3,-3){0.7071}{-45}{45} 
\psarc{-}(6,-4){0.7071}{135}{225}
\psarc{-}(4,-4){0.7071}{-45}{45} 
\psarc{-}(7,-5){0.7071}{135}{225}
\psarc{-}(5,-5){0.7071}{-45}{45} 
\psset{linecolor=myc2}\unlw
\psline{-}(0.5,0.5)(0.5,6.25)
\psline{-}(1.5,1.5)(1.5,6.25)
\psline{-}(2.5,2.5)(2.5,6.25)
\psline{-}(3.5,3.5)(3.5,6.25)
\psline{-}(4.5,4.5)(4.5,6.25)
\psline{-}(5.5,5.5)(5.5,6.25)
\psline{-}(6.5,5.5)(6.5,6.25)
\psset{linecolor=myc3,linestyle=dashed,dash=2pt 2pt}
\psarc{-}(1,-1){0.7071}{135}{315} 
\psline{-}(2.5,-2.5)(2.5,-6.25)
\psline{-}(3.5,-3.5)(3.5,-6.25)
\psline{-}(4.5,-4.5)(4.5,-6.25)
\psline{-}(5.5,-5.5)(5.5,-6.25)
\psline{-}(6.5,-5.5)(6.5,-6.25)
\end{pspicture}
\end{equation*}
\\
\noindent We now sum the boxes of the upper diagonal row. Only $r-1$ of the $2^{r-2}$ terms contribute, namely:
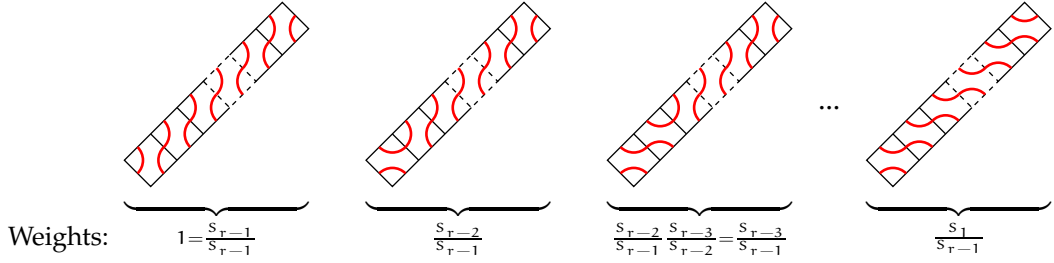
\begin{equation*}
\psset{unit=0.35}
\underbrace{\begin{pspicture}(0,-1.5)(7,6.0)
\lw
\rput(-2.5,-3){Weights:}
\psline{-}(3,3)(0,0)(1,-1)(4,2)
\psline{-}(4,4)(6,6)(7,5)(5,3)
\psline[linestyle=dashed,dash=2pt 2pt]{-}(3,3)(4,4)(5,3)(4,2)(3,3)
\psline{-}(1,1)(2,0)
\psline{-}(2,2)(3,1)
\psline{-}(5,5)(6,4)
\psset{linecolor=myc}\unlw
\psarc{-}(0,0){0.7071}{-45}{45}\psarc{-}(2,0){0.7071}{135}{225}
\psarc{-}(1,1){0.7071}{-45}{45}\psarc{-}(3,1){0.7071}{135}{225}
\psarc{-}(2,2){0.7071}{-45}{45}\psarc{-}(4,2){0.7071}{135}{225}
\psarc{-}(3,3){0.7071}{-45}{45}\psarc{-}(5,3){0.7071}{135}{225}
\psarc{-}(4,4){0.7071}{-45}{45}\psarc{-}(6,4){0.7071}{135}{225}
\psarc{-}(5,5){0.7071}{-45}{45}\psarc{-}(7,5){0.7071}{135}{225} 
\end{pspicture}}_{1 = \frac{S_{r-1}}{S_{r-1}}} \qquad
\underbrace{\begin{pspicture}(0,-1.5)(7,6.0)
\lw
\psline{-}(3,3)(0,0)(1,-1)(4,2)
\psline{-}(4,4)(6,6)(7,5)(5,3)
\psline[linestyle=dashed,dash=2pt 2pt]{-}(3,3)(4,4)(5,3)(4,2)(3,3)
\psline{-}(1,1)(2,0)
\psline{-}(2,2)(3,1)
\psline{-}(5,5)(6,4)
\psset{linecolor=myc}\unlw
\psarc{-}(1,-1){0.7071}{45}{135}\psarc{-}(1,1){0.7071}{225}{315}
\psarc{-}(1,1){0.7071}{-45}{45}\psarc{-}(3,1){0.7071}{135}{225}
\psarc{-}(2,2){0.7071}{-45}{45}\psarc{-}(4,2){0.7071}{135}{225}
\psarc{-}(3,3){0.7071}{-45}{45}\psarc{-}(5,3){0.7071}{135}{225}
\psarc{-}(4,4){0.7071}{-45}{45}\psarc{-}(6,4){0.7071}{135}{225}
\psarc{-}(5,5){0.7071}{-45}{45}\psarc{-}(7,5){0.7071}{135}{225} 
\end{pspicture}}_{\frac{S_{r-2}}{S_{r-1}}}  \qquad
\underbrace{\begin{pspicture}(0,-1.5)(7,6.0)
\lw
\psline{-}(3,3)(0,0)(1,-1)(4,2)
\psline{-}(4,4)(6,6)(7,5)(5,3)
\psline[linestyle=dashed,dash=2pt 2pt]{-}(3,3)(4,4)(5,3)(4,2)(3,3)
\psline{-}(1,1)(2,0)
\psline{-}(2,2)(3,1)
\psline{-}(5,5)(6,4)
\psset{linecolor=myc}\unlw
\psarc{-}(1,-1){0.7071}{45}{135}\psarc{-}(1,1){0.7071}{225}{315}
\psarc{-}(2,0){0.7071}{45}{135}\psarc{-}(2,2){0.7071}{225}{315}
\psarc{-}(2,2){0.7071}{-45}{45}\psarc{-}(4,2){0.7071}{135}{225}
\psarc{-}(3,3){0.7071}{-45}{45}\psarc{-}(5,3){0.7071}{135}{225}
\psarc{-}(4,4){0.7071}{-45}{45}\psarc{-}(6,4){0.7071}{135}{225}
\psarc{-}(5,5){0.7071}{-45}{45}\psarc{-}(7,5){0.7071}{135}{225} 
\end{pspicture}}_{\frac{S_{r-2}}{S_{r-1}}\frac{S_{r-3}}{S_{r-2}}=\frac{S_{r-3}}{S_{r-1}}}
\ \ \
\begin{pspicture}(0,-1.5)(1,6.0)
\rput(0.5,2.0){$...$}
\end{pspicture} \ \ \
\underbrace{\begin{pspicture}(0,-1.5)(7,6.0)
\lw
\psline{-}(3,3)(0,0)(1,-1)(4,2)
\psline{-}(4,4)(6,6)(7,5)(5,3)
\psline[linestyle=dashed,dash=2pt 2pt]{-}(3,3)(4,4)(5,3)(4,2)(3,3)
\psline{-}(1,1)(2,0)
\psline{-}(2,2)(3,1)
\psline{-}(5,5)(6,4)
\psset{linecolor=myc}\unlw
\psarc{-}(1,-1){0.7071}{45}{135}\psarc{-}(1,1){0.7071}{225}{315}
\psarc{-}(2,0){0.7071}{45}{135}\psarc{-}(2,2){0.7071}{225}{315}
\psarc{-}(3,1){0.7071}{45}{135}\psarc{-}(3,3){0.7071}{225}{315}
\psarc{-}(4,2){0.7071}{45}{135}\psarc{-}(4,4){0.7071}{225}{315}
\psarc{-}(5,3){0.7071}{45}{135}\psarc{-}(5,5){0.7071}{225}{315}
\psarc{-}(6,4){0.7071}{45}{135}\psarc{-}(6,6){0.7071}{225}{315}
\end{pspicture}}_{\frac{S_1}{S_{r-1}}}
\end{equation*}
Other box configurations produce half arcs, which propagate down and eventually annihilate with
\psset{linewidth=1pt}
\begin{pspicture}(-0.2,-0.1)(0.15,0.1)\lw
\psline{-}(-0.2,0)(0,0.2)(0.2,0)(0,-0.2)(-0.2,0)
\psset{linecolor=myc}
\rput(0,0){$_1$}
\end{pspicture}
. Therefore the sums of the upper diagonal row lead to the downward propagation, by one box, of all incoming defects and to an overall weight of
\begin{equation*}\sum_{i=1}^{r-1} \frac{S_i}{S_{r-1}} = \frac{S_{(r-1)/2}S_{r/2}}{S_{r-1}S_{1/2}}.\end{equation*}
Finally,
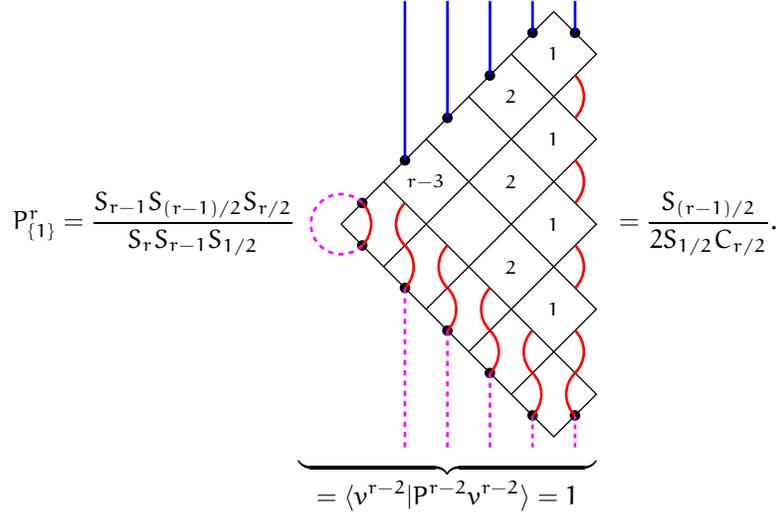
\begin{equation*}
\psset{unit=0.565685} \qquad \qquad
\underbrace{\begin{pspicture}(-0.0,-6.5)(7,4.0)
\psdots(1.5,-0.5)(2.5,0.5)(3.5,1.5)(4.5,2.5)(5.5,3.5)(6.5,3.5)
\psdots(1.5,-1.5)(2.5,-2.5)(3.5,-3.5)(4.5,-4.5)(5.5,-5.5)(6.5,-5.5)
\lw
\rput(-3.4,-1.0){$\displaystyle{ P^r_{\{1\}} = \frac{S_{r-1}S_{(r-1)/2}S_{r/2}}{S_rS_{r-1}S_{1/2}}}$}
\rput(9.4,-1.0){$\displaystyle{ = \frac{S_{(r-1)/2}}{2 S_{1/2} C_{r/2}} .}$}
\psline{-}(2,-2)(1,-1)
\psline{-}(6,4)(7,3)(6,2)(7,1)(6,0)(7,-1)(6,-2)(7,-3)(6,-4)(7,-5)(6,-6)(4,-4)
\psline{-}(4,-4)(5,-3)(4,-2)(5,-1)(4,0)(5,1)(4,2)
\psline{-}(6,4)(5,3)(6,2)(5,1)(6,0)(5,-1)(6,-2)(5,-3)(6,-4)(5,-5)
\psline{-}(2,0)(1,-1)
\psline{-}(4,2)(5,3)
\psline{-}(2,-2)(3,-3)
\psline{-}(4,-4)(3,-3)(4,-2)(3,-1)(4,0)(3,1)(4,2)
\psline{-}(3,1)(2,0)(3,-1)(2,-2)
\rput(6,3){$_1$}
\rput(6,1){$_1$}
\rput(6,-1){$_1$}
\rput(6,-3){$_1$}
\rput(5,2){$_2$}
\rput(5,0){$_2$}
\rput(5,-2){$_2$}
\rput(3,0){$_{r-3}$}
\psset{linecolor=myc}\unlw
\psarc{-}(6,-4){0.7071}{-45}{45}
\psarc{-}(6,-2){0.7071}{-45}{45}
\psarc{-}(6,0){0.7071}{-45}{45}
\psarc{-}(6,2){0.7071}{-45}{45}
\psarc{-}(3,-1){0.7071}{135}{225}
\psarc{-}(1,-1){0.7071}{-45}{45} 
\psarc{-}(4,-2){0.7071}{135}{225}
\psarc{-}(2,-2){0.7071}{-45}{45} 
\psarc{-}(5,-3){0.7071}{135}{225}
\psarc{-}(3,-3){0.7071}{-45}{45} 
\psarc{-}(6,-4){0.7071}{135}{225}
\psarc{-}(4,-4){0.7071}{-45}{45} 
\psarc{-}(7,-5){0.7071}{135}{225}
\psarc{-}(5,-5){0.7071}{-45}{45} 
\psset{linecolor=myc2}\unlw
\psline{-}(2.5,0.5)(2.5,4.25)
\psline{-}(3.5,1.5)(3.5,4.25)
\psline{-}(4.5,2.5)(4.5,4.25)
\psline{-}(5.5,3.5)(5.5,4.25)
\psline{-}(6.5,3.5)(6.5,4.25)
\psset{linecolor=myc3,linestyle=dashed,dash=2pt 2pt}
\psarc{-}(1,-1){0.7071}{45}{315} 
\psline{-}(2.5,-2.5)(2.5,-6.25)
\psline{-}(3.5,-3.5)(3.5,-6.25)
\psline{-}(4.5,-4.5)(4.5,-6.25)
\psline{-}(5.5,-5.5)(5.5,-6.25)
\psline{-}(6.5,-5.5)(6.5,-6.25)
\end{pspicture}}_{\displaystyle{= \langle v^{r-2} | P^{r-2} v^{r-2} \rangle =1 }}
\end{equation*}
\hfill$\square$
\begin{Lemme} The single bubble in general position:
\begin{equation*}
P^r_{\{n\}} = \frac{S_{(r-n)/2}S_{n/2}}{2 \left(S_{1/2}\right)^2C_{r/2}} \qquad \textrm{for\ } n = 1, ..., r-1.
\end{equation*}
\label{sec: singlebubble}
\end{Lemme}
\noindent{\scshape Proof\ \ } The expression given for $P^r_{\{n\}}$ certainly works for $n=1$ and all $r$. The rest of the proof will be by induction. Using the same strategy as before, we build a recursion relation for $P^r_{\{n\}}$: 
\begin{equation*}
\psset{unit=0.565685}\begin{pspicture}(-2.0,-8.3)(9,8.3)
\psdots(0.5,0.5)(1.5,1.5)(2.5,2.5)(3.5,3.5)(4.5,4.5)(5.5,5.5)(6.5,6.5)(7.5,7.5)(8.5,7.5)
\psdots(0.5,-0.5)(1.5,-1.5)(2.5,-2.5)(3.5,-3.5)(4.5,-4.5)(5.5,-5.5)(6.5,-6.5)(7.5,-7.5)(8.5,-7.5)
\lw
\rput(-1.4,-0.0){$ P^r_{\{n\}}=$}

\psline{-}(1,1)(0,0)(1,-1)
\psline{-}(2,2)(3,3)(4,2)(3,1)(4,0)(3,-1)(4,-2)(3,-3)(2,-2)
\psline{-}(3,3)(4,4)(5,3)(4,2)(5,1)(4,0)(5,-1)(4,-2)(5,-3)(4,-4)(3,-3)
\psline{-}(4,4)(5,5)
\psline{-}(5,5)(6,4)(5,3)(6,2)(5,1)(6,0)(5,-1)(6,-2)(5,-3)(6,-4)(5,-5)(4,-4)
\psline{-}(5,-5)(4,-4)
\psline{-}(5,5)(6,6)(7,5)(6,4)(7,3)(6,2)(7,1)(6,0)(7,-1)(6,-2)(7,-3)(6,-4)(7,-5)(6,-6)(5,-5)
\psline{-}(6,6)(7,7)(8,6)(7,5)(8,4)(7,3)(8,2)(7,1)(8,0)(7,-1)(8,-2)(7,-3)(8,-4)(7,-5)(8,-6)(7,-7)(6,-6)
\psline{-}(7,7)(8,8)(9,7)(8,6)(9,5)(8,4)(9,3)(8,2)(9,1)(8,0)(9,-1)(8,-2)(9,-3)(8,-4)(9,-5)(8,-6)(9,-7)(8,-8)(7,-7)
\psline{-}(2,-2)(1,-1)(2,0)(1,1)(2,2)(3,1)(2,0)(3,-1)(2,-2)
\rput(8,7){$_1$}
\rput(8,5){$_1$}
\rput(8,3){$_1$}
\rput(8,1){$_1$}
\rput(8,-1){$_1$}
\rput(8,-3){$_1$}
\rput(8,-5){$_1$}
\rput(8,-7){$_1$}
\rput(5,4){$_{r-n-1}$}
\rput(5,2){$_{r-n-1}$}
\rput(5,0){$_{r-n-1}$}
\rput(5,-2){$_{r-n-1}$}
\rput(5,-4){$_{r-n-1}$}
\rput(4,3){$_{r-n}$}
\rput(4,1){$_{r-n}$}
\rput(4,-1){$_{r-n}$}
\rput(4,-3){$_{r-n}$}
\rput(1,0){$_{r-1}$}
\psset{linecolor=myc}\unlw
\psarc{-}(8,-6){0.7071}{-45}{45}
\psarc{-}(8,-4){0.7071}{-45}{45}
\psarc{-}(8,-2){0.7071}{-45}{45}
\psarc{-}(8,0){0.7071}{-45}{45}
\psarc{-}(8,2){0.7071}{-45}{45}
\psarc{-}(8,4){0.7071}{-45}{45} 
\psarc{-}(8,6){0.7071}{-45}{45} 
\psset{linecolor=myc2}\unlw
\psline{-}(0.5,0.5)(0.5,8.25)
\psline{-}(1.5,1.5)(1.5,8.25)
\psline{-}(2.5,2.5)(2.5,8.25)
\psline{-}(3.5,3.5)(3.5,8.25)
\psline{-}(4.5,4.5)(4.5,8.25)
\psline{-}(5.5,5.5)(5.5,8.25)
\psline{-}(6.5,6.5)(6.5,8.25)
\psline{-}(7.5,7.5)(7.5,8.25)
\psline{-}(8.5,7.5)(8.5,8.25)
\psset{linecolor=myc3,linestyle=dashed,dash=2pt 2pt}
\psline{-}(0.5,-0.5)(0.5,-8.25)
\psline{-}(1.5,-1.5)(1.5,-8.25)
\psline{-}(2.5,-2.5)(2.5,-8.25)
\psarc{-}(4,-4){0.7071}{135}{315} 
\psline{-}(5.5,-5.5)(5.5,-8.25)
\psline{-}(6.5,-6.5)(6.5,-8.25)
\psline{-}(7.5,-7.5)(7.5,-8.25)
\psline{-}(8.5,-7.5)(8.5,-8.25)
\end{pspicture}
\begin{pspicture}(-2.0,-8.3)(9,8.3)
\psdots(0.5,0.5)(1.5,1.5)(2.5,2.5)(3.5,3.5)(4.5,4.5)(5.5,5.5)(6.5,6.5)(7.5,7.5)(8.5,7.5)
\psdots(0.5,-0.5)(1.5,-1.5)(2.5,-2.5)(3.5,-3.5)(4.5,-4.5)(5.5,-5.5)(6.5,-6.5)(7.5,-7.5)(8.5,-7.5)
\lw
\rput(-1.1,-0.0){$ =$}
\psline{-}(1,1)(0,0)(1,-1)
\psline{-}(2,2)(3,3)(4,2)(3,1)(4,0)(3,-1)(4,-2)(3,-3)(2,-2)
\psline{-}(3,3)(4,4)(5,3)(4,2)(5,1)(4,0)(5,-1)(4,-2)(5,-3)(4,-4)(3,-3)
\psline{-}(4,4)(5,5)
\psline{-}(5,5)(6,4)(5,3)(6,2)(5,1)(6,0)(5,-1)(6,-2)(5,-3)(6,-4)(5,-5)(4,-4)
\psline{-}(5,-5)(4,-4)
\psline{-}(5,5)(6,6)(7,5)(6,4)(7,3)(6,2)(7,1)(6,0)(7,-1)(6,-2)(7,-3)(6,-4)(7,-5)(6,-6)(5,-5)
\psline{-}(6,6)(7,7)(8,6)(7,5)(8,4)(7,3)(8,2)(7,1)(8,0)(7,-1)(8,-2)(7,-3)(8,-4)(7,-5)(8,-6)(7,-7)(6,-6)
\psline{-}(7,7)(8,8)(9,7)(8,6)(9,5)(8,4)(9,3)(8,2)(9,1)(8,0)(9,-1)(8,-2)(9,-3)(8,-4)(9,-5)(8,-6)(9,-7)(8,-8)(7,-7)
\psline{-}(2,-2)(1,-1)(2,0)(1,1)(2,2)(3,1)(2,0)(3,-1)(2,-2)
\rput(8,7){$_1$}
\rput(8,5){$_1$}
\rput(8,3){$_1$}
\rput(8,1){$_1$}
\rput(8,-1){$_1$}
\rput(8,-3){$_1$}
\rput(8,-5){$_1$}
\rput(5,4){$_{r-n-1}$}
\rput(5,2){$_{r-n-1}$}
\rput(5,0){$_{r-n-1}$}
\rput(5,-2){$_{r-n-1}$}
\rput(4,3){$_{r-n}$}
\rput(4,1){$_{r-n}$}
\rput(4,-1){$_{r-n}$}
\rput(4,-3){$_{r-n}$}
\rput(1,0){$_{r-1}$}
\psset{linecolor=myc}\unlw
\psarc{-}(8,-6){0.7071}{-45}{45}
\psarc{-}(8,-4){0.7071}{-45}{45}
\psarc{-}(8,-2){0.7071}{-45}{45}
\psarc{-}(8,0){0.7071}{-45}{45}
\psarc{-}(8,2){0.7071}{-45}{45}
\psarc{-}(8,4){0.7071}{-45}{45} 
\psarc{-}(8,6){0.7071}{-45}{45} 
\psarc{-}(4,-4){0.7071}{-45}{45}
\psarc{-}(5,-5){0.7071}{-45}{45}
\psarc{-}(6,-6){0.7071}{-45}{45}
\psarc{-}(7,-7){0.7071}{-45}{45}
\psarc{-}(6,-4){0.7071}{135}{-135}
\psarc{-}(7,-5){0.7071}{135}{-135}
\psarc{-}(8,-6){0.7071}{135}{-135}
\psarc{-}(9,-7){0.7071}{135}{-135}
\psset{linecolor=myc2}\unlw
\psline{-}(0.5,0.5)(0.5,8.25)
\psline{-}(1.5,1.5)(1.5,8.25)
\psline{-}(2.5,2.5)(2.5,8.25)
\psline{-}(3.5,3.5)(3.5,8.25)
\psline{-}(4.5,4.5)(4.5,8.25)
\psline{-}(5.5,5.5)(5.5,8.25)
\psline{-}(6.5,6.5)(6.5,8.25)
\psline{-}(7.5,7.5)(7.5,8.25)
\psline{-}(8.5,7.5)(8.5,8.25)
\psset{linecolor=myc3,linestyle=dashed,dash=2pt 2pt}
\psline{-}(0.5,-0.5)(0.5,-8.25)
\psline{-}(1.5,-1.5)(1.5,-8.25)
\psline{-}(2.5,-2.5)(2.5,-8.25)
\psarc{-}(4,-4){0.7071}{135}{315} 
\psline{-}(5.5,-5.5)(5.5,-8.25)
\psline{-}(6.5,-6.5)(6.5,-8.25)
\psline{-}(7.5,-7.5)(7.5,-8.25)
\psline{-}(8.5,-7.5)(8.5,-8.25)
\end{pspicture}
\end{equation*}
\begin{equation*}
\psset{unit=0.565685}\begin{pspicture}(-2.0,-8.3)(9,8.3)
\psdots(0.5,0.5)(1.5,1.5)(2.5,2.5)(3.5,3.5)(4.5,4.5)(5.5,5.5)(6.5,6.5)(7.5,7.5)(8.5,7.5)
\psdots(0.5,-0.5)(1.5,-1.5)(2.5,-2.5)(3.5,-3.5)(4.5,-4.5)(5.5,-5.5)(6.5,-6.5)(7.5,-7.5)(8.5,-7.5)
\lw
\rput(-1.1,-0.0){$ =$}
\psline{-}(1,1)(0,0)(1,-1)
\psline{-}(2,2)(3,3)(4,2)(3,1)(4,0)(3,-1)(4,-2)(3,-3)(2,-2)
\psline{-}(3,3)(4,4)(5,3)(4,2)(5,1)(4,0)(5,-1)(4,-2)(5,-3)(4,-4)(3,-3)
\psline{-}(4,4)(5,5)
\psline{-}(5,5)(6,4)(5,3)(6,2)(5,1)(6,0)(5,-1)(6,-2)(5,-3)(6,-4)(5,-5)(4,-4)
\psline{-}(5,-5)(4,-4)
\psline{-}(5,5)(6,6)(7,5)(6,4)(7,3)(6,2)(7,1)(6,0)(7,-1)(6,-2)(7,-3)(6,-4)(7,-5)(6,-6)(5,-5)
\psline{-}(6,6)(7,7)(8,6)(7,5)(8,4)(7,3)(8,2)(7,1)(8,0)(7,-1)(8,-2)(7,-3)(8,-4)(7,-5)(8,-6)(7,-7)(6,-6)
\psline{-}(7,7)(8,8)(9,7)(8,6)(9,5)(8,4)(9,3)(8,2)(9,1)(8,0)(9,-1)(8,-2)(9,-3)(8,-4)(9,-5)(8,-6)(9,-7)(8,-8)(7,-7)
\psline{-}(2,-2)(1,-1)(2,0)(1,1)(2,2)(3,1)(2,0)(3,-1)(2,-2)
\rput(8,7){$_1$}
\rput(8,5){$_1$}
\rput(8,3){$_1$}
\rput(8,1){$_1$}
\rput(8,-1){$_1$}
\rput(8,-3){$_1$}
\rput(8,-5){$_1$}
\rput(5,4){$_{r-n-1}$}
\rput(5,2){$_{r-n-1}$}
\rput(5,0){$_{r-n-1}$}
\rput(5,-2){$_{r-n-1}$}
\rput(4,3){$_{r-n}$}
\rput(4,1){$_{r-n}$}
\rput(4,-1){$_{r-n}$}
\rput(1,0){$_{r-1}$}
\psset{linecolor=myc}\unlw
\psarc{-}(8,-6){0.7071}{-45}{45}
\psarc{-}(8,-4){0.7071}{-45}{45}
\psarc{-}(8,-2){0.7071}{-45}{45}
\psarc{-}(8,0){0.7071}{-45}{45}
\psarc{-}(8,2){0.7071}{-45}{45}
\psarc{-}(8,4){0.7071}{-45}{45} 
\psarc{-}(8,6){0.7071}{-45}{45} 
\psarc{-}(3,-3){0.7071}{-45}{45}
\psarc{-}(4,-4){0.7071}{-45}{45}
\psarc{-}(5,-5){0.7071}{-45}{45}
\psarc{-}(6,-6){0.7071}{-45}{45}
\psarc{-}(7,-7){0.7071}{-45}{45}
\psarc{-}(5,-3){0.7071}{135}{-135}
\psarc{-}(6,-4){0.7071}{135}{-135}
\psarc{-}(7,-5){0.7071}{135}{-135}
\psarc{-}(8,-6){0.7071}{135}{-135}
\psarc{-}(9,-7){0.7071}{135}{-135}
\psset{linecolor=myc2}\unlw
\psline{-}(0.5,0.5)(0.5,8.25)
\psline{-}(1.5,1.5)(1.5,8.25)
\psline{-}(2.5,2.5)(2.5,8.25)
\psline{-}(3.5,3.5)(3.5,8.25)
\psline{-}(4.5,4.5)(4.5,8.25)
\psline{-}(5.5,5.5)(5.5,8.25)
\psline{-}(6.5,6.5)(6.5,8.25)
\psline{-}(7.5,7.5)(7.5,8.25)
\psline{-}(8.5,7.5)(8.5,8.25)
\psset{linecolor=myc3,linestyle=dashed,dash=2pt 2pt}
\psline{-}(0.5,-0.5)(0.5,-8.25)
\psline{-}(1.5,-1.5)(1.5,-8.25)
\psline{-}(2.5,-2.5)(2.5,-8.25)
\psarc{-}(4,-4){0.7071}{135}{315} 
\psline{-}(5.5,-5.5)(5.5,-8.25)
\psline{-}(6.5,-6.5)(6.5,-8.25)
\psline{-}(7.5,-7.5)(7.5,-8.25)
\psline{-}(8.5,-7.5)(8.5,-8.25)
\end{pspicture}
\begin{pspicture}(-2.9,-8.3)(9,8.3)
\psdots(0.5,0.5)(1.5,1.5)(2.5,2.5)(3.5,3.5)(4.5,4.5)(5.5,5.5)(6.5,6.5)(7.5,7.5)(8.5,7.5)
\psdots(0.5,-0.5)(1.5,-1.5)(2.5,-2.5)(3.5,-3.5)(4.5,-4.5)(5.5,-5.5)(6.5,-6.5)(7.5,-7.5)(8.5,-7.5)
\lw
\rput(-1.4,-0.0){$ + \frac{S_{r-n}}{S_{r-n+1}}$}
\psline{-}(1,1)(0,0)(1,-1)
\psline{-}(2,2)(3,3)(4,2)(3,1)(4,0)(3,-1)(4,-2)(3,-3)(2,-2)
\psline{-}(3,3)(4,4)(5,3)(4,2)(5,1)(4,0)(5,-1)(4,-2)(5,-3)(4,-4)(3,-3)
\psline{-}(4,4)(5,5)
\psline{-}(5,5)(6,4)(5,3)(6,2)(5,1)(6,0)(5,-1)(6,-2)(5,-3)(6,-4)(5,-5)(4,-4)
\psline{-}(5,-5)(4,-4)
\psline{-}(5,5)(6,6)(7,5)(6,4)(7,3)(6,2)(7,1)(6,0)(7,-1)(6,-2)(7,-3)(6,-4)(7,-5)(6,-6)(5,-5)
\psline{-}(6,6)(7,7)(8,6)(7,5)(8,4)(7,3)(8,2)(7,1)(8,0)(7,-1)(8,-2)(7,-3)(8,-4)(7,-5)(8,-6)(7,-7)(6,-6)
\psline{-}(7,7)(8,8)(9,7)(8,6)(9,5)(8,4)(9,3)(8,2)(9,1)(8,0)(9,-1)(8,-2)(9,-3)(8,-4)(9,-5)(8,-6)(9,-7)(8,-8)(7,-7)
\psline{-}(2,-2)(1,-1)(2,0)(1,1)(2,2)(3,1)(2,0)(3,-1)(2,-2)
\rput(8,7){$_1$}
\rput(8,5){$_1$}
\rput(8,3){$_1$}
\rput(8,1){$_1$}
\rput(8,-1){$_1$}
\rput(8,-3){$_1$}
\rput(8,-5){$_1$}
\rput(5,4){$_{r-n-1}$}
\rput(5,2){$_{r-n-1}$}
\rput(5,0){$_{r-n-1}$}
\rput(5,-2){$_{r-n-1}$}
\rput(4,3){$_{r-n}$}
\rput(4,1){$_{r-n}$}
\rput(4,-1){$_{r-n}$}
\rput(1,0){$_{r-1}$}
\psset{linecolor=myc}\unlw
\psarc{-}(8,-6){0.7071}{-45}{45}
\psarc{-}(8,-4){0.7071}{-45}{45}
\psarc{-}(8,-2){0.7071}{-45}{45}
\psarc{-}(8,0){0.7071}{-45}{45}
\psarc{-}(8,2){0.7071}{-45}{45}
\psarc{-}(8,4){0.7071}{-45}{45} 
\psarc{-}(8,6){0.7071}{-45}{45} 
\psarc{-}(4,-4){0.7071}{-45}{135}
\psarc{-}(5,-5){0.7071}{-45}{45}
\psarc{-}(6,-6){0.7071}{-45}{45}
\psarc{-}(7,-7){0.7071}{-45}{45}
\psarc{-}(4,-2){0.7071}{-135}{-45}
\psarc{-}(6,-4){0.7071}{135}{-135}
\psarc{-}(7,-5){0.7071}{135}{-135}
\psarc{-}(8,-6){0.7071}{135}{-135}
\psarc{-}(9,-7){0.7071}{135}{-135}
\psset{linecolor=myc2}\unlw
\psline{-}(0.5,0.5)(0.5,8.25)
\psline{-}(1.5,1.5)(1.5,8.25)
\psline{-}(2.5,2.5)(2.5,8.25)
\psline{-}(3.5,3.5)(3.5,8.25)
\psline{-}(4.5,4.5)(4.5,8.25)
\psline{-}(5.5,5.5)(5.5,8.25)
\psline{-}(6.5,6.5)(6.5,8.25)
\psline{-}(7.5,7.5)(7.5,8.25)
\psline{-}(8.5,7.5)(8.5,8.25)
\psset{linecolor=myc3,linestyle=dashed,dash=2pt 2pt}
\psline{-}(0.5,-0.5)(0.5,-8.25)
\psline{-}(1.5,-1.5)(1.5,-8.25)
\psline{-}(2.5,-2.5)(2.5,-8.25)
\psarc{-}(4,-4){0.7071}{135}{315} 
\psline{-}(5.5,-5.5)(5.5,-8.25)
\psline{-}(6.5,-6.5)(6.5,-8.25)
\psline{-}(7.5,-7.5)(7.5,-8.25)
\psline{-}(8.5,-7.5)(8.5,-8.25)
\end{pspicture}
\end{equation*}
Summing the lower diagonal row of boxes has yielded two terms. For the second, the remaining $n-1$ boxes of the first lower diagonal row must all be
\psset{linewidth=1pt}
\begin{pspicture}(-0.2,-0.1)(0.2,0.2)\lw
\psline{-}(-0.2,0)(0,0.2)(0.2,0)(0,-0.2)(-0.2,0)\unlw
\psset{linecolor=myc}
\psarc{-}(0,0.2){0.14}{225}{315}
\psarc{-}(0,-0.2){0.14}{45}{135}
\end{pspicture}
. Otherwise, a half-arc is formed, propagates in the upper right direction and annihilates.
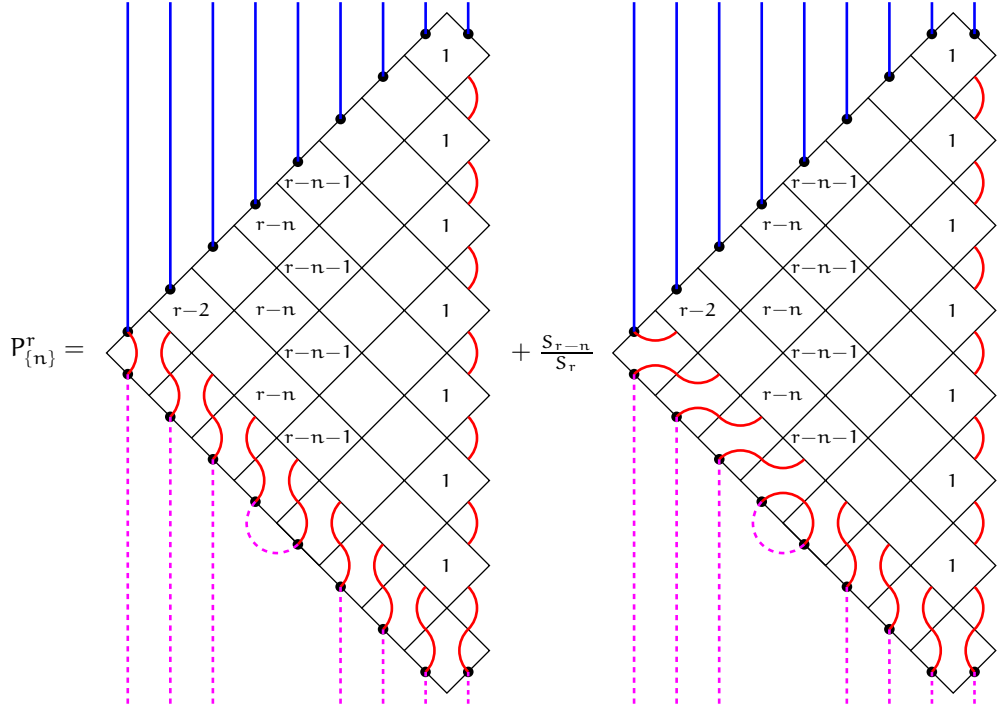
\begin{equation*}
\psset{unit=0.565685}\begin{pspicture}(-2.0,-8.3)(9,8.3)
\psdots(0.5,0.5)(1.5,1.5)(2.5,2.5)(3.5,3.5)(4.5,4.5)(5.5,5.5)(6.5,6.5)(7.5,7.5)(8.5,7.5)
\psdots(0.5,-0.5)(1.5,-1.5)(2.5,-2.5)(3.5,-3.5)(4.5,-4.5)(5.5,-5.5)(6.5,-6.5)(7.5,-7.5)(8.5,-7.5)
\lw
\rput(-1.4,-0.0){$P^r_{\{n\}} =$}
\psline{-}(1,1)(0,0)(1,-1)
\psline{-}(2,2)(3,3)(4,2)(3,1)(4,0)(3,-1)(4,-2)(3,-3)(2,-2)
\psline{-}(3,3)(4,4)(5,3)(4,2)(5,1)(4,0)(5,-1)(4,-2)(5,-3)(4,-4)(3,-3)
\psline{-}(4,4)(5,5)
\psline{-}(5,5)(6,4)(5,3)(6,2)(5,1)(6,0)(5,-1)(6,-2)(5,-3)(6,-4)(5,-5)(4,-4)
\psline{-}(5,-5)(4,-4)
\psline{-}(5,5)(6,6)(7,5)(6,4)(7,3)(6,2)(7,1)(6,0)(7,-1)(6,-2)(7,-3)(6,-4)(7,-5)(6,-6)(5,-5)
\psline{-}(6,6)(7,7)(8,6)(7,5)(8,4)(7,3)(8,2)(7,1)(8,0)(7,-1)(8,-2)(7,-3)(8,-4)(7,-5)(8,-6)(7,-7)(6,-6)
\psline{-}(7,7)(8,8)(9,7)(8,6)(9,5)(8,4)(9,3)(8,2)(9,1)(8,0)(9,-1)(8,-2)(9,-3)(8,-4)(9,-5)(8,-6)(9,-7)(8,-8)(7,-7)
\psline{-}(2,-2)(1,-1)(2,0)(1,1)(2,2)(3,1)(2,0)(3,-1)(2,-2)
\rput(8,7){$_1$}
\rput(8,5){$_1$}
\rput(8,3){$_1$}
\rput(8,1){$_1$}
\rput(8,-1){$_1$}
\rput(8,-3){$_1$}
\rput(8,-5){$_1$}
\rput(5,4){$_{r-n-1}$}
\rput(5,2){$_{r-n-1}$}
\rput(5,0){$_{r-n-1}$}
\rput(5,-2){$_{r-n-1}$}
\rput(4,3){$_{r-n}$}
\rput(4,1){$_{r-n}$}
\rput(4,-1){$_{r-n}$}
\rput(2,1){$_{r-2}$}
\psset{linecolor=myc}\unlw
\psarc{-}(8,-6){0.7071}{-45}{45}
\psarc{-}(8,-4){0.7071}{-45}{45}
\psarc{-}(8,-2){0.7071}{-45}{45}
\psarc{-}(8,0){0.7071}{-45}{45}
\psarc{-}(8,2){0.7071}{-45}{45}
\psarc{-}(8,4){0.7071}{-45}{45} 
\psarc{-}(8,6){0.7071}{-45}{45} 
\psarc{-}(0,-0){0.7071}{-45}{45}
\psarc{-}(1,-1){0.7071}{-45}{45}
\psarc{-}(2,-2){0.7071}{-45}{45}
\psarc{-}(3,-3){0.7071}{-45}{45}
\psarc{-}(4,-4){0.7071}{-45}{45}
\psarc{-}(5,-5){0.7071}{-45}{45}
\psarc{-}(6,-6){0.7071}{-45}{45}
\psarc{-}(7,-7){0.7071}{-45}{45}
\psarc{-}(2,-0){0.7071}{135}{-135}
\psarc{-}(3,-1){0.7071}{135}{-135}
\psarc{-}(4,-2){0.7071}{135}{-135}
\psarc{-}(5,-3){0.7071}{135}{-135}
\psarc{-}(6,-4){0.7071}{135}{-135}
\psarc{-}(7,-5){0.7071}{135}{-135}
\psarc{-}(8,-6){0.7071}{135}{-135}
\psarc{-}(9,-7){0.7071}{135}{-135}
\psset{linecolor=myc2}\unlw
\psline{-}(0.5,0.5)(0.5,8.25)
\psline{-}(1.5,1.5)(1.5,8.25)
\psline{-}(2.5,2.5)(2.5,8.25)
\psline{-}(3.5,3.5)(3.5,8.25)
\psline{-}(4.5,4.5)(4.5,8.25)
\psline{-}(5.5,5.5)(5.5,8.25)
\psline{-}(6.5,6.5)(6.5,8.25)
\psline{-}(7.5,7.5)(7.5,8.25)
\psline{-}(8.5,7.5)(8.5,8.25)
\psset{linecolor=myc3,linestyle=dashed,dash=2pt 2pt}
\psline{-}(0.5,-0.5)(0.5,-8.25)
\psline{-}(1.5,-1.5)(1.5,-8.25)
\psline{-}(2.5,-2.5)(2.5,-8.25)
\psarc{-}(4,-4){0.7071}{135}{315} 
\psline{-}(5.5,-5.5)(5.5,-8.25)
\psline{-}(6.5,-6.5)(6.5,-8.25)
\psline{-}(7.5,-7.5)(7.5,-8.25)
\psline{-}(8.5,-7.5)(8.5,-8.25)
\end{pspicture}
\begin{pspicture}(-2.9,-8.3)(9,8.3)
\psdots(0.5,0.5)(1.5,1.5)(2.5,2.5)(3.5,3.5)(4.5,4.5)(5.5,5.5)(6.5,6.5)(7.5,7.5)(8.5,7.5)
\psdots(0.5,-0.5)(1.5,-1.5)(2.5,-2.5)(3.5,-3.5)(4.5,-4.5)(5.5,-5.5)(6.5,-6.5)(7.5,-7.5)(8.5,-7.5)
\lw
\rput(-1.4,-0.0){$ + \ \frac{S_{r-n}}{S_{r}}$}
\psline{-}(1,1)(0,0)(1,-1)
\psline{-}(2,2)(3,3)(4,2)(3,1)(4,0)(3,-1)(4,-2)(3,-3)(2,-2)
\psline{-}(3,3)(4,4)(5,3)(4,2)(5,1)(4,0)(5,-1)(4,-2)(5,-3)(4,-4)(3,-3)
\psline{-}(4,4)(5,5)
\psline{-}(5,5)(6,4)(5,3)(6,2)(5,1)(6,0)(5,-1)(6,-2)(5,-3)(6,-4)(5,-5)(4,-4)
\psline{-}(5,-5)(4,-4)
\psline{-}(5,5)(6,6)(7,5)(6,4)(7,3)(6,2)(7,1)(6,0)(7,-1)(6,-2)(7,-3)(6,-4)(7,-5)(6,-6)(5,-5)
\psline{-}(6,6)(7,7)(8,6)(7,5)(8,4)(7,3)(8,2)(7,1)(8,0)(7,-1)(8,-2)(7,-3)(8,-4)(7,-5)(8,-6)(7,-7)(6,-6)
\psline{-}(7,7)(8,8)(9,7)(8,6)(9,5)(8,4)(9,3)(8,2)(9,1)(8,0)(9,-1)(8,-2)(9,-3)(8,-4)(9,-5)(8,-6)(9,-7)(8,-8)(7,-7)
\psline{-}(2,-2)(1,-1)(2,0)(1,1)(2,2)(3,1)(2,0)(3,-1)(2,-2)
\rput(8,7){$_1$}
\rput(8,5){$_1$}
\rput(8,3){$_1$}
\rput(8,1){$_1$}
\rput(8,-1){$_1$}
\rput(8,-3){$_1$}
\rput(8,-5){$_1$}
\rput(5,4){$_{r-n-1}$}
\rput(5,2){$_{r-n-1}$}
\rput(5,0){$_{r-n-1}$}
\rput(5,-2){$_{r-n-1}$}
\rput(4,3){$_{r-n}$}
\rput(4,1){$_{r-n}$}
\rput(4,-1){$_{r-n}$}
\rput(2,1){$_{r-2}$}
\psset{linecolor=myc}\unlw
\psarc{-}(8,-6){0.7071}{-45}{45}
\psarc{-}(8,-4){0.7071}{-45}{45}
\psarc{-}(8,-2){0.7071}{-45}{45}
\psarc{-}(8,0){0.7071}{-45}{45}
\psarc{-}(8,2){0.7071}{-45}{45}
\psarc{-}(8,4){0.7071}{-45}{45} 
\psarc{-}(8,6){0.7071}{-45}{45} 
\psarc{-}(4,-4){0.7071}{-45}{135}
\psarc{-}(5,-5){0.7071}{-45}{45}
\psarc{-}(6,-6){0.7071}{-45}{45}
\psarc{-}(7,-7){0.7071}{-45}{45}
\psarc{-}(4,-2){0.7071}{-135}{-45}
\psarc{-}(6,-4){0.7071}{135}{-135}
\psarc{-}(7,-5){0.7071}{135}{-135}
\psarc{-}(8,-6){0.7071}{135}{-135}
\psarc{-}(9,-7){0.7071}{135}{-135}
\psarc{-}(3,-1){0.7071}{-135}{-45}\psarc{-}(3,-3){0.7071}{45}{135}
\psarc{-}(2,0){0.7071}{-135}{-45}\psarc{-}(2,-2){0.7071}{45}{135}
\psarc{-}(1,1){0.7071}{-135}{-45}\psarc{-}(1,-1){0.7071}{45}{135}
\psset{linecolor=myc2}\unlw
\psline{-}(0.5,0.5)(0.5,8.25)
\psline{-}(1.5,1.5)(1.5,8.25)
\psline{-}(2.5,2.5)(2.5,8.25)
\psline{-}(3.5,3.5)(3.5,8.25)
\psline{-}(4.5,4.5)(4.5,8.25)
\psline{-}(5.5,5.5)(5.5,8.25)
\psline{-}(6.5,6.5)(6.5,8.25)
\psline{-}(7.5,7.5)(7.5,8.25)
\psline{-}(8.5,7.5)(8.5,8.25)
\psset{linecolor=myc3,linestyle=dashed,dash=2pt 2pt}
\psline{-}(0.5,-0.5)(0.5,-8.25)
\psline{-}(1.5,-1.5)(1.5,-8.25)
\psline{-}(2.5,-2.5)(2.5,-8.25)
\psarc{-}(4,-4){0.7071}{135}{315} 
\psline{-}(5.5,-5.5)(5.5,-8.25)
\psline{-}(6.5,-6.5)(6.5,-8.25)
\psline{-}(7.5,-7.5)(7.5,-8.25)
\psline{-}(8.5,-7.5)(8.5,-8.25)
\end{pspicture}
\end{equation*}
The first term is $P^{r-1}_{\{n-1\}}$. For the second, the upper diagonal of boxes is summed in the same fashion as in the proof of Lemma \ref{sec:Pr1}. The desired recurrence relation is:
\begin{equation}
\label{eq:rec1}
P^r_{\{n\}} = P^{r-1}_{\{n-1\}} + \frac{S_{r-n}}{4 S_{1/2}C_{r/2}C_{(r-1)/2}}.
\end{equation} 
To complete the proof, we suppose the proposition is true for $r-1$ and verify that the recurrence relation yields the correct answer for $r$:
\begin{align*}
P^{r-1}_{\{n-1\}} + \frac{S_{r-n}}{4 S_{1/2}C_{r/2}C_{(r-1)/2}} &=  \frac{S_{(r-n)/2}S_{(n-1)/2}}{2 \left(S_{1/2}\right)^2C_{(r-1)/2}} + \frac{S_{r-n}}{4 S_{1/2}C_{r/2}C_{(r-1)/2}} \\ &=  \frac{S_{(r-n)/2}}{2 \left(S_{1/2}\right)^2C_{r/2}C_{(r-1)/2}} \left( S_{(n-1)/2}C_{r/2}+S_{1/2}C_{(r-n)/2}\right) \\ &=  \frac{S_{(r-n)/2}S_{n/2}}{2 \left(S_{1/2}\right)^2C_{r/2}}.
\end{align*}  \hfill$\square$ 

\medskip
\noindent The invariance of $P^r$ under reflection appears through $P^r_{\{n\}} = P^r_{\{r-n\}}$. Note also that (\ref{eq:rec1}) holds for $n=1$ if we impose the unphysical condition $P^{r-1}_{\{0 \}}=0$. This will become the initial condition in recursion arguments in propositions to come. 

The presence of $C_{r/2}$ in the denominator will cause $P^r_{\{n\}}$ to diverge for specific values of $\Lambda$, but for all $n$. We now direct our interest to $m$ concentrical bubbles, starting with all the defects sitting at the right end of the link state.
\begin{Lemme}\label{lem:concentric} Concentric bubbles in leftmost position:
\begin{equation*}
P^r_{\{m^m\}} = \frac1{ \left( 2 S_{1/2} \right)^m} \prod_{i=0}^{m-1} \frac{S_{(r-m-i)/2}}{C_{(r-i)/2}}.
\end{equation*}
\end{Lemme}
\noindent{\scshape Proof\ \ } The proof does not require new ideas. Using the graphical representation, one finds the following recurrence relation:
\begin{equation}
\label{eq:rec2}
P^r_{\{m^m\}} = \frac{S_{r-m}}{4C_{r/2}C_{(r-1)/2}S_{1/2}}P^{r-2}_{\{(m-1)^{m-1}\}}.
\end{equation}
The proposed $P^r_{\{m^m\}}$ certainly fits the initial condition,  $m=1$, given in Lemma \ref{sec:Pr1}. Proving the induction is straightforward.  \hfill$\square$ 

\medskip
\noindent A little more work would also yield:
\begin{equation}
P^r_{\{n^m\}} = \frac1{ \left( 2 S_{1/2} \right)^m} \prod_{i=0}^{m-1} \frac{S_{(r-n-i)/2}S_{(n-i)/2}}{C_{(r-i)/2}S_{(i+1)/2}}.\label{eq:encoreUneAutre}
\end{equation}
Since the $S_{(r-n-i)/2} / S_{1/2}$ are analytic in $\Lambda$ on $ \mathbb{R}$, the set of singularities of $P^r_{\{m^m\}}$ is contained in the set $\cup_{i=0}^{m-1} \{ \Lambda\, |\, C_{(r-i)/2 } = 0\}$. The singularities of $P^r_{\{n^m\}}$ could be worked out from the above expression, but the analysis would be more tedious because zeroes of the denominator could be cancelled by zeroes of the numerator. Similar difficulties arise when one studies $w$ with more complicated patterns of arcs. This is why the analysis in the next subsection turns to expressing the general matrix elements in terms of some already computed, instead of giving new explicit expressions.

\subsection{$P^r_{\{n_1,n_2,...,n_k\}}$ with non trivial bubble patterns}
\label{sec:nontrivialB}

This section shows how to express the matrix elements $P^r_w$ for any link state $w$ in terms of a sum of $P^r_{\{m^m\}}$. This is done by two basic operations: the removal of the leftmost defects and the replacement of a cluster of arcs by a cluster of concentric ones.

The operation of shifting a pattern of arcs by removing the defect at position ``$1$'' will be denoted by ``$\leftarrow$''. It is an operation from $B^r$ to $B^{r-1}\cup\{0\}$. If $w$ does not start with a defect, the result $\overleftarrow{w}$ is zero. For example, $\overleftarrow{v^{10}_{2,6,8,7}} = v^9_{1,5,7,6}$  but $\overleftarrow{v^{10}_{2,2,7,7}} = 0$.

We also define the operation, from $B^r$ to $B^{r-2}$, of removing the $j$-th $1$-bubble from $w$. The result is noted $w\setminus \{ j \}$. For example, $v^{10}_{2,6,8,7}\setminus\{1\} = v^8_{4,6,5}$ and $v^{10}_{2,6,8,7} \setminus\{2\} = v^8_{2,6,6} = v^{10}_{2,6,8,7} \setminus\{3\} $.

\begin{Lemme}Left shift of arcs.
Let $w \in B^r$ any link state, with $b$ $1$-bubbles, labelled by $j$ and centered at $n_j$ (see section \ref{sub:notation}). The matrix element $P^r_w$ satisfies the following recursion equation:
\begin{equation}
P^r_w = P^{r-1}_{\overleftarrow{w}} + \frac{1}{4S_{1/2}C_{r/2}C_{(r-1)/2}} \sum_{j=1}^b S_{r-n_j} P^{r-2}_{w\setminus\{j\}} 
\label{eq: generalrec}
\end{equation}
It also holds when $\overleftarrow{w} =0$ if we define $P^{r-1}_{\overleftarrow{w}=0} =0$.
\label{sec: recurgenerale}
\end{Lemme}
\noindent{\scshape Proof\ \ } This is a generalization of (\ref{eq:rec1}) and (\ref{eq:encoreUneAutre}). Summing the first lower diagonal row of boxes of $P^r_w$, only $b+1$ configurations have non zero contributions. For example,
\begin{equation*}
\psset{unit=0.365685}\begin{pspicture}(-2.0,-8.3)(9,8.3)
\lw
\psline{-}(1,1)(0,0)(1,-1)
\psline{-}(2,2)(3,3)(4,2)(3,1)(4,0)(3,-1)(4,-2)(3,-3)(2,-2)
\psline{-}(3,3)(4,4)(5,3)(4,2)(5,1)(4,0)(5,-1)(4,-2)(5,-3)(4,-4)(3,-3)
\psline{-}(4,4)(5,5)
\psline{-}(5,5)(6,4)(5,3)(6,2)(5,1)(6,0)(5,-1)(6,-2)(5,-3)(6,-4)(5,-5)(4,-4)
\psline{-}(5,-5)(4,-4)
\psline{-}(5,5)(6,6)(7,5)(6,4)(7,3)(6,2)(7,1)(6,0)(7,-1)(6,-2)(7,-3)(6,-4)(7,-5)(6,-6)(5,-5)
\psline{-}(6,6)(7,7)(8,6)(7,5)(8,4)(7,3)(8,2)(7,1)(8,0)(7,-1)(8,-2)(7,-3)(8,-4)(7,-5)(8,-6)(7,-7)(6,-6)
\psline{-}(7,7)(8,8)(9,7)(8,6)(9,5)(8,4)(9,3)(8,2)(9,1)(8,0)(9,-1)(8,-2)(9,-3)(8,-4)(9,-5)(8,-6)(9,-7)(8,-8)(7,-7)
\psline{-}(2,-2)(1,-1)(2,0)(1,1)(2,2)(3,1)(2,0)(3,-1)(2,-2)
\psline[linewidth=1.5pt]{-}(0,0)(1,1)(9,-7)(8,-8)(0,0)
\psset{linecolor=myc}\unlw
\psarc{-}(8,-6){0.7071}{-45}{45}
\psarc{-}(8,-4){0.7071}{-45}{45}
\psarc{-}(8,-2){0.7071}{-45}{45}
\psarc{-}(8,0){0.7071}{-45}{45}
\psarc{-}(8,2){0.7071}{-45}{45}
\psarc{-}(8,4){0.7071}{-45}{45} 
\psarc{-}(8,6){0.7071}{-45}{45} 
\psset{linecolor=myc2}\unlw
\psline{-}(0.5,0.5)(0.5,8.25)
\psline{-}(1.5,1.5)(1.5,8.25)
\psline{-}(2.5,2.5)(2.5,8.25)
\psline{-}(3.5,3.5)(3.5,8.25)
\psline{-}(4.5,4.5)(4.5,8.25)
\psline{-}(5.5,5.5)(5.5,8.25)
\psline{-}(6.5,6.5)(6.5,8.25)
\psline{-}(7.5,7.5)(7.5,8.25)
\psline{-}(8.5,7.5)(8.5,8.25)
\psset{linecolor=myc3,linestyle=dashed,dash=2pt 2pt}
\psarc{-}(2,-2){0.7071}{135}{315} 
\psbezier{-}(0.5,-0.5)(-0.5,-1.5)(2.5,-4.5)(3.5,-3.5)
\psarc{-}(5,-5){0.7071}{135}{315} 
\psline{-}(8.5,-7.5)(8.5,-8.25)
\psline{-}(7.5,-7.5)(7.5,-8.25)
\psline[linestyle=dotted,dotsep=1pt]{-}(5.7,-6.7)(6.3,-7.3)
\end{pspicture}
\qquad \qquad \qquad \qquad
\begin{pspicture}(0,0)(19,1)\lw
\psline[linewidth=1.2pt]{-}(0,0)(16,0)(16,1)(0,1)(0,0)
\psline[linewidth=1.2pt]{-}(0,16)(16,16)(16,17)(0,17)(0,16)
\psline[linewidth=1.2pt]{-}(0,12.5)(16,12.5)(16,13.5)(0,13.5)(0,12.5)
\psline[linewidth=1.2pt]{-}(0,7)(16,7)(16,8)(0,8)(0,7)
\psline[linewidth=1.2pt]{-}(0,3.5)(16,3.5)(16,4.5)(0,4.5)(0,3.5)
\psline{-}(1,0)(1,1)\psline{-}(1,3.5)(1,4.5)\psline{-}(1,7)(1,8)\psline{-}(1,12.5)(1,13.5)\psline{-}(1,16)(1,17)
\psline{-}(2,0)(2,1)\psline{-}(2,3.5)(2,4.5)\psline{-}(2,7)(2,8)\psline{-}(2,12.5)(2,13.5)\psline{-}(2,16)(2,17)
\psline{-}(3,0)(3,1)\psline{-}(3,3.5)(3,4.5)\psline{-}(3,7)(3,8)\psline{-}(3,12.5)(3,13.5)\psline{-}(3,16)(3,17)
\psline{-}(4,0)(4,1)\psline{-}(4,3.5)(4,4.5)\psline{-}(4,7)(4,8)\psline{-}(4,12.5)(4,13.5)\psline{-}(4,16)(4,17)
\psline{-}(5,0)(5,1)\psline{-}(5,3.5)(5,4.5)\psline{-}(5,7)(5,8)\psline{-}(5,12.5)(5,13.5)\psline{-}(5,16)(5,17)
\psline{-}(6,0)(6,1)\psline{-}(6,3.5)(6,4.5)\psline{-}(6,7)(6,8)\psline{-}(6,12.5)(6,13.5)\psline{-}(6,16)(6,17)
\psline{-}(7,0)(7,1)\psline{-}(7,3.5)(7,4.5)\psline{-}(7,7)(7,8)\psline{-}(7,12.5)(7,13.5)\psline{-}(7,16)(7,17)
\psline{-}(8,0)(8,1)\psline{-}(8,3.5)(8,4.5)\psline{-}(8,7)(8,8)\psline{-}(8,12.5)(8,13.5)\psline{-}(8,16)(8,17)
\psline{-}(9,0)(9,1)\psline{-}(9,3.5)(9,4.5)\psline{-}(9,7)(9,8)\psline{-}(9,12.5)(9,13.5)\psline{-}(9,16)(9,17)
\psline{-}(10,0)(10,1)\psline{-}(10,3.5)(10,4.5)\psline{-}(10,7)(10,8)\psline{-}(10,12.5)(10,13.5)\psline{-}(10,16)(10,17)
\psline{-}(11,0)(11,1)\psline{-}(11,3.5)(11,4.5)\psline{-}(11,7)(11,8)\psline{-}(11,12.5)(11,13.5)\psline{-}(11,16)(11,17)
\psline{-}(12,0)(12,1)\psline{-}(12,3.5)(12,4.5)\psline{-}(12,7)(12,8)\psline{-}(12,12.5)(12,13.5)\psline{-}(12,16)(12,17)
\psline{-}(13,0)(13,1)\psline{-}(13,3.5)(13,4.5)\psline{-}(13,7)(13,8)\psline{-}(13,12.5)(13,13.5)\psline{-}(13,16)(13,17)
\psline{-}(14,0)(14,1)\psline{-}(14,3.5)(14,4.5)\psline{-}(14,7)(14,8)\psline{-}(14,12.5)(14,13.5)\psline{-}(14,16)(14,17)
\psline{-}(15,0)(15,1)\psline{-}(15,3.5)(15,4.5)\psline{-}(15,7)(15,8)\psline{-}(15,12.5)(15,13.5)\psline{-}(15,16)(15,17)
\psset{linecolor=myc}
\psarc{-}(0,0){0.5}{0}{90}\psarc{-}(1,1){0.5}{180}{270}
\psarc{-}(1,0){0.5}{0}{90}\psarc{-}(2,1){0.5}{180}{270}
\psarc{-}(2,0){0.5}{0}{90}\psarc{-}(3,1){0.5}{180}{270}
\psarc{-}(3,0){0.5}{0}{90}\psarc{-}(4,1){0.5}{180}{270}
\psarc{-}(4,0){0.5}{0}{90}\psarc{-}(5,1){0.5}{180}{270}
\psarc{-}(5,0){0.5}{0}{90}\psarc{-}(6,1){0.5}{180}{270}
\psarc{-}(6,0){0.5}{0}{90}\psarc{-}(7,1){0.5}{180}{270}
\psarc{-}(7,0){0.5}{0}{90}\psarc{-}(8,1){0.5}{180}{270}
\psarc{-}(8,0){0.5}{0}{90}\psarc{-}(9,1){0.5}{180}{270}
\psarc{-}(9,0){0.5}{0}{90}\psarc{-}(10,1){0.5}{180}{270}
\psarc{-}(10,0){0.5}{0}{90}\psarc{-}(11,1){0.5}{180}{270}
\psarc{-}(11,0){0.5}{0}{90}\psarc{-}(12,1){0.5}{180}{270}
\psarc{-}(12,0){0.5}{0}{90}\psarc{-}(13,1){0.5}{180}{270}
\psarc{-}(13,0){0.5}{0}{90}\psarc{-}(14,1){0.5}{180}{270}
\psarc{-}(14,0){0.5}{0}{90}\psarc{-}(15,1){0.5}{180}{270}
\psarc{-}(15,0){0.5}{0}{90}\psarc{-}(16,1){0.5}{180}{270}
\psarc{-}(0,4.5){0.5}{270}{360}\psarc{-}(1,3.5){0.5}{90}{180}
\psarc{-}(1,4.5){0.5}{270}{360}\psarc{-}(2,3.5){0.5}{90}{180}
\psarc{-}(2,3.5){0.5}{0}{90}\psarc{-}(3,4.5){0.5}{180}{270}
\psarc{-}(3,3.5){0.5}{0}{90}\psarc{-}(4,4.5){0.5}{180}{270}
\psarc{-}(4,3.5){0.5}{0}{90}\psarc{-}(5,4.5){0.5}{180}{270}
\psarc{-}(5,3.5){0.5}{0}{90}\psarc{-}(6,4.5){0.5}{180}{270}
\psarc{-}(6,3.5){0.5}{0}{90}\psarc{-}(7,4.5){0.5}{180}{270}
\psarc{-}(7,3.5){0.5}{0}{90}\psarc{-}(8,4.5){0.5}{180}{270}
\psarc{-}(8,3.5){0.5}{0}{90}\psarc{-}(9,4.5){0.5}{180}{270}
\psarc{-}(9,3.5){0.5}{0}{90}\psarc{-}(10,4.5){0.5}{180}{270}
\psarc{-}(10,3.5){0.5}{0}{90}\psarc{-}(11,4.5){0.5}{180}{270}
\psarc{-}(11,3.5){0.5}{0}{90}\psarc{-}(12,4.5){0.5}{180}{270}
\psarc{-}(12,3.5){0.5}{0}{90}\psarc{-}(13,4.5){0.5}{180}{270}
\psarc{-}(13,3.5){0.5}{0}{90}\psarc{-}(14,4.5){0.5}{180}{270}
\psarc{-}(14,3.5){0.5}{0}{90}\psarc{-}(15,4.5){0.5}{180}{270}
\psarc{-}(15,3.5){0.5}{0}{90}\psarc{-}(16,4.5){0.5}{180}{270}
\psarc{-}(0,8){0.5}{270}{360}\psarc{-}(1,7){0.5}{90}{180}
\psarc{-}(1,8){0.5}{270}{360}\psarc{-}(2,7){0.5}{90}{180}
\psarc{-}(2,8){0.5}{270}{360}\psarc{-}(3,7){0.5}{90}{180}
\psarc{-}(3,8){0.5}{270}{360}\psarc{-}(4,7){0.5}{90}{180}
\psarc{-}(4,8){0.5}{270}{360}\psarc{-}(5,7){0.5}{90}{180}
\psarc{-}(5,7){0.5}{0}{90}\psarc{-}(6,8){0.5}{180}{270}
\psarc{-}(6,7){0.5}{0}{90}\psarc{-}(7,8){0.5}{180}{270}
\psarc{-}(7,7){0.5}{0}{90}\psarc{-}(8,8){0.5}{180}{270}
\psarc{-}(8,7){0.5}{0}{90}\psarc{-}(9,8){0.5}{180}{270}
\psarc{-}(9,7){0.5}{0}{90}\psarc{-}(10,8){0.5}{180}{270}
\psarc{-}(10,7){0.5}{0}{90}\psarc{-}(11,8){0.5}{180}{270}
\psarc{-}(11,7){0.5}{0}{90}\psarc{-}(12,8){0.5}{180}{270}
\psarc{-}(12,7){0.5}{0}{90}\psarc{-}(13,8){0.5}{180}{270}
\psarc{-}(13,7){0.5}{0}{90}\psarc{-}(14,8){0.5}{180}{270}
\psarc{-}(14,7){0.5}{0}{90}\psarc{-}(15,8){0.5}{180}{270}
\psarc{-}(15,7){0.5}{0}{90}\psarc{-}(16,8){0.5}{180}{270}
\psarc{-}(0,13.5){0.5}{270}{360}\psarc{-}(1,12.5){0.5}{90}{180}
\psarc{-}(1,13.5){0.5}{270}{360}\psarc{-}(2,12.5){0.5}{90}{180}
\psarc{-}(2,13.5){0.5}{270}{360}\psarc{-}(3,12.5){0.5}{90}{180}
\psarc{-}(3,13.5){0.5}{270}{360}\psarc{-}(4,12.5){0.5}{90}{180}
\psarc{-}(4,13.5){0.5}{270}{360}\psarc{-}(5,12.5){0.5}{90}{180}
\psarc{-}(5,13.5){0.5}{270}{360}\psarc{-}(6,12.5){0.5}{90}{180}
\psarc{-}(6,13.5){0.5}{270}{360}\psarc{-}(7,12.5){0.5}{90}{180}
\psarc{-}(7,13.5){0.5}{270}{360}\psarc{-}(8,12.5){0.5}{90}{180}
\psarc{-}(8,13.5){0.5}{270}{360}\psarc{-}(9,12.5){0.5}{90}{180}
\psarc{-}(9,13.5){0.5}{270}{360}\psarc{-}(10,12.5){0.5}{90}{180}
\psarc{-}(10,13.5){0.5}{270}{360}\psarc{-}(11,12.5){0.5}{90}{180}
\psarc{-}(11,12.5){0.5}{0}{90}\psarc{-}(12,13.5){0.5}{180}{270}
\psarc{-}(12,12.5){0.5}{0}{90}\psarc{-}(13,13.5){0.5}{180}{270}
\psarc{-}(13,12.5){0.5}{0}{90}\psarc{-}(14,13.5){0.5}{180}{270}
\psarc{-}(14,12.5){0.5}{0}{90}\psarc{-}(15,13.5){0.5}{180}{270}
\psarc{-}(15,12.5){0.5}{0}{90}\psarc{-}(16,13.5){0.5}{180}{270}
\psarc{-}(0,17){0.5}{270}{360}\psarc{-}(1,16){0.5}{90}{180}
\psarc{-}(1,17){0.5}{270}{360}\psarc{-}(2,16){0.5}{90}{180}
\psarc{-}(2,17){0.5}{270}{360}\psarc{-}(3,16){0.5}{90}{180}
\psarc{-}(3,17){0.5}{270}{360}\psarc{-}(4,16){0.5}{90}{180}
\psarc{-}(4,17){0.5}{270}{360}\psarc{-}(5,16){0.5}{90}{180}
\psarc{-}(5,17){0.5}{270}{360}\psarc{-}(6,16){0.5}{90}{180}
\psarc{-}(6,17){0.5}{270}{360}\psarc{-}(7,16){0.5}{90}{180}
\psarc{-}(7,17){0.5}{270}{360}\psarc{-}(8,16){0.5}{90}{180}
\psarc{-}(8,17){0.5}{270}{360}\psarc{-}(9,16){0.5}{90}{180}
\psarc{-}(9,17){0.5}{270}{360}\psarc{-}(10,16){0.5}{90}{180}
\psarc{-}(10,17){0.5}{270}{360}\psarc{-}(11,16){0.5}{90}{180}
\psarc{-}(11,17){0.5}{270}{360}\psarc{-}(12,16){0.5}{90}{180}
\psarc{-}(12,17){0.5}{270}{360}\psarc{-}(13,16){0.5}{90}{180}
\psarc{-}(13,17){0.5}{270}{360}\psarc{-}(14,16){0.5}{90}{180}
\psarc{-}(14,16){0.5}{0}{90}\psarc{-}(15,17){0.5}{180}{270}
\psarc{-}(15,16){0.5}{0}{90}\psarc{-}(16,17){0.5}{180}{270}
\unlw
\psset{linecolor=myc3,linestyle=dashed,dash=1pt 1pt}
\psarc{-}(2,0){0.5}{180}{360}\psarc{-}(2,3.5){0.5}{180}{360}\psarc{-}(2,7){0.5}{180}{360}\psarc{-}(2,12.5){0.5}{180}{360}\psarc{-}(2,16){0.5}{180}{360}
\psbezier{-}(0.5,0)(0.5,-1)(3.5,-1)(3.5,0)\psbezier{-}(0.5,3.5)(0.5,2.5)(3.5,2.5)(3.5,3.5)\psbezier{-}(0.5,7)(0.5,6)(3.5,6)(3.5,7)\psbezier{-}(0.5,12.5)(0.5,11.5)(3.5,11.5)(3.5,12.5)\psbezier{-}(0.5,16)(0.5,15)(3.5,15)(3.5,16)
\psarc{-}(5,0){0.5}{180}{360}\psarc{-}(5,3.5){0.5}{180}{360}\psarc{-}(5,7){0.5}{180}{360}\psarc{-}(5,12.5){0.5}{180}{360}\psarc{-}(5,16){0.5}{180}{360}
\psline{-}(6.5,0)(6.5,-0.8)\psline{-}(6.5,3.5)(6.5,2.7)\psline{-}(6.5,7)(6.5,6.2)\psline{-}(6.5,12.5)(6.5,11.3)\psline{-}(6.5,15.2)(6.5,16)
\psline[linestyle=dotted,dotsep=1pt]{-}(7.5,-0.5)(9.5,-0.5)
\psline[linestyle=dotted,dotsep=1pt]{-}(7.5,3)(9.5,3)
\psline[linestyle=dotted,dotsep=1pt]{-}(7.5,6.5)(9.5,6.5)
\psline[linestyle=dotted,dotsep=1pt]{-}(7.5,12)(9.5,12)
\psline[linestyle=dotted,dotsep=1pt]{-}(7.5,15.5)(9.5,15.5)
\psarc{-}(11,0){0.5}{180}{360}\psarc{-}(11,3.5){0.5}{180}{360}\psarc{-}(11,7){0.5}{180}{360}\psarc{-}(11,12.5){0.5}{180}{360}\psarc{-}(11,16){0.5}{180}{360}
\psline{-}(12.5,-0.0)(12.5,-0.8)\psline{-}(12.5,3.5)(12.5,2.7)\psline{-}(12.5,7)(12.5,6.2)\psline{-}(12.5,12.5)(12.5,11.3)\psline{-}(12.5,16)(12.5,15.2)
\psarc{-}(14,0){0.5}{180}{360}\psarc{-}(14,3.5){0.5}{180}{360}\psarc{-}(14,7){0.5}{180}{360}\psarc{-}(14,12.5){0.5}{180}{360}\psarc{-}(14,16){0.5}{180}{360}
\psline{-}(15.5,0.0)(15.5,-0.8)\psline{-}(15.5,3.5)(15.5,2.7)\psline{-}(15.5,7)(15.5,6.2)\psline{-}(15.5,12.5)(15.5,11.7)\psline{-}(15.5,16)(15.5,15.2)
\psarc{-}(16,0){0.5}{0}{90}\psarc{-}(16,3.5){0.5}{0}{90}\psarc{-}(16,7){0.5}{0}{90}\psarc{-}(16,12.5){0.5}{0}{90}\psarc{-}(16,16){0.5}{0}{90}
\psline{-}(16.5,-0.1)(16.5,-0.8)
\psline{-}(16.5,3.4)(16.5,2.7)
\psline{-}(16.5,6.9)(16.5,6.2)
\psline{-}(16.5,12.4)(16.5,11.7)
\psline{-}(16.5,15.9)(16.5,15.2)
\rput(4,10){$\vdots$}
\rput(8,10){$\vdots$}
\rput(12,10){$\vdots$}
\end{pspicture}
\end{equation*}
All other configurations have zero weight, either because of (\ref{eq:dotprod}) or because bubbles formed in this lower diagonal will propagate in the upper right direction to form an arc attached to a braid box labelled by a ``$1$'', leading to zero contribution by \eqref{eq:diamant}. In the configurations drawn above, the last one has weight $1$ and is simply a translation of the link state towards the left, giving rise to $P^{r-1}_{\overleftarrow{w}}$. (When the link state has a bubble in first position, it has weight zero.) The other configurations all remove a $1$-bubble from $w$. Removing the $j$-th $1$-bubble yields a weight $S_{r-n_j}/S_{r}$. Since the first box is always\begin{pspicture}(-0.25,-0.1)(0.25,0.2)
\psset{unit=0.3}
\psline[linewidth=0.3pt]{-}(-0.5,-0.5)(0.5,-0.5)(0.5,0.5)(-0.5,0.5)(-0.5,-0.5)
\psset{linecolor=myc}
\psarc(-0.5,0.5){0.5}{270}{360}
\psarc(0.5,-0.5){0.5}{90}{180}
\end{pspicture}, the first upper right row can be summed, as was done in the proof of Lemma \ref{sec:Pr1}. The result is 
\begin{equation*}
\frac{S_{(r-1)/2}S_{r/2}}{S_{r-1}S_{1/2}} \frac{S_{r-n_j}}{S_r} P^{r-2}_{w/ \{j\}} = \frac{S_{r-n_j}  P^{r-2}_{w/ \{j\}}} {4 S_{1/2} C_{r/2} C_{(r-1)/2}},
\end{equation*}
finishing the proof.
\hfill$\square$

\medskip

Let $w \in B_N^d$ be a link state starting with $a_0$ defects, followed by one cluster of arcs (uninterrupted by defects) and then followed by $a_1$ defects. The patterns of arcs in $w$ is described by $\zeta(w)=[a_0](k_1k_2...k_m)[a_1]$ where $k_i$ is half the number of points under the arc $i$ (including the starting and ending points). The arcs are added at the first free point available, starting from the left. The number $m=(N-a_0-a_1)/2$ is the number of arcs in $w$. Again, an example is useful:
\begin{equation*} \zeta \left(
\begin{pspicture}(0.35,0.1)(9.25,0.55)
\psset{unit=0.45}
\psdots(1,0)(2,0)(3,0)(4,0)(5,0)(6,0)(7,0)(8,0)(9,0)(10,0)(11,0)(12,0)(13,0)(14,0)(15,0)(16,0)(17,0)(18,0)(19,0)(20,0)
\psset{linecolor=myc2}
\psline{-}(1,0)(1,1)
\psline{-}(2,0)(2,1)
\psline{-}(3,0)(3,1)
\psline{-}(4,0)(4,1)
\psline{-}(19,0)(19,1)
\psline{-}(20,0)(20,1)
\psarc(6.5,0){0.5}{0}{180}
\psbezier(5,0)(5,1)(8,1)(8,0)
\psarc(10.5,0){0.5}{0}{180}
\psarc(13.5,0){0.5}{0}{180}
\psarc(15.5,0){0.5}{0}{180}
\psbezier{-}(12,0)(12,1.4)(17,1.4)(17,0)
\psbezier{-}(9,0)(9,1.8)(18,1.8)(18,0)
\rput(4.8,0.3){$_2$}
\rput(5.8,0.3){$_1$}
\rput(8.8,0.3){$_5$}
\rput(9.8,0.3){$_1$}
\rput(11.8,0.3){$_3$}
\rput(12.8,0.3){$_1$}
\rput(14.8,0.3){$_1$}
\end{pspicture} \right) = [4](2151311)[2]
\end{equation*}

\begin{Lemme}\label{sec:multiB} Replacing a cluster of arcs by concentric ones.
Let $w \in B^r$ a link state with pattern $\zeta(w) = [a_0](k_1k_2...k_m)[a_1]$.  Then,
\begin{equation}
P^r_w = \left(\prod_{i=1}^m \frac{S_i}{S_{k_i}}\right)P^r_u
\label{eq: multbubble}
\end{equation} 
where $u$, given by $\zeta(u)=[a_0](m \ m-1...21)[a_1]$, has a single cluster of $m$ concentric arcs.
\end{Lemme}
\noindent{\scshape Proof\ \ } The proposition is trivially true if $\zeta(w) =[a_0](m \ m-1...21)[a_1]$. The first step is to show that it is true for
$\zeta(w) = [a_0](\underbrace{11...1}_{m})[a_1]$. From Lemma \ref{sec: recurgenerale}, $u$ and $w$ satisfy the following relations:
\begin{equation*}
P^r_u = P^{r-1}_{\overleftarrow{u}} +  \frac{S_{r-m-a_0} P^{r-2}_{u\setminus\{1\}}}{4S_{1/2}C_{r/2}C_{(r-1)/2}} \end{equation*}
and
\begin{alignat*}{3}
P^r_w &= P^{r-1}_{\overleftarrow{w}} + \frac{\sum_{j=1}^m S_{r-n_j} P^{r-2}_{w\setminus\{j\}}}{4S_{1/2}C_{r/2}C_{(r-1)/2}}  \\
            &= P^{r-1}_{\overleftarrow{w}} + \frac{ \left(\sum_{j=1}^m S_{r-(a_0+2j-1)}\right) P^{r-2}_{w\setminus\{1\}}}{4S_{1/2}C_{r/2}C_{(r-1)/2}}\\
            &= P^{r-1}_{\overleftarrow{w}} + \frac{ S_m S_{r-m-a_0} P^{r-2}_{w\setminus\{1\}}}{4S_{1/2}C_{r/2}C_{(r-1)/2}S_1}.
\end{alignat*}
Equation~(\ref{eq: multbubble}) is valid for $\zeta(w)=[a_0](11\dots1)[a_1]$ with $r=2$ and $r=3$ for all $m\le r/2$, that is, for $m=1$. Suppose it is valid for $r-1$ and $r-2$, then
\begin{alignat*}{3}
P^r_w &= \left(\prod_{i=1}^m  \frac{S_i}{S_1} \right) P^{r-1}_{\overleftarrow{u}} + \frac{S_{r-m-a_0}}{4S_{1/2}C_{r/2}C_{(r-1)/2}} \frac{S_m }{S_1} \left(\prod_{i=1}^{m-1}  \frac{S_i}{S_1} \right) P^{r-2}_{u\setminus\{1\}} \\
            &=  \left(\prod_{i=1}^m  \frac{S_i}{S_1} \right) \left( P^{r-1}_{\overleftarrow{u}} + \frac{S_{r-m-a_0}  P^{r-2}_{u\setminus\{1\}}}{4S_{1/2}C_{r/2}C_{(r-1)/2}} \right)\\
           &=  \left(\prod_{i=1}^m  \frac{S_i}{S_1} \right) P^r_u
\end{alignat*} 
as required. Since (\ref{eq: multbubble}) is true for the fictitious initial conditions introduced earlier, after Lemma \ref{sec: singlebubble}, the proof is complete.

We now define the operation $\mathcal{O}_{b,x}$ that acts on link state $w$ by removing $b$ $1$-bubbles centered on $x$ and replacing them with $(b-1)$ $1$-bubbles, circumscribed by a large
 $2$-bubble:
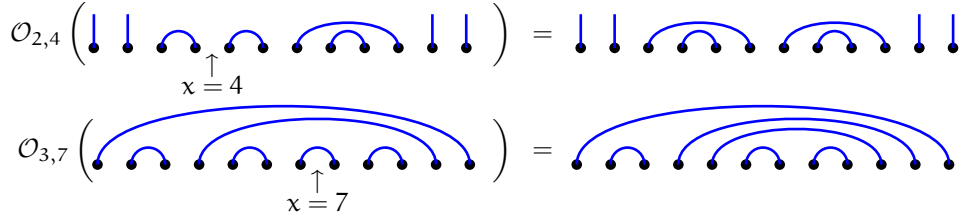
\begin{alignat*}{3}
\mathcal{O}_{2,4}\left(
\begin{pspicture}(0.35,0.1)(5.75,0.55)
\psset{unit=0.45}
\psdots(1,0)(2,0)(3,0)(4,0)(5,0)(6,0)(7,0)(8,0)(9,0)(10,0)(11,0)(12,0)
\rput(4.5,-0.5){$\uparrow$}
\rput(4.5,-1.1){$x=4$}
\psset{linecolor=myc2}
\psline{-}(1,0)(1,1)
\psline{-}(2,0)(2,1)
\psarc(3.5,0){0.5}{0}{180}
\psarc(5.5,0){0.5}{0}{180}
\psbezier(7,0)(7,1)(10,1)(10,0)
\psarc(8.5,0){0.5}{0}{180}
\psline{-}(11,0)(11,1)
\psline{-}(12,0)(12,1)
\end{pspicture} \right)\ \ &= \ \
\begin{pspicture}(0.35,0.1)(6.15,0.55)
\psset{unit=0.45}
\psdots(1,0)(2,0)(3,0)(4,0)(5,0)(6,0)(7,0)(8,0)(9,0)(10,0)(11,0)(12,0)
\psset{linecolor=myc2}
\psline{-}(1,0)(1,1)
\psline{-}(2,0)(2,1)
\psbezier(3,0)(3,1)(6,1)(6,0)
\psarc(4.5,0){0.5}{0}{180}
\psbezier(7,0)(7,1)(10,1)(10,0)
\psarc(8.5,0){0.5}{0}{180}
\psline{-}(11,0)(11,1)
\psline{-}(12,0)(12,1)
\end{pspicture}
\\
& \\
\mathcal{O}_{3,7}\left(
\begin{pspicture}(0.85,0.1)(6.15,0.55)
\psset{unit=0.45}
\psdots(2,0)(3,0)(4,0)(5,0)(6,0)(7,0)(8,0)(9,0)(10,0)(11,0)(12,0)(13,0)
\rput(8.5,-0.5){$\uparrow$}
\rput(8.5,-1.1){$x=7$}
\psset{linecolor=myc2}
\psarc(3.5,0){0.5}{0}{180}
\psbezier(2.0,0)(2,2.3)(13,2.3)(13,0)
\psbezier(5,0)(5,1.8)(12,1.8)(12,0)
\psarc(6.5,0){0.5}{0}{180}
\psarc(8.5,0){0.5}{0}{180}
\psarc(10.5,0){0.5}{0}{180}
\end{pspicture} \right)\ \ &= \ \
\begin{pspicture}(0.85,0.1)(6.65,0.55)
\psset{unit=0.45}
\psdots(2,0)(3,0)(4,0)(5,0)(6,0)(7,0)(8,0)(9,0)(10,0)(11,0)(12,0)(13,0)
\psset{linecolor=myc2}
\psarc(3.5,0){0.5}{0}{180}
\psbezier(2.0,0)(2,2.3)(13,2.3)(13,0)
\psbezier(5,0)(5,1.8)(12,1.8)(12,0)
\psbezier(6,0)(6,1.4)(11,1.4)(11,0)
\psarc(7.5,0){0.5}{0}{180}
\psarc(9.5,0){0.5}{0}{180}
\end{pspicture}
\end{alignat*}
\\
The action of $\mathcal{O}_{b,x}$ is not defined for all $w$, but any $w \in B_N $ with $\zeta(w)=[a_0](k_1k_2...k_m)[a_1]$ can be written as 
\begin{equation*}
[a_0](k_1k_2...k_m)[a_1] = \left( \prod_{i} \mathcal{O}_{b_i,x_i}  \right) [a_0](\underbrace{11...1}_{m})[a_1]
\end{equation*}
for some $b_i$ and $x_i$. For instance, the right-hand
sides of the above examples are 
$$[2](2,1,2,1)[2]=  \mathcal{O}_{2,8}\mathcal{O}_{2,4} \ [2](1111)[2]\qquad\text{\rm and}\qquad[0](6,1,4,3,1,1)[0]=  \mathcal{O}_{3,7} \mathcal{O}_{4,7}  \mathcal{O}_{6,6}  \ [0](111111)[0].$$

Let $w_1$ and $w_2$ be two link states such that $w_2 = \mathcal{O}_{b,x}w_1$. If we can show that 
\begin{equation}
\label{eq:multeq}
P^r_{w_1} = \frac{S_b}{S_1} P^r_{w_2},
\end{equation}
then the proof of the proposition will be complete. Note first that this relation holds (somewhat trivially) for $r=3$. For $r=4$ the relation is non-trivial only for $w_1=v^4_{1,3}$, $w_2=v^4_{2^2}$ and $\mathcal{O}_{2,2}$. The element $P^4_{w_2}$ has been calculated in Lemma \ref{lem:concentric}; $P^4_{w_1}$ can be obtained from \eqref{eq: generalrec} and it agrees with \eqref{eq:multeq}.
We now use (\ref{eq: generalrec}) for such a pair $w_1$ and $w_2=\mathcal{O}_{b,x}w_1$. For $w_1$ we partition the sum over $j$ into the set $G_1$ of $b$ $1$-bubbles modified by the action $\mathcal{O}_{b,x}$ and its complement:
\begin{align*}
P^r_{w_1} &= P^{r-1}_{\overleftarrow{w_1}} + \frac{1}{4S_{1/2}C_{r/2}C_{(r-1)/2}} \left( \sum_{j \in G_1^c} S_{r-n_j} P^{r-2}_{w_1\setminus\{j\}} +  \sum_{j \in G_1} S_{r-n_j} P^{r-2}_{w_1\setminus\{j\}} \right) \\
&= P^{r-1}_{\overleftarrow{w_1}} + \frac{1}{4S_{1/2}C_{r/2}C_{(r-1)/2}} \left( \sum_{j \in G_1^c} S_{r-n_j} P^{r-2}_{w_1\setminus\{j\}} + \Big(\sum_{j = 1}^{b} S_{r-(x-b+2j-1)} \Big) P^{r-2}_{w_1\setminus\{g_1\}} \right) \\
&= P^{r-1}_{\overleftarrow{w_1}} + \frac{1}{4S_{1/2}C_{r/2}C_{(r-1)/2}} \left( \sum_{j \in G_1^c} S_{r-n_j} P^{r-2}_{w_1\setminus\{j\}} + \Big( \frac{ S_b S_{r-x}}{S_1} \Big) P^{r-2}_{w_1\setminus\{g_1\}} \right)
\end{align*}
where $g_1$ is the first $1$-bubble in $G_1$. The same can be carried out for $w_2$:
\begin{align*}
P^r_{w_2} &= P^{r-1}_{\overleftarrow{w_2}} + \frac{1}{4S_{1/2}C_{r/2}C_{(r-1)/2}} \left( \sum_{j \in G_1^c} S_{r-n_j} P^{r-2}_{w_2\setminus\{j\}} +  \Big(\sum_{j = 1}^{b-1} S_{r-(x-b+2j)} \Big) P^{r-2}_{w_2\setminus\{g_2\}} \right) \\
&= P^{r-1}_{\overleftarrow{w_2}} + \frac{1}{4S_{1/2}C_{r/2}C_{(r-1)/2}} \left( \sum_{j \in G_1^c} S_{r-n_j} P^{r-2}_{w_2\setminus\{j\}} + \Big(  \frac{S_{b-1} S_{r-x}}{S_1} \Big) P^{r-2}_{w_2\setminus\{g_2\}} \right)
\end{align*}
where again $g_2$ is the position of the first $1$-bubble in the complement in $w_2$ of $G_1^c$. If (\ref{eq:multeq}) is true for $r-1$ and $r-2$,
\begin{equation*}
P^{r-1}_{\overleftarrow{w_1}} = \frac{S_b}{S_1} P^{r-1}_{\overleftarrow{w_2}}, \qquad P^{r-2}_{w_1\setminus\{j\}} = \frac{S_b}{S_1}  P^{r-2}_{w_2\setminus\{j\}} \ \forall \ j \in G^c_1, \qquad P^{r-2}_{w_1\setminus\{g_1\}} = \frac{S_{b-1}}{S_1}  P^{r-2}_{w_2\setminus\{g_2\}}, 
\end{equation*}
and then for $r$:
\begin{align*}
P^r_{w_1} &= \frac{S_b}{S_1} P^{r-1}_{\overleftarrow{w_2}} + \frac{1}{4S_{1/2}C_{r/2}C_{(r-1)/2}} \left( \sum_{j \in G_1^c} \frac{S_bS_{r-n_j}}{S_1} P^{r-2}_{w_2\setminus\{j\}} + \Big( \frac{ S_b S_{r-x}}{S_1} \Big)  \frac{S_{b-1}}{S_1} P^{r-2}_{w_2\setminus\{g_2\}} \right) \\
&= \frac{S_b}{S_1} \left( P^{r-1}_{\overleftarrow{w_2}} + \frac{1}{4S_{1/2}C_{r/2}C_{(r-1)/2}} \left( \sum_{j \in G_1^c} S_{r-n_j} P^{r-2}_{w_2\setminus\{j\}} + \Big(  \frac{S_{b-1} S_{r-x}}{S_1} \Big) P^{r-2}_{w_2\setminus\{g_2\}} \right)  \right)\\
&=  \frac{S_b}{S_1}P^r_{w_2}.
\end{align*}\hfill$\square$

\medskip

The definition of the pattern $\zeta(w)$ can be extended to link states $w$ with more than one cluster of bubbles. Clusters of defects and clusters of bubbles are noted, respectively, in $[ \ ]$ and $( \ )$:
\begin{equation*}
\zeta(w) = [a_0] \Big( \prod_{j}(k_{1,j}k_{2,j}...k_{m_j,j})[a_j]\Big)
\end{equation*}
where the product over $j$ is over the clusters of bubbles. 
\begin{Lemme}For a general link state $w$, the result of the previous lemma extends to 
\begin{equation*}
P^r_w = \Big(\prod_j \prod_{i=1}^{m_j} \frac{S_i}{S_{k_{i,j}}}\Big)P^r_u
\end{equation*}
where
\begin{equation*}
\zeta(u) = [a_0] \Big( \prod_{j}(m_j \ m_j-1... 21)[a_j]\Big)
\end{equation*}
is a state that has only clusters of concentrical bubbles.
\label{sec: tresgeneral}
\end{Lemme}
\noindent{\scshape Proof\ \ } The definition of  $\mathcal{O}_{b,x}$ was not restricted to link states with only one cluster of defects. Multiple-cluster link states can then be created by successive applications of $\mathcal{O}_{b_i,x_i}$ on link states with only $1$-bubbles, changing one cluster at a time. The result will follow if (\ref{eq:multeq}) still holds for states with multiple clusters. This is indeed the case because, in the previous proof, the fact that the $1$-bubbles of $G^c_1$ were in a single cluster was not used. The rest of the argument goes through. \hfill$\square$ \\

\subsection{Singular points of $P^r_{\{n_1,n_2,...,n_k\}}$}
\label{sec:generalP}

The goal of this section is to show that
\begin{equation*}
\{\Lambda \in \mathbb{R} \, |\, P^r_{\{n_1,n_2,...,n_k \}} \ \textrm{diverges at }\Lambda\}  \subset  \cup_{i=1}^k \{\Lambda \in \mathbb{R} \, |\, P^r_{\{i^i\}} \ \textrm{diverges at }\Lambda\}  \subset \cup_{i=0}^{k-1} \{ \Lambda \in \mathbb{R} \, |\, C_{(r-i)/2} =0   \}.
\end{equation*}
We start by investigating the singularities of $P^r_{\{n_1,n_2,...,n_k\}}$, when $w = v^r_{n_1,n_2,...,n_k}$ has only $1$-bubbles, and will show that 
\begin{equation}
\label{eq:lincomb}
P^r_{\{n_1,n_2,...,n_k\}} = \sum_{i=0}^k a_iP^r_{\{i^i\}}
\end{equation}
where $a_i$'s are analytic functions of $\Lambda \in \mathbb{R}$ and the term $i=0$ is for $P^r_{\{\}}$.

\begin{Lemme} Moving a single bubble to position ``$1$''.
\begin{equation}\label{eq:Prnbis}
P^r_{\{n\}} = \frac{S_n}{S_1}P^r_{\{1\}} - \frac{S_{n/2}S_{(n-1)/2}}{S_{1/2}S_{2/2}} P^r_{\{\}}.
\end{equation}
\end{Lemme}
\noindent{\scshape Proof\ \ } The trick is to start by calculating $\langle v^r_{1} | e_1 P^r v^r\rangle$. The only link states $w \in B^r$ satisfying $\langle v^r_{1} | e_1 w\rangle \neq 0$ are $v^r$, $v^r_{1}$ and $v^r_{2}$ and 
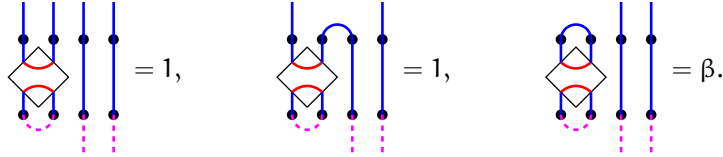
\begin{equation*}
\begin{pspicture}(0,-0.5)(1.2,0.5)
\psdots(0.0,0)(0.4,0)(0.8,0)(1.2,0)
\psdots(0.0,-1)(0.4,-1)(0.8,-1)(1.2,-1)\lw
\psline{-}(0.2,-0.9)(0.6,-0.5)(0.2,-0.1)(-0.2,-0.5)(0.2,-0.9)\unlw
\psset{linecolor=myc}
\psarc{-}(0.2,-0.1){0.282843}{225}{315}
\psarc{-}(0.2,-0.9){0.282843}{45}{135}
\psset{linecolor=myc2}
\psline{-}(0,0)(0,0.5)
\psline{-}(0.4,0)(0.4,0.5)
\psline{-}(0.8,0)(0.8,0.5)
\psline{-}(1.2,0)(1.2,0.5)
\psline{-}(0,0)(0,-0.3)
\psline{-}(0.4,0)(0.4,-0.3)
\psline{-}(0,-0.7)(0,-1)
\psline{-}(0.4,-0.7)(0.4,-1)
\psline{-}(0.8,0)(0.8,-1)
\psline{-}(1.2,0)(1.2,-1)
\psset{linecolor=myc3,linestyle=dashed,dash=2pt 2pt}
\psarc{-}(0.2,-1){0.2}{180}{360}
\psline{-}(0.8,-1)(0.8,-1.5)
\psline{-}(1.2,-1)(1.2,-1.5)
\end{pspicture} \ \ = 1, \qquad \qquad
\begin{pspicture}(0,-0.5)(1.2,0.5)
\psdots(0.0,0)(0.4,0)(0.8,0)(1.2,0)
\psdots(0.0,-1)(0.4,-1)(0.8,-1)(1.2,-1)\lw
\psline{-}(0.2,-0.9)(0.6,-0.5)(0.2,-0.1)(-0.2,-0.5)(0.2,-0.9)\unlw
\psset{linecolor=myc}
\psarc{-}(0.2,-0.1){0.282843}{225}{315}
\psarc{-}(0.2,-0.9){0.282843}{45}{135}
\psset{linecolor=myc2}
\psline{-}(0,0)(0,0.5)
\psarc{-}(0.6,0){0.2}{0}{180}
\psline{-}(1.2,0)(1.2,0.5)
\psline{-}(0,0)(0,-0.3)
\psline{-}(0.4,0)(0.4,-0.3)
\psline{-}(0,-0.7)(0,-1)
\psline{-}(0.4,-0.7)(0.4,-1)
\psline{-}(0.8,0)(0.8,-1)
\psline{-}(1.2,0)(1.2,-1)
\psset{linecolor=myc3,linestyle=dashed,dash=2pt 2pt}
\psarc{-}(0.2,-1){0.2}{180}{360}
\psline{-}(0.8,-1)(0.8,-1.5)
\psline{-}(1.2,-1)(1.2,-1.5)
\end{pspicture} \ \ = 1, \qquad \qquad
\begin{pspicture}(0,-0.5)(1.2,0.5)
\psdots(0.0,0)(0.4,0)(0.8,0)(1.2,0)
\psdots(0.0,-1)(0.4,-1)(0.8,-1)(1.2,-1)\lw
\psline{-}(0.2,-0.9)(0.6,-0.5)(0.2,-0.1)(-0.2,-0.5)(0.2,-0.9)\unlw
\psset{linecolor=myc}
\psarc{-}(0.2,-0.1){0.282843}{225}{315}
\psarc{-}(0.2,-0.9){0.282843}{45}{135}
\psset{linecolor=myc2}
\psarc{-}(0.2,0){0.2}{0}{180}
\psline{-}(0.8,0)(0.8,0.5)
\psline{-}(1.2,0)(1.2,0.5)
\psline{-}(0,0)(0,-0.3)
\psline{-}(0.4,0)(0.4,-0.3)
\psline{-}(0,-0.7)(0,-1)
\psline{-}(0.4,-0.7)(0.4,-1)
\psline{-}(0.8,0)(0.8,-1)
\psline{-}(1.2,0)(1.2,-1)
\psset{linecolor=myc3,linestyle=dashed,dash=2pt 2pt}
\psarc{-}(0.2,-1){0.2}{180}{360}
\psline{-}(0.8,-1)(0.8,-1.5)
\psline{-}(1.2,-1)(1.2,-1.5)
\end{pspicture} \ \ = \beta.
\end{equation*} 
\\ \\
Since $e_1 P^r$ is zero whenever $r\ge 2$,
\begin{equation*}
0 = \langle v^r_{1} | e_1 P^r v^r\rangle = P^r_{\{\}} \langle v^r_{1} | e_1 v^r\rangle + P^r_{\{1\}}  \langle v^r_{1} | e_1 v^r_1\rangle + P^r_{\{2\}}  \langle v^r_{1} | e_1 v^r_2\rangle = P^r_{\{\}} - \frac{S_2}{S_1} P^r_{\{1\}} + P^r_{\{2\}},
\end{equation*}
that is
\begin{equation*}
P^r_{\{2\}} = \frac{S_2}{S_1} P^r_{\{1\}} -  P^r_{\{\}}
\end{equation*}
and the proposed $P^r_{\{n\}}$ is correct for $n=1$ and $2$. For $n>2$, we calculate $\langle v^r_{n-1} | e_{n-1} P^r v^r\rangle$ and use induction. Four link states carry non-zero contributions:
\begin{alignat*}{3}
0 &= \langle v^r_{n-1} | e_{n-1} P^r v^r\rangle \\& = P^r_{\{\}} \langle v^r_{n-1} | e_{n-1} v^r\rangle + P^r_{\{n-2\}}  \langle v^r_{n-1} | e_{n-1} v^r_{n-2}\rangle + P^r_{\{n-1\}}  \langle v^r_{n-1} | e_{n-1} v^r_{n-1}\rangle+ P^r_{\{n\}}  \langle v^r_{n-1} | e_{n-1} v^r_{n}\rangle \\ &= P^r_{\{\}} +  P^r_{\{n-2\}}  -\frac{S_2}{S_1} P^r_{\{n-1\}} + P^r_{\{n\}}.
\end{alignat*}
If eq.~(\ref{eq:Prnbis}) is valid for $n-1$ and $n-2$, then
\begin{alignat*}{3}
P^r_{\{n\}} &= \frac{S_2}{S_1} P^r_{\{n-1\}} - P^r_{\{n-2\}} - P^r_{\{\}} \\
&=  \frac{S_2}{S_1} \left( \frac{S_{n-1}}{S_1} P^r_1 -  \frac{S_{(n-1)/2} S_{(n-2)/2}}{S_{1/2}S_{2/2}}P^r_{\{\}} \right) - \left( \frac{S_{n-2}}{S_1}P^r_1 - \frac{S_{(n-2)/2} S_{(n-3)/2}}{S_{1/2}S_{2/2}}P^r_{\{\}} \right) - P^r_{\{\}} \\ 
&= Y_n P^r_{\{1\}} -  Z_n P^r_{\{\}}
\end{alignat*}
where
\begin{equation*}
Y_n = \frac{S_2S_{n-1} - S_{n-2}S_1}{S_1^2}, \qquad Z_n =  \frac{S_2 S_{(n-1)/2} S_{(n-2)/2}}{S_{1/2}S_{2/2}^2} - \frac{S_{(n-2)/2}S_{(n-3)/2}}{S_{1/2}S_{2/2}} + 1.
\end{equation*}
Trigonometric identities can be used to bring $Y_n$ and $Z_n$ onto the forms proposed for the coefficients of $P_{\{1\}}$ and $P_{\{\}}$ respectively.\hfill$\square$

\begin{Lemme} Let $\{n,X\}$ be a set of $1$-bubbles such that the positions $m_1,m_2,\dots,m_k$ represented by $X$ are all to the right of the first bubble at $n$. Then,
\begin{equation*}
P^r_{\{n,X\}} = \frac{S_n}{S_1}P^r_{\{1,X\}} - \frac{S_{n/2}S_{(n-1)/2}}{S_{1/2}S_{2/2}} P^r_{\{X\}}.
\end{equation*}
\end{Lemme}
\noindent{\scshape Proof\ \ } Reading the previous proof, one sees that $X$ acts as a spectator. For instance, the same four terms will appear in $\langle v^r_{n-1} | e_{n-1} P^r v^r\rangle$ and none others, even when the bubbles at $n$ and $m_1$ are immediate neighbors. In other words, the proof is identical. \hfill$\square$

\medskip

The ultimate objective is to write $P^r_{\{n_1,...,n_k\}}$ as a linear combination of the form (\ref{eq:lincomb}). The previous lemma allows to move the first bubble to the extreme left, or make it disappear. We now show how this can be done with the remaining bubbles.
\begin{Lemme}\label{lem:A9} Moving bubbles to the leftmost available position.
\begin{equation}
P^r_{\{1,3,...,2b-3,n,X\}} = \frac{S_{n-b+1}}{S_b}P^r_{\{1,3,...,2b-3,2b-1,X\}} -  \frac{S_{(n-2b+2)/2}S_{(n-2b+1)/2}}{S_{1/2}S_{2/2}}P^r_{\{1,3,...,2b-3,X\}}.
\label{eq: generalmove}
\end{equation}
\end{Lemme}
The previous lemma is just the case $b=1$.

\noindent{\scshape Proof\ \ } In this proof, $X$ plays no role and will be omitted. We start with the case of two $1$-bubbles. The element $\langle v^r_{1,3} | e_3 P^r v^r\rangle$ has four contributions:
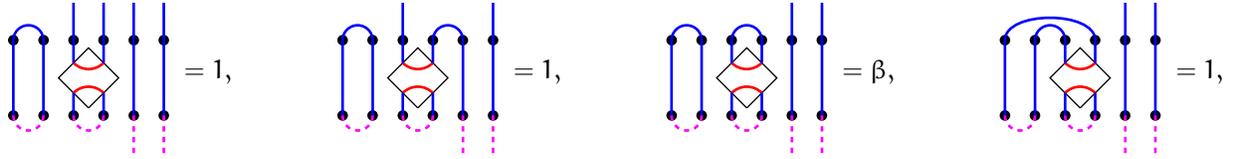
\begin{equation*}
\begin{pspicture}(-0.8,-0.5)(1.2,0.5)
\psdots(-0.8,0)(-0.4,0)(0.0,0)(0.4,0)(0.8,0)(1.2,0)
\psdots(-0.8,-1)(-0.4,-1)(0.0,-1)(0.4,-1)(0.8,-1)(1.2,-1)\lw
\psline{-}(0.2,-0.9)(0.6,-0.5)(0.2,-0.1)(-0.2,-0.5)(0.2,-0.9)\unlw
\psset{linecolor=myc}
\psarc{-}(0.2,-0.1){0.282843}{225}{315}
\psarc{-}(0.2,-0.9){0.282843}{45}{135}
\psset{linecolor=myc2}
\psline{-}(-0.4,0)(-0.4,-1)
\psline{-}(-0.8,0)(-0.8,-1)
\psarc{-}(-0.6,0){0.2}{0}{180}
\psline{-}(0,0)(0,0.5)
\psline{-}(0.4,0)(0.4,0.5)
\psline{-}(0.8,0)(0.8,0.5)
\psline{-}(1.2,0)(1.2,0.5)
\psline{-}(0,0)(0,-0.3)
\psline{-}(0.4,0)(0.4,-0.3)
\psline{-}(0,-0.7)(0,-1)
\psline{-}(0.4,-0.7)(0.4,-1)
\psline{-}(0.8,0)(0.8,-1)
\psline{-}(1.2,0)(1.2,-1)
\psset{linecolor=myc3,linestyle=dashed,dash=2pt 2pt}
\psarc{-}(-0.6,-1){0.2}{180}{360}
\psarc{-}(0.2,-1){0.2}{180}{360}
\psline{-}(0.8,-1)(0.8,-1.5)
\psline{-}(1.2,-1)(1.2,-1.5)
\end{pspicture} \ \ = 1, \qquad \qquad
\begin{pspicture}(-0.8,-0.5)(1.2,0.5)
\psdots(-0.8,0)(-0.4,0)(0.0,0)(0.4,0)(0.8,0)(1.2,0)
\psdots(-0.8,-1)(-0.4,-1)(0.0,-1)(0.4,-1)(0.8,-1)(1.2,-1)\lw
\psline{-}(0.2,-0.9)(0.6,-0.5)(0.2,-0.1)(-0.2,-0.5)(0.2,-0.9)\unlw
\psset{linecolor=myc}
\psarc{-}(0.2,-0.1){0.282843}{225}{315}
\psarc{-}(0.2,-0.9){0.282843}{45}{135}
\psset{linecolor=myc2}
\psline{-}(-0.4,0)(-0.4,-1)
\psline{-}(-0.8,0)(-0.8,-1)
\psarc{-}(-0.6,0){0.2}{0}{180}
\psline{-}(0,0)(0,0.5)
\psarc{-}(0.6,0){0.2}{0}{180}
\psline{-}(1.2,0)(1.2,0.5)
\psline{-}(0,0)(0,-0.3)
\psline{-}(0.4,0)(0.4,-0.3)
\psline{-}(0,-0.7)(0,-1)
\psline{-}(0.4,-0.7)(0.4,-1)
\psline{-}(0.8,0)(0.8,-1)
\psline{-}(1.2,0)(1.2,-1)
\psset{linecolor=myc3,linestyle=dashed,dash=2pt 2pt}
\psarc{-}(-0.6,-1){0.2}{180}{360}
\psarc{-}(0.2,-1){0.2}{180}{360}
\psline{-}(0.8,-1)(0.8,-1.5)
\psline{-}(1.2,-1)(1.2,-1.5)
\end{pspicture} \ \ = 1, \qquad \qquad
\begin{pspicture}(-0.8,-0.5)(1.2,0.5)
\psdots(-0.8,0)(-0.4,0)(0.0,0)(0.4,0)(0.8,0)(1.2,0)
\psdots(-0.8,-1)(-0.4,-1)(0.0,-1)(0.4,-1)(0.8,-1)(1.2,-1)\lw
\psline{-}(0.2,-0.9)(0.6,-0.5)(0.2,-0.1)(-0.2,-0.5)(0.2,-0.9)\unlw
\psset{linecolor=myc}
\psarc{-}(0.2,-0.1){0.282843}{225}{315}
\psarc{-}(0.2,-0.9){0.282843}{45}{135}
\psset{linecolor=myc2}
\psline{-}(-0.4,0)(-0.4,-1)
\psline{-}(-0.8,0)(-0.8,-1)
\psarc{-}(-0.6,0){0.2}{0}{180}
\psarc{-}(0.2,0){0.2}{0}{180}
\psline{-}(0.8,0)(0.8,0.5)
\psline{-}(1.2,0)(1.2,0.5)
\psline{-}(0,0)(0,-0.3)
\psline{-}(0.4,0)(0.4,-0.3)
\psline{-}(0,-0.7)(0,-1)
\psline{-}(0.4,-0.7)(0.4,-1)
\psline{-}(0.8,0)(0.8,-1)
\psline{-}(1.2,0)(1.2,-1)
\psset{linecolor=myc3,linestyle=dashed,dash=2pt 2pt}
\psarc{-}(-0.6,-1){0.2}{180}{360}
\psarc{-}(0.2,-1){0.2}{180}{360}
\psline{-}(0.8,-1)(0.8,-1.5)
\psline{-}(1.2,-1)(1.2,-1.5)
\end{pspicture} \ \ = \beta, \qquad \qquad
\begin{pspicture}(-0.8,-0.5)(1.2,0.5)
\psdots(-0.8,0)(-0.4,0)(0.0,0)(0.4,0)(0.8,0)(1.2,0)
\psdots(-0.8,-1)(-0.4,-1)(0.0,-1)(0.4,-1)(0.8,-1)(1.2,-1)\lw
\psline{-}(0.2,-0.9)(0.6,-0.5)(0.2,-0.1)(-0.2,-0.5)(0.2,-0.9)\unlw
\psset{linecolor=myc}
\psarc{-}(0.2,-0.1){0.282843}{225}{315}
\psarc{-}(0.2,-0.9){0.282843}{45}{135}
\psset{linecolor=myc2}
\psline{-}(-0.4,0)(-0.4,-1)
\psline{-}(-0.8,0)(-0.8,-1)
\psarc{-}(-0.2,0){0.2}{0}{180}
\psbezier{-}(-0.8,0)(-0.8,0.4)(0.4,0.4)(0.4,0)
\psline{-}(0.8,0)(0.8,0.5)
\psline{-}(1.2,0)(1.2,0.5)
\psline{-}(0,0)(0,-0.3)
\psline{-}(0.4,0)(0.4,-0.3)
\psline{-}(0,-0.7)(0,-1)
\psline{-}(0.4,-0.7)(0.4,-1)
\psline{-}(0.8,0)(0.8,-1)
\psline{-}(1.2,0)(1.2,-1)
\psset{linecolor=myc3,linestyle=dashed,dash=2pt 2pt}
\psarc{-}(-0.6,-1){0.2}{180}{360}
\psarc{-}(0.2,-1){0.2}{180}{360}
\psline{-}(0.8,-1)(0.8,-1.5)
\psline{-}(1.2,-1)(1.2,-1.5)
\end{pspicture} \ \ = 1,
\end{equation*} \\
\\
so that
\begin{equation*}
 P^r_{\{1,4\}} = \frac{S_2}{S_1} P^r_{\{1,3\}} - P^r_{\{2,2\}} - P^r_{\{1\}}  
 = \left( \frac{S_2}{S_1} - \frac{S_1}{S_2}  \right)  P^r_{\{1,3\}} - P^r_{\{1\}}
 = \frac{S_3}{S_2} P^r_{\{1,3\}} - P^r_{\{1\}}, 
\end{equation*}
where Lemma \ref{sec:multiB} was used for $P^r_{\{2,2\}}$. When the second bubble is at $n>4$, the equation $0=\langle v^r_{1,n-1} | e_{n-1} P^r v^r\rangle$ and induction yield 
\begin{equation*}
P^r_{\{1,n\}} = \frac{S_{2}}{S_1} P^r_{\{1,n-1\}} - P^r_{\{1,n-2\}} - P^r_{\{1\}} = \frac{S_{n-1}}{S_2} P^r_{\{1,3\}}  - \frac{S_{(n-2)/2}S_{(n-3)/2}}{S_{1/2}S_{2/2}} P^r_{\{1\}}.
\end{equation*}

More generally, when the $b$-th bubble is at position $2b$, we compute $\langle v^r_{1, 3, ..., 2b-1} | e_{2b-1} P^r v^r\rangle$. The following contributions
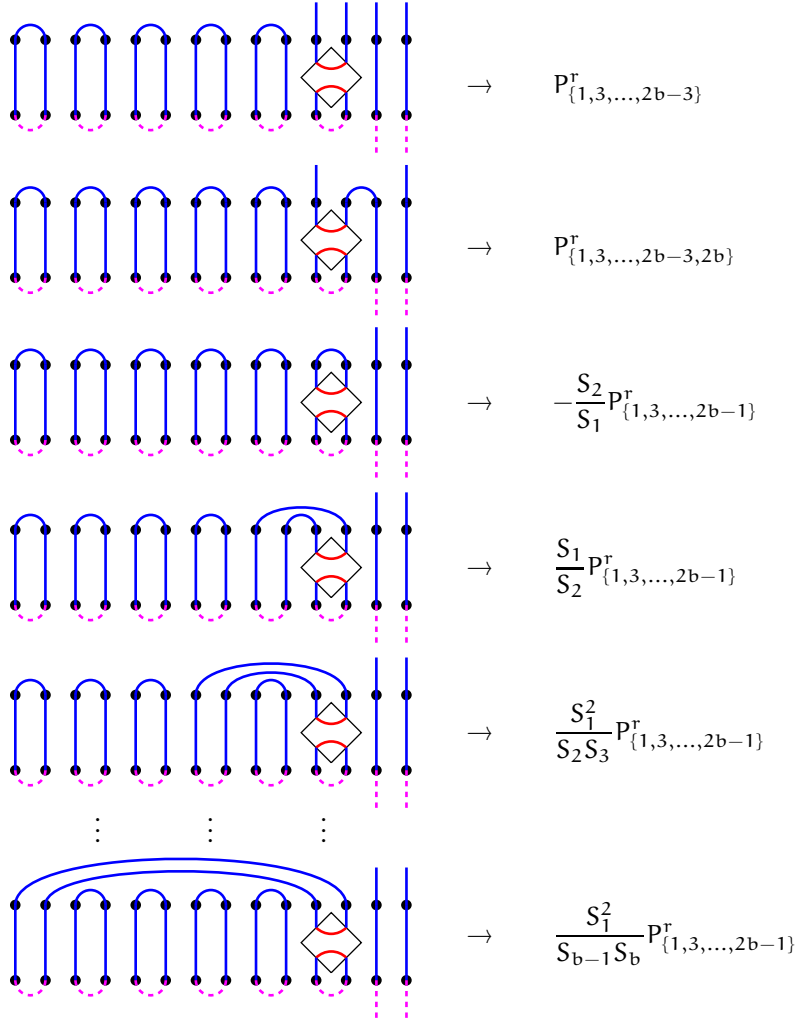
\begin{alignat*}{3}
\begin{pspicture}(-4.0,-0.7)(1.2,0.5)
\psdots(-4.0,0)(-3.6,0)(-3.2,0)(-2.8,0)(-2.4,0)(-2.0,0)(-1.6,0)(-1.2,0)(-0.8,0)(-0.4,0)(0.0,0)(0.4,0)(0.8,0)(1.2,0)
\psdots(-4.0,-1)(-3.6,-1)(-3.2,-1)(-2.8,-1)(-2.4,-1)(-2.0,-1)(-1.6,-1)(-1.2,-1)(-0.8,-1)(-0.4,-1)(0.0,-1)(0.4,-1)(0.8,-1)(1.2,-1)\lw
\psline{-}(0.2,-0.9)(0.6,-0.5)(0.2,-0.1)(-0.2,-0.5)(0.2,-0.9)\unlw
\psset{linecolor=myc}
\psarc{-}(0.2,-0.1){0.282843}{225}{315}
\psarc{-}(0.2,-0.9){0.282843}{45}{135}
\psset{linecolor=myc2}
\psline{-}(-0.4,0)(-0.4,-1)
\psline{-}(-0.8,0)(-0.8,-1)
\psline{-}(-1.2,0)(-1.2,-1)
\psline{-}(-1.6,0)(-1.6,-1)
\psline{-}(-2.0,0)(-2.0,-1)
\psline{-}(-2.4,0)(-2.4,-1)
\psline{-}(-2.8,0)(-2.8,-1)
\psline{-}(-3.2,0)(-3.2,-1)
\psline{-}(-3.6,0)(-3.6,-1)
\psline{-}(-4.0,0)(-4.0,-1)
\psarc{-}(-3.8,0){0.2}{0}{180}
\psarc{-}(-3.0,0){0.2}{0}{180}
\psarc{-}(-2.2,0){0.2}{0}{180}
\psarc{-}(-1.4,0){0.2}{0}{180}
\psarc{-}(-0.6,0){0.2}{0}{180}
\psline{-}(0,0)(0,0.5)
\psline{-}(0.4,0)(0.4,0.5)
\psline{-}(0.8,0)(0.8,0.5)
\psline{-}(1.2,0)(1.2,0.5)
\psline{-}(0,0)(0,-0.3)
\psline{-}(0.4,0)(0.4,-0.3)
\psline{-}(0,-0.7)(0,-1)
\psline{-}(0.4,-0.7)(0.4,-1)
\psline{-}(0.8,0)(0.8,-1)
\psline{-}(1.2,0)(1.2,-1)
\psset{linecolor=myc3,linestyle=dashed,dash=2pt 2pt}
\psarc{-}(-3.8,-1){0.2}{180}{360}
\psarc{-}(-3.0,-1){0.2}{180}{360}
\psarc{-}(-2.2,-1){0.2}{180}{360}
\psarc{-}(-1.4,-1){0.2}{180}{360}
\psarc{-}(-0.6,-1){0.2}{180}{360}
\psarc{-}(0.2,-1){0.2}{180}{360}
\psline{-}(0.8,-1)(0.8,-1.5)
\psline{-}(1.2,-1)(1.2,-1.5)
\end{pspicture} \qquad &\rightarrow  \qquad P^r_{\{1,3,...,2b-3 \}} 
\\ &  \\
\begin{pspicture}(-4.0,-0.7)(1.2,0.6)
\psdots(-4.0,0)(-3.6,0)(-3.2,0)(-2.8,0)(-2.4,0)(-2.0,0)(-1.6,0)(-1.2,0)(-0.8,0)(-0.4,0)(0.0,0)(0.4,0)(0.8,0)(1.2,0)
\psdots(-4.0,-1)(-3.6,-1)(-3.2,-1)(-2.8,-1)(-2.4,-1)(-2.0,-1)(-1.6,-1)(-1.2,-1)(-0.8,-1)(-0.4,-1)(0.0,-1)(0.4,-1)(0.8,-1)(1.2,-1)\lw
\psline{-}(0.2,-0.9)(0.6,-0.5)(0.2,-0.1)(-0.2,-0.5)(0.2,-0.9)\unlw
\psset{linecolor=myc}
\psarc{-}(0.2,-0.1){0.282843}{225}{315}
\psarc{-}(0.2,-0.9){0.282843}{45}{135}
\psset{linecolor=myc2}
\psline{-}(-0.4,0)(-0.4,-1)
\psline{-}(-0.8,0)(-0.8,-1)
\psline{-}(-1.2,0)(-1.2,-1)
\psline{-}(-1.6,0)(-1.6,-1)
\psline{-}(-2.0,0)(-2.0,-1)
\psline{-}(-2.4,0)(-2.4,-1)
\psline{-}(-2.8,0)(-2.8,-1)
\psline{-}(-3.2,0)(-3.2,-1)
\psline{-}(-3.6,0)(-3.6,-1)
\psline{-}(-4.0,0)(-4.0,-1)
\psarc{-}(-3.8,0){0.2}{0}{180}
\psarc{-}(-3.0,0){0.2}{0}{180}
\psarc{-}(-2.2,0){0.2}{0}{180}
\psarc{-}(-1.4,0){0.2}{0}{180}
\psarc{-}(-0.6,0){0.2}{0}{180}
\psarc{-}(0.6,0){0.2}{0}{180}
\psline{-}(0.0,0)(0.0,0.5)
\psline{-}(1.2,0)(1.2,0.5)
\psline{-}(0,0)(0,-0.3)
\psline{-}(0.4,0)(0.4,-0.3)
\psline{-}(0,-0.7)(0,-1)
\psline{-}(0.4,-0.7)(0.4,-1)
\psline{-}(0.8,0)(0.8,-1)
\psline{-}(1.2,0)(1.2,-1)
\psset{linecolor=myc3,linestyle=dashed,dash=2pt 2pt}
\psarc{-}(-3.8,-1){0.2}{180}{360}
\psarc{-}(-3.0,-1){0.2}{180}{360}
\psarc{-}(-2.2,-1){0.2}{180}{360}
\psarc{-}(-1.4,-1){0.2}{180}{360}
\psarc{-}(-0.6,-1){0.2}{180}{360}
\psarc{-}(0.2,-1){0.2}{180}{360}
\psline{-}(0.8,-1)(0.8,-1.5)
\psline{-}(1.2,-1)(1.2,-1.5)
\end{pspicture} \qquad &\rightarrow  \qquad  P^r_{\{1,3,...,2b-3,2b \}} 
\\ &  \\
\begin{pspicture}(-4.0,-0.6)(1.2,0.6)
\psdots(-4.0,0)(-3.6,0)(-3.2,0)(-2.8,0)(-2.4,0)(-2.0,0)(-1.6,0)(-1.2,0)(-0.8,0)(-0.4,0)(0.0,0)(0.4,0)(0.8,0)(1.2,0)
\psdots(-4.0,-1)(-3.6,-1)(-3.2,-1)(-2.8,-1)(-2.4,-1)(-2.0,-1)(-1.6,-1)(-1.2,-1)(-0.8,-1)(-0.4,-1)(0.0,-1)(0.4,-1)(0.8,-1)(1.2,-1)\lw
\psline{-}(0.2,-0.9)(0.6,-0.5)(0.2,-0.1)(-0.2,-0.5)(0.2,-0.9)\unlw
\psset{linecolor=myc}
\psarc{-}(0.2,-0.1){0.282843}{225}{315}
\psarc{-}(0.2,-0.9){0.282843}{45}{135}
\psset{linecolor=myc2}
\psline{-}(-0.4,0)(-0.4,-1)
\psline{-}(-0.8,0)(-0.8,-1)
\psline{-}(-1.2,0)(-1.2,-1)
\psline{-}(-1.6,0)(-1.6,-1)
\psline{-}(-2.0,0)(-2.0,-1)
\psline{-}(-2.4,0)(-2.4,-1)
\psline{-}(-2.8,0)(-2.8,-1)
\psline{-}(-3.2,0)(-3.2,-1)
\psline{-}(-3.6,0)(-3.6,-1)
\psline{-}(-4.0,0)(-4.0,-1)
\psarc{-}(-3.8,0){0.2}{0}{180}
\psarc{-}(-3.0,0){0.2}{0}{180}
\psarc{-}(-2.2,0){0.2}{0}{180}
\psarc{-}(-1.4,0){0.2}{0}{180}
\psarc{-}(-0.6,0){0.2}{0}{180}
\psarc{-}(0.2,0){0.2}{0}{180}
\psline{-}(0.8,0)(0.8,0.5)
\psline{-}(1.2,0)(1.2,0.5)
\psline{-}(0,0)(0,-0.3)
\psline{-}(0.4,0)(0.4,-0.3)
\psline{-}(0,-0.7)(0,-1)
\psline{-}(0.4,-0.7)(0.4,-1)
\psline{-}(0.8,0)(0.8,-1)
\psline{-}(1.2,0)(1.2,-1)
\psset{linecolor=myc3,linestyle=dashed,dash=2pt 2pt}
\psarc{-}(-3.8,-1){0.2}{180}{360}
\psarc{-}(-3.0,-1){0.2}{180}{360}
\psarc{-}(-2.2,-1){0.2}{180}{360}
\psarc{-}(-1.4,-1){0.2}{180}{360}
\psarc{-}(-0.6,-1){0.2}{180}{360}
\psarc{-}(0.2,-1){0.2}{180}{360}
\psline{-}(0.8,-1)(0.8,-1.5)
\psline{-}(1.2,-1)(1.2,-1.5)
\end{pspicture} \qquad &\rightarrow  \qquad -\frac{S_2}{S_1} P^r_{\{1, 3, ..., 2b-1\}} \\ &  \\
\begin{pspicture}(-4.0,-0.6)(1.2,0.6)
\psdots(-4.0,0)(-3.6,0)(-3.2,0)(-2.8,0)(-2.4,0)(-2.0,0)(-1.6,0)(-1.2,0)(-0.8,0)(-0.4,0)(0.0,0)(0.4,0)(0.8,0)(1.2,0)
\psdots(-4.0,-1)(-3.6,-1)(-3.2,-1)(-2.8,-1)(-2.4,-1)(-2.0,-1)(-1.6,-1)(-1.2,-1)(-0.8,-1)(-0.4,-1)(0.0,-1)(0.4,-1)(0.8,-1)(1.2,-1)\lw
\psline{-}(0.2,-0.9)(0.6,-0.5)(0.2,-0.1)(-0.2,-0.5)(0.2,-0.9)\unlw
\psset{linecolor=myc}
\psarc{-}(0.2,-0.1){0.282843}{225}{315}
\psarc{-}(0.2,-0.9){0.282843}{45}{135}
\psset{linecolor=myc2}
\psline{-}(-0.4,0)(-0.4,-1)
\psline{-}(-0.8,0)(-0.8,-1)
\psline{-}(-1.2,0)(-1.2,-1)
\psline{-}(-1.6,0)(-1.6,-1)
\psline{-}(-2.0,0)(-2.0,-1)
\psline{-}(-2.4,0)(-2.4,-1)
\psline{-}(-2.8,0)(-2.8,-1)
\psline{-}(-3.2,0)(-3.2,-1)
\psline{-}(-3.6,0)(-3.6,-1)
\psline{-}(-4.0,0)(-4.0,-1)
\psarc{-}(-3.8,0){0.2}{0}{180}
\psarc{-}(-3.0,0){0.2}{0}{180}
\psarc{-}(-2.2,0){0.2}{0}{180}
\psarc{-}(-1.4,0){0.2}{0}{180}
\psarc{-}(-0.2,0){0.2}{0}{180}
\psbezier{-}(-0.8,0)(-0.8,0.4)(0.4,0.4)(0.4,0)
\psline{-}(0.8,0)(0.8,0.5)
\psline{-}(1.2,0)(1.2,0.5)
\psline{-}(0,0)(0,-0.3)
\psline{-}(0.4,0)(0.4,-0.3)
\psline{-}(0,-0.7)(0,-1)
\psline{-}(0.4,-0.7)(0.4,-1)
\psline{-}(0.8,0)(0.8,-1)
\psline{-}(1.2,0)(1.2,-1)
\psset{linecolor=myc3,linestyle=dashed,dash=2pt 2pt}
\psarc{-}(-3.8,-1){0.2}{180}{360}
\psarc{-}(-3.0,-1){0.2}{180}{360}
\psarc{-}(-2.2,-1){0.2}{180}{360}
\psarc{-}(-1.4,-1){0.2}{180}{360}
\psarc{-}(-0.6,-1){0.2}{180}{360}
\psarc{-}(0.2,-1){0.2}{180}{360}
\psline{-}(0.8,-1)(0.8,-1.5)
\psline{-}(1.2,-1)(1.2,-1.5)
\end{pspicture} \qquad  &\rightarrow  \qquad \frac{S_1}{S_2} P^r_{\{1,3,...,2b-1 \}} 
\\ &  \\
\begin{pspicture}(-4.0,-0.6)(1.2,0.6)
\psdots(-4.0,0)(-3.6,0)(-3.2,0)(-2.8,0)(-2.4,0)(-2.0,0)(-1.6,0)(-1.2,0)(-0.8,0)(-0.4,0)(0.0,0)(0.4,0)(0.8,0)(1.2,0)
\psdots(-4.0,-1)(-3.6,-1)(-3.2,-1)(-2.8,-1)(-2.4,-1)(-2.0,-1)(-1.6,-1)(-1.2,-1)(-0.8,-1)(-0.4,-1)(0.0,-1)(0.4,-1)(0.8,-1)(1.2,-1)\lw
\psline{-}(0.2,-0.9)(0.6,-0.5)(0.2,-0.1)(-0.2,-0.5)(0.2,-0.9)\unlw
\psset{linecolor=myc}
\psarc{-}(0.2,-0.1){0.282843}{225}{315}
\psarc{-}(0.2,-0.9){0.282843}{45}{135}
\psset{linecolor=myc2}
\psline{-}(-0.4,0)(-0.4,-1)
\psline{-}(-0.8,0)(-0.8,-1)
\psline{-}(-1.2,0)(-1.2,-1)
\psline{-}(-1.6,0)(-1.6,-1)
\psline{-}(-2.0,0)(-2.0,-1)
\psline{-}(-2.4,0)(-2.4,-1)
\psline{-}(-2.8,0)(-2.8,-1)
\psline{-}(-3.2,0)(-3.2,-1)
\psline{-}(-3.6,0)(-3.6,-1)
\psline{-}(-4.0,0)(-4.0,-1)
\psarc{-}(-3.8,0){0.2}{0}{180}
\psarc{-}(-3.0,0){0.2}{0}{180}
\psarc{-}(-2.2,0){0.2}{0}{180}
\psarc{-}(-0.6,0){0.2}{0}{180}
\psbezier{-}(-1.2,0)(-1.2,0.4)(0,0.4)(0,0)
\psbezier{-}(-1.6,0)(-1.6,0.56)(0.4,0.56)(0.4,0)
\psline{-}(0.8,0)(0.8,0.5)
\psline{-}(1.2,0)(1.2,0.5)
\psline{-}(0,0)(0,-0.3)
\psline{-}(0.4,0)(0.4,-0.3)
\psline{-}(0,-0.7)(0,-1)
\psline{-}(0.4,-0.7)(0.4,-1)
\psline{-}(0.8,0)(0.8,-1)
\psline{-}(1.2,0)(1.2,-1)
\psset{linecolor=myc3,linestyle=dashed,dash=2pt 2pt}
\psarc{-}(-3.8,-1){0.2}{180}{360}
\psarc{-}(-3.0,-1){0.2}{180}{360}
\psarc{-}(-2.2,-1){0.2}{180}{360}
\psarc{-}(-1.4,-1){0.2}{180}{360}
\psarc{-}(-0.6,-1){0.2}{180}{360}
\psarc{-}(0.2,-1){0.2}{180}{360}
\psline{-}(0.8,-1)(0.8,-1.5)
\psline{-}(1.2,-1)(1.2,-1.5)
\end{pspicture} \qquad  &\rightarrow  \qquad \frac{S_1^2}{S_2S_3} P^r_{\{1,3,...,2b-1 \}}   
\\ &  \\
\begin{pspicture}(-4.0,-0.5)(1.2,1.2)
\rput(-2.9,1.1){$\vdots$}
\rput(-1.4,1.1){$\vdots$}
\rput(0.1,1.1){$\vdots$}
\psdots(-4.0,0)(-3.6,0)(-3.2,0)(-2.8,0)(-2.4,0)(-2.0,0)(-1.6,0)(-1.2,0)(-0.8,0)(-0.4,0)(0.0,0)(0.4,0)(0.8,0)(1.2,0)
\psdots(-4.0,-1)(-3.6,-1)(-3.2,-1)(-2.8,-1)(-2.4,-1)(-2.0,-1)(-1.6,-1)(-1.2,-1)(-0.8,-1)(-0.4,-1)(0.0,-1)(0.4,-1)(0.8,-1)(1.2,-1)\lw
\psline{-}(0.2,-0.9)(0.6,-0.5)(0.2,-0.1)(-0.2,-0.5)(0.2,-0.9)\unlw
\psset{linecolor=myc}
\psarc{-}(0.2,-0.1){0.282843}{225}{315}
\psarc{-}(0.2,-0.9){0.282843}{45}{135}
\psset{linecolor=myc2}
\psline{-}(-0.4,0)(-0.4,-1)
\psline{-}(-0.8,0)(-0.8,-1)
\psline{-}(-1.2,0)(-1.2,-1)
\psline{-}(-1.6,0)(-1.6,-1)
\psline{-}(-2.0,0)(-2.0,-1)
\psline{-}(-2.4,0)(-2.4,-1)
\psline{-}(-2.8,0)(-2.8,-1)
\psline{-}(-3.2,0)(-3.2,-1)
\psline{-}(-3.6,0)(-3.6,-1)
\psline{-}(-4.0,0)(-4.0,-1)
\psarc{-}(-3.0,0){0.2}{0}{180}
\psarc{-}(-2.2,0){0.2}{0}{180}
\psarc{-}(-1.4,0){0.2}{0}{180}
\psarc{-}(-0.6,0){0.2}{0}{180}
\psbezier{-}(-3.6,0)(-3.6,0.60)(0,0.60)(0,0)
\psbezier{-}(-4.0,0)(-4.0,0.82)(0.4,0.82)(0.4,0)
\psline{-}(0.8,0)(0.8,0.5)
\psline{-}(1.2,0)(1.2,0.5)
\psline{-}(0,0)(0,-0.3)
\psline{-}(0.4,0)(0.4,-0.3)
\psline{-}(0,-0.7)(0,-1)
\psline{-}(0.4,-0.7)(0.4,-1)
\psline{-}(0.8,0)(0.8,-1)
\psline{-}(1.2,0)(1.2,-1)
\psset{linecolor=myc3,linestyle=dashed,dash=2pt 2pt}
\psarc{-}(-3.8,-1){0.2}{180}{360}
\psarc{-}(-3.0,-1){0.2}{180}{360}
\psarc{-}(-2.2,-1){0.2}{180}{360}
\psarc{-}(-1.4,-1){0.2}{180}{360}
\psarc{-}(-0.6,-1){0.2}{180}{360}
\psarc{-}(0.2,-1){0.2}{180}{360}
\psline{-}(0.8,-1)(0.8,-1.5)
\psline{-}(1.2,-1)(1.2,-1.5)
\end{pspicture} \qquad  &\rightarrow  \qquad \frac{S_1^2}{S_{b-1}S_{b}} P^r_{\{1,3,...,2b-1 \}} 
\end{alignat*} \\
\\
are obtained using, among others, \eqref{eq:multeq}. Overall,
\begin{equation*}
P^r_{\{1,3,...,2b-3,2b \}} =  \Big( \underbrace{\frac{S_2}{S_1} - \sum_{i=1}^{b-1} \frac{S_1^2}{S_iS_{i+1}}}_{A_b} \Big) P^r_{\{1, 3, ..., 2b-1\}} - P^r_{\{1, 3, ..., 2b-3\}}.
\end{equation*}
The factor $A_b$ satisfies $A_{b} = A_{b-1} - S_1^2/(S_b S_{b-1})$ and the initial condition $A_1 = S_2/S_1$. An induction confirms that $A_{b} =S_{b+1}/S_b$. When the $b$-th bubble is placed in $n > 2b$, only four terms contribute to $\langle v^r_{1, 3, ..., 2b-3, n-1} | e_{n-1} P^r v^r\rangle$, and
\begin{alignat*}{3}
P^r_{\{1, 3, ..., 2b-3, n\}} &= \frac{S_2}{S_1}P^r_{\{1, 3, ..., 2b-3, n-1\}} - P^r_{\{1, 3, ..., 2b-3, n-2\}} - P^r_{\{1, 3, ..., 2b-3\}} \\ &= \frac{S_{n-b+1}}{S_b}P^r_{\{1,3,...,2b-3,2b-1\}} -  \frac{S_{(n-2b+2)/2}S_{(n-2b+1)/2}}{S_{1/2}S_{2/2}}P^r_{\{1,3,...,2b-3\}}
\end{alignat*}
where a last induction, on $n$, was used to obtain the last line. \hfill$\square$

\begin{Lemme} Let $w= v^r_{n_1,n_2,...,n_k}$ a link state with only $1$-bubbles. Then 
\begin{equation*}
\{\Lambda\in \mathbb{R}\, |\, P^r_w \ \textrm{diverges at }\Lambda\}  \subset  \cup_{i=1}^k \{\Lambda\in \mathbb{R}\, |\, P^r_{\{i^i\}} \ \textrm{diverges at }\Lambda\}.
\end{equation*}
\end{Lemme}
\noindent{\scshape Proof\ \ } By repeated use of (\ref{eq: generalmove}), it is certainly possible to express $P^r_w$ as:
$$
P^r_{\{n_1,n_2,...,n_k\}} = \sum_{i=0}^k \alpha_iP^r_{\{1, 3, ..., 2i-1\}}
$$
where the $\alpha_i$ are functions independent of $r$ and the term $i=0$ is for $P^r_{\{\}}$. Due to the $S_b$ in the denominator of the first term of (\ref{eq: generalmove}), the $\alpha_i$'s may be divergent for some values of $\Lambda$. More precisely,  $\alpha_i$ may carry at most one factor $S_j$, for any $j$ in $1<j \leq i $, in its denominator. (The coefficient of the second term of \eqref{eq: generalmove} is analytic for $\Lambda\in\mathbb R$.) However, from (\ref{eq: multbubble}),
\begin{equation*}
P^r_{\{n_1,n_2,...,n_k\}} = \sum_{i=0}^k \Big( \alpha_i \prod_{j=1}^i \frac{S_j}{S_1} \Big) P^r_{\{i^i\}}
\end{equation*}
and the terms in the denominator of $\alpha_i$ are cancelled by the product $\prod_{1\le j\le i} S_j/S_1$. The coefficients of $P^r_{\{i^i\}}$ are therefore analytic for $\Lambda\in \mathbb{R}$. \hfill$\square$

\begin{Lemme}
Let $w= v^r_{n_1,n_2,...,n_k}$ be any link state. Then 
$$P^r_w=\sum_{i=0}^k\alpha_iP^r_{\{i^i\}}$$
for some functions $\alpha_i$ analytic in $\Lambda$. Therefore
\begin{equation}
\{\Lambda \in \mathbb{R} \, |\, P^r_{w} \ \textrm{diverges at }\Lambda\}  \subset  \cup_{i=1}^k \{\Lambda \in \mathbb{R}\, |\, P^r_{\{i^i\}} \ \textrm{diverges at } \Lambda\}  \subset \cup_{i=0}^{k-1} \{\Lambda \in \mathbb{R}\, |\, C_{(r-i)/2}=0 \}.
\tag{29}
\end{equation}
\end{Lemme}

\noindent{\scshape Proof\ \ } Let $w_1,w_2\in B^d_r$ and let $\Delta_{1,2}$ and $\Delta_{2,1}$ be the set of bubbles of $w_1$ and $w_2$ respectively that are not shared by both. These sets are noted in the form $\Delta_{1,2}=\{p_1,p_2, \dots, p_k\}$ and $\Delta_{2,1}=\{q_1,q_2,\dots, q_k\}$ where $k$ is the number of distinct bubbles and the $p_i$'s ($p_i<p_{i+1}$) are the end points of the bubbles in $w_1$ that do not appear in $w_2$. For example 
\begin{align*}
w_1 & = 
\begin{pspicture}(0.2,0.)(7.8,0.2)
\psset{unit=0.4}
\psdots[dotsize=2.5pt](1,0)(2,0)(3,0)(4,0)(5,0)(6,0)(7,0)(8,0)(9,0)(10,0)(11,0)(12,0)(13,0)(14,0)(15,0)(16,0)(17,0)(18,0)(19,0)
\psset{linecolor=myc2}
\psline{-}(3,0)(3,1)
\psline{-}(4,0)(4,1)
\psline{-}(9,0)(9,1)
\psline{-}(16,0)(16,1)
\psline{-}(17,0)(17,1)
\psarc(1.5,0){0.5}{0}{180}
\psarc(5.5,0){0.5}{0}{180}
\psarc(7.5,0){0.5}{0}{180}
\psarc(11.5,0){0.5}{0}{180}
\psarc(13.5,0){0.5}{0}{180}
\psarc(18.5,0){0.5}{0}{180}
\psbezier{-}(10,0)(10,1.5)(15,1.5)(15,0)
\end{pspicture}, &\quad  \Delta_{1,2} = \{6,8,15 \} \\
w_2 & = 
\begin{pspicture}(0.2,0.)(7.8,0.5)
\psset{unit=0.4}
\psdots[dotsize=2.5pt](1,0)(2,0)(3,0)(4,0)(5,0)(6,0)(7,0)(8,0)(9,0)(10,0)(11,0)(12,0)(13,0)(14,0)(15,0)(16,0)(17,0)(18,0)(19,0)
\psset{linecolor=myc2}
\psline{-}(3,0)(3,1)
\psline{-}(4,0)(4,1)
\psline{-}(15,0)(15,1)
\psline{-}(16,0)(16,1)
\psline{-}(17,0)(17,1)
\psarc(1.5,0){0.5}{0}{180}
\psarc(6.5,0){0.5}{0}{180}
\psarc(9.5,0){0.5}{0}{180}
\psarc(11.5,0){0.5}{0}{180}
\psarc(13.5,0){0.5}{0}{180}
\psarc(18.5,0){0.5}{0}{180}
\psbezier{-}(5,0)(5,1.5)(8,1.5)(8,0)
\end{pspicture},  &\quad \Delta_{2,1} = \{7,8,10 \}.
\end{align*}
A partial order ``$<$'' is defined on the link basis $B_r$ by setting $w_1<w_2$ if $w_1$ has less bubbles than $w_2$ and, if $w_1, w_2$ have the same number of bubbles, then $w_1<w_2$ if $p_k<q_k$. For the above example, $w_2<w_1$.

Each subset $B_r^d$ contains several minimal elements for this order, namely all elements whose arcs are all bunched up on the left. (The number of minimal elements in $B_r^d$ is $\dim V_d^0$.) The proof of the lemma is by induction on this order. More precisely, we first prove the statement for all minimal elements. Then we construct an iterative procedure that expresses $P^r_w$ as a sum $\sum_{y<w}\beta_yP_y^r$ with $\beta_y$'s analytic in $\Lambda$. 

The first step is actually given by Lemma \ref{sec:multiB}. If $w\in B^d_r$ is minimal, then $\zeta(w)=[0](k_1, k_2, \dots, k_m)[d]$ for $m=(r-d)/2$ and 
$$P^r_w=P^r_{\{m^m\}}\Big(\prod_{1\le i\le m}\frac{S_i}{S_{k_i}}\Big).$$
The analyticity of $\prod_iS_i/S_{k_i}$ will follow if all zeroes of the denominator are canceled by zeroes in the numerator. Those of the denominator originate from $S_k=0$, that is $e^{2ik\Lambda}=1$, and they are therefore all of the form $\pi a/b$, with $1\le b\le m$ and $a,b$ coprimes. Because all $S_k$ have $2\pi$ as a period, the integer $a$ can be restricted to $0\le a<2b$. Moreover the multiplicities of the zero at $\Lambda=0$ of both numerator and denominator are identical and the case $a=0$ can be forgotten. Fix now a zero $\pi a/b$ of the denominator, $a\neq 0$. Its multiplicity will depend of the $k_i$'s. This multiplicity is never larger than when the $k_i$'s contain as many $b$ as possible. The largest multiplicity of $\pi a/b$ in the denominator is therefore $\lfloor m/b\rfloor$. But each term in the numerator whose index is a multiple of $b$ has also $\pi a/b$ among its zeroes and there are precisely $\lfloor m/b\rfloor$ of those. So $\prod_iS_i/S_{k_i}$ is not singular at $\pi a/b$ and is therefore analytic.

Suppose now that $y$ is not minimal. This means that there are some defects on the left of some bubbles. We study the leftmost patch of defects and divide the remaining argument depending on whether this patch contains a single defect or many. We assume first that this patch contains more than one defect, as in 
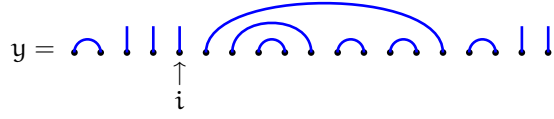
\begin{equation*}\begin{pspicture}(-1.,-0.4)(8.0,0.5)
\rput(0.5,0){$y =$}
\psset{unit=0.35}
\psdots[dotsize=2.5pt](3,0)(4,0)(5,0)(6,0)(7,0)(8,0)(9,0)(10,0)(11,0)(12,0)(13,0)(14,0)(15,0)(16,0)(17,0)(18,0)(19,0)(20,0)(21,0)
\psset{linecolor=myc2}
\psline{-}(7,0)(7,1)
\psline{-}(20,0)(20,1)
\psline{-}(21,0)(21,1)
\psline{-}(5,0)(5,1)
\psline{-}(6,0)(6,1)
\rput(7,-0.8){$\uparrow$}
\rput(7,-1.8){$i$}
\psarc(3.5,0){0.5}{0}{180}
\psarc(10.5,0){0.5}{0}{180}
\psarc(13.5,0){0.5}{0}{180}
\psarc(15.5,0){0.5}{0}{180}
\psarc(18.5,0){0.5}{0}{180}
\psbezier{-}(9,0)(9,1.5)(12,1.5)(12,0)
\psbezier{-}(8,0)(8,2.5)(17,2.5)(17,0)
\end{pspicture}
\end{equation*}
Note that the patch of defects might not be preceded on the left by any arcs. Let $i$ be the position of the last defect on the right of this patch and consider the matrix element $\langle e_i y|e_i P^rv^r\rangle$ which is zero because $e_iP^r=0$. Several terms of the linear combination $P^rv^r$ will contribute. First the following coefficients will appear: $P^r_y$, $P^r_{e_iy}$ and $P^r_{e_iy/\{i\}}$ where $e_iy/\{i\}$ stands for the link state obtained from $e_iy$ by replacing the $1$-bubble starting at $i$ by two defects. For example, for the $y$ above
\begin{equation*}e_iy=\begin{pspicture}(1.,-0.0)(7.5,0.4)
\psset{unit=0.35}
\psdots[dotsize=2.5pt](3,0)(4,0)(5,0)(6,0)(7,0)(8,0)(9,0)(10,0)(11,0)(12,0)(13,0)(14,0)(15,0)(16,0)(17,0)(18,0)(19,0)(20,0)(21,0)
\psset{linecolor=myc2}
\psarc(7.5,0){0.5}{0}{180}
\psline{-}(20,0)(20,1)
\psline{-}(21,0)(21,1)
\psline{-}(5,0)(5,1)
\psline{-}(6,0)(6,1)
\psarc(3.5,0){0.5}{0}{180}
\psarc(10.5,0){0.5}{0}{180}
\psarc(13.5,0){0.5}{0}{180}
\psarc(15.5,0){0.5}{0}{180}
\psline{-}(17,0)(17,1)
\psarc(18.5,0){0.5}{0}{180}
\psbezier{-}(9,0)(9,1.5)(12,1.5)(12,0)
\end{pspicture}\quad \text{\rm and}\quad
e_iy/\{i\}=\begin{pspicture}(1.,-0.0)(7.5,0.4)
\psset{unit=0.35}
\psdots[dotsize=2.5pt](3,0)(4,0)(5,0)(6,0)(7,0)(8,0)(9,0)(10,0)(11,0)(12,0)(13,0)(14,0)(15,0)(16,0)(17,0)(18,0)(19,0)(20,0)(21,0)
\psset{linecolor=myc2}
\psline{-}(7,0)(7,1)
\psline{-}(8,0)(8,1)
\psline{-}(20,0)(20,1)
\psline{-}(21,0)(21,1)
\psline{-}(5,0)(5,1)
\psline{-}(6,0)(6,1)
\psarc(3.5,0){0.5}{0}{180}
\psarc(10.5,0){0.5}{0}{180}
\psarc(13.5,0){0.5}{0}{180}
\psarc(15.5,0){0.5}{0}{180}
\psline{-}(17,0)(17,1)
\psarc(18.5,0){0.5}{0}{180}
\psbezier{-}(9,0)(9,1.5)(12,1.5)(12,0)
\end{pspicture}
\end{equation*}
Second, because there is a defect at position $i-1$, the generator $e_i$ can be used to exchange this defect with a $1$-bubble starting at $i$ and $P^r_{e_{i-1}e_iy}$ will also contribute. Finally, some link states with a pattern of bubbles on the right of the defect $i$ will also contribute. If the notation $P^r(w)$ is used for $P^r_w$, the contributions for the example are
\begin{align*}
0 & =  \langle e_iy\,|\, e_iP^rv^r \rangle \\
& = P^{r}_y + \beta P^{r}(\begin{pspicture}(0.5,-0.0)(4.30,0.2)
\psset{unit=0.2}
\psdots[dotsize=2.5pt](3,0)(4,0)(5,0)(6,0)(7,0)(8,0)(9,0)(10,0)(11,0)(12,0)(13,0)(14,0)(15,0)(16,0)(17,0)(18,0)(19,0)(21,0)(20,0)
\psset{linecolor=myc2}
\psline{-}(17,0)(17,1)
\psline{-}(20,0)(20,1)
\psline{-}(21,0)(21,1)
\psarc(18.5,0){0.5}{0}{180}
\psarc(3.5,0){0.5}{0}{180}
\psline{-}(5,0)(5,1)
\psline{-}(6,0)(6,1)
\psarc(7.5,0){0.5}{0}{180}
\psarc(10.5,0){0.5}{0}{180}
\psarc(13.5,0){0.5}{0}{180}
\psarc(15.5,0){0.5}{0}{180}
\psbezier{-}(9,0)(9,1.5)(12,1.5)(12,0)
\end{pspicture})  + P^{r}(\begin{pspicture}(0.5,-0.0)(4.30,0.2)
\psset{unit=0.2}
\psdots[dotsize=2.5pt](3,0)(4,0)(5,0)(6,0)(7,0)(8,0)(9,0)(10,0)(11,0)(12,0)(13,0)(14,0)(15,0)(16,0)(17,0)(18,0)(19,0)(21,0)(20,0)
\psset{linecolor=myc2}
\psline{-}(17,0)(17,1)
\psline{-}(20,0)(20,1)
\psline{-}(21,0)(21,1)
\psarc(18.5,0){0.5}{0}{180}
\psarc(3.5,0){0.5}{0}{180}
\psline{-}(5,0)(5,1)
\psline{-}(6,0)(6,1)
\psline{-}(7,0)(7,1)
\psline{-}(8,0)(8,1)
\psarc(10.5,0){0.5}{0}{180}
\psarc(13.5,0){0.5}{0}{180}
\psarc(15.5,0){0.5}{0}{180}
\psbezier{-}(9,0)(9,1.5)(12,1.5)(12,0)
\end{pspicture}) + P^{r}(\begin{pspicture}(0.5,-0.0)(4.30,0.2)
\psset{unit=0.2}
\psdots[dotsize=2.5pt](3,0)(4,0)(5,0)(6,0)(7,0)(8,0)(9,0)(10,0)(11,0)(12,0)(13,0)(14,0)(15,0)(16,0)(17,0)(18,0)(19,0)(21,0)(20,0)
\psset{linecolor=myc2}
\psline{-}(17,0)(17,1)
\psline{-}(20,0)(20,1)
\psline{-}(21,0)(21,1)
\psarc(18.5,0){0.5}{0}{180}
\psarc(3.5,0){0.5}{0}{180}
\psline{-}(5,0)(5,1)
\psline{-}(8,0)(8,1)
\psarc(6.5,0){0.5}{0}{180}
\psarc(10.5,0){0.5}{0}{180}
\psarc(13.5,0){0.5}{0}{180}
\psarc(15.5,0){0.5}{0}{180}
\psbezier{-}(9,0)(9,1.5)(12,1.5)(12,0)
\end{pspicture})\\ & \quad+\Big( P^{r}(\begin{pspicture}(0.5,-0.0)(4.3,0.2)
\psset{unit=0.2}
\psdots[dotsize=2.5pt](21,0)(20,0)(3,0)(4,0)(5,0)(6,0)(7,0)(8,0)(9,0)(10,0)(11,0)(12,0)(13,0)(14,0)(15,0)(16,0)(17,0)(18,0)(19,0)
\psset{linecolor=myc2}
\psline{-}(17,0)(17,1)
\psline{-}(20,0)(20,1)
\psline{-}(21,0)(21,1)
\psarc(18.5,0){0.5}{0}{180}
\psarc(3.5,0){0.5}{0}{180}
\psline{-}(5,0)(5,1)
\psline{-}(6,0)(6,1)
\psarc(8.5,0){0.5}{0}{180}
\psarc(10.5,0){0.5}{0}{180}
\psarc(13.5,0){0.5}{0}{180}
\psarc(15.5,0){0.5}{0}{180}
\psbezier{-}(7,0)(7,2.0)(12,2.0)(12,0)
\end{pspicture})  +P^{r}(\begin{pspicture}(0.5,-0.0)(4.3,0.2)
\psset{unit=0.2}
\psdots[dotsize=2.5pt](21,0)(20,0)(3,0)(4,0)(5,0)(6,0)(7,0)(8,0)(9,0)(10,0)(11,0)(12,0)(13,0)(14,0)(15,0)(16,0)(17,0)(18,0)(19,0)
\psset{linecolor=myc2}
\psline{-}(17,0)(17,1)
\psline{-}(20,0)(20,1)
\psline{-}(21,0)(21,1)
\psarc(18.5,0){0.5}{0}{180}
\psarc(3.5,0){0.5}{0}{180}
\psline{-}(5,0)(5,1)
\psline{-}(6,0)(6,1)
\psarc(10.5,0){0.5}{0}{180}
\psarc(15.5,0){0.5}{0}{180}
\psbezier{-}(8,0)(8,2.5)(13,2.5)(13,0)
\psbezier{-}(7,0)(7,3.5)(14,3.5)(14,0)
\psbezier{-}(9,0)(9,1.5)(12,1.5)(12,0)
\end{pspicture})+P^{r}(\begin{pspicture}(0.5,-0.0)(4.3,0.2)
\psset{unit=0.2}
\psdots[dotsize=2.5pt](21,0)(20,0)(3,0)(4,0)(5,0)(6,0)(7,0)(8,0)(9,0)(10,0)(11,0)(12,0)(13,0)(14,0)(15,0)(16,0)(17,0)(18,0)(19,0)
\psset{linecolor=myc2}
\psline{-}(17,0)(17,1)
\psline{-}(20,0)(20,1)
\psline{-}(21,0)(21,1)
\psarc(18.5,0){0.5}{0}{180}
\psarc(3.5,0){0.5}{0}{180}
\psline{-}(5,0)(5,1)
\psline{-}(6,0)(6,1)
\psarc(10.5,0){0.5}{0}{180}
\psarc(13.5,0){0.5}{0}{180}
\psbezier{-}(9,0)(9,1.5)(12,1.5)(12,0)
\psbezier{-}(8,0)(8,2.5)(15,2.5)(15,0)
\psbezier{-}(7,0)(7,3.5)(16,3.5)(16,0)
\end{pspicture})
\Big) 
\end{align*}
The terms in the last line are reminiscent of those appearing in the proof of Lemma \ref{lem:A9}. Each term in this line is accounted as follows. Let a closed cluster be a cluster that contains $n$ bubbles of which one contains all the others. All $1$-bubbles are closed clusters and, in the above $y$, a closed cluster with $n=5$ starts at position $i+1=6$. Consider the closed cluster to the right of the patch of defects. If this is not a $1$-bubble, it contains itself a certain number $n_c$ of closed clusters. (In the example, $n_c=3$.) The state $e_iy$ breaks this closed cluster and introduces an arc joining $i\leftrightarrow i+1$, itself followed by the $n_c$ smaller closed clusters that were included in the larger original one. Choose one of these $n_c$ closed clusters and let $p_1$ and $p_2$ be the beginning and end positions of its exterior bubble. In $e_iy$, replace the arcs $i\leftrightarrow i+1$ and $p_1\leftrightarrow p_2$ by the arcs $i\leftrightarrow p_2$ and $i+1\leftrightarrow p_1$. The resulting state $y'$ is such that $e_iy'=e_iy$; it is one of the $n_c$ terms in the last line. Note that if the closed cluster at the right of the defect in $y$ is a $1$-bubble, then $n_c=0$ and the last line of the above example would have been empty.
One more feature of $e_iy$ needs to be underlined. The breaking of the exterior loop of the closed cluster has left a new defect at the end point $i'=i+2n$ of this loop. Any arc or defect to the right of this new defect appears without change in each term contributing to the matrix element under study. The matrix element is thus
\begin{equation}0 = P^{r}_y + \beta P^{r}_{e_iy} + P^{r}_{e_iy / \{i\}} +  P^{r}_{e_{i-1}e_iy} + \sum_{y'}P^{i}_{y'}\label{eq:laPetite}\end{equation}
and allows to express $P^r_y$ in terms of the other $P^r_w$'s. Note that $e_iy$, $e_iy/\{i\}$, $e_{i-1}e_iy$ and the $n_c$ terms $y'$ of the sum all share the new defect at position $i'$ and they are therefore all smaller than $y$ for the partial order. Since $\beta=-S_2/S_1=-2\cos\Lambda$ is analytic, the statement follows. Note, before closing this case, that the expression \eqref{eq:laPetite} also holds for link vectors $y$ that have a single defect at position $1$ followed by a bubble starting at position $2$ (setting $P^{r}_{e_{i-1}e_iy}$ to $0$).

We now study the case where the leftmost patch of defects is actually a single defect at position $i$. (Note that this $i$ is always odd.) Because the previous case covered the possibility of $i=1$, we can now assume that a cluster starts at position $1$. Given such a link state $Y$, we replace the leftmost cluster by  $1$-bubbles, using Lemma \ref{sec:multiB}. The factor $\prod_i S_i/S_{k_i}$ that this lemma introduces is set aside and will be dealt with later. Here is an example of the resulting state $y$:
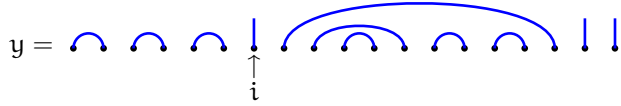
\begin{equation*}\begin{pspicture}(-1,-0.4)(8,0.6)
\psset{unit=0.4}
\psdots[dotsize=2.5pt](1,0)(2,0)(3,0)(4,0)(5,0)(6,0)(7,0)(8,0)(9,0)(10,0)(11,0)(12,0)(13,0)(14,0)(15,0)(16,0)(17,0)(18,0)(19,0)
\psset{linecolor=myc2}
\rput(-0.4,0){$y=$}
\rput(7,-0.6){$\uparrow$}
\rput(7,-1.4){$i$}
\psline{-}(7,0)(7,1)
\psline{-}(18,0)(18,1)
\psline{-}(19,0)(19,1)
\psarc(1.5,0){0.5}{0}{180}
\psarc(3.5,0){0.5}{0}{180}
\psarc(5.5,0){0.5}{0}{180}
\psarc(10.5,0){0.5}{0}{180}
\psarc(13.5,0){0.5}{0}{180}
\psarc(15.5,0){0.5}{0}{180}
\psbezier{-}(9,0)(9,1.)(12,1.)(12,0)
\psbezier{-}(8,0)(8,2.0)(17,2.0)(17,0)
\end{pspicture}
\end{equation*}
with
\begin{equation*}\begin{pspicture}(-1,-0)(8,0.4)
\psset{unit=0.4}
\psdots[dotsize=2.5pt](1,0)(2,0)(3,0)(4,0)(5,0)(6,0)(7,0)(8,0)(9,0)(10,0)(11,0)(12,0)(13,0)(14,0)(15,0)(16,0)(17,0)(18,0)(19,0)
\psset{linecolor=myc2}
\rput(-0.5,0){$e_iy=$}
\psline{-}(17,0)(17,1)
\psline{-}(18,0)(18,1)
\psline{-}(19,0)(19,1)
\psarc(1.5,0){0.5}{0}{180}
\psarc(3.5,0){0.5}{0}{180}
\psarc(5.5,0){0.5}{0}{180}
\psarc(7.5,0){0.5}{0}{180}
\psarc(10.5,0){0.5}{0}{180}
\psarc(13.5,0){0.5}{0}{180}
\psarc(15.5,0){0.5}{0}{180}
\psbezier{-}(9,0)(9,1.)(12,1.)(12,0)
\end{pspicture}
\end{equation*}
Again a vanishing matrix element is computed: $\langle e_iy|e_iP^rv^r\rangle$. The components of $P^rv^r$ that contribute to this matrix element are of several types. Again $P^r_y$, $P^r_{e_iy}$ and $P^r_{e_iy/\{i\}}$ will appear, as well as a sum $\sum_{y'}$ over states constructed as previously from the $n_c$ closed clusters inside the large closed cluster to the right of the defect. (Again the present $y$ has $n_c=3$.) Besides these, a sum appears over states $y''$ obtained from $e_iy$ by replacing the arcs $i\leftrightarrow i+1$ and $2j-1\leftrightarrow 2j$ with $2j<i$ by arcs $i\leftrightarrow 2j$ and $i+1\leftrightarrow 2j-1$. There are $(i-1)/2$ terms of this type. Here are all the terms for the above $y$. The three terms $y''$ are contained in the first parenthesis.
\begin{align*}
0 & =  \langle \begin{pspicture}(0.1,-0.0)(3.90,0.2)
\psset{unit=0.2}
\psdots[dotsize=2.5pt](1,0)(2,0)(3,0)(4,0)(5,0)(6,0)(7,0)(8,0)(9,0)(10,0)(11,0)(12,0)(13,0)(14,0)(15,0)(16,0)(17,0)(18,0)(19,0)
\psset{linecolor=myc2}
\psline{-}(17,0)(17,1)
\psline{-}(18,0)(18,1)
\psline{-}(19,0)(19,1)
\psarc(1.5,0){0.5}{0}{180}
\psarc(3.5,0){0.5}{0}{180}
\psarc(5.5,0){0.5}{0}{180}
\psarc(7.5,0){0.5}{0}{180}
\psarc(10.5,0){0.5}{0}{180}
\psarc(13.5,0){0.5}{0}{180}
\psarc(15.5,0){0.5}{0}{180}
\psbezier{-}(9,0)(9,1.5)(12,1.5)(12,0)
\end{pspicture}
| e_i P^r v^r \rangle \\
& = P^r_y +  \beta P^r(e_iy=\begin{pspicture}(0.1,-0.0)(3.90,0.2)
\psset{unit=0.2}
\psdots[dotsize=2.5pt](1,0)(2,0)(3,0)(4,0)(5,0)(6,0)(7,0)(8,0)(9,0)(10,0)(11,0)(12,0)(13,0)(14,0)(15,0)(16,0)(17,0)(18,0)(19,0)
\psset{linecolor=myc2}
\psline{-}(17,0)(17,1)
\psline{-}(18,0)(18,1)
\psline{-}(19,0)(19,1)
\psarc(1.5,0){0.5}{0}{180}
\psarc(3.5,0){0.5}{0}{180}
\psarc(5.5,0){0.5}{0}{180}
\psarc(7.5,0){0.5}{0}{180}
\psarc(10.5,0){0.5}{0}{180}
\psarc(13.5,0){0.5}{0}{180}
\psarc(15.5,0){0.5}{0}{180}
\psbezier{-}(9,0)(9,1.5)(12,1.5)(12,0)
\end{pspicture}) \\ &\quad +\Big( 
P^r(\begin{pspicture}(0.1,-0.0)(3.90,0.2)
\psset{unit=0.2}
\psdots[dotsize=2.5pt](1,0)(2,0)(3,0)(4,0)(5,0)(6,0)(7,0)(8,0)(9,0)(10,0)(11,0)(12,0)(13,0)(14,0)(15,0)(16,0)(17,0)(18,0)(19,0)
\psset{linecolor=myc2}
\psline{-}(17,0)(17,1)
\psline{-}(18,0)(18,1)
\psline{-}(19,0)(19,1)
\psarc(1.5,0){0.5}{0}{180}
\psarc(3.5,0){0.5}{0}{180}
\psarc(6.5,0){0.5}{0}{180}
\psbezier{-}(5,0)(5,1.5)(8,1.5)(8,0)
\psarc(10.5,0){0.5}{0}{180}
\psarc(13.5,0){0.5}{0}{180}
\psarc(15.5,0){0.5}{0}{180}
\psbezier{-}(9,0)(9,1.5)(12,1.5)(12,0)
\end{pspicture})+
P^r(\begin{pspicture}(0.1,-0.0)(3.90,0.2)
\psset{unit=0.2}
\psdots[dotsize=2.5pt](1,0)(2,0)(3,0)(4,0)(5,0)(6,0)(7,0)(8,0)(9,0)(10,0)(11,0)(12,0)(13,0)(14,0)(15,0)(16,0)(17,0)(18,0)(19,0)
\psset{linecolor=myc2}
\psline{-}(17,0)(17,1)
\psline{-}(18,0)(18,1)
\psline{-}(19,0)(19,1)
\psarc(1.5,0){0.5}{0}{180}
\psbezier{-}(3,0)(3,2.5)(8,2.5)(8,0)
\psbezier{-}(4,0)(4,1.5)(7,1.5)(7,0)
\psarc(5.5,0){0.5}{0}{180}
\psarc(10.5,0){0.5}{0}{180}
\psarc(13.5,0){0.5}{0}{180}
\psarc(15.5,0){0.5}{0}{180}
\psbezier{-}(9,0)(9,1.5)(12,1.5)(12,0)
\end{pspicture})  +
P^r(\begin{pspicture}(0.1,-0.0)(3.90,0.2)
\psset{unit=0.2}
\psdots[dotsize=2.5pt](1,0)(2,0)(3,0)(4,0)(5,0)(6,0)(7,0)(8,0)(9,0)(10,0)(11,0)(12,0)(13,0)(14,0)(15,0)(16,0)(17,0)(18,0)(19,0)
\psset{linecolor=myc2}
\psline{-}(17,0)(17,1)
\psline{-}(18,0)(18,1)
\psline{-}(19,0)(19,1)
\psarc(3.5,0){0.5}{0}{180}
\psarc(5.5,0){0.5}{0}{180}
\psarc(10.5,0){0.5}{0}{180}
\psarc(13.5,0){0.5}{0}{180}
\psarc(15.5,0){0.5}{0}{180}
\psbezier{-}(1,0)(1,2.5)(8,2.5)(8,0)
\psbezier{-}(2,0)(2,1.5)(7,1.5)(7,0)
\psbezier{-}(9,0)(9,1.5)(12,1.5)(12,0)
\end{pspicture})
\Big) \\ &\quad + P^r(e_iy/\{i\}=\begin{pspicture}(0.1,-0.0)(3.90,0.2)
\psset{unit=0.2}
\psdots[dotsize=2.5pt](1,0)(2,0)(3,0)(4,0)(5,0)(6,0)(7,0)(8,0)(9,0)(10,0)(11,0)(12,0)(13,0)(14,0)(15,0)(16,0)(17,0)(18,0)(19,0)
\psset{linecolor=myc2}
\psline{-}(17,0)(17,1)
\psline{-}(18,0)(18,1)
\psline{-}(19,0)(19,1)
\psline{-}(7,0)(7,1)
\psline{-}(8,0)(8,1)
\psarc(1.5,0){0.5}{0}{180}
\psarc(3.5,0){0.5}{0}{180}
\psarc(5.5,0){0.5}{0}{180}
\psarc(10.5,0){0.5}{0}{180}
\psarc(13.5,0){0.5}{0}{180}
\psarc(15.5,0){0.5}{0}{180}
\psbezier{-}(9,0)(9,1.5)(12,1.5)(12,0)
\end{pspicture})\\ &\quad +\Big( P^r(\begin{pspicture}(0.1,-0.0)(3.90,0.2)
\psset{unit=0.2}
\psdots[dotsize=2.5pt](1,0)(2,0)(3,0)(4,0)(5,0)(6,0)(7,0)(8,0)(9,0)(10,0)(11,0)(12,0)(13,0)(14,0)(15,0)(16,0)(17,0)(18,0)(19,0)
\psset{linecolor=myc2}
\psline{-}(17,0)(17,1)
\psline{-}(18,0)(18,1)
\psline{-}(19,0)(19,1)
\psarc(1.5,0){0.5}{0}{180}
\psarc(3.5,0){0.5}{0}{180}
\psarc(5.5,0){0.5}{0}{180}
\psarc(8.5,0){0.5}{0}{180}
\psarc(10.5,0){0.5}{0}{180}
\psarc(13.5,0){0.5}{0}{180}
\psarc(15.5,0){0.5}{0}{180}
\psbezier{-}(7,0)(7,2.0)(12,2.0)(12,0)
\end{pspicture}) +P^r(\begin{pspicture}(0.1,-0.0)(3.90,0.2)
\psset{unit=0.2}
\psdots[dotsize=2.5pt](1,0)(2,0)(3,0)(4,0)(5,0)(6,0)(7,0)(8,0)(9,0)(10,0)(11,0)(12,0)(13,0)(14,0)(15,0)(16,0)(17,0)(18,0)(19,0)
\psset{linecolor=myc2}
\psline{-}(17,0)(17,1)
\psline{-}(18,0)(18,1)
\psline{-}(19,0)(19,1)
\psarc(1.5,0){0.5}{0}{180}
\psarc(3.5,0){0.5}{0}{180}
\psarc(5.5,0){0.5}{0}{180}
\psarc(10.5,0){0.5}{0}{180}
\psarc(15.5,0){0.5}{0}{180}
\psbezier{-}(8,0)(8,2.5)(13,2.5)(13,0)
\psbezier{-}(7,0)(7,3.5)(14,3.5)(14,0)
\psbezier{-}(9,0)(9,1.5)(12,1.5)(12,0)
\end{pspicture})+P^r(\begin{pspicture}(0.1,-0.0)(3.90,0.2)
\psset{unit=0.2}
\psdots[dotsize=2.5pt](1,0)(2,0)(3,0)(4,0)(5,0)(6,0)(7,0)(8,0)(9,0)(10,0)(11,0)(12,0)(13,0)(14,0)(15,0)(16,0)(17,0)(18,0)(19,0)
\psset{linecolor=myc2}
\psline{-}(17,0)(17,1)
\psline{-}(18,0)(18,1)
\psline{-}(19,0)(19,1)
\psarc(1.5,0){0.5}{0}{180}
\psarc(3.5,0){0.5}{0}{180}
\psarc(5.5,0){0.5}{0}{180}
\psarc(10.5,0){0.5}{0}{180}
\psarc(13.5,0){0.5}{0}{180}
\psbezier{-}(9,0)(9,1.5)(12,1.5)(12,0)
\psbezier{-}(8,0)(8,2.5)(15,2.5)(15,0)
\psbezier{-}(7,0)(7,3.5)(16,3.5)(16,0)
\end{pspicture})
\Big) 
\end{align*}
Using Lemma \ref{sec:multiB}, all terms of type $y''$ can be brought down to $P^r_{e_iy}$. Their sum, with $\beta P^r_{e_iy}$ included, is 
$$P^r_{e_iy}\left(-\frac{S_2}{S_1}+\frac{S_1^2}{S_1S_2}+\dots+\frac{S_1^2}{S_{(i-1)/2}S_{(i+1)/2}}\right)=-P^r_{e_iy}A_{(i+1)/2}$$
where the coefficients $A_b=S_{b+1}/S_b$ was introduced in Lemma \ref{lem:A9}. The equation for the matrix elements is therefore
$$0=P^r_y-P^r_{e_iy}A_{(i+1)/2}+P^r_{e_iy/\{i\}}+\sum_{y'}P^r_{y'}$$
and allows to express $P^r_y$ as a function of coefficients in $P^rv^r$ of the states $e_iy$, $e_iy/\{i\}$ and $y'$ which are all smaller than $y$ for the partial order $<$.

Now is the time to address the factor $\prod_iS_i/S_{k_i}$ left aside at the beginning. The states $e_iy/\{i\}$ and $y'$ start on the left with the same number of bubbles as $y$ does. Through the use of Lemma \ref{sec:multiB}, the left cluster of the original state $Y$ can be reconstructed using this product; the coefficients of the corresponding terms will be $1$ and the new states are smaller than $Y$. The state $P^r_{e_iy}$ contains an additional $1$-bubble joining $i$ and $i+1$. A large bubble from $1$ to $i+1$ is first drawn to get rid of the denominator $S_{(i+1)/2}$ in $A_{(i+1)/2}$. Then, within it, the original left cluster of the state $Y$ is reconstructed, taking care of $\prod_iS_i/S_{k_i}$. The corresponding state has an analytic factor, namely $S_{(i+1)/2+1}$, and is smaller than $Y$. So, as claimed, $P^r_Y=\sum_{w<Y}\beta_w P^r_w$ form some analytic $\beta_w$'s.

The example given above has only two clusters separated by one defect. What if, to the right of the second cluster, there are others? As for the case with a patch with more than one defects, the other clusters are left untouched by the procedure above. Only the cluster starting at $1$ and the {\em closed} cluster starting at $i+1$ are transformed by the computation of $\langle e_iy|e_iP^rv^r\rangle$. Because the other clusters remain as they are in the starting state $y$, all new terms produced during the computation are smaller than $y$ as required. \hfill$\square$

%
%

\section{Matrix elements of $\rho(F_N(\lambda))$}\label{app:b}

In this appendix, we describe a method for calculating any matrix element of $\rho(F_N(\lambda))$. From the simple graphical identity \eqref{eq:semicircle}, we know that every half arc in $u = v^N_{n_1, n_2, ..., n_k}$ propagates down through $F_N(\lambda)$. The only $w$'s for which the matrix element $\langle w|\,\rho(F_N(\lambda)) u \rangle$ does not vanish are those that contain all arcs in $u$ and, maybe, additional ones. They are then of the form $w=v^N_{n_1, n_2, ..., n_k, m_1, m_2, ..., m_j}$. (The notation introduced in \ref{sub:notation} is used throughout this appendix.) Calculating off-diagonal elements of $\rho(F_N(\lambda))$ amounts to calculating the last column of $\rho(F_r(\lambda))$ for all $r$ smaller or equal to $N$, but with the same parity, or more specifically, $\langle v^r_{\ell_1, \ell_2, ..., \ell_n}|\,\rho(F_r(\lambda)) v^{r} \rangle$ where, again, $v^r$ stands for the link vector with $r$ defects ($\in B_r^r$). 

The result is expressed in terms of another notation for the link vector. To each element $w \in B_N$ is associated a set $\eta(w)$ of $N$ integers $\in \{ -1,0,1\}$. Going from left to right along the $N$ entries of the link state $w$, we assign a ``$0$'' for a defect, a ``$1$'' for the beginning of a bubble and a ``$-1$'' for the end of bubble. The examples given in section \ref{sub:notation} have the following notation $\eta$:
\begin{alignat*}{3}
\eta (v^{10}_{3,7}) & =  \{ 0,0,1,-1,0,0,1,-1,0,0 \}, & \qquad \eta(v^{10}_{2,2,7,7}) & = \{1,1,-1,-1,0,1,1,-1,-1,0\},\\ 
\eta(v^{10}_{2,6,8,7})  & =  \{0,1,-1,0,1,1,-1,1,-1,-1\}, & \qquad \eta(v^{10}_{3,5,4,8,5}) & = \{1,1,1,-1,1,-1,-1,1,-1,-1\} .
\end{alignat*}

\begin{Proposition} Let $w \in B^r$. Then the matrix element of $\rho(F_r(\lambda))$ between $w$ and $v^r$ is
\begin{equation} 
\langle w | \rho(F_r(\lambda)) v^r \rangle = I ^\dagger \ \Big( \prod_{1\le k\le r} N_{\eta(w)_k} \Big) \ G \ I \label{eq:8par8}\end{equation}
where the matrices in the product are ordered from left to right as $k$ increases and the matrices $G$, $N_i$, $i\in\{-1,0,1\}$, and vector $I$ are
\begin{equation*}N_0=\begin{pmatrix} 
		 -e^{i\lambda} & 1-e^{-2i\lambda} & 0 & 0 & 0 & 0 & 0 & 0 \\
                    0 & -e^{-i\lambda} & 0 & 1 & 0 & 0 & 0 & 0 \\
                    -e^{i\lambda} & 1-e^{-i\lambda} & 1 & 0 & 0 & 0 & 0 & 0 \\
                    0 & 0 & 0 & 1 & 0 & 0 & 0 & 0 \\
                    0 & 0 & 0 & 0 & 0 & 0 & 0 & 0 \\
                    0 & 0 & 0 & 0 & 0 & 0 & 0 & 0 \\
                    0 & 0 & 0 & 0 & 0 & 0 & 0 & 0 \\
                    0 & -e^{-i\lambda} & 0 & 0 & 0 & 0 & 0 & 0 \\
\end{pmatrix},
\qquad
N_1=\begin{pmatrix} 
		 0 & 0 & 0 & 0 & 1-e^{-2i\lambda} & 0 & 0 & 0 \\
                    0 & 0 & 0 & 0 & -e^{-i\lambda} & 1 & 0 & 0 \\
                    0 & 0 & 0 & 0 & 1-e^{-i\lambda} & 0 & 0 & 0 \\
                    0 & 0 & 0 & 0 & 0 & 1 & 0 & 0 \\
                    0 & 0 & 0 & 0 & 0 & 0 & 1 & 0 \\
                    0 & 0 & 0 & 0 & 0 & 0 & 1 & 0 \\
                    0 & 0 & 0 & 0 & 0 & 0 & 0 & 0 \\
                    0 & 0 & 0 & 0 & -e^{-i\lambda} & 0 & 0 & 0 \\
\end{pmatrix},
\end{equation*}
\begin{equation*}N_{-1}=\begin{pmatrix} 0 & 0 & 0 & 0 & 0 & 0 & 0 & 0 \\
                     0 & 0 & 0 & 0 & 0 & 0 & 0 & 0 \\
                    0 & 0 & 0 & 0 & 0 & 0 & 0 & 0 \\
                    0 & 0 & 0 & 0 & 0 & 0 & 0 & 0 \\
                    -e^{i\lambda} & 0 & 1 & 0 & 0 & 0 & 0 & 1 \\
                    -e^{i\lambda} & 1 & 0 & 0 & 0 & 0 & 0 & 1 \\
                    0 & 0 & 0 & 0 & 1 & 0 & 0 & 0 \\
                    0 & 0 & 0 & 0 & 0 & 0 & 0 & 0 \\
\end{pmatrix},
\qquad
G=\begin{pmatrix} 2 \cos(\lambda) & 1 & 1 & 0 & 0 & 0 & 0 & 1 \\
                     1 & 0 & 1 & 0 & 0 & 0 & 0 & 0 \\
                    1 & 1 & 1 & 1 & 0 & 0 & 0 & 0 \\
                    0 & 0 & 1 & 0 & 0 & 0 & 0 & 0 \\
                    0 & 0 & 0 & 0 & 1 & 1 & 0 & 0 \\
                    0 & 0 & 0 & 0 & 1 & 0 & 0 & 0 \\
                    0 & 0 & 0 & 0 & 0 & 0 & 1 & 0 \\
                    1 & 0 & 0 & 0 & 0 & 0 & 0 & 0 \\
\end{pmatrix},
\qquad
I=\begin{pmatrix} 1 \\ 0 \\ 0 \\ 0 \\ 0 \\ 0 \\ 0 \\ 0 \end{pmatrix}.
\end{equation*}
\label{sec:elementFN}
\end{Proposition}
\noindent{\scshape Proof\ \ }The strategy will be to sum the braid boxes in $F_r$ from left to right, using this identity:
\begin{equation*}
\psset{unit=0.8}\begin{pspicture}(-0.6,-0.13)(1.2,1.0)\lw
\psline{-}(0,-1)(0,1)(1,1)(1,-1)(0,-1)
\psline{-}(0,0)(1,0)
\psset{linecolor=myc}\unlw
\psline{-}(0,-0.5)(1,-0.5)
\psline{-}(0,0.5)(0.35,0.5)\psline{-}(0.65,0.5)(1,0.5)
\psline{-}(0.5,0)(0.5,1)
\psline{-}(0.5,0)(0.5,-0.35)\psline{-}(0.5,-0.65)(0.5,-1)
\end{pspicture}
= (-e^{-i \lambda})
\begin{pspicture}(-0.2,-0.13)(1.2,1.0)\lw
\psline{-}(0,-1)(0,1)(1,1)(1,-1)(0,-1)
\psline{-}(0,0)(1,0)
\psset{linecolor=myc}\unlw
\psarc{-}(0,0){0.5}{270}{90}
\psarc{-}(1,1){0.5}{180}{270}
\psarc{-}(1,-1){0.5}{90}{180}
\end{pspicture}
+(-e^{i \lambda})
\begin{pspicture}(-0.2,-0.13)(1.2,1.0)\lw
\psline{-}(0,-1)(0,1)(1,1)(1,-1)(0,-1)
\psline{-}(0,0)(1,0)
\psset{linecolor=myc}\unlw
\psarc{-}(1,0){0.5}{90}{270}
\psarc{-}(0,1){0.5}{270}{360}
\psarc{-}(0,-1){0.5}{0}{90}
\end{pspicture}
\  + \ 1 \
\begin{pspicture}(-0.2,-0.13)(1.2,1.0)\lw
\psline{-}(0,-1)(0,1)(1,1)(1,-1)(0,-1)
\psline{-}(0,0)(1,0)
\psset{linecolor=myc}\unlw
\psarc{-}(0,0){0.5}{0}{90}
\psarc{-}(0,-1){0.5}{0}{90}
\psarc{-}(1,0){0.5}{180}{270}
\psarc{-}(1,1){0.5}{180}{270}
\end{pspicture}
\ + \ 1 \
\begin{pspicture}(-0.2,-0.13)(1.2,1.2)\lw
\psline{-}(0,-1)(0,1)(1,1)(1,-1)(0,-1)
\psline{-}(0,0)(1,0)
\psset{linecolor=myc}\unlw
\psarc{-}(0,0){0.5}{270}{360}
\psarc{-}(0,1){0.5}{270}{360}
\psarc{-}(1,0){0.5}{90}{180}
\psarc{-}(1,-1){0.5}{90}{180}
\end{pspicture}
\end{equation*}
\\The matrix element $\langle v^4_{2} |\,\rho(F_4(\lambda)) v^4 \rangle$ will serve as an example. Using the previous identity on the first two boxes, we obtain:
\begin{equation}
\psset{unit=0.8}\begin{pspicture}(-0.6,-0.1)(4.7,1.8)
\psdots(0.5,1)(1.5,1)(2.5,1)(3.5,1)
\psdots(0.5,-1)(1.5,-1)(2.5,-1)(3.5,-1)
\lw
\psline{-}(0,-1)(0,1)(4,1)(4,-1)(0,-1)
\psline{-}(0,0)(4,0)
\psline{-}(1,-1)(1,1)\psline{-}(2,-1)(2,1)\psline{-}(3,-1)(3,1)
\psline{-}(3,1)(4,1)\psline{-}(3,0)(4,0)\psline{-}(3,-1)(4,-1)
\psset{linecolor=myc}\unlw
\psarc{-}(0,0){0.5}{90}{270}
\psarc{-}(4,0){0.5}{270}{90}
\psline{-}(0,-0.5)(4,-0.5)
\psline{-}(0,0.5)(0.35,0.5)\psline{-}(0.65,0.5)(1,0.5)
\psline{-}(1.0,0.5)(1.35,0.5)\psline{-}(1.65,0.5)(2,0.5)
\psline{-}(2.0,0.5)(2.35,0.5)\psline{-}(2.65,0.5)(3,0.5)
\psline{-}(3.0,0.5)(3.35,0.5)\psline{-}(3.65,0.5)(4,0.5)
\psline{-}(0.5,0)(0.5,1)\psline{-}(1.5,0)(1.5,1)
\psline{-}(2.5,0)(2.5,1)\psline{-}(3.5,0)(3.5,1)
\psline{-}(0.5,0)(0.5,-0.35)\psline{-}(0.5,-0.65)(0.5,-1)
\psline{-}(1.5,0)(1.5,-0.35)\psline{-}(1.5,-0.65)(1.5,-1)
\psline{-}(2.5,0)(2.5,-0.35)\psline{-}(2.5,-0.65)(2.5,-1)
\psline{-}(3.5,0)(3.5,-0.35)\psline{-}(3.5,-0.65)(3.5,-1)
\psset{linecolor=myc2}
\psline{-}(0.5,1)(0.5,1.8)
\psline{-}(1.5,1)(1.5,1.8)
\psline{-}(2.5,1)(2.5,1.8)
\psline{-}(3.5,1)(3.5,1.8)
\psset{linecolor=myc3,linestyle=dashed,dash=2pt 2pt}
\psline{-}(0.5,-1)(0.5,-1.8)
\psline{-}(3.5,-1)(3.5,-1.8)
\psarc{-}(2,-1){0.5}{180}{360}
\end{pspicture}
= -e^{i \lambda}
\begin{pspicture}(0.3,-0.1)(4.7,1.8)
\psdots(1.5,1)(2.5,1)(3.5,1)
\psdots(1.5,-1)(2.5,-1)(3.5,-1)
\lw
\psline{-}(1,-1)(1,1)(4,1)(4,-1)(1,-1)
\psline{-}(1,0)(4,0)
\psline{-}(2,-1)(2,1)\psline{-}(3,-1)(3,1)
\psline{-}(3,1)(4,1)\psline{-}(3,0)(4,0)\psline{-}(3,-1)(4,-1)
\psset{linecolor=myc}\unlw
\psarc{-}(1,0){0.5}{90}{270}
\psarc{-}(4,0){0.5}{270}{90}
\psline{-}(1,-0.5)(4,-0.5)
\psline{-}(1.0,0.5)(1.35,0.5)\psline{-}(1.65,0.5)(2,0.5)
\psline{-}(2.0,0.5)(2.35,0.5)\psline{-}(2.65,0.5)(3,0.5)
\psline{-}(3.0,0.5)(3.35,0.5)\psline{-}(3.65,0.5)(4,0.5)
\psline{-}(1.5,0)(1.5,1)
\psline{-}(2.5,0)(2.5,1)\psline{-}(3.5,0)(3.5,1)
\psline{-}(1.5,0)(1.5,-0.35)\psline{-}(1.5,-0.65)(1.5,-1)
\psline{-}(2.5,0)(2.5,-0.35)\psline{-}(2.5,-0.65)(2.5,-1)
\psline{-}(3.5,0)(3.5,-0.35)\psline{-}(3.5,-0.65)(3.5,-1)
\psset{linecolor=myc2}
\psline{-}(1.5,1)(1.5,1.8)
\psline{-}(2.5,1)(2.5,1.8)
\psline{-}(3.5,1)(3.5,1.8)
\psset{linecolor=myc3,linestyle=dashed,dash=2pt 2pt}
\psline{-}(3.5,-1)(3.5,-1.8)
\psarc{-}(2,-1){0.5}{180}{360}
\end{pspicture}
+ (\underbrace{ 2 - \beta e^{-i \lambda}}_{1-e^{-2i \lambda}} )
\begin{pspicture}(0.3,-0.1)(4.7,1.8)
\psdots(1.5,1)(2.5,1)(3.5,1)
\psdots(1.5,-1)(2.5,-1)(3.5,-1)
\lw
\psline{-}(1,-1)(1,1)(4,1)(4,-1)(1,-1)
\psline{-}(1,0)(4,0)
\psline{-}(2,-1)(2,1)\psline{-}(3,-1)(3,1)
\psline{-}(3,1)(4,1)\psline{-}(3,0)(4,0)\psline{-}(3,-1)(4,-1)
\psset{linecolor=myc}\unlw
\psarc{-}(4,0){0.5}{270}{90}
\psline{-}(1,-0.5)(4,-0.5)
\psline{-}(1.0,0.5)(1.35,0.5)\psline{-}(1.65,0.5)(2,0.5)
\psline{-}(2.0,0.5)(2.35,0.5)\psline{-}(2.65,0.5)(3,0.5)
\psline{-}(3.0,0.5)(3.35,0.5)\psline{-}(3.65,0.5)(4,0.5)
\psline{-}(1.5,0)(1.5,1)
\psline{-}(2.5,0)(2.5,1)\psline{-}(3.5,0)(3.5,1)
\psline{-}(1.5,0)(1.5,-0.35)\psline{-}(1.5,-0.65)(1.5,-1)
\psline{-}(2.5,0)(2.5,-0.35)\psline{-}(2.5,-0.65)(2.5,-1)
\psline{-}(3.5,0)(3.5,-0.35)\psline{-}(3.5,-0.65)(3.5,-1)
\psset{linecolor=myc2}
\psline{-}(0.5,1)(0.5,1.8)
\psline{-}(1.5,1)(1.5,1.8)
\psline{-}(2.5,1)(2.5,1.8)
\psline{-}(3.5,1)(3.5,1.8)
\psarc{-}(1,1){0.5}{180}{270}
\psset{linecolor=myc3,linestyle=dashed,dash=2pt 2pt}
\psline{-}(0.5,-1)(0.5,-1.8)
\psline{-}(3.5,-1)(3.5,-1.8)
\psarc{-}(2,-1){0.5}{180}{360}
\psarc{-}(1,-1){0.5}{90}{180}
\end{pspicture}\label{eq:exemple1}
\end{equation}
\\
\\
Summing the first two boxes of $F_r(\lambda)$, we obtain a linear combination of new objects made of $(r-1) \times 2$ boxes, with different boundary condition on the left side. If we repeat the process on the second term, two terms are found to be zero due to eq.~(\ref{eq:dotprod}) and we get: 
\begin{equation}
\psset{unit=0.8}
\begin{pspicture}(0.3,-0.1)(4.7,1.8)
\psdots(1.5,1)(2.5,1)(3.5,1)
\psdots(1.5,-1)(2.5,-1)(3.5,-1)
\lw
\psline{-}(1,-1)(1,1)(4,1)(4,-1)(1,-1)
\psline{-}(1,0)(4,0)
\psline{-}(2,-1)(2,1)\psline{-}(3,-1)(3,1)
\psline{-}(3,1)(4,1)\psline{-}(3,0)(4,0)\psline{-}(3,-1)(4,-1)
\psset{linecolor=myc}\unlw
\psarc{-}(4,0){0.5}{270}{90}
\psline{-}(1,-0.5)(4,-0.5)
\psline{-}(1.0,0.5)(1.35,0.5)\psline{-}(1.65,0.5)(2,0.5)
\psline{-}(2.0,0.5)(2.35,0.5)\psline{-}(2.65,0.5)(3,0.5)
\psline{-}(3.0,0.5)(3.35,0.5)\psline{-}(3.65,0.5)(4,0.5)
\psline{-}(1.5,0)(1.5,1)
\psline{-}(2.5,0)(2.5,1)\psline{-}(3.5,0)(3.5,1)
\psline{-}(1.5,0)(1.5,-0.35)\psline{-}(1.5,-0.65)(1.5,-1)
\psline{-}(2.5,0)(2.5,-0.35)\psline{-}(2.5,-0.65)(2.5,-1)
\psline{-}(3.5,0)(3.5,-0.35)\psline{-}(3.5,-0.65)(3.5,-1)
\psset{linecolor=myc2}
\psline{-}(0.5,1)(0.5,1.8)
\psline{-}(1.5,1)(1.5,1.8)
\psline{-}(2.5,1)(2.5,1.8)
\psline{-}(3.5,1)(3.5,1.8)
\psarc{-}(1,1){0.5}{180}{270}
\psset{linecolor=myc3,linestyle=dashed,dash=2pt 2pt}
\psline{-}(0.5,-1)(0.5,-1.8)
\psline{-}(3.5,-1)(3.5,-1.8)
\psarc{-}(2,-1){0.5}{180}{360}
\psarc{-}(1,-1){0.5}{90}{180}
\end{pspicture} = (-e^{-i\lambda})
\begin{pspicture}(1.3,-0.1)(4.7,1.8)
\psdots(2.5,1)(3.5,1)
\psdots(2.5,-1)(3.5,-1)
\lw
\psline{-}(2,-1)(2,1)(4,1)(4,-1)(2,-1)
\psline{-}(2,0)(4,0)
\psline{-}(2,-1)(2,1)\psline{-}(3,-1)(3,1)
\psline{-}(3,1)(4,1)\psline{-}(3,0)(4,0)\psline{-}(3,-1)(4,-1)
\psset{linecolor=myc}\unlw
\psarc{-}(4,0){0.5}{270}{90}
\psline{-}(2,-0.5)(4,-0.5)
\psline{-}(2.0,0.5)(2.35,0.5)\psline{-}(2.65,0.5)(3,0.5)
\psline{-}(3.0,0.5)(3.35,0.5)\psline{-}(3.65,0.5)(4,0.5)
\psline{-}(2.5,0)(2.5,1)\psline{-}(3.5,0)(3.5,1)
\psline{-}(2.5,0)(2.5,-0.35)\psline{-}(2.5,-0.65)(2.5,-1)
\psline{-}(3.5,0)(3.5,-0.35)\psline{-}(3.5,-0.65)(3.5,-1)
\psset{linecolor=myc2}
\psline{-}(1.5,1)(1.5,1.8)
\psline{-}(2.5,1)(2.5,1.8)
\psline{-}(3.5,1)(3.5,1.8)
\psarc{-}(2,1){0.5}{180}{270}
\psset{linecolor=myc3,linestyle=dashed,dash=2pt 2pt}
\psline{-}(3.5,-1)(3.5,-1.8)
\psarc{-}(2,-1){0.5}{90}{360}
\end{pspicture}+
\begin{pspicture}(0.3,-0.1)(4.7,1.8)
\psdots(2.5,1)(3.5,1)
\psdots(2.5,-1)(3.5,-1)
\lw
\psline{-}(2,-1)(2,1)(4,1)(4,-1)(2,-1)
\psline{-}(2,0)(4,0)
\psline{-}(2,-1)(2,1)\psline{-}(3,-1)(3,1)
\psline{-}(3,1)(4,1)\psline{-}(3,0)(4,0)\psline{-}(3,-1)(4,-1)
\psset{linecolor=myc}\unlw
\psarc{-}(4,0){0.5}{270}{90}
\psline{-}(2,-0.5)(4,-0.5)
\psline{-}(2.0,0.5)(2.35,0.5)\psline{-}(2.65,0.5)(3,0.5)
\psline{-}(3.0,0.5)(3.35,0.5)\psline{-}(3.65,0.5)(4,0.5)
\psline{-}(2.5,0)(2.5,1)\psline{-}(3.5,0)(3.5,1)
\psline{-}(2.5,0)(2.5,-0.35)\psline{-}(2.5,-0.65)(2.5,-1)
\psline{-}(3.5,0)(3.5,-0.35)\psline{-}(3.5,-0.65)(3.5,-1)
\psset{linecolor=myc2}
\psline{-}(2.5,1)(2.5,1.8)
\psline{-}(3.5,1)(3.5,1.8)
\psset{linecolor=myc3,linestyle=dashed,dash=2pt 2pt}
\psline{-}(0.5,-1.1)(0.5,-1.8)
\psline{-}(3.5,-1)(3.5,-1.8)
\psarc{-}(2,-1){0.5}{90}{360}
\psarc{-}(2,-1){1.5}{90}{180}
\end{pspicture}
\label{eq:exemple2}
\end{equation}
\\
\\
\\
We are now left with a combination of $(r-2) \times 2$ boxes. The incoming straight lines (or curves) represent the defects of $v^r$ and the dashed ones represent where the defects and arcs of $w$ are. The same procedure can be repeated until all the sums in the braid boxes are performed. At each step, the pattern of straight and dashed lines (or curves) entering the leftmost pair of braid boxes is one of the following eight ``vectors'':
\begin{equation*}
\mathcal L=\Bigg\{
\psset{unit=0.7}
\begin{pspicture}(-0.75,-0.13)(0.4,1.8)\lw
\psline{-}(0,-1)(0,1)\unlw
\psset{linecolor=myc}
\psarc{-}(0,0){0.5}{90}{270}
\psset{linecolor=myc2}
\psset{linecolor=myc3,linestyle=dashed,dash=2pt 2pt}
\end{pspicture}\textrm{,}
\begin{pspicture}(-0.75,-0.13)(0.4,1.8)\lw
\psline{-}(0,-1)(0,1)\unlw
\psset{linecolor=myc}
\psset{linecolor=myc2}
\psarc{-}(0,1){0.5}{180}{270}
\psline{-}(-0.5,1)(-0.5,1.8)
\psset{linecolor=myc3,linestyle=dashed,dash=2pt 2pt}
\psarc{-}(0,-1){0.5}{90}{180}
\psline{-}(-0.5,-1.1)(-0.5,-1.8)
\end{pspicture}\textrm{,}
\begin{pspicture}(-1.75,-0.13)(0.4,1.8)\lw
\psline{-}(0,-1)(0,1)\unlw
\psset{linecolor=myc}
\psset{linecolor=myc2}
\psarc{-}(0,1){0.5}{180}{270}
\psline{-}(-0.5,1)(-0.5,1.8)
\psarc{-}(0,1){1.5}{180}{270}
\psline{-}(-1.5,1)(-1.5,1.8)
\psset{linecolor=myc3,linestyle=dashed,dash=2pt 2pt}
\end{pspicture}\textrm{,}
\begin{pspicture}(-1.75,-0.13)(0.4,1.8)
\lw
\psline{-}(0,-1)(0,1)\unlw
\psset{linecolor=myc}
\psset{linecolor=myc2}
\psset{linecolor=myc3,linestyle=dashed,dash=2pt 2pt}
\psarc{-}(0,-1){0.5}{90}{180}
\psline{-}(-0.5,-1.1)(-0.5,-1.8)
\psarc{-}(0,-1){1.5}{90}{180}
\psline{-}(-1.5,-1.1)(-1.5,-1.8)
\end{pspicture}\textrm{,}
\begin{pspicture}(-0.75,-0.13)(0.4,1.8)\lw
\psline{-}(0,-1)(0,1)\unlw
\psset{linecolor=myc}
\psset{linecolor=myc2}
\psarc{-}(0,1){0.5}{180}{270}
\psline{-}(-0.5,1)(-0.5,1.8)
\psset{linecolor=myc3,linestyle=dashed,dash=2pt 2pt}
\psarc{-}(0,-1){0.5}{90}{270}
\end{pspicture}\textrm{,}
\begin{pspicture}(-1.75,-0.13)(0.4,1.8)
\lw
\psline{-}(0,-1)(0,1)\unlw
\psset{linecolor=myc}
\psset{linecolor=myc2}
\psset{linecolor=myc3,linestyle=dashed,dash=2pt 2pt}
\psarc{-}(0,-1){0.5}{90}{270}
\psarc{-}(0,-1){1.5}{90}{180}
\psline{-}(-1.5,-1.1)(-1.5,-1.8)
\end{pspicture}\textrm{,}
\begin{pspicture}(-1.75,-0.13)(0.4,1.8)
\lw
\psline{-}(0,-1)(0,1)\unlw
\psset{linecolor=myc}
\psset{linecolor=myc2}
\psset{linecolor=myc3,linestyle=dashed,dash=2pt 2pt}
\psarc{-}(0,-1){0.5}{90}{270}
\psarc{-}(0,-1){1.5}{90}{270}
\end{pspicture}\textrm{,}
\begin{pspicture}(-0.75,-0.13)(0.4,1.8)
\lw
\psline{-}(0,-1)(0,1)\unlw
\psset{linecolor=myc}
\psset{linecolor=myc2}
\psset{linecolor=myc3,linestyle=dashed,dash=2pt 2pt}
\psarc{-}(0,0){0.5}{90}{270}
\end{pspicture}
\Bigg\}\textrm{.}
\end{equation*}
\\ \\ \\
The dashed curves represent the states of $w$, either a defect if it ends as a line or an arc if it is a half-circle. One can understand the leftmost pair of braid boxes as acting linearly on one of these eight basis vectors of the basis $\mathcal L$. This action depends on the states entering this new pair of braids and three linear transformations must be distinguished: 
\begin{equation*}
\psset{unit=0.8}
\begin{pspicture}(-1,-0.13)(1.4,1.8)
\rput(-1,0){$N_0=$}
\psdots(0.5,1)(0.5,-1)\lw
\psline{-}(0,-1)(0,1)(1,1)(1,-1)(0,-1)
\psline{-}(0,0)(1,0)
\psset{linecolor=myc}\unlw
\psline{-}(0,-0.5)(1,-0.5)
\psline{-}(0,0.5)(0.35,0.5)\psline{-}(0.65,0.5)(1,0.5)
\psline{-}(0.5,0)(0.5,1)
\psline{-}(0.5,0)(0.5,-0.35)\psline{-}(0.5,-0.65)(0.5,-1)
\psset{linecolor=myc2}
\psline{-}(0.5,1)(0.5,1.8)
\psset{linecolor=myc3,linestyle=dashed,dash=2pt 2pt}
\psline{-}(0.5,-1)(0.5,-1.8)
\end{pspicture}
\begin{pspicture}(-1.5,-0.13)(1.6,1.8)
\rput(-1,0){, $N_1=$}
\psdots(0.5,1)(0.5,-1)\lw
\psline{-}(0,-1)(0,1)(1,1)(1,-1)(0,-1)
\psline{-}(0,0)(1,0)
\psset{linecolor=myc}\unlw
\psline{-}(0,-0.5)(1,-0.5)
\psline{-}(0,0.5)(0.35,0.5)\psline{-}(0.65,0.5)(1,0.5)
\psline{-}(0.5,0)(0.5,1)
\psline{-}(0.5,0)(0.5,-0.35)\psline{-}(0.5,-0.65)(0.5,-1)
\psset{linecolor=myc2}
\psline{-}(0.5,1)(0.5,1.8)
\psset{linecolor=myc3,linestyle=dashed,dash=2pt 2pt}
\psarc{-}(1,-1){0.5}{180}{270}
\end{pspicture}
\begin{pspicture}(-2.5,-0.13)(1.4,1.8)
\rput(-1.5,0){and $N_{-1}=$}
\psdots(0.5,1)(0.5,-1)\lw
\psline{-}(0,-1)(0,1)(1,1)(1,-1)(0,-1)
\psline{-}(0,0)(1,0)
\psset{linecolor=myc}\unlw
\psline{-}(0,-0.5)(1,-0.5)
\psline{-}(0,0.5)(0.35,0.5)\psline{-}(0.65,0.5)(1,0.5)
\psline{-}(0.5,0)(0.5,1)
\psline{-}(0.5,0)(0.5,-0.35)\psline{-}(0.5,-0.65)(0.5,-1)
\psset{linecolor=myc2}
\psline{-}(0.5,1)(0.5,1.8)
\psset{linecolor=myc3,linestyle=dashed,dash=2pt 2pt}
\psarc{-}(0,-1){0.5}{270}{360}
\end{pspicture}
\textrm{.}
\end{equation*}
\\ \\ \\
The three linear transformations can be expressed in the basis $\mathcal L$, with the action given by right multiplication on a row vector of components. The two examples \eqref{eq:exemple1} and \eqref{eq:exemple2} have computed one line of each $N_0$ and $N_1$ in this basis:
\begin{itemize}
\item eq.~(\ref{eq:exemple1}) is $\begin{pmatrix} 1 & 0 & 0 & 0 & 0 & 0 & 0 & 0 \end{pmatrix} \cdot N_0 = \begin{pmatrix} -e^{i \lambda} & 1-e^{-2i \lambda} & 0 & 0 & 0 & 0 & 0 & 0 \end{pmatrix}$;
\item eq.~(\ref{eq:exemple2}) is $\begin{pmatrix} 0 & 1 & 0 & 0 & 0 & 0 & 0 & 0 \end{pmatrix} \cdot N_1 = \begin{pmatrix} 0 & 0 & 0 & 0 & -e^{-i \lambda} & 1 & 0 & 0 \end{pmatrix}$.
\end{itemize}
A direct computation gives the three matrices in the statement.
After the procedure has been applied $r$ times, the row vector is $ I^\dagger  \ \left( \prod_{1\le k\le r} N_{\eta(w)_k} \right)$. All boxes have been summed and it remains to connect the state on left and with the half-circle on the right, using the rules \eqref{eq:dotprod} as in
\begin{equation*}
\psset{unit=0.7}
\begin{pspicture}(-0.75,-0.13)(0.6,1.8)\lw
\psline{-}(0,-1)(0,1)\unlw
\psset{linecolor=myc}
\psarc{-}(0,0){0.5}{0}{360}
\psset{linecolor=myc2}
\psset{linecolor=myc3,linestyle=dashed,dash=2pt 2pt}
\end{pspicture} = 2 \cos(\lambda) \  \textrm{,} \qquad
\begin{pspicture}(-0.75,-0.13)(0.5,1.8)\lw
\psline{-}(0,-1)(0,1)\unlw
\psset{linecolor=myc}
\psarc{-}(0,0){0.5}{-90}{90}
\psset{linecolor=myc2}
\psarc{-}(0,1){0.5}{180}{270}
\psline{-}(-0.5,1)(-0.5,1.8)
\psset{linecolor=myc3,linestyle=dashed,dash=2pt 2pt}
\psarc{-}(0,-1){0.5}{90}{180}
\psline{-}(-0.5,-1.1)(-0.5,-1.8)
\end{pspicture}=1 \ \textrm{,}\qquad
\begin{pspicture}(-1.5,-0.13)(0.75,1.8)\lw
\psline{-}(0,-1)(0,1)\unlw
\psset{linecolor=myc}
\psarc{-}(0,0){0.5}{-90}{90}
\psset{linecolor=myc2}
\psarc{-}(0,1){0.5}{180}{270}
\psline{-}(-0.5,1)(-0.5,1.8)
\psarc{-}(0,1){1.5}{180}{270}
\psline{-}(-1.5,1)(-1.5,1.8)
\psset{linecolor=myc3,linestyle=dashed,dash=2pt 2pt}
\end{pspicture}=1 \ \textrm{,} \qquad
\begin{pspicture}(-1.5,-0.13)(0.75,1.8)\lw
\psline{-}(0,-1)(0,1)\unlw
\psset{linecolor=myc}
\psarc{-}(0,0){0.5}{-90}{90}
\psset{linecolor=myc2}
\psset{linecolor=myc3,linestyle=dashed,dash=2pt 2pt}
\psarc{-}(0,-1){0.5}{90}{180}
\psline{-}(-0.5,-1.1)(-0.5,-1.8)
\psarc{-}(0,-1){1.5}{90}{180}
\psline{-}(-1.5,-1.1)(-1.5,-1.8)
\end{pspicture}=0 \ \textrm{,}\qquad \textrm{...}
\end{equation*}
\\ \\ \\
\noindent This pairing of elements in the basis elements $\mathcal L$ with the mirror image of the first element can be extended to the mirror images of all elements, thus defining the symmetric bilinear form $G$ given in the proposition. But for the specific boundary condition represented by the first vector of $\mathcal L$, only the first column of $G$ is needed and one gets  $\langle w | \rho(F_r(\lambda)) v^r \rangle = I ^\dagger \ ( \prod_{1\le k\le r} N_{\eta(w)_k}) \ G \ I$ as claimed. \hfill$\square$

\medskip
\noindent To end the example started in the proof, we can use the formula to express the matrix element $\langle v^4_{2} |\,\rho(F_4(\lambda)) v^4 \rangle$ as $I^\dagger \ N_0 \ N_1 \ N_{-1} \ N_0 \ G \ I=-2^5\cos\lambda\, \sin^2\lambda\,\sin^2\lambda/2$. A few remarks are in order. It is impossible to create $3$-bubbles with only two rows of braid-boxes. Consequently $\langle w|\,\rho(F_N(\lambda)) v^N \rangle$ is always zero if $w$ has $n$-bubbles, with $n \ge 3$ and the two matrices $N_1$ and $N_{-1}$ are nilpotent of degree $3$: $(N_{-1})^3 = (N_{1})^3 = 0$. The previous proposition allows for another proof of the eigenvalue \eqref{eq:cmaxdiag} of $\rho(F_N)$ in the sector $d$ (Lemma \ref{lem:fn}). This eigenvalue is $\langle v^d |\,\rho(F_d(\lambda)) v^d \rangle = I^{\dagger} \, N_0^d \, G \, I $. The powers of $N_0$ are obtained by diagonalisation: $N_0 = S \, n_0 \, S^{-1}$ with
\begin{equation}
n_0=\begin{pmatrix} 
		 0 & 0 & 0 & 0 & 0 & 0 & 0 & 0 \\
                    0 & 0 & 0 & 0 & 0 & 0 & 0 & 0 \\
                    0 & 0 & 0 & 0 & 0 & 0 & 0 & 0 \\
                    0 & 0 & 0 & 0 & 0 & 0 & 0 & 0 \\
                    0 & 0 & 0 & 0 & 1 & 0 & 0 & 0 \\
                    0 & 0 & 0 & 0 & 0 & 1 & 0 & 0 \\
                    0 & 0 & 0 & 0 & 0 & 0 & -e^{-i \lambda} & 0 \\
                   0 & 0 & 0 & 0 & 0 & 0 & 0 & -e^{i \lambda} \\
\end{pmatrix},
\qquad
S=\begin{pmatrix} 
                    0 & 0 & 0 & 0 & -1+e^{-i \lambda} & 0 & e^{-i \lambda} & 1+e^{-i \lambda} \\
                    0 & 0 & 0 & 0 & -e^{i \lambda} & 0 & 1 & 0 \\
                    0 & 0 & 0 & 0 & 0 & 1 & \frac1{1+e^{i \lambda}} & 1 \\
                    0 & 0 & 0 & 0 & -1-e^{i \lambda} & 0 & 0 & 0 \\
                    0 & 0 & 0 & 1 & 0 & 0 & 0 & 0 \\
                    0 & 0 & 1 & 0 & 0 & 0 & 0 & 0 \\
                    0 & 1 & 0 & 0 & 0 & 0 & 0 & 0 \\
                    1 & 0 & 0 & 0 & 0 & 0 & 1 & 0 \\
\end{pmatrix}.
\label{eq:N0diag}
\end{equation}
\noindent With these, it is now trivial to verify that $I^{\dagger} \, S \, (n_0)^d \, S^{-1} \, G \, I  = 2 (-1)^d \cos(\lambda(d+1))$. 
As a last remark, note that in the proof of the previous proposition, we have summed $2$-boxes from left to right. We could very well have started from the right, summing $2$-boxes from right to left. Simple matrix products yield $N_k  G = G  N_{-k}^\dagger$. This allows to move the matrix $G$ in \eqref{eq:8par8} anywhere within $\prod_{k} N_{\eta(w)_k}$. In particular,  
\begin{equation*} 
\langle w | F_r(\lambda) v^r \rangle = I ^\dagger \ G \ \Big( \prod_{1\le k \le r} N^\dagger_{-\eta(w)_k} \Big)  \ I .\end{equation*}

Our last expression for the matrix elements of $\rho(F_N)$ is given in yet another notation.
Let $w \in B_N^d$ be a link state with only $1$-bubbles and let $n_1 = (N-d)/2$ be their number. We associate with $w$ a set of $(N-d)/2+1$ positive integers $\mu(w)=[m_0,m_1,...,m_{n_1}]$, where $m_i$ is the number of points separating the middle of bubbles $i$ and $(i+1)$-th, for $i = 1, ..., n_1-1$. Then $m_0$ counts the points before the middle of the first bubble and $m_{n_1}$ those after the middle of the last.
For instance, $\mu(v^{20}_{3,11,13,19}) = [3,8,2,6,1]$:
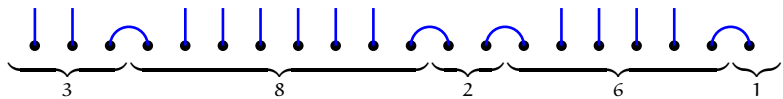
\begin{equation*}
\underbrace{
\begin{pspicture}(0.15,-0.2)(1.72,0.5)
\psset{unit=0.5}
\psdots(1,0)(2,0)(3,0)(4,0)(5,0)(6,0)(7,0)(8,0)(9,0)(10,0)(11,0)(12,0)(13,0)(14,0)(15,0)(16,0)(17,0)(18,0)(19,0)(20,0)
\psset{linecolor=myc2}
\psline{-}(1,0)(1,1)
\psline{-}(2,0)(2,1)
\psline{-}(5,0)(5,1)
\psline{-}(6,0)(6,1)
\psline{-}(7,0)(7,1)
\psline{-}(8,0)(8,1)
\psline{-}(9,0)(9,1)
\psline{-}(10,0)(10,1)
\psline{-}(15,0)(15,1)
\psline{-}(16,0)(16,1)
\psline{-}(17,0)(17,1)
\psline{-}(18,0)(18,1)
\psarc(3.5,0){0.5}{0}{180}
\psarc(11.5,0){0.5}{0}{180}
\psarc(13.5,0){0.5}{0}{180}
\psarc(19.5,0){0.5}{0}{180}
\end{pspicture}}_{3}
\underbrace{
\begin{pspicture}(0.30,-0.1)(4.25,0.5)
\psset{unit=0.5}
\end{pspicture}}_{8}
\underbrace{
\begin{pspicture}(0.30,-0.1)(1.24,0.5)
\psset{unit=0.5}
\end{pspicture}}_{2}
\underbrace{
\begin{pspicture}(0.30,-0.1)(3.24,0.5)
\psset{unit=0.5}
\end{pspicture}}_{6}
\underbrace{
\begin{pspicture}(0.30,-0.1)(0.32,0.5)
\psset{unit=0.5}
\end{pspicture}}_{1}
\end{equation*}
With this definition, $\sum_i m_i = N$. 
Now let $w$ be a link state with $1$-bubbles and $2$-bubbles. Again we associate with $w$ the set $\mu(w)=[m_0,m_1,...,m_{L}]$, with $L = n_1 + 2 n_2$, where $n_2$ is the number of $2$-bubbles and $n_1$, the number of $1$-bubbles that are not circumscribed by $2$-bubbles. This time, $m$ can either be an integer (and counts defects between bubbles including the first or last point of delimiting arcs, as before), or an asterisk to denote all points within a $2$-bubble. For instance, $\mu(v^{20}_{3,5,4,12,12,19}) = [2,*,5,*,6,1]$:
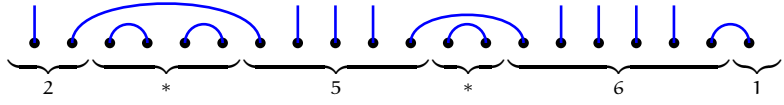
\begin{equation*}
\underbrace{
\begin{pspicture}(0.15,-0.2)(1.22,0.5)
\psset{unit=0.5}
\psdots(1,0)(2,0)(3,0)(4,0)(5,0)(6,0)(7,0)(8,0)(9,0)(10,0)(11,0)(12,0)(13,0)(14,0)(15,0)(16,0)(17,0)(18,0)(19,0)(20,0)
\psset{linecolor=myc2}
\psline{-}(1,0)(1,1)
\psline{-}(8,0)(8,1)
\psline{-}(9,0)(9,1)
\psline{-}(10,0)(10,1)
\psline{-}(15,0)(15,1)
\psline{-}(16,0)(16,1)
\psline{-}(17,0)(17,1)
\psline{-}(18,0)(18,1)
\psarc(5.5,0){0.5}{0}{180}
\psarc(3.5,0){0.5}{0}{180}
\psbezier{-}(2,0)(2,1.4)(7,1.4)(7,0)
\psarc(12.5,0){0.5}{0}{180}
\psbezier{-}(11,0)(11,1)(14,1)(14,0)
\psarc(19.5,0){0.5}{0}{180}
\end{pspicture}}_{2}
\underbrace{
\begin{pspicture}(0.30,-0.1)(2.25,0.5)
\psset{unit=0.5}
\end{pspicture}}_{*}
\underbrace{
\begin{pspicture}(0.30,-0.1)(2.75,0.5)
\psset{unit=0.5}
\end{pspicture}}_{5}
\underbrace{
\begin{pspicture}(0.30,-0.1)(1.24,0.5)
\psset{unit=0.5}
\end{pspicture}}_{*}
\underbrace{
\begin{pspicture}(0.30,-0.1)(3.24,0.5)
\psset{unit=0.5}
\end{pspicture}}_{6}
\underbrace{
\begin{pspicture}(0.30,-0.1)(0.32,0.5)
\psset{unit=0.5}
\end{pspicture}}_{1}
\end{equation*}
\begin{Lemme}\label{lem:finalExpression}
Let $w \in B_r$ be a link state with $\mu(w)=[m_0,m_1,...,m_{L}]$. Then 
\begin{equation*}
\langle w | \rho(F_r(\lambda)) v^r \rangle = V(m_0)^T \  \left(\prod_{i=1}^{L-1} W(m_i)\right) \ G' \  V(m_{L})
\end{equation*}
with
\begin{equation}
V(m)= 2 \begin{pmatrix} 
                    S_m  \\
                    S_{\frac{m-1}{2}} S_{\frac{m}{2}}/S_{\frac12}  \\ \end{pmatrix},
\quad
W(m)=\begin{pmatrix} 
                    S_{\frac{2m-1}{2}}/S_{\frac12} &  S_{\frac{m}{2}}S_{\frac{m-2}{2}}/S_{\frac12}^2  \\
                    2 \ C_{m-1} &  S_{\frac{2m-3}{2}}/S_{\frac12}  \\ \end{pmatrix},
\quad
W(*)=\begin{pmatrix} 
                    1 & 0  \\
                    1 & 0  \\
\end{pmatrix},
\quad
G'=\begin{pmatrix} 
                    1 & 1  \\
                    1 & 0  \\
\end{pmatrix}
\label{eq:2par2}
\end{equation}
where the notation of Definition \ref{def:fN} was used.
\end{Lemme}
\noindent{\scshape Proof\ \ } We start with $w$'s that have no $2$-bubbles.
The product of eq.~(\ref{eq:8par8}) can be rearranged by displacing $G$ to the left until it comes across the first bubble, that is, until it passes the rightmost $N_{-1}$. The matrices on the left of $G$ can then be gathered in groups labelled by the subscript $i = 1, ..., n_1 -1$. Each group starts with $N_{-1}$, followed by $m_i-2$ occurrences of $N_0$, and ends with $N_{-1}$. The result is
\begin{equation}
\langle w | F_N(\lambda) v^N \rangle = \left( I^\dagger  \left(N_0\right)^{m_0-1}  N_1 \right)  \left( \prod_{i=1}^{n_1-1} \left( N_{-1} \left(N_0\right)^{m_i-2} N{_1} \right) \right)  G  (N^\dagger_1  (N^\dagger_0)^{m_{n_1}-1} I).
\label{eq:intermed}\end{equation} 
Using the diagonal form \eqref{eq:N0diag} of $N_0$, we can compute $I^\dagger \left(N_0\right)^{m-1} N_1$ and $N_{-1} \left(N_0\right)^{m-2} N{_1}$:
\begin{equation*}
I^\dagger  \ S \  n_0^{m-1} \ S^{-1} \ N_1  = -2 i e^{i \Lambda}\begin{pmatrix} 
		 0  &
                    0  &
                    0  &
                    0  &
                    S_{m}  &
                    S_{\frac{m}{2}} S_{\frac{m-1}{2}} / S_{\frac{1}{2}}  &
                    0  &
                   0  
\end{pmatrix}
\end{equation*}
\begin{equation*}
N_{-1} S  \ n_0^{m-2} \ S^{-1}  N{_1}=\begin{pmatrix} 
                    0 & 0 & 0 & 0 & 0 & 0 & 0 & 0 \\
                    0 & 0 & 0 & 0 & 0 & 0 & 0 & 0 \\
                    0 & 0 & 0 & 0 & 0 & 0 & 0 & 0 \\
                    0 & 0 & 0 & 0 & 0 & 0 & 0 & 0 \\
                    0 & 0 & 0 & 0 & S_{\frac{2m-1}{2}}/S_{\frac12} & S_{\frac{m}{2}}S_{\frac{m-2}{2}}/S_{\frac12}^2 & 0 & 0 \\
                    0 & 0 & 0 & 0 & 2 \ C_{m-1} & S_{\frac{2m-3}{2}}/S_{\frac12} & 0 & 0 \\
                    0 & 0 & 0 & 0 & 0 & 0 & 0 & 0 \\
                    0 & 0 & 0 & 0 & 0 & 0 & 0 & 0 \\
\end{pmatrix}
\end{equation*}
Hence, $I^\dagger \left(N_0\right)^{m_0} N_1$ belongs to the subspace spanned by $\{ (0,0,0,0,1,0,0,0) , (0,0,0,0,0,1,0,0) \}$, which is also stable under the action of both $N_{-1} \left(N_0\right)^{m_i} N{_1}$ and $G$. The products in \eqref{eq:intermed} can therefore be restricted to this subspace. Note that the two factors $(-ie^{i\Lambda})$ and $(ie^{-i\Lambda})$ coming from the extreme factors cancel.

The same rearrangement can be carried out on the product of (\ref{eq:8par8}) when link states carry $2$-bubbles. When they do, $(N_1N_{-1})^p$ appears between $N_{-1}(N_0)^{m_i}N_1$ and $N_{-1}(N_0)^{m_{i+2}}N_1$. Here $p$ is the number of $1$-bubbles enclosed in the $2$-bubble. But the $2$-dimensional subspace used above is also stable under
\begin{equation*}
N_{1} N{_{-1}}=\begin{pmatrix} 
                    e^{-i \lambda} - e^{i \lambda} & 0 & 1-e^{-2i \lambda} & 0 & 0 & 0 & 0 & 1-e^{-2 i \lambda} \\
                    1-e^{i \lambda} & 1 & -e^{-i \lambda} & 0 & 0 & 0 & 0 & 1-e^{-i \lambda} \\
                    1-e^{i \lambda} & 0 & 1-e^{-i \lambda} & 0 & 0 & 0 & 0 & 1-e^{-i \lambda} \\
                    -e^{i \lambda} & 1 & 0 & 0 & 0 & 0 & 0 & 1 \\
                    0 & 0 & 0 & 0 & 1 & 0 & 0 & 0 \\
                    0 & 0 & 0 & 0 & 1 & 0 & 0 & 0 \\
                    0 & 0 & 0 & 0 & 0 & 0 & 0 & 0 \\
                    1 & 0 & -e^{-i \lambda} & 0 & 0 & 0 & 0 & -e^{-i \lambda}
\end{pmatrix}.
\end{equation*}
The fact that the restriction $W(*)$ of this matrix to the subspace is idempotent shows that the contribution of $2$-bubbles is independent of $p$. This concludes the proof.\hfill$\square$
\medskip

%
%


\end{document}